## ABSTRACT

ZHANG, XIAO. Capacitive Micromachined Ultrasonic Transducers (CMUTs) on Glass Substrates for Next-Generation Medical Imaging and Beyond. (Under the direction of Dr. Ömer Oralkan).

Capacitive Micromachined Ultrasonic Transducers (CMUTs) have been widely researched for the last two decades. One of the most immediate applications is ultrasound medical imaging. Comparing to the traditional medical ultrasound transducer technology based on piezoelectric materials, CMUT technology offers benefits including fabrication of large transducer arrays, close integration with electronics, and broad bandwidth in immersion. CMUT technology is especially competitive for fabrication of 2D transducer arrays and their close integration with electronics for real-time 3D imaging and for fabrication of high-frequency broadband transducer arrays for high-resolution imaging.

CMUT technology was invented and has been widely researched on silicon substrates. In recent years, fabrication of CMUT on glass substrates has raised significant interests because a number of advantages it can offer. First, parasitic capacitance could be easily reduced since glass is an insulating material. Second, insulation steps required in the silicon-based CMUT fabrication process could be avoided, leading to a reduced fabrication complexity. Third, anodic bonding can be used when the substrate is borosilicate glass, which is a low-temperature bonding technique and has a high tolerance to bonding surface area and roughness in addition to the advantages of silicon wafer bonding. Moreover, glass transparency can enable novel applications beyond pulse-echo ultrasound medical imaging and enlarge the potential markets of CMUTs.

This dissertation investigates the design and fabrication of several CMUT-based key components on glass substrates to provide a tool kit for the development of the next-generation ultrasound system frontends. On one side, a densely populated 2D transducer array with through-wafer interconnects and a MEMS transmit/receive (T/R) switch were developed to overcome the transducer integration challenges. On the other side, by taking the advantage of glass substrate transparency, a transparent CMUT was developed for both medical applications and consumer electronics applications.

We first developed a platform process to fabricate vacuum-sealed CMUTs on a glass substrate with anodic bonding. The CMUT performance and the array uniformity were characterized. Low parasitics, good process control, and good array uniformity were achieved by this platform technology. We further demonstrated a high-frequency (30-MHz), broadband (100%) 1D CMUT array based on this process.

Furthermore, we extended the platform process and built a 16×16 2D CMUT array by incorporating through-glass-via (TGV) interconnects. The parasitic capacitance of a via pair with a 250-μm pitch was measured as 21 fF. The resistance of a single via plus via-to-electrode contact resistance was $2\,\Omega$. The 2D CMUT array element showed a low parasitic capacitance.

In addition, we developed a MEMS switch co-fabricated with a CMUT. Static and dynamic characterization was performed in air and in immersion, respectively. The MEMS switch showed a dc switching voltage of 68 V and could be operated using a control voltage of 2.5 V when biased at 67 V. The optimum switching time is 1.34 μs and the optimum release time is 80 ns.

Last but not least, we demonstrated a CMUT with improved transparency for backward-mode PAI and discussed the effect of silicon plate absorption. Second, a fully transparent CMUT was fabricated with two ITO coated glass wafers by adhesive bonding. The fabricated device shows 60%-80% optical transmittance in the full visible wavelength range.

Capacitive Micromachined Ultrasonic Transducers (CMUTs) on Glass Substrates for Next-Generation Medical Imaging and Beyond

by
Xiao Zhang

A dissertation submitted to the Graduate Faculty of
North Carolina State University
in partial fulfillment of the
requirements for the degree of
Doctor of Philosophy

Electrical and Computer Engineering

Raleigh, North Carolina

2017

APPROVED BY:

| | |
|---|---|
| Dr. Ömer Oralkan<br>Committee Chair | Dr. Alper Bozkurt |
| Dr. Troy Nagle | Dr. Yun Jing |

## BIOGRAPHY

Xiao Zhang received the B.S. degree from Xi'an Jiaotong University, Xi'an, China, in 2012. He was then admitted to the direct doctoral program in the Department of Electrical and Computer Engineering at North Carolina State University, Raleigh, NC.

His research focuses on design and fabrication of 1D and 2D capacitive micromachined ultrasonic transducer (CMUT) arrays and associated MEMS transmit/receive switches on glass substrates to overcome the transducer integration challenges with the front-end electronics. He also developed transparent CMUTs for medical applications and consumer electronics applications.

He has authored or co-authored over 20 scientific publications. He received student travel awards in the 2015, 2016, and 2017 IEEE International Ultrasonics Symposia. He was a student paper competition finalist in the 2016 and 2017 IEEE International Ultrasonics Symposia. He won the first place in 2017 NCSU ECE Research Symposium and he was also a recipient of the UGSA Conference Award at NCSU in 2015.

# ACKNOWLEDGMENTS

First of all, I would like to express my sincere gratitude to my Ph.D. advisor Dr. Ömer Oralkan. I was fortunate to be the first student in his research group and also the first Ph.D. student to graduate. Being the first is a big challenge but it is also the most rewarding experience through my five years of study. During the years I worked with Dr. Oralkan, he has been not only my academic advisor but also my mentor. What I learned from him is not only the knowledge about ultrasound and MEMS but also the approach and attitude to address a research problem with little or no references. He not only helped me to make breakthroughs on my research projects but also guided me to be a successful researcher and engineer who could be positive and do not quit even facing the most challenging technical problems.

I would also like to acknowledge the other professors in my Ph.D. committee, Dr. Alper Bozkurt, Dr. Yun Jing, and Dr. Troy Nagle. The knowledge I acquired on their courses about human-machine interfaces, bio-electronics, and acoustics is very helpful for my research and my understanding of electrical and computer engineering. I would also like to thank them for working with me through all the milestones in my Ph.D. program from qualification exam, preliminary exam, and to the final defense.

During my years of study at NC State, I'm very glad to know and become friends with my lab mates and colleagues. Dr. Feysel Yalcin Yamaner joined the group on the same day as me. He taught me the basics of CMUT, finite-element modeling, and helped me make my first CMUT in the cleanroom. Xun, I enjoyed our useful discussions towards the medical imaging algorithms and also our fun chats about Japanese music and animation. Jeanne, thank you for correcting my grammar and pronunciation before my conference presentations. Marzana, thanks for inviting us to your baby's events and I wish your daughter happiness every day. Femi, I won't forget the countless hours we spent together in the cleanroom. CK, thank you for praying for me during my difficult times. Besides, I would also like acknowledge other colleagues and close friends: Dr. Xiaoning Jiang, Dr. Sibo Li, Dr. Chen Shen, Mr. Weiyi Chang, Dr. Zhenhao Li, Dr. Szuheng Ho, Dr. Xu Zhang, Dr. Murat Yokus, Dr. Lujun Huang, Mr. Peter Sotory, Mr. Rupak Roy, Mr. Mohit Kumar, Dr. Bhoj Gautam, Mr. Robert Younts, Ms. Feiyan Lin, Mr. Dingjie Suo, Mr, Steve Lipa, Mr. Joe Matthews, etc.

In addition, I would like to thank all the current and previous NNF staff members (Marcio, Nicole, Henry, Bruce, and Jeff) and AIF staff members (Chuck and Roberto) for their detailed instructions on microfabrication tools and microscopy techniques, as well as the fun chats we had while waiting for the process to complete.

Finally, I want to express my deepest gratitude to my family and my beloved wife. To my parents and my elder brother: Your unconditional love started from the very beginning of time. You made everything possible for me and I'm blessed to be born in this family. To my wife Hui: Meeting, dating, marrying you were the most exciting moments in my life. Thank you for always standing by my side and encouraging me even during our most difficult times.

This dissertation is dedicated to my parents, my elder brother, and my wife.

# TABLE OF CONTENTS

# LIST OF TABLES

## LIST OF FIGURES

# Chapter 1 INTRODUCTION

## 1.1 Overview of Ultrasound Transducers

Ultrasound is defined as sound waves that have a frequency above 20 kHz. An ultrasound transducer is a device that could transmit ultrasound by converting electrical energy into acoustical energy, and could also receive ultrasound by converting acoustic energy into electrical energy. The history of modern ultrasound can be traced back to the discovery of piezoelectric effect in 1880 by French physicists Jacques and Pierre Curie [1]. Piezoelectric ceramics were first introduced as ultrasonic transducers during the 1940s. Later, piezoelectric materials, including polyvinylidene fluoride (PVDF), were introduced in the 1960s, piezo-composites in 1980s, and piezoelectric single crystals in the 1990s.

Since the late 1960s, the exponential growth of the microelectronic industry has largely accelerated the application of ultrasonic devices in different areas, such as ultrasound medical imaging, nondestructive testing (NDT), acoustic microscopy, flow metering, surface acoustic wave (SAW), and bulk acoustic wave (BAW) devices. Figure 1-1 summarizes the ultrasound applications at different frequency ranges. The application market for ultrasound still grows steadily. More recently, ultrasound fingerprint sensors, ultrasound wireless power transfer, etc. are being widely investigated and pushed towards commercialization.

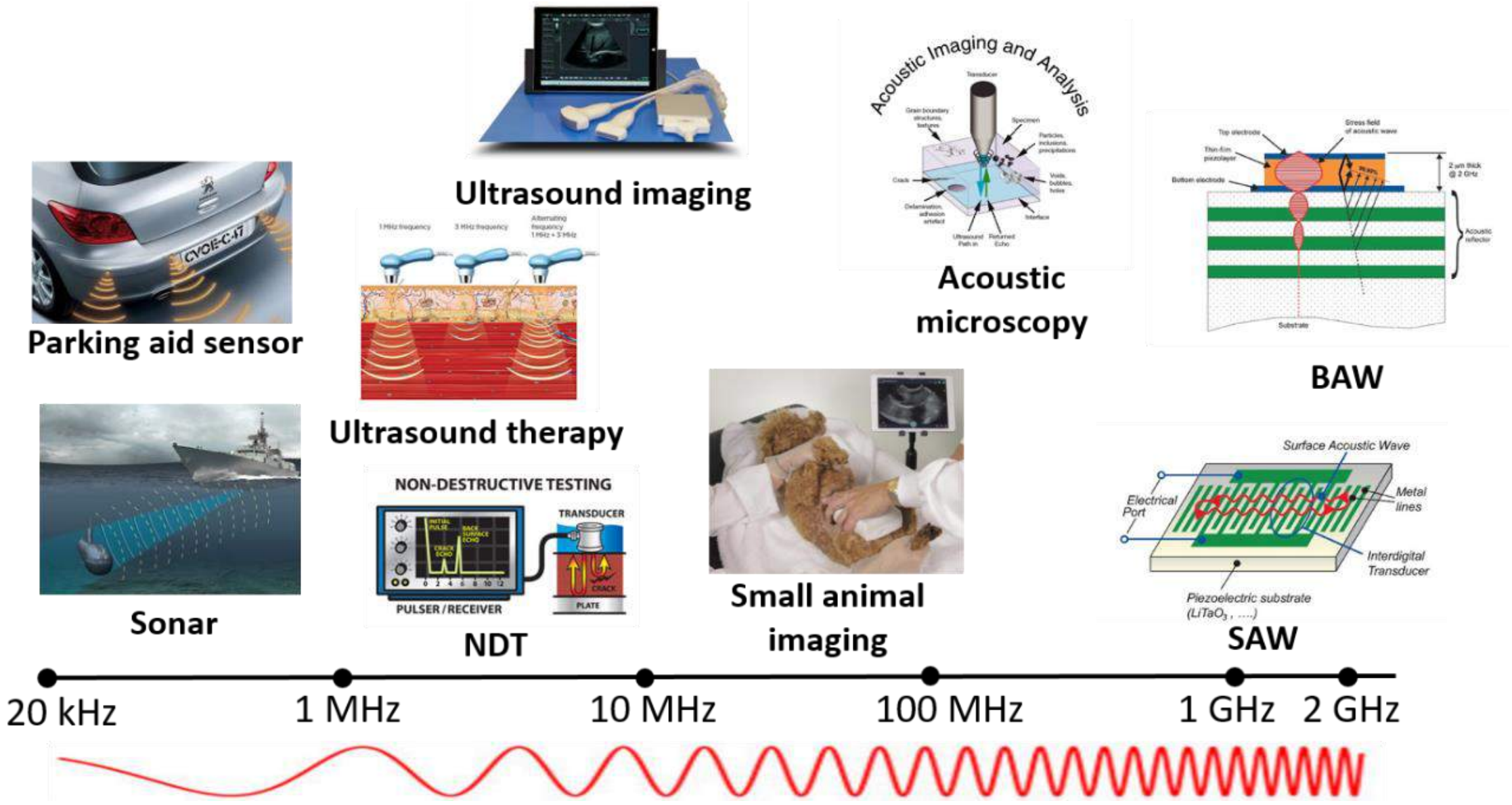

**Figure 1-1: Examples of ultrasound applications at various frequencies.**

## 1.2 Capacitive Micromachined Ultrasonic Transducers (CMUTs)

CMUTs were made practical in the mid-1990s [2], although the idea of capacitive ultrasound transducers is as old as early piezoelectric transducers. The reason why capacitive transducers had not been popular is that a high electric field strength of the order of a million volt per centimeter ($10^6$ V/cm) is required [3], which is critical for delivering a reasonable transducer efficiency. The explosive growth of microelectronics and MEMS technology in the late 1960s and 1970s has enabled the definition of a device dimension with an unprecedented accuracy. With the advanced micro-fabrication and micromachining techniques, a very shallow (sub-micron) cavity could be defined in a capacitive transducer, which translates to a relatively low voltage (<100 V) to reach the required electric field strength, making electrostatic ultrasound transducers possible.

The first generation CMUTs were built on silicon wafers with no vacuum-sealing for air-coupled applications [2]. Since then, extensive research effort has been conducted to improve the device performance and reliability as well as to exploit the potential applications.

On the device aspect, various design and fabrication approaches have been investigated, including vacuum sealing of the cavity, dual-electrode designs, various plate geometries, reduction of electric charging, etc. In terms of device fabrication, CMUT was originally invented on silicon substrates using the sacrificial release process. Later in 2003, the CMUT fabrication was demonstrated by direct wafer-to-wafer fusion bonding technique [4], which offers better dimensional and stress control compared to surface micromachining. During the same period, 2D CMUT arrays with through-silicon-via (TSV) interconnects were developed for 3D real-time ultrasound imaging [5]. Later trench-isolated through-wafer interconnects were developed which eliminate the complicated process to make TSVs and are also compatible with wafer bonding [6]. High-frequency CMUT arrays were also demonstrated for high-resolution imaging where deep penetration is not required [7]. Later in 2009, row-column addressed CMUT (RC-CMUT) arrays were demonstrated and investigated for 3D ultrasound imaging [8]. In recent years, CMUT has been made on insulating substrates in order to further reduce the device parasitics and the process complexity [9]–[11]. Also, adhesive bonding has been proposed as an alternative way to build CMUTs [12]. More recently, transparent CMUT has been investigated for applications where optics and acoustics need to be combined [13]. An overview of the above-described progress regarding CMUT fabrication is depicted below.

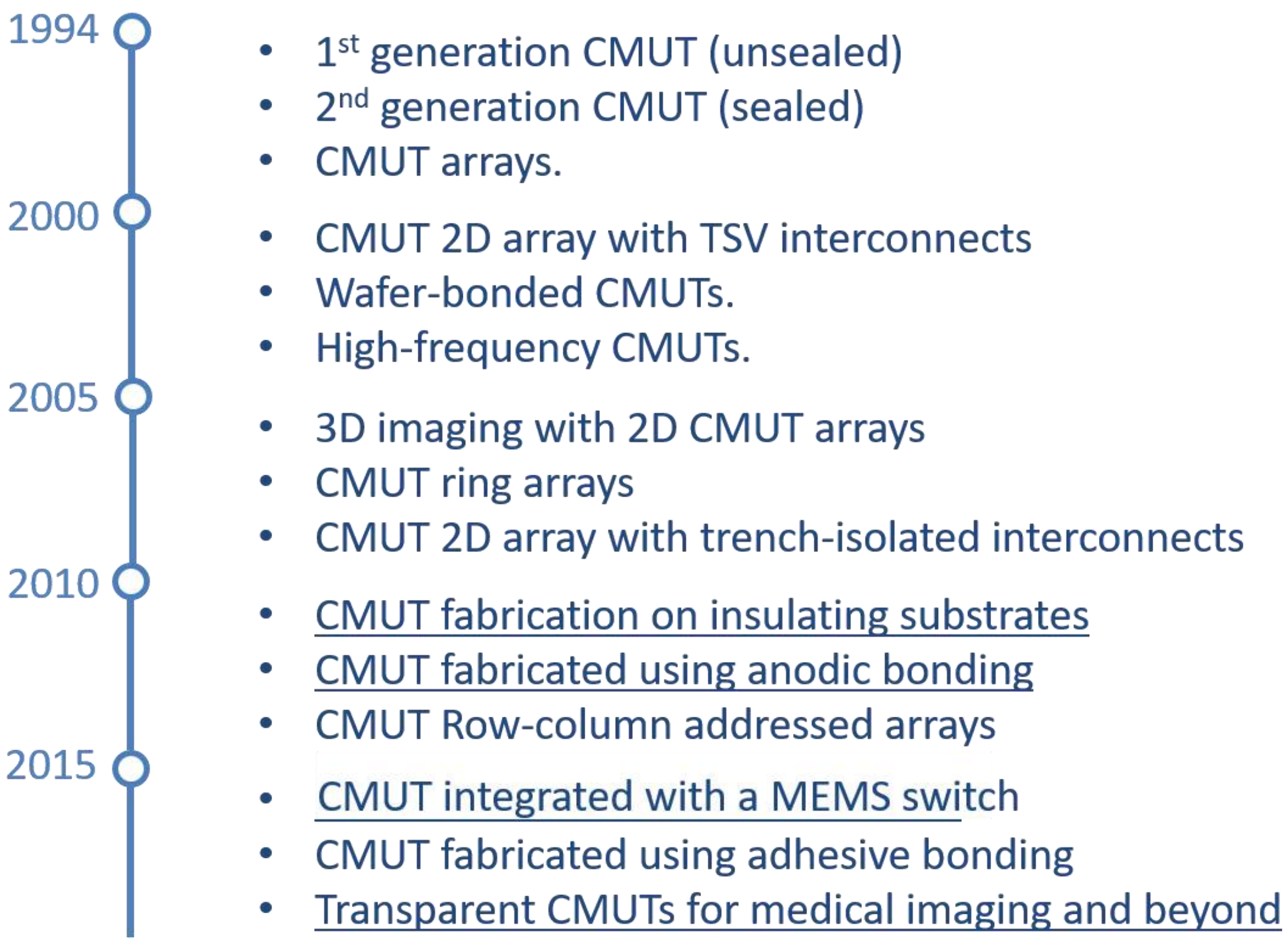

**Figure 1-2: A limited overview of CMUT technology progress. The underlined items indicate author's contribution.**

In terms of CMUT application, although first developed for air-coupled applications, CMUT later found its most promising application in ultrasound medical imaging. The tremendous interest of using CMUT technology as an alternative or complement for piezoelectric medical transducers stems from the advantages offered by CMUT, such as ease of fabricating large arrays and integration with electronics, as well as the wide bandwidth in immersion. CMUT is especially attractive for high-frequency ultrasound imaging and 3D real-time ultrasound imaging, where transducer arrays are difficult to fabricate and integrate with electronics using the traditional piezoelectric technology and its associated fabrication approach. Table 1-1 is a summary of the important advantages of CMUT technology compared to the piezoelectric transducer technology for medical imaging applications.

**Table 1-1: Comparison of transducer technologies for ultrasound imaging applications**

| | Piezoelectric transducers | CMUT |
|---|---|---|
| Fabrication method | Ceramic technology | MEMS technology |
| Array fabrication | Difficult for 2D and high-frequency arrays | 2D arrays enabled by through-wafer interconnects. Enable high-frequency broadband transducer. |
| Transducer bandwidth | Moderate, matching layer required | Wide, no matching layer needed |
| Array uniformity | Moderate | High |
| Thermal stability | Low | High |
| IC integration | Difficult | Monolithic or hybrid integration |
| Output pressure | High | Relatively low but improving |

Researchers have progressed from early demonstration of the CMUT operation to the full integration of 2D [14], 3D [15], [16], intravascular [17]–[19], and photoacoustic imaging systems [20], [21]. CMUT for medical imaging has reached the commercial market in 2009 by Hitachi, Ltd. [22] and the first practical CMUT 1D linear array probe was presented in 2015 by KOLO

Medical Inc. [23], which is originated from the Stanford research group where CMUT was invented.

In addition, CMUTs have been demonstrated for therapeutic ultrasound, CMUTs can provide MR-compatible HIFU [24], [25]. Apart from medical ultrasound, CMUT has also been widely researched for applications such as chemical sensors [26]–[28], parametric speaker [29], and biometric applications such as ultrasonic fingerprint sensor, ultrasonic touchless interactive panels [30]–[32], etc.

## 1.3 Summary of Contributions

The first objective of this dissertation is to develop a new approach to fabricate CMUT 1D and 2D arrays on glass substrates. The CMUTs fabricated on glass substrates simplifies the fabrication flow and reduce the device parasitics and thus improve the transducer efficiency. The second objective is to facilitate the CMUT integration with electronics. On one side, we demonstrate the CMUT 2D arrays with through-glass-via interconnects that are feasible for hybrid integration. On the other side, a MEMS switch that could be embedded in a CMUT structure is developed. The significance of the MEMS switch is that it could eliminate the high-voltage process required to implement the front-end electronics that therefore could improve the performance of the front-end electronics of the system. The third objective is to develop transparent CMUTs on glass substrates and demonstrate its potential for applications where acoustics and optics are combined.

The contributions of this dissertation are listed below.

1. This dissertation will demonstrate a novel CMUT fabrication method that is versatile to make different types of CMUT arrays and enable novel applications.

2. A 16×16 CMUT 2D array on a glass substrate with through-glass-via (TGV) interconnects was fabricated. A single via resistance plus the contact resistance is 2 Ω. The capacitance of a 250-µm pitch via pair was measured as 21 fF. A 2D array element was characterized in the air.

3. A finite-element model was developed for a MEMS switch that could be embedded and co-fabricated with CMUT on glass substrates using anodic bonding. The switching mechanism and switching speed were studied using the model and verified by characterizing the actual fabricated device.

4. The MEMS switch was fabricated and characterized by static characterization in the air and dynamic characterization under oil. The actual device performance agrees well with the FEM model. A switching time of 1.34 µs and a releasing time of 80 ns were demonstrated. The control voltage of the switch is 2.5 V.

5. A CMUT element with improved transparency was presented by using ITO as the bottom electrode. The fabricated CMUT was functional and shows 30%-70% optical transmission in 600-800 nm wavelength range, which is suitable for photoacoustic imaging applications.

6. The CMUT with improved transparency was used for backward photoacoustic imaging. A graphite target and a tube filled with ICG solution were used as two imaging targets. The effect of silicon plate absorption was not significant (30 dB less than the graphite target signal at 830 nm wavelength).

7. An optically transparent CMUT is developed for integrating ultrasound with flat panel displays. The process only requires two masks and the maximum processing temperature is 250ºC. The fabricated device has an optical transparency of 60%-80% in the full visible wavelength range.

Most of the works related to this dissertation have been previously published or presented.

**The publications related to Chapter 3 are:**

[1] F. Y. Yamaner, **X. Zhang**, and Ö. Oralkan, "A three-mask process for fabricating vacuum-sealed capacitive micromachined ultrasonic transducers using anodic bonding," *IEEE Trans. Ultrason. Ferroelect., Freq. Contr.*, vol. 62, no. 5, pp. 972–982, 2015.

[2] F. Y. Yamaner, **X. Zhang**, and Ö. Oralkan, "Fabrication of anodically bonded capacitive micromachined ultrasonic transducers with vacuum-sealed cavities," *Proc. IEEE Ultrason. Symp.*, 2014, pp. 604–607.

[3] **X. Zhang,** F. Y. Yamaner, O. Adelegan, and Ö. Oralkan, "Design of high-frequency broadband CMUT arrays," in *Proc. IEEE Ultrason. Symp.*, 2015, pp. 1–4.

## The publications related to Chapter 4 are:

[1] **X. Zhang**, F. Y. Yamaner, and Ö. Oralkan, "Fabrication of Vacuum-Sealed Capacitive Micromachined Ultrasonic Transducers With Through-Glass-Via Interconnects Using Anodic Bonding," *J. Microelectromech. Syst.*, vol. 26, no. 1, pp. 226–234, 2017.

[2] **X. Zhang**, F. Y. Yamaner, and Ö. Oralkan, "Fabrication of capacitive micromachined ultrasonic transducers with through-glass-via interconnects," *Proc. IEEE Ultrason. Symp.*, 2015, pp. 1–4.

## The publications related to Chapter 5 are:

[1] **X. Zhang**, F. Y. Yamaner, and Ö. Oralkan, "*A Fast-Switching (1.35-μs) Low-Control-Voltage (2.5-V) MEMS T/R Switch Monolithically Integrated With a Capacitive Micromachined Ultrasonic Transducer (CMUT)," J. Microelectromech. Syst. (Accepted)*

[2] **X. Zhang**, A. Zeshan, O. J. Adelegan, F. Y. Yamaner, and Ö. Oralkan, "A MEMS T/R switch embedded in CMUT structure for ultrasound imaging frontends," *Proc. IEEE Ultrason. Symp.*, pp. 1–4, 2016.

## The publications related to Chapter 6 are:

[1] **X. Zhang**, X. Wu, O. J. Adelegan, F. Y. Yamaner, and Ö. Oralkan, "Backward-mode photoacoustic imaging using illumination through a CMUT with improved transparency," *IEEE Trans. Ultrason., Ferroelect., Freq. Contr.* DOI 10.1109/TUFFC.2017.2774283 *(IEEE early access article available).*

[2] **X. Zhang**, O. Adelegan, F. Y. Yamaner, and Ö. Oralkan, "CMUTs on the glass with ITO bottom electrodes for improved transparency," *Proc. IEEE Ultrason. Symp.*, pp. 1–4, 2016.

[3] **X. Zhang**, O. Adelegan, F. Y. Yamaner, and Ö. Oralkan, "An Optically Transparent Air-Coupled Capacitive Micromachined Ultrasonic Transducer (CMUT) Fabricated Using Adhesive Bonding" pp. 1–4, 2017.

## 1.4 Thesis Organization

Chapter 2 introduces the background and modeling of CMUTs. Chapter 3 explains the platform process, which requires only three masks and three photolithographic steps to fabricate CMUT single transducers and 1D arrays on glass substrates using anodic bonding.

In Chapter 4, we incorporate through-glass-via (TGV) interconnects into the platform process to make 2D CMUT arrays. The TGV parasitic resistance and capacitance were measured and an element of a 16×16 2D CMUT array was characterized in the air.

Chapter 5 introduces a MEMS switch that can be embedded and co-fabricated in the CMUT structure that is fabricated on glass substrates using anodic bonding. The MEMS switch is characterized in air for static performance and characterized under oil for dynamic performance. The results demonstrated that the co-fabricated MEMS switch is suitable for the front-end electronics of a medical ultrasound imaging system.

In Chapter 6, we first demonstrate CMUT with improved transparency that could be used for backward-mode photoacoustic imaging applications. Then we replace the thin silicon plate with a glass plate. By using adhesive bonding technique, we demonstrate a 2-mask process to make fully transparent air-coupled CMUT that could be integrated with a flat-panel display. Such applications include a parametric array for generating uni-directional sound, fingerprint sensors, and ultrasound gesture sensors, etc.

Chapter 7 is the summary of the dissertation.

# Chapter 2 CAPACITIVE MICROMACHINED ULTRASONIC TRANSDUCERS (CMUTs)

## 2.1 CMUT Structure and Operation Principle

The basic building block of a CMUT array is a capacitor cell consisting of a movable plate suspended above a vacuum gap. An insulation layer is incorporated between the two electrodes to prevent the two electrodes shorting in case of contact during the CMUT operation. The insulation layer could be implemented either under the top electrode or on the bottom electrode. The rule of thumb for CMUT design is to maximize the electric field strength in the cavity with a practical dc bias voltage so that the CMUT could be operated with sufficient efficiency. Therefore it is desired to have the top electrode at the bottom of the plate.

The substrate for fabricating CMUT can be a conductive substrate, such as a heavily doped silicon substrate, or a dielectric substrate, such as a glass substrate. CMUT plate can be implemented either using a conductive material, such as conductive silicon, or a dielectric material.

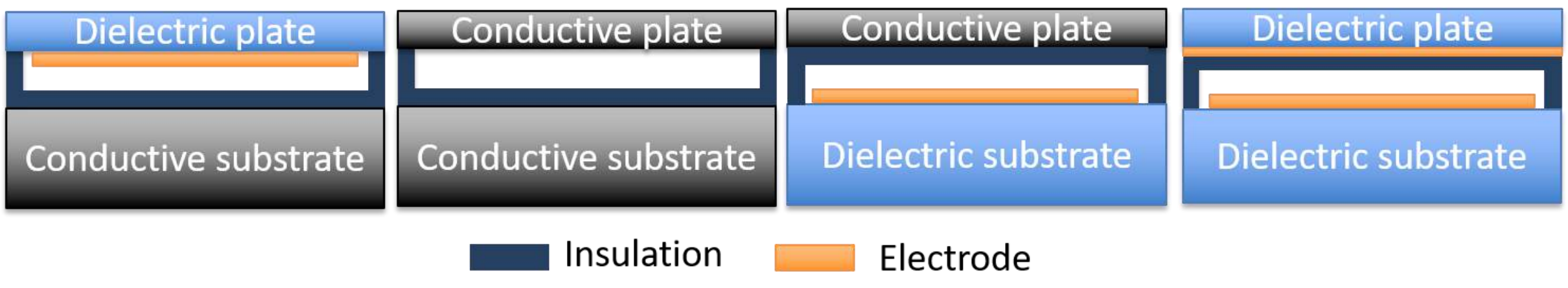

**Figure 2-1: Different implementations of a CMUT cell.**

A single CMUT element consists of many capacitor cells connected in parallel, and multiple elements form an array. Arrays of different geometries can be fabricated on the same wafer simultaneously (Figure 2-2).

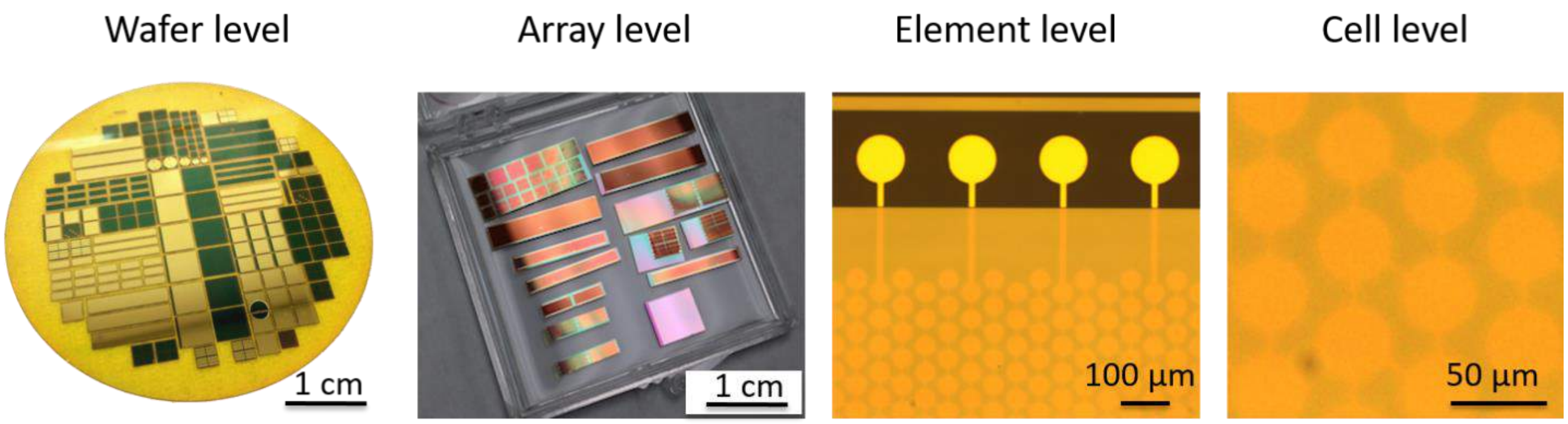

**Figure 2-2: CMUT arrays of different geometries fabricated on the same wafer. The four pictures shows the CMUTs at wafer level, array level, element level, and cell level.**

During CMUT operation, a dc voltage is applied to an element and the movable top plates are attracted towards the bottom electrodes due to electrostatic force, which is resisted by the mechanical restoring force due to the stiffness of the plate. The plate reaches a steady-state when the two force balances. Driving the plate with an alternating voltage (ac) modulates the electrostatic force and therefore generates ultrasound.

On the receive side, if the biased plate is subjected to ultrasound pressure, the stored electric charge on the electrodes will change due to the capacitance change under the constant bias voltage.

$$\frac{dQ}{dt} = V_{dc}\frac{dC}{dt} \tag{2.1}$$

Therefore, a current is produced with an amplitude that is a function of the bias voltage $V_{dc}$, the frequency of the incident wave $\omega_0$, the steady-state capacitance of the device $C_0$, and the fractional displacement $\frac{\Delta x}{x_0}$ due to the incident wave, which can be mathematically expressed as:

$$I_{ac} = V_{dc} \cdot \omega_0 \cdot C_0 \cdot \frac{\Delta x}{x_0} \qquad (2.2)$$

## 2.2 CMUT Performance Attributes

### 2.2.1 CMUT Modeling

In order to analyze the CMUT performance attributes, we need to understand the CMUT modeling. There are mainly two approaches used to simulate the performance attributes of a CMUT, which are the analytical electrical equivalent circuit model (ECM), and the finite element model (FEM). In this chapter, we will review the equivalent circuit model. The finite-element model will be discussed in Chapter 5.

Mason’s equivalent circuit model is shown in Figure 2-3, with an inset showing the CMUT parallel plate capacitor model (spring-mass-damper model) to derive some important parameters of CMUT that is needed to understand the equivalent circuit model.

In this parallel plate capacitor model, the CMUT top plate is constrained so that it moves like a piston, which forms a parallel plate capacitor with the fixed bottom electrode. The piston with a mass $m$ is suspended over an effective electrostatic gap $g_0$ by a spring constant $k$, which derives from the stiffness of the plate and results in the mechanical restoring force during the CMUT operation. The damper with a damping factor $b$ represents the mechanical loss during the piston movement. $V$ is the voltage applied between the top and bottom electrode to introduce electrostatic force. It should be noted that this model deviates from the reality since CMUT plate is edge-clamped with a maximum displacement in the center. In addition, the piston-like motion also neglects the higher order vibration modes of the plate that are inherent in the actual devices. Furthermore, the assumption of a linear spring differs from the actual restoring force especially when the displacement is significant. Despite the limitations, the model can still provide a good prediction of important CMUT parameters such as pull-in voltage, center frequency, fractional bandwidth, etc.

The Mason's equivalent circuit model is a small-signal model and therefore is only suitable for the analysis of the receiver, or when the ac signal is small compared to the dc bias during transmit. The model can be considered as a two-port network composed of an electrical domain and a mechanical domain. On the electrical port, the input of the circuit is a voltage source with an internal resistance of $R_s$. $C_0$ is the clamped device capacitance at the bias voltage. $C_p$ is the parasitic capacitance. On the mechanical port of the circuit, voltage corresponds to force and current corresponds to the plate average velocity with a transformer ratio of $n$, which is also related to the electromechanical coupling coefficient ($Kt^2$). $Z_{plate}$ represents the mechanical impedance of the CMUT plate, which could be modeled using the spring constant $k$ and the mass $m$ of the plate in the spring-mass-damper model. $Z_{medium}$ is the radiation impedance of the surrounding medium. The spring softening behavior is modeled using a spring softening capacitor $\frac{-C_0}{n^2}$.

The equation of motion is expressed as:

$$m\frac{d^2x}{dt^2}+b\frac{dx}{dt}+kx+f_e=0 \tag{2.3}$$

where $x$ is the displacement of the plate, m is the mass of the piston, $b$ is the damping coefficient, $k$ is the spring constant, and $f_e$ is the electrostatic force.

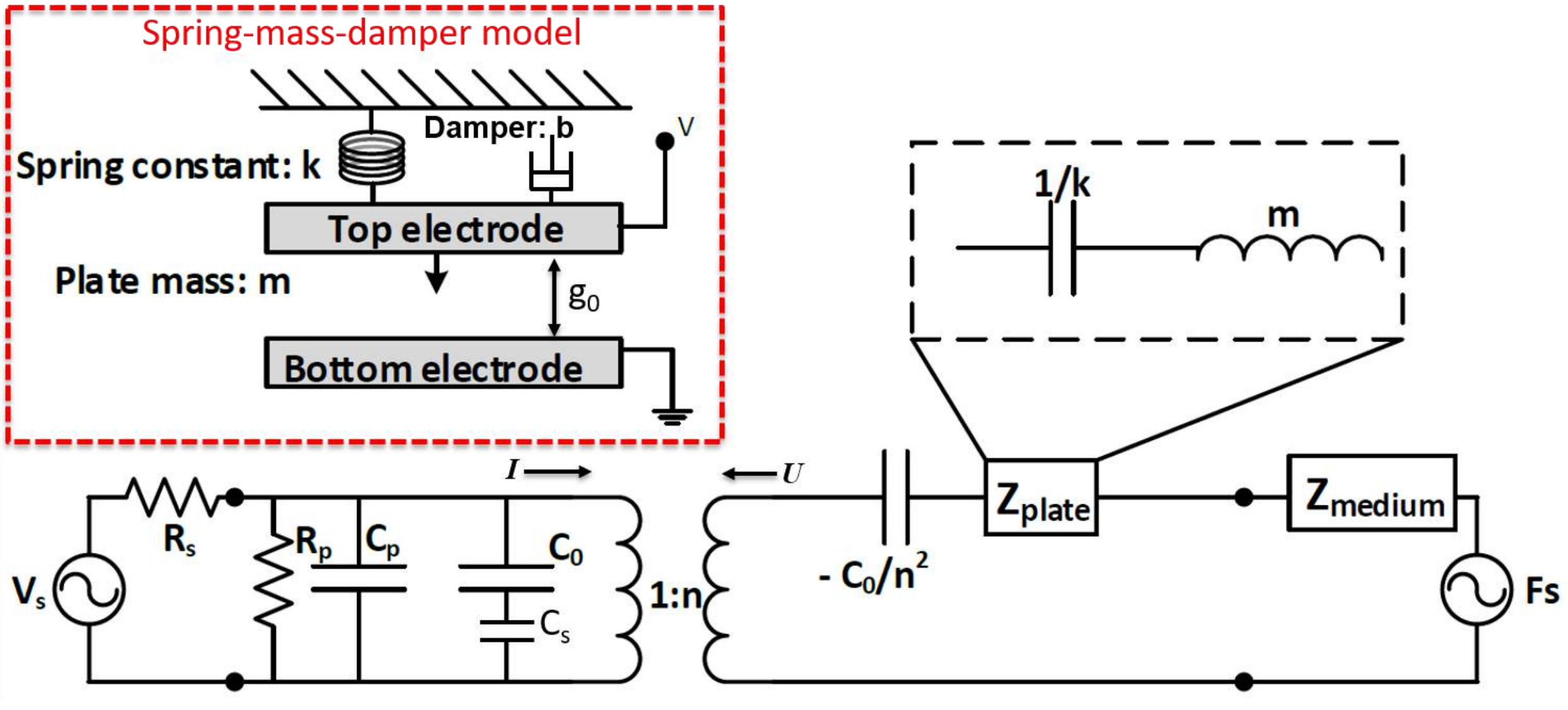

**Figure 2-3: Equivalent circuit model of CMUT that relates voltage and current (V and I) to force and velocity (F and U). (For TX, Fs=0; For RX, Vs=0)**

## 2.2.2 Center frequency ($f_c$), fractional bandwidth (FBW), and quality factor (Q)

### 2.2.2.1 Center Frequency

When the CMUT operates in air or vacuum, the center frequency of a mass-spring-damper system is determined by the equivalent mass $m_{eq}$ and the equivalent spring constant $k_{eq}$. Since the medium damping is low it can be ignored in such case. The center frequency is equivalent to the natural resonant frequency $f_r$ of the plate and can be written as:

$$\omega_c = \omega_r = \frac{2.95h}{a^2}\sqrt{\frac{E}{\rho_p(1-\nu^2)}} = \sqrt{\frac{k_{eq}}{m_{eq}}} \,. \tag{2.4}$$

where $E$, $\rho_p$ ,and $\nu$ denote Young's modulus, density, and Poisson's ratio of the plate material, respectively. $a$ represents the radius of the plate and $h$ is the plate thickness [33].

When the CMUT operates in immersion, the center frequency will also be influenced by the complex damping factor $b = b_r + b_i$, the resonant frequency of the system becomes:

$$\omega_c = \frac{\dfrac{2.95h}{a^2}\sqrt{\dfrac{E}{\rho(1-\upsilon^2)}}}{\sqrt{1+0.67\dfrac{\rho_m a}{\rho_p h}}} = \alpha\sqrt{\frac{k_{eq}}{m_{eq}}} \tag{2.5}$$

where $\rho_m$ is the density of the medium. The imaginary part of the loading impedance of the medium contributes to the additional mass to the plate while the spring constant of the plate remains the same. As a result, the center frequency shifts down. This equation is a more general form of the center frequency, which indicates the center frequency shift down with the presence of medium damping ($\alpha = 1$ in the air and $\alpha < 1$ in immersion).

2.2.2.2 Fractional Bandwidth (FBW) and Quality Factor (Q)

As seen in Figure 2-3, the second order spring-mass-damper model can be regarded as a second-order RLC resonant network as an electrical equivalent. Therefore, one can identify the lower 3-dB and higher 3-dB cutoff frequencies which can be written as the following two equations respectively [34]:

$$\omega_H = \omega_0 (\frac{1}{2Q} + \sqrt{(\frac{1}{2Q})^2 + 1}) \tag{2.6}$$

$$\omega_H = \omega_0 (-\frac{1}{2Q} + \sqrt{(\frac{1}{2Q})^2 + 1}) \tag{2.7}$$

where $Q = \frac{X_L}{R} = \frac{\sqrt{k(m+b_i)}}{b_r}$ is the quality factor of the system, which is proportional to the ratio of the stored energy over dissipated energy at the operating frequency. The fractional bandwidth can therefore be obtained as:

$$FBW = \frac{\omega_H - \omega_L}{\omega_0} = \frac{1}{Q} = \frac{b_r}{\sqrt{k(m+b_i)}} \tag{2.8}$$

The complex damping factor $b$ is contributed by the mechanical loss within the CMUT structure as well as the energy dissipated into the medium. The medium loss can be expressed as:

$$Z_{medium} = \frac{K^2a^2 + jKa}{1 + K^2a^2} Z_0 \qquad (2.9)$$

where $K = \frac{2\pi}{\lambda}$ is the wave number. It can be seen that $Ka$ needs to be large so that a wider FBW could be obtained.

#### 2.2.2.3 Discussion

The operating frequency and the bandwidth of the transducer are mainly determined by the dimensions, the shape, and the mechanical properties of the thin plate. A wide bandwidth is desired for better axial resolution in ultrasound imaging. One of the major advantages of CMUTs compared to piezoelectric transducers is the wide bandwidth when operated in immersion. Compared to a PZT transducer, the vibrating plate of a CMUT is substantially thinner and therefore the stiffness or the spring constant of the plate is low. The transducer's plate mechanical impedance is much smaller than the loading impedance over a large frequency range. As a result, a CMUT experiences a much higher damping which leads to a wider bandwidth, which translates into less ringing in the time domain signal and therefore leads to a higher axial resolution in ultrasound imaging. The higher cutoff frequency is determined by the plate mass and the mass loading of the immersion medium as well as the higher order resonant modes of the plate. The lower cutoff frequency is determined by the spring constant. As the spring constant increases, the bandwidth of the transducer decreases.

### 2.2.3 Pull-in Voltage ($V_{pull-in}$), Coupling Coefficient ($k_T^2$), T/R Sensitivity ($S_{TX}, S_{RX}$)

#### 2.2.3.1 Pull-in Voltage ($V_{pull-in}$)

Pull-in of the plate occurs when electrostatic force overcomes the mechanical restoring force, and the plate abruptly collapses down to the bottom electrode.

Before the plate pulls in, the equilibrium exists between the electrostatic force $F_e$ and the mechanical restoring force $F_m$:

$$\sum F = F_m + F_e = 0 \tag{2.10}$$

where $F_m = k(g_{eff} - g)$, $F_e = -\frac{\varepsilon_0 A V_{dc}^2}{2g^2}$, $A$ is the piston electrode area, and $\varepsilon_0$ is the permittivity of the free space. Then the pull-in voltage can be obtained when electrostatic force gradient overcomes the gradient of the mechanical restoring force:

$$\left.\frac{\partial(\sum F)}{\partial g}\right|_{g_{pi},V_{pi}} = \frac{\varepsilon_0 A V_{pi}^2}{g_{pi}^3} - k = 0 \Rightarrow V_{pi} = \sqrt{\frac{k g_{pi}^3}{\varepsilon_0 A}} \tag{2.11}$$

It can be seen that $g_{pi}$ is 2/3 of the effective electrostatic gap:

$$k = \frac{\varepsilon_0 A V_{pi}^2}{g_{pi}^3} \Rightarrow g_{pi} = g_{eff} - \frac{\varepsilon_0 A V_{pi}^2}{2k g_{pi}^2} = g_{eff} - \frac{g_{pi}}{2} \Rightarrow g_{pi} = \frac{2}{3} g_{eff} \tag{2.12}$$

Therefore, the pull-in voltage can be expressed as:

$$V_{pi} = \sqrt{\frac{8k g_{eff}^3}{27 A \varepsilon_0}} \tag{2.13}$$

2.2.3.2 Electromechanical Coupling Coefficient ( $k_T^2$ )

The electromechanical coupling coefficient is an important figure of merit of ultrasonic transducers. The coupling coefficient is, by definition, the ratio of delivered mechanical energy to the stored total energy in the transducer, as expressed by equation,

$$k_T^2 = \frac{E_{mech}}{E_{total}} = \frac{1}{1 + \frac{E_{elec}}{E_{mech}}} \tag{2.14}$$

where $E_{total} = E_{mech} + E_{elec}$. For the parallel plate structure, the coupling coefficient is

$$k_T^2 = \frac{2x}{g_0 - x} \tag{2.15}$$

Initially, $k_T^2$ is zero and increases as the displacement increases. When the displacement equals to one-third of the initial gap distance the electrostatic force gradient is larger than that of mechanical force and the top plate collapses on the bottom electrode; at this point $k_T^2$ is equal to 1 [3].

It should be understood that the coupling coefficient has a direct impact on the device efficiency for piezoelectric transducers. However, this is not the case for CMUTs. Although the $k_T^2$ can be as high as 0.85 when CMUT is biased close to $V_{pull-in}$ [35], it is not desired in actual CMUT as a transmitter because in such case there will not be enough gap for the plate to vibrate. Typically, a CMUT device is biased approximately 80% of the $V_{pull-in}$ for the optimum balance between output pressure and coupling coefficient.

#### 2.2.3.3 Transmit Sensitivity ( $S_{TX}$ )

During the transmission, the input is the electrical actuation voltage and the output is the acoustic pressure delivered to the surrounding medium. The transmission efficiency ( $S_{Tx}$ ), is expressed as the ratio of the output pressure delivered to the medium to the electrical actuation voltage. The stored electric energy inside the CMUT is:

$$W = \frac{CV^2}{2} = \frac{\varepsilon A V^2}{2(d+x)} \tag{2.16}$$

where $C$ is the device capacitance, $V$ is the voltage difference between the top and bottom electrodes, $A$ is the area of the piston, and $\varepsilon$ is the permittivity of the medium between the electrodes. Therefore, the electrostatic force can be represented as:

$$f_e = -\frac{\partial W}{\partial x} = -\frac{\varepsilon A V^2}{2(d+x)^2} \tag{2.17}$$

where $V^2 = V_{dc}{}^2 + V_{ac}(t)^2 + 2V_{dc}V_{ac}(t)$ . For the fundamental frequency, $2V_{dc}V_{ac}(t)$ is the only term contributes to the output pressure. Therefore, at a given dc operation point, the acoustic output pressure could be expressed as:

$$p(t)\cdot A = F(t) = -\frac{\varepsilon A V_{dc}V_{ac}(t)}{2(d+x)^2} = CEV_{ac}(t) \equiv nV_{ac}(t) \qquad (2.18)$$

where $C = \frac{\varepsilon A}{d+x}$ and $E = \frac{V_{dc}}{d+x}$ are the capacitance of the device and the electrical field strength in the gap at the dc operating point, respectively. As a result, the transmission efficiency can be represented as:

$$S_{Tx} = \frac{P(t)}{V_{ac}(t)} = \frac{n}{A}(Pa/V) \qquad (2.19)$$

The above equation indicates that CMUTs with larger capacitance and stronger electrical field (i.e. a smaller gap and larger dc bias) have better coupling efficiency between electrical and acoustic domains during transmission.

The maximum output pressure at a frequency of $\omega$ can be written as:

$$P = \mathrm{Re}(Z_m \cdot \frac{dx}{dt}) = \mathrm{Re}(Z_m)\omega x \qquad (2.20)$$

where x is the amplitude of the acoustic wave, which is equivalent to the CMUT plate displacement, which is determined by the gap height and the dc bias. As a result, a larger electrode gap is desired for higher output pressure, while a smaller gap is required for a better transmit efficiency at a certain dc bias. Therefore the transmitting efficiency and the maximum output pressure is a trade-off.

2.2.3.4 Receive Sensitivity ($S_{RX}$):

On the receive side, the input of the CMUT is the acoustic pressure from the surrounding medium and the output of the device is a current which resulted from the device capacitance change caused by plate vibration under a constant dc bias voltage.

$$I_{ac} = \frac{\partial Q}{\partial t} = V_{dc}\frac{\partial C}{\partial t} \tag{2.21}$$

Therefore, a current is produced with an amplitude which is a function of the bias voltage, $V_{dc}$, the frequency of the incident wave $\omega_0$, the steady-state capacitance of the device $C_0$, and the fractional displacement $\frac{\Delta x}{x_0}$ due to the incident wave, which can be mathematically expressed as:

$$I_{ac} = V_{dc} \cdot \omega_0 \cdot C_0 \cdot \frac{\Delta x}{x_0} \tag{2.22}$$

By plugging in the $C = \frac{\varepsilon A}{x_0}$, $E = \frac{V_{dc}}{x_0}$, and $P = \mathrm{Re}(Z_m \cdot \frac{\partial x}{\partial t})$ into above equation, we can rewrite the received current as:

$$I = CE \cdot \frac{\partial x}{\partial t} \tag{2.23}$$

Therefore, the receive sensitivity can be expressed as:

$$S_{RX} = \frac{I}{P} = \frac{CE}{Z_m} \cdot A = \frac{nA}{Z_m} \tag{2.24}$$

It is seen that both $S_{TX}$ and $S_{RX}$ are proportional to the electromechanical transformer ratio $n$. We can also define the two-way sensitivity, $S_{TX/RX}$, with the unit A/V by the product of $S_{TX}$ and $S_{RX}$:

$$S_{TX/RX} = \frac{n^2}{Z_m} \tag{2.25}$$

We can derive the transfer function of the CMUT in TX mode by setting $F = 0$ and in RX mode by setting the $V = 0$. The voltage across $R_m$ represents the force on the top plate. Assuming the parasitics and $R_{loss}$ and $m_m$ can be ignored, the magnitude of TX and RX transfer function can be expressed as:

$$S_{TX} = \left|\frac{P}{V}\right| = \frac{n}{A}\left[1 + (\frac{\omega\sum m}{R_m})^2(1 - \frac{\omega_0^2}{\omega^2})^2\right]^{-\frac{1}{2}} \tag{2.26}$$

$$S_{RX} = \left|\frac{I}{V}\right| = \frac{An}{R_m}\left[1 + (\frac{\omega\sum m}{R_m})^2(1 - \frac{\omega_0^2}{\omega^2})^2\right]^{-\frac{1}{2}} \tag{2.27}$$

where $\omega_0 = \sqrt{\frac{\sum k}{\sum m}}$, $\sum k = k_p + k_{soft}$, $k_{soft} = \frac{n^2}{C_0}$ $\sum m = m_{piston} + m_{medium}$. It can be concluded that the maximum sensitivity in TX and RX both occur at the center frequency $\omega = \omega_0$, where the TX, RX, and two-way sensitivity can be expressed as:

$$S_{TX,\max} = \frac{n}{A} \tag{2.28}$$

$$S_{RX,\max} = \frac{nA}{R_m} \tag{2.29}$$

$$S_{TX/RX,\max} = \frac{n^2}{R_m} \tag{2.30}$$

2.2.3.5 Discussion

Overall, it is observed that the vacuum gap height design and control is one of the most critical aspects to achieve a balance between transmit and receive performances. A larger gap is desired for higher output pressure, while a smaller gap is desired for optimum sensitivity at a practical dc bias voltage.

Parasitics can heavily affect CMUT efficiency. In the Mason's equivalent circuit model, the parasitics derive from the following aspects. First, the parallel parasitic capacitance $C_p$ is the most dominant parasitic component because it sinks the current in the electrical port and reduces the available current that could be converted into the mechanical domain. Essentially, $C_p$ exists where the actuation voltage is applied but there is no plate movement, which is where not electrically active. It includes the overlapped post region, cable capacitance, top-to-bottom via capacitance when TSVs are incorporated, and $C_p$ also exists because the plate average movement deviates from the ideal piston movement. Having an insulating substrate is beneficial for reducing the parallel capacitance resulting from the CMUT structure.

Another parasitic capacitance is the series parasitic capacitance $C_s$, which comes from the insulating layer embedded inside the CMUT structure. A high-k dielectric insulation layer and a thinner dielectric layer thickness are desired to reduce the effect of the series parasitic capacitance.

## 2.3 CMUT Fabrication Process

### 2.3.1 CMUT on Silicon Substrates

#### 2.3.1.1 Sacrificial-Release Process

CMUT technology was invented on silicon substrates and extensive research has been conducted to improve the device performance and reliability in terms of the device structure design and fabrication process flow.

Surface micromachining with sacrificial release process is the first technology used to fabricate CMUTs. The basic principle of sacrificial release process is to first deposit a sacrificial layer and then selectively remove it using an appropriate etchant after depositing the plate layer on top. An extension of this process includes through-silicon via interconnects from the front side of the substrate to the backside which can be used for making 2D CMUT arrays. A general method for making CMUTs through surface micromachining method, or sacrificial releasing method is described as following. (a) First, an insulation is deposited; it also acts as an etch stop layer. (b) Next, a sacrificial layer is deposited. (c) Channels are etched define etch channels. (d) Membrane material is deposited. (e) Release holes are etched. (f) Membrane is released by wet-etching of the sacrificial layer. (g) Etch holes are filled. (h) Electrodes are deposited.

There are several disadvantages associated with this process. First of all, the fabrication process is relatively complex. Second, the deposited layers have poor uniformity and could introduce stress. Finally, it introduces additional parasitics because the silicon substrate is a conductive material and requires additional insulation steps.

#### 2.3.1.2 Wafer Bonding Process

The second method, typically referred to as wafer bonding fabrication, involves fabricating the cavities/bottom electrode on one wafer and the membrane/top electrode on a second wafer [3.6]. The wafers are then bonded together with high force and elevated temperature. The fabrication process is described as following. (a) First, an insulation layer is grown on a highly doped substrate. (b) Cavities are then etched on the insulation layer. (c) A separate silicon-on-insulator (SOI) wafer is bonded together with the (b) wafer. (d) An etch-back process is done to release the substrate and buried-oxide (BOX) layer of the SOI wafer. (e) A two-step etching process is done to expose the bottom substrate. (f) Metal electrodes are deposited. Because the cavities and membranes can be fabricated separately, it decouples design constraints so each structure can be optimized individually. With wafer bonding, sacrificial layers, and thus holes, are not required hence the fill-factor is improved. Also, without the constraints of releasing, cavity depths can be made shallower which further improves electrical stability as actuation voltages can be reduced and insulating thickness can be increased instead. Furthermore, fabrication repeatability and controllability is generally better with the wafer bonding process as the number of lithography steps and masks are typically fewer.

Wafer bonding method is demonstrated to overcome the above shortcomings. Wafer bonding process begins with two wafers: a substrate wafer and a silicon-on-insulator (SOI) wafer. For silicon substrates, the cavities are defined on the substrate wafer. Then the SOI wafer and the substrate are brought together. Wafer bonding has also been implemented using eutectic bonding and more recently adhesive bonding [12] to make CMUTs on silicon substrates.

#### 2.3.1.3 Insulation on Silicon Substrate

In CMUT operation, it is desired to have the electric field applied only where it is required, which is the vacuum gap. However, the commonly used silicon substrate is a semiconductor and therefore result in the parallel parasitic capacitance. Several strategies have been investigated to address the problem on the silicon substrate.

One way is to use an extended insulation layer structure in the post area to address the low breakdown voltage and high parasitic capacitance using a LOCOS process [36][37]. However, this process is not suitable for 2D arrays.

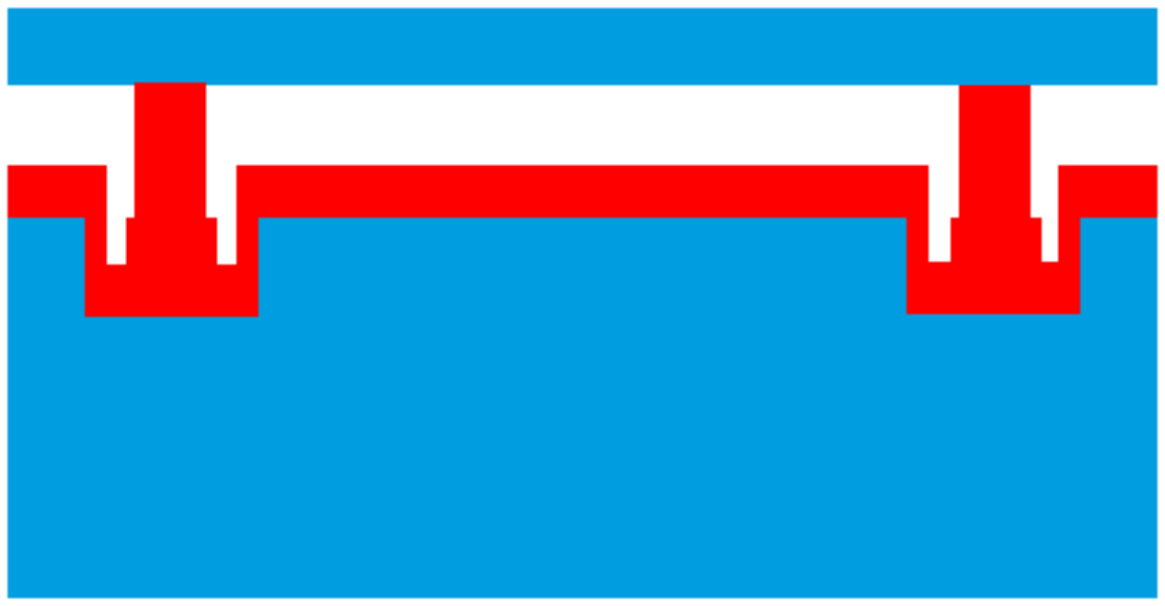

**Figure 2-4: CMUT with an extended insulation layer.**

Another structure is to isolate the CMUT bottom electrode only to the region under the gap where a high electric field is desired so that the possibility of dielectric breakdown and parasitic capacitance in the post region can be minimized [38][37]. This structure also includes TSV interconnects for accessing each element from the backside. The key component in this approach is an SOI wafer with a thick buried oxide layer to completely insulate silicon bottom electrodes below the active plate region in each CMUT cell. The hot electrode is provided through an opening in the thick buried oxide layer filled with TSV.

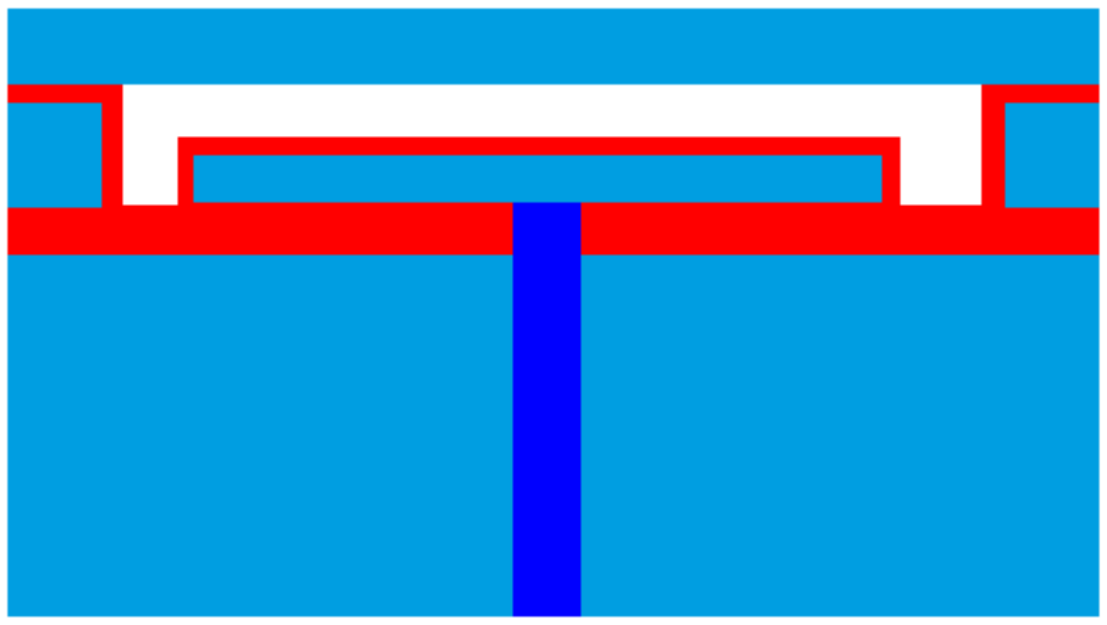

**Figure 2-5: CMUT with a thick buried oxide layer.**

### 2.3.2 CMUT on Insulating Substrates

Building MEMS devices on insulating substrates have raised tremendous research interest [39]–[41]. Using an electrically insulating substrate to fabricate CMUTs can reduce the device parasitics and simplify the fabrication flow by eliminating the complicated isolation steps. Furthermore, insulating substrates such as glass, fused silica, quartz, can provide excellent optical transparency and therefore can enable novel applications where optics and acoustics are combined. Fabrication of CMUT on insulating substrates is relatively new and is attracting more and more attention.

It is desired to fabricate CMUTs on insulating substrates using the wafer bonding technique, although it has also been shown that sacrificial release process could be employed and result in a reduced parasitic capacitance [9]. Anodic bonding comes across as an attractive method to combine the benefit of wafer bonding and an insulating substrate. The bottom electrode could be patterned and deposited in the etched glass cavities and a silicon or SOI wafer could be anodically bonded as the plate to realize a CMUT structure. However, one challenge with anodic bonding is the outgassing during bonding. In the previously demonstrated CMUTs on glass substrates with anodic bonding, either the cavities were pressurized with trapped oxygen gas under a thick plate [42], or the cavities were exposed to outside for gas evacuation, thus making the transducer not suitable for immersion operation [10]. This question needs to be addressed to fully unleash the potential to make CMUTs on glass substrates using anodic bonding. Apart from anodic bonding, adhesive bonding (with SU8, BCB) or eutectic bonding can also be combined with insulating substrates.

Below is a summary of the CMUT fabrication technologies categorized by the substrate type. The topics in the black box will be discussed in this dissertation.

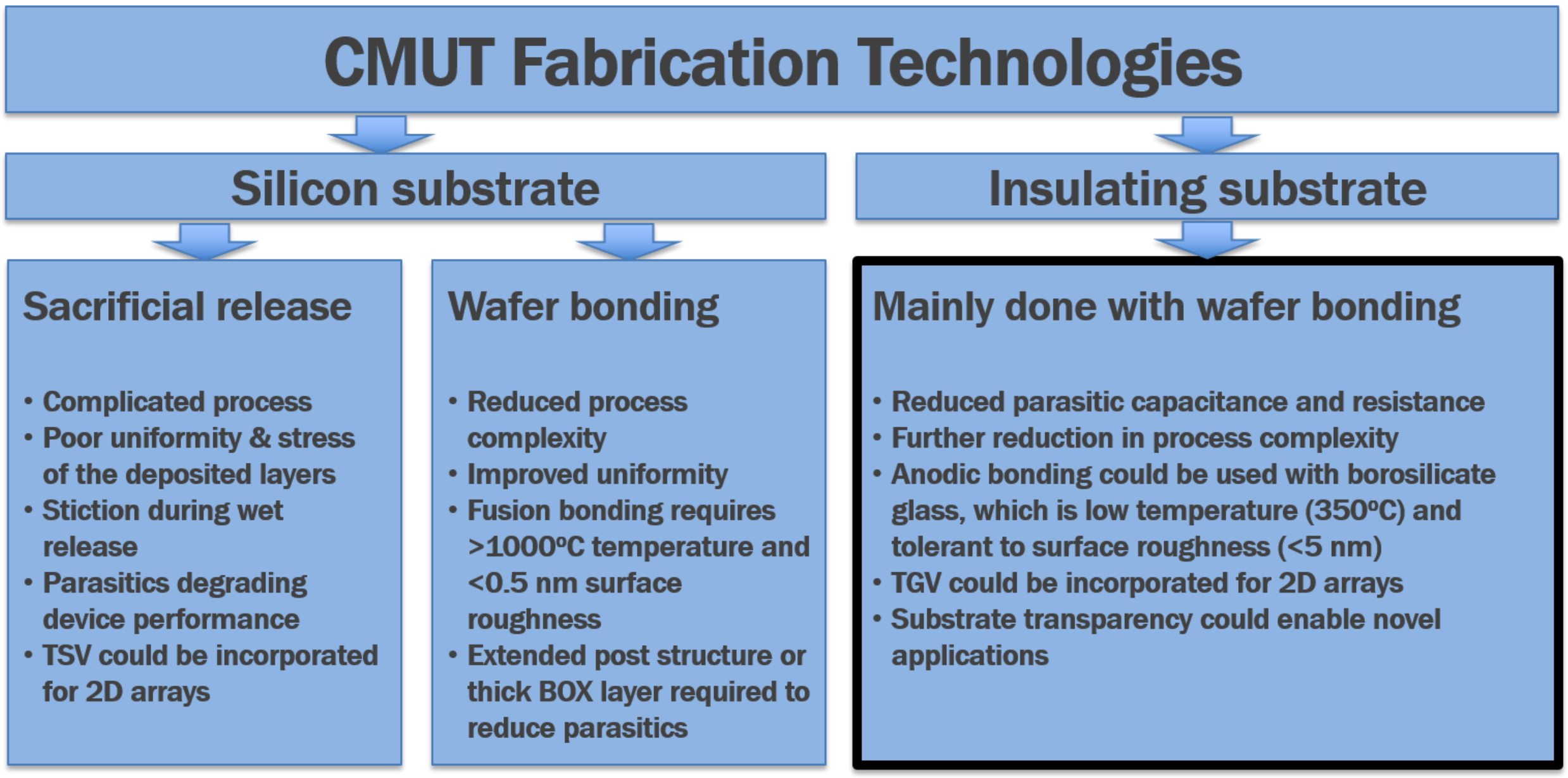

**Figure 2-6: CMUT fabrication technologies categorized by substrate types.**

## 2.4 CMUT Integration with Electronics

Close integration of CMUT and the electronics are highly desired, especially for transducer arrays with small elements, thus small capacitance, such as 2D arrays and the arrays designed for use at the end of a catheter. In the current conventional ultrasonic imaging systems, the array is located in a hand-held probe, which is connected to the main processing unit via a cable bundle. Transmit and receive electronics are located in the main processing unit. In order to avoid the additional parasitic capacitance introduced by the cable degrading the signal quality, the receiver electronics must be close to the transducer array. Furthermore, multiplexing and beamforming circuitry can also be integrated with the array to minimize the number of active electronic channels and the number of physical connections between the probe and the backend system.

For integrating CMUT arrays with electronic circuits, mainly two categories of methods have been adapted.

### 2.4.1 Monolithic Integration

The first approach is monolithic integration, which is to build CMUTs and CMOS concurrently (co-processing), or build the CMUTs on top of the finished electronic wafer (post-processing).

#### 2.4.1.1 Co-processing

One method to monolithically integrate CMUTs and electronics is to fabricate both components concurrently using a standard or minimally modified process. A BiCMOS process using 16 masks has been used with only minor modifications including an additional photolithography step and sacrificial layer etching to fabricate CMUTs side-by-side with electronic circuits on the same substrate [43]. Although this method offers a cost-effective solution for electronics integration, it suffers from two major limitations: (1) the transducer area is shared by electronic circuits or interconnections. (2) The device layer thickness are limited by the process used for electronic circuits.

#### 2.4.1.2 Post-process (CMUT-on-CMOS)

The second monolithic integration technique is to fabricate the electronics first using a standard foundry process and then build the CMUT on top of the finished electronics. This post-processing is augmenting a standard foundry process only with two blanket post-process steps for sacrificial etching and cavity sealing [44]. This approach has good area utilization and more control over device dimensions, but it still suffers from the limitations on the device dimensions in the vertical direction due to the layers available in the standard foundry process.

Low-temperature wafer bonding can also be used in post-process. It brings the advantages of wafer bonding process such as control over the plate thickness, process simplicity, and provides a monolithic integration solution without going through the complexity of sacrificial release-process [45]. In this low-temperature bonding process, a thin titanium layer is used for electrical via-contact to CMUTs substrate as well as an adhesion layer for wafer bonding. The gap height is set by the total thickness of the titanium adhesion layer and the passivation layer on the CMOS wafer.

### 2.4.2 Hybrid Integration

#### 2.4.2.1 Chip-to-chip bonding

The fabrication processes for both CMUT and CMOS could be optimized if the two components can be fabricated on separate substrates. In this case, CMUTs need through-wafer-interconnections from the front side of the substrate to the backside. The CMUT chip can then be directly bonded on the electronics using the well-established flip-chip bonding process [46]. Direct bonding requires that the electronic die area be greater than the CMUT die area so that the peripheral pads on the electronic die will be accessible to provide connections between the front-end circuits and the backend system.

#### 2.4.2.2 Bonding Through Intermediate Substrates

Using an intermediate substrate will make the size of the CMUT and the IC be independent and it is desirable in the following cases: (1) To implement a very large CMUT array, electronics or both electronics and CMUT array should be implemented by tiling several unit blocks. This approach also helps improve the overall yield one can achieve [47]. (2) Using a flexible intermediate substrate, the overall form factor of an integrated ultrasound probe can be minimized. This approach has been used to demonstrate a 64-element CMUT ring array for forward-looking catheter-based intracardiac imaging [48].

## 2.5 Summary

In this chapter, we reviewed the CMUT basics and the device modeling (parallel plate capacitor model and equivalent circuit model). The critical performance attributes are derived and the design trade-offs are discussed. CMUTs have a wider bandwidth immersion derived from the thin plate and low plate mechanical impedance. Therefore it holds great potential for ultrasound medical imaging as a complement to piezoelectric transducers. A good device design and minimized parasitics are required for optimized CMUT device efficiency.

The CMUT fabrication process and integration methods are also discussed. We categorized the CMUT fabrication technology based on substrate types. CMUTs are traditionally fabricated on silicon substrates. However, fabricating CMUT on insulating substrates can reduce device parasitics and simplify the fabrication process flow. Furthermore, insulating substrates such as glass can offer the benefit of optical transparency and help innovate for new applications.

# Chapter 3 PLATFORM TECHNOLOGY: VACUUM-SEALED CMUTS FABRICATED ON GLASS SUBSTRATE USING ANODIC BONDING

## 3.1 Introduction

CMUT technology has demonstrated great promise for next-generation ultrasound applications. Wafer-bonding technology has largely simplified the fabrication of CMUTs by eliminating the requirement for a sacrificial layer and increases control over device parameters. Fabrication of CMUTs on glass substrates using anodic bonding has many advantages over other bonding methods, such as low-temperature compatibility, high bond strength, high tolerance to particle contamination and surface roughness, and cost savings. Furthermore, the glass substrates lower the parasitic capacitance and improve reliability. The major drawback is the trapped gas inside the cavities, which occurs during bonding. Earlier CMUT fabrication efforts using anodic bonding failed to demonstrate a vacuum-sealed cavity.

In this Chapter, we developed a fabrication scheme to overcome this issue and demonstrated vacuum-backed CMUTs using anodic bonding. This new approach also simplifies the overall fabrication process for CMUTs. We demonstrated a CMUT fabrication process with three lithography steps. A vibrating plate is formed by bonding the device layer of a silicon-on-insulator (SOI) wafer on top of submicron cavities defined on a borosilicate glass wafer. The cavities and the bottom electrodes are created on the borosilicate glass wafer with a single lithography step. The recessed bottom metal layer over the glass surface allows bonding the plate directly on glass posts and therefore helps reduce the parasitic capacitance and improve the breakdown reliability.

## 3.2 Fabrication Process Development

The realized CMUT structure consists of a thin single-crystal silicon plate with a thin silicon nitride insulation layer and a patterned metal bottom electrode deposited inside a vacuum-sealed, sub-micrometer cavity (Figure 3-1). The vibrating plate is formed by depositing a layer of silicon nitride on the device layer of an SOI wafer before bonding. The silicon nitride layer is mainly to prevent electrical shorting when the plate comes in contact with the metal bottom electrode following pull-in and also acts as an intermediate bonding layer. Anodic bonding of borosilicate glass to thin-film coated silicon wafers was demonstrated previously [49]–[51]. The device layer is chosen to be highly doped conductive silicon as it is used as the top electrode. The bottom electrode is deposited on the surface of the cavities so that the overall parasitic capacitance of the device is reduced and the dielectric reliability is improved as the post region in this approach does not experience any significant electric field. During the formation of bond pads to provide electrical access to bottom electrodes of individual transducer elements, a metal layer is also deposited on top of the silicon plate to further increase the conductivity of the top electrode and to provide a suitable layer for wire bonding.

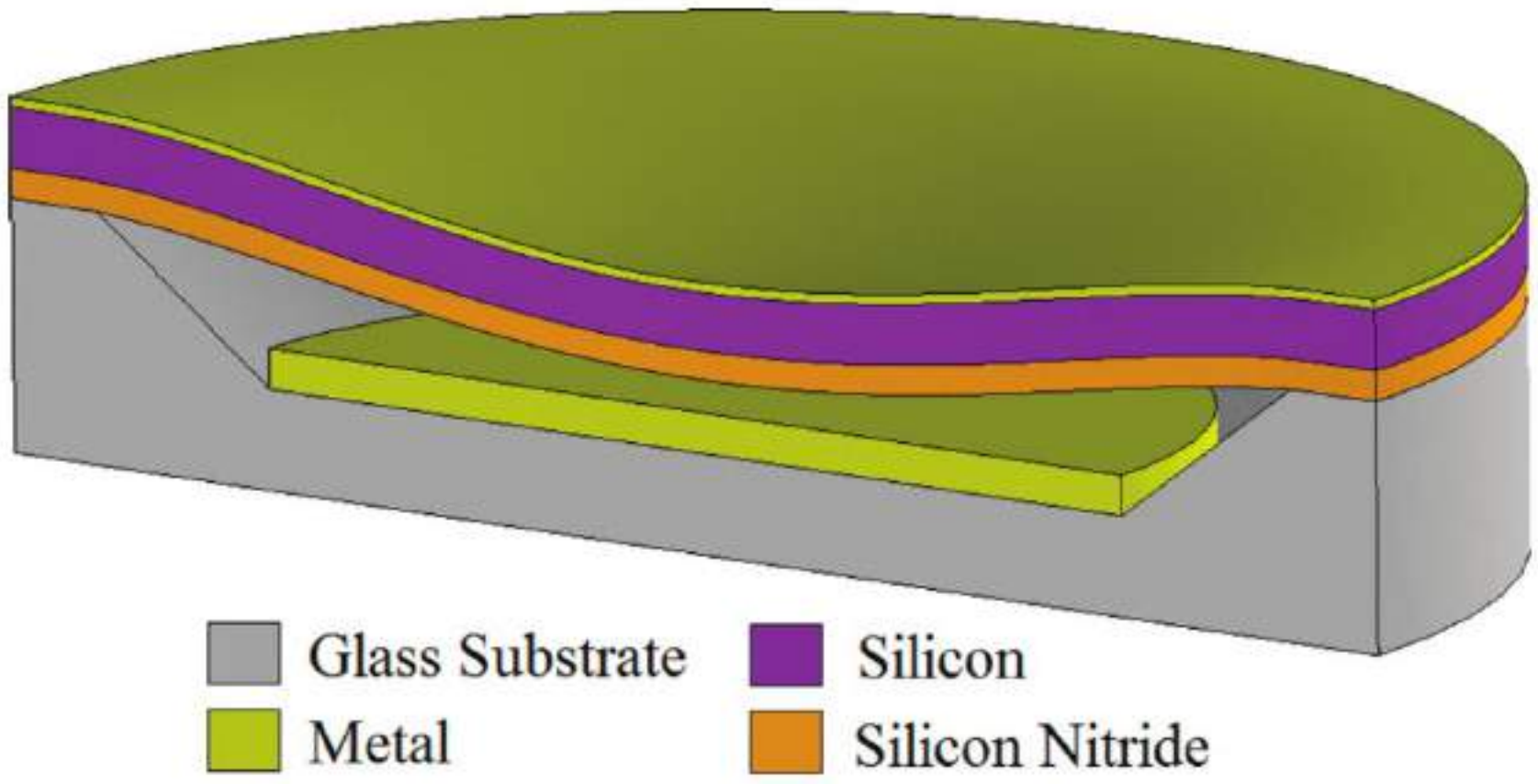

**Figure 3-1: 3D cross-sectional model of a completed CMUT cell.**

The initial substrate was a standard 0.7-mm-thick, 100-mm-diameter borosilicate glass wafer (Borofloat33, Schott AG, Jena, Germany) that has a high surface quality with an RMS roughness ($R_q$) of 0.7 nm and a good flatness with a warp that is less than 0.05%. The thermal expansion coefficient of the borosilicate glass substrate is 3.25 ppm/°C, close to that of silicon (3.2 ppm/°C) preventing stress in the silicon plate after anodic bonding. The SOI wafer that we used for fabrication has a 2 ± 0.5-µm-thick, n-type device layer with 0.001 to 0.005 Ω·cm resistivity, a 0.5-µm-thick BOX layer, and a 500-µm-thick handle wafer with 1 to 10 Ω·cm resistivity. Before the process, the borosilicate glass substrate was cleaned for 15 minutes in a Piranha solution for removal of organics and other gross particle contaminants from the surface. The cavity pattern was defined using 2-µm-thick negative photoresist (AZ-5214E IR, Clariant, Wiesbaden, Germany), which is suitable for lift-off [Figure 3-2 (a)]. The patterned wafer was hard-baked for 2 h at an elevated temperature of 125°C [Figure 3-2 (b)]. This step promotes the adhesion between the photoresist and the substrate and makes the photoresist a better mask for the etching. The cavities were created in 10:1 buffered oxide etch (BOE) solution [Figure 3-2 (c)]. Wet etching was preferred for a uniform etching and minimal surface roughness in the cavities. We measured the BOE lateral etch as 10 times faster than its vertical etch. It has been reported that a water-rich interface layer between the wafer and resist causes the etchant to penetrate very fast laterally [21]. As a result, faster lateral than vertical etching is often seen in isotropic etching. This lateral etch must be considered in the mask design to achieve the target cavity size after etching. 230-nm-deep cavities were etched in 15-min total time with 5 cycles of BOE etching of 3 min each. The photoresist was hard-baked for 10 min between each cycle to prevent peeling. After the cavity etching was completed, the wafer was transferred to the evaporation chamber without removing the resist. The gap height of the CMUTs was defined by the difference of the etch depth and the thickness of the metal deposited in the cavities. Thus a metal stack that consists of 20 nm of chromium as an adhesion layer and 90 nm of gold was deposited to obtain the 120-nm gap height [Figure 3-2 (d)]. The undercut that was formed during the wet etch helps to confine the metal electrode to the bottom surface of the cavity and also makes the lift-off process easier.

Prior to bonding, we deposited 200-nm silicon nitride on top of the device layer of the SOI wafer by using plasma-enhanced, chemical-vapor deposition (PECVD) at 1000-mTorr chamber pressure and 350°C temperature. This silicon nitride layer serves as an insulation layer between the conductive silicon plate (top electrode) and the metal in the cavity (bottom electrode) during device operation. The glass wafer and the SOI wafers were cleaned using solvents and Piranha solution, respectively. The borosilicate glass surface and the nitride surface were anodically bonded together at 350°C under 2.5-kN down force in vacuum ($10^{-4}$ Torr) in a semi-automatic bonding system [model EVG510, EVG Group, St. Florian, Austria; Figure 3-2 (e)]. Typically, it is recommended to limit the current during the bonding [22]. The setup that we use does not have a current limited bonding option. Thus, the voltage was ramped up to its final value at a rate of 20 V/min not to cause a breakdown in the silicon nitride layer and kept at the target value for 30 min. The metal is exposed to a high electrical field during bonding, thus various bonding voltages were evaluated to maintain high bond yield without damaging the floating bottom electrode inside the cavities. The bonding was tested at 1000, 700, 600, and 500 V. A high bonding yield with no damage on the bottom electrodes was observed at 600 V for the presented process and wafer parameters. After bonding, the handle wafer was ground down to 100 µm. We used a heated tetramethylammonium hydroxide (TMAH) solution (10% TMAH at 80°C) to selectively remove the remaining handle wafer over the BOX layer, which was subsequently removed using 10:1 BOE solution [Figure 3-2 (f)].

The borosilicate glass substrate is exposed to a high electrostatic field, which causes outgassing during bonding [52]. We proposed to evacuate the gas inside the cavities and seal them in vacuum. To access the bottom electrode for forming bond pads, the plate at the pad location has to be etched. When the plate over the metal pad region is etched, the channel over the metal surface is exposed and allows the gas to escape [Figure 3-2 (g)]. We used reactive ion etching with SF6 gas to etch silicon. Different arrays implemented on the same wafer are also separated during this step by etching the conductive silicon between different arrays. After evacuating the trapped gas in the cavities, oxygen plasma is used to remove the photoresist. Because there is no wet cleaning at this step, no liquid reaches inside the cavities. Avoiding wet processing is important at this step as it can lead to stiction and consequently collapsed cells in the drying stage. To seal the cavities we deposited PECVD silicon nitride [Figure 3-2 (h)]. The thickness of the silicon nitride was chosen to be more than three times the cavity height for a proper sealing [9].

After the sealing step, the wafer surface is completely covered by silicon nitride. To create electrical contacts, the silicon nitride layer on the bond pads has to be removed. At this point, the silicon nitride deposited on the conductive silicon plate is also removed leaving the silicon nitride only on the locations where sealing is required. For etching the silicon nitride we used reactive ion etching where the AZ5214E IR photoresist was used as a mask. The photoresist was hard-baked for 5 min at 125°C before etching. After removing the nitride layer on the pads and on the silicon plates [Figure 3-2 (i)], 20-nm chromium and 130-nm gold were deposited. The chromium-gold metal stack was then lifted off in N-methyl-2-pyrrolidone (NMP) solvent [Figure 3-2 (j)]. At this step, the device fabrication is completed.

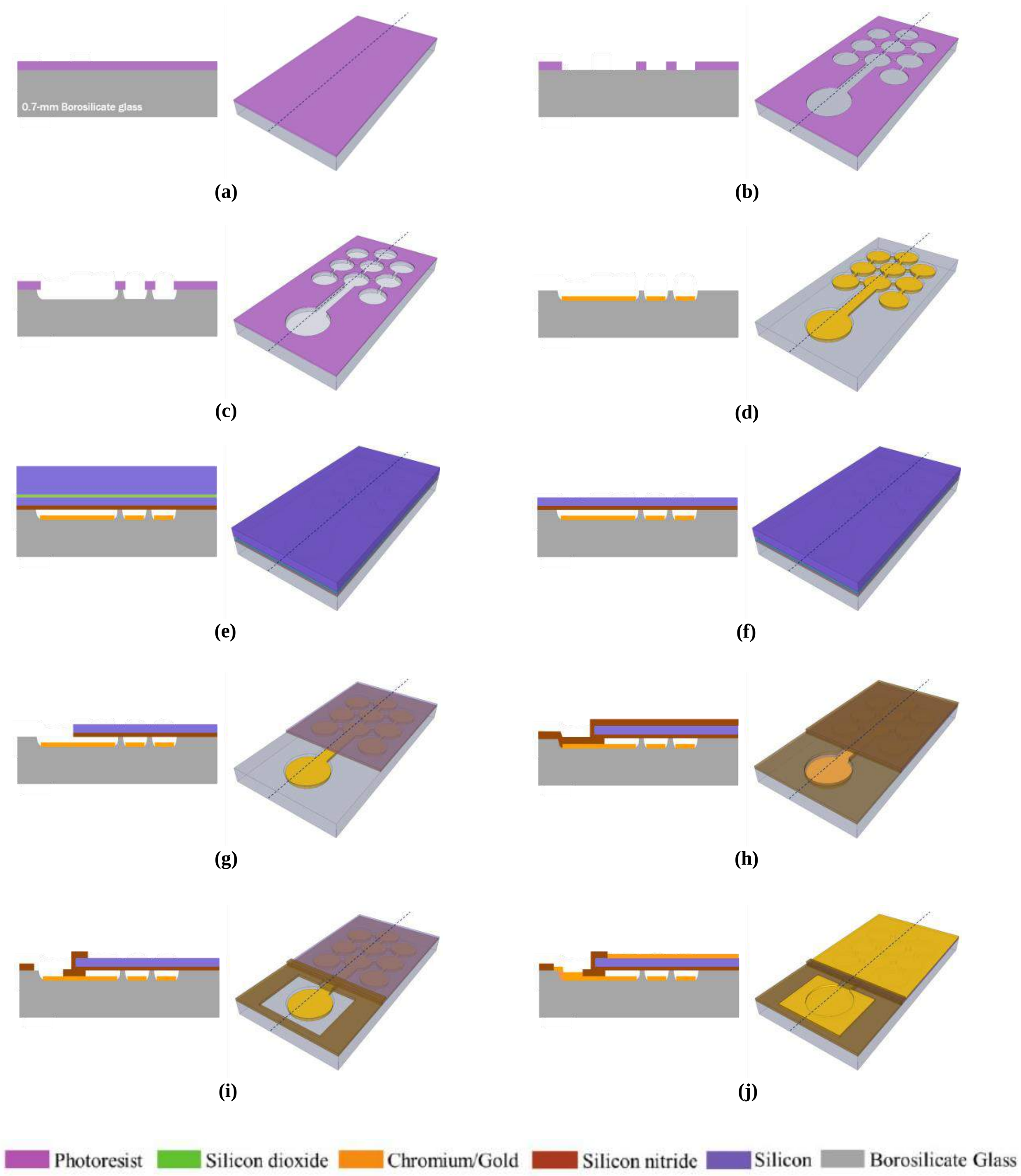

**Figure 3-2: Fabrication process flow**

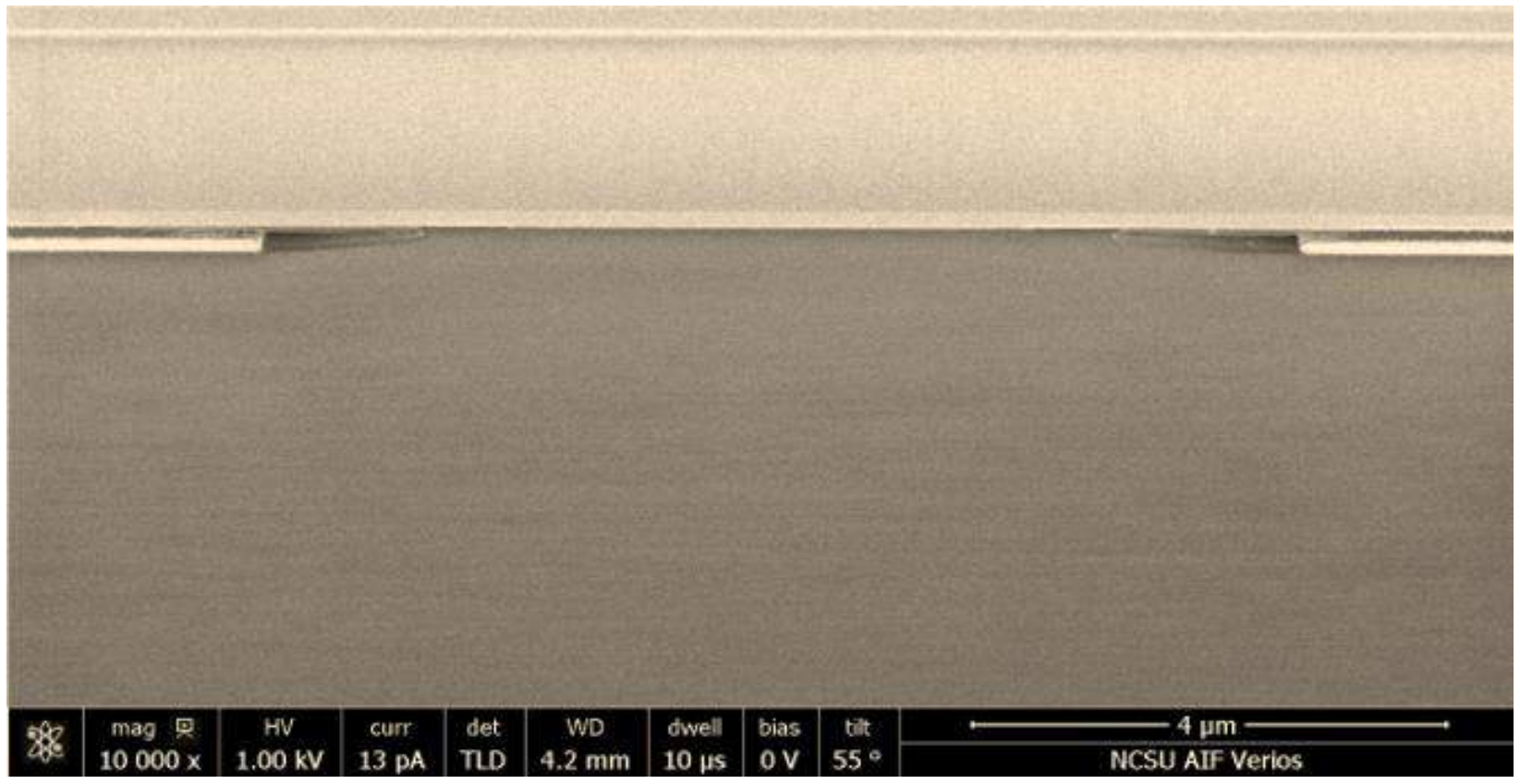

**Figure 3-3: SEM cross-sectional image of a completed CMUT cell.**

## 3.3 Characterization

### 3.3.1 Characterization in Air

The success of the sealing process was confirmed by measuring the deflection profile of the plate after the completion of the entire process (Figure 3-4). The upward plate deflection due to gas trapped inside an array of cells of design #1 was measured using an optical surface profilometer (model NewView 5000, Zygo Corporation, Middlefield, CT, USA). After evacuating the gas and sealing the channels, the deflection under atmospheric pressure was measured as 28 nm. Finite element analysis (FEA; ANSYS v.14, ANSYS Inc., Canonsburg, PA, USA) for this cell predicts the atmospheric deflection as 26.8 nm, which also proves that there is no significant stress due to anodic bonding and the cavity is under vacuum.

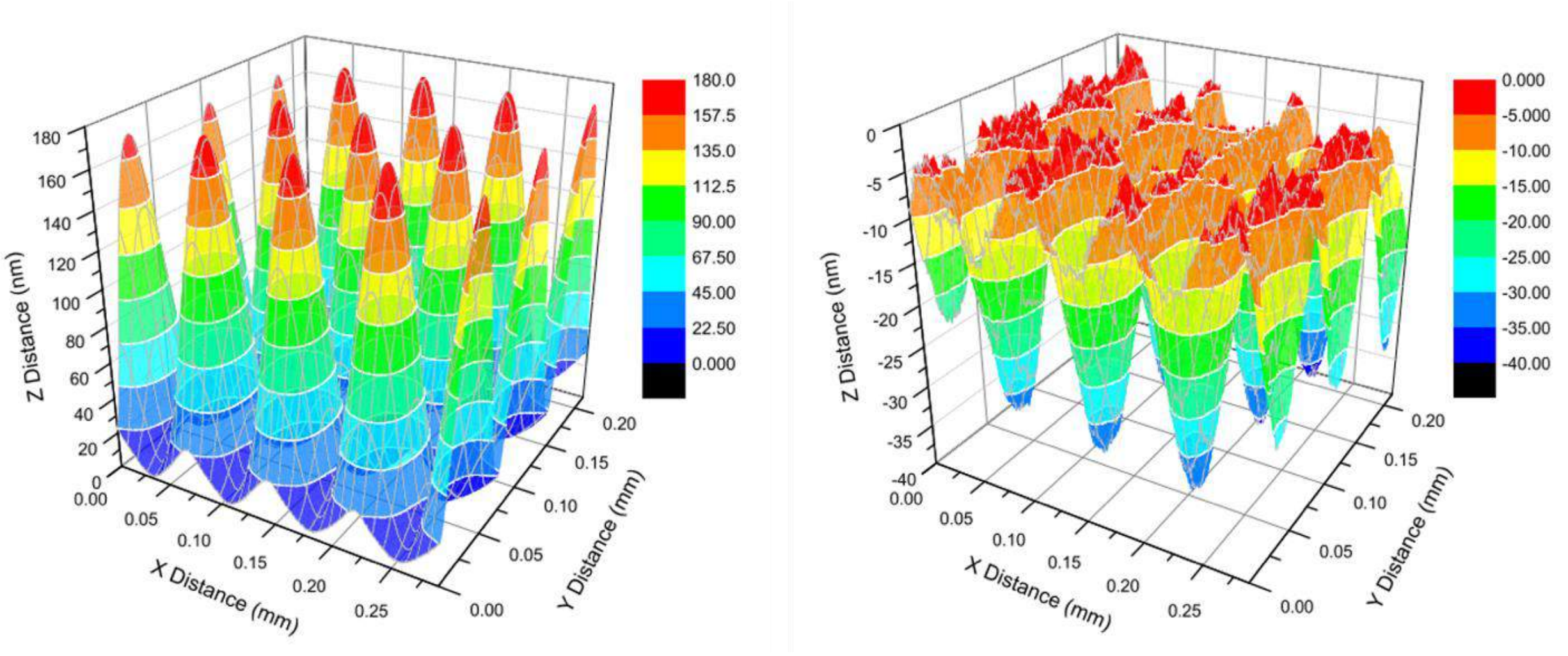

**Figure 3-4: Zygo interferometer measurement. Before gas evacuation (left). After gas evacuation and re-seal (right).**

The device was tested in air using a network analyzer (model E5061B, Agilent Technologies Inc., Santa Clara, CA, USA) with an internal dc supply available up to 40 V. The measured real and imaginary parts of electrical input impedance in air are shown for bias voltages of 10, 15, and 20 V in Figure 3-5. The baseline in the real part corresponds to the series resistance of the device, which is measured as 21 Ω. To measure the collapse voltage, the resonant frequency is observed while increasing the bias voltage by 1-V steps. The collapse voltage is determined by the sudden resonant frequency jump at the collapse.

The collapse voltages and the resonant frequencies of these three designs were also simulated using FEA. TRANS 126, electromechanical transducer elements are used for the direct coupling of electrostatic and structural domains. The element is capable of handling the spring softening effect in the simulations. First, the static analysis was performed to find the collapse voltages.

Second, the prestressed harmonic analysis was carried out to find the resonant frequency at the 70% of the collapse voltage. The results show that the fabricated CMUTs operate as predicted by the finite element model. We have measured the electromechanical coupling coefficient ($k_T^2$) as 0.1 at 15-V dc bias (75% of the collapse voltage) and 0.3 at 20-V dc bias (90% of the collapse voltage), which are consistent with results reported earlier [35]. The single cell capacitance of the same design under atmospheric pressure is calculated as 0.23 pF by using FEA, which corresponds to a total capacitance of 64.17 pF when multiplied by the total number of cells. We measured the total capacitance of the device as 67.84 pF, indicating that the external parasitic capacitance is less than 6%.

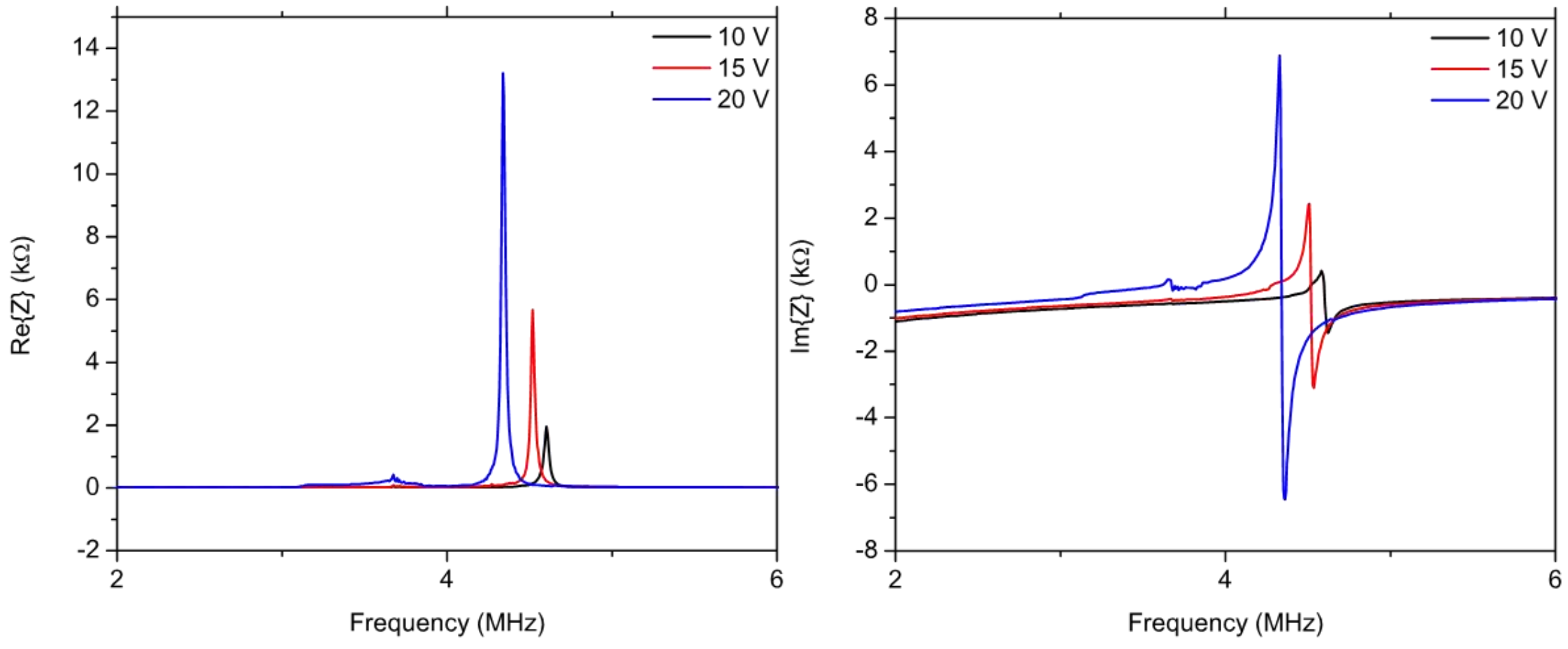

**Figure 3-5: Electrical input impedance measurement. Real part (left). Imaginary part (right).**

### 3.3.2 Characterization in Immersion

For immersion tests, vegetable oil was used because the transducer surface was not electrically insulated. A small tank was built over the transducer element that was wire-bonded to a chip carrier. A calibrated hydrophone (model HGL-0200, Onda Corporation, Sunnyvale, CA, USA) connected to a preamplifier (model AH-2010, Onda Corporation) was placed at 14-mm distance from the transducer surface on the central axis of the transducer. A 20-V unipolar pulse was superimposed on the dc bias voltage through a bias-T circuit. The conductive plate layer was grounded and the bottom electrode was used as the active electrode. A 12-V dc voltage was applied and the element was driven by a 110-ns wide pulse. The signal received by the hydrophone is shown in Figure 3-6. The notches at 3.8 MHz and its higher harmonics correspond to the ringing in the substrate [53] as it was also evident in the time domain data as a tail following the main pulse. Substrate ringing can be pushed out of the frequency band of interest by choosing the substrate thickness accordingly.

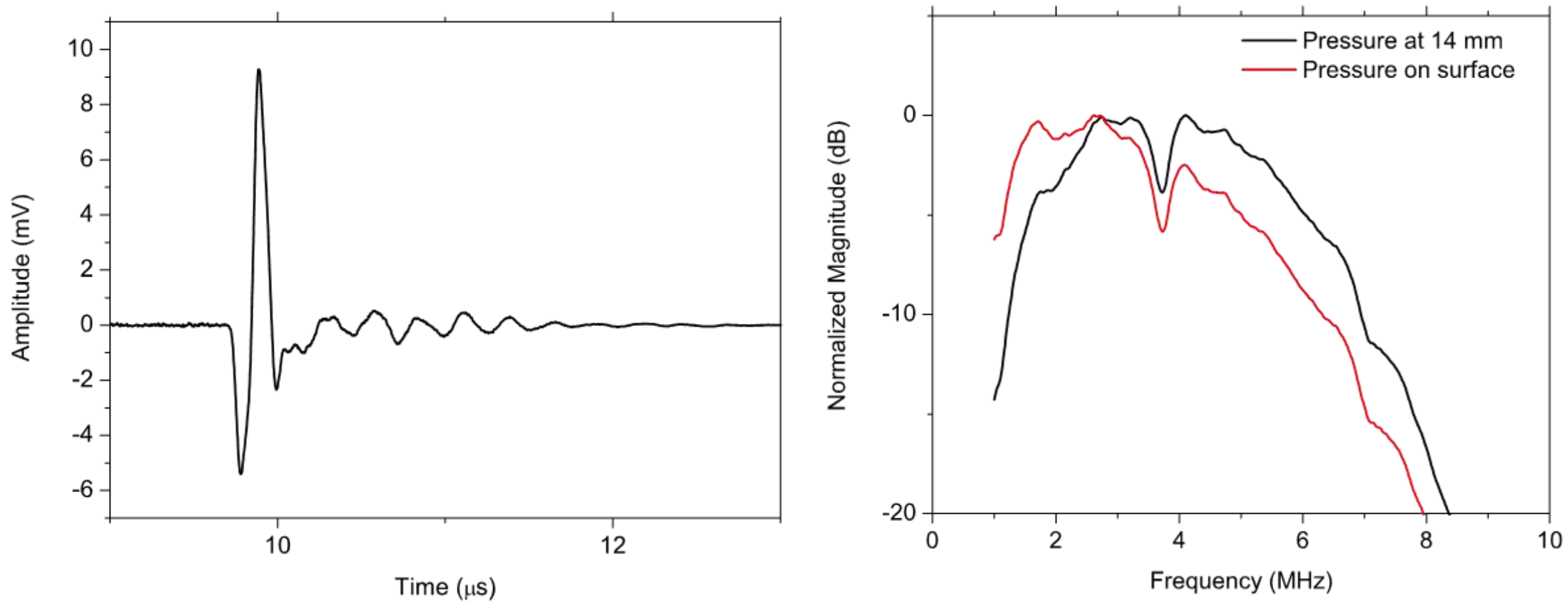

(a) (b)

**Figure 3-6: Immersion measurement using a calibrated hydrophone. (a) Received time-domain signal. (b) Fourier transform.**

### 3.3.3 Array Characterization

We checked the resonant frequency of each individual element in an imaging array that we fabricated (Figure 3-7). The results show that the standard deviation is 0.11 MHz with a mean value of 12.5 MHz in resonant frequency for a 66-element array.

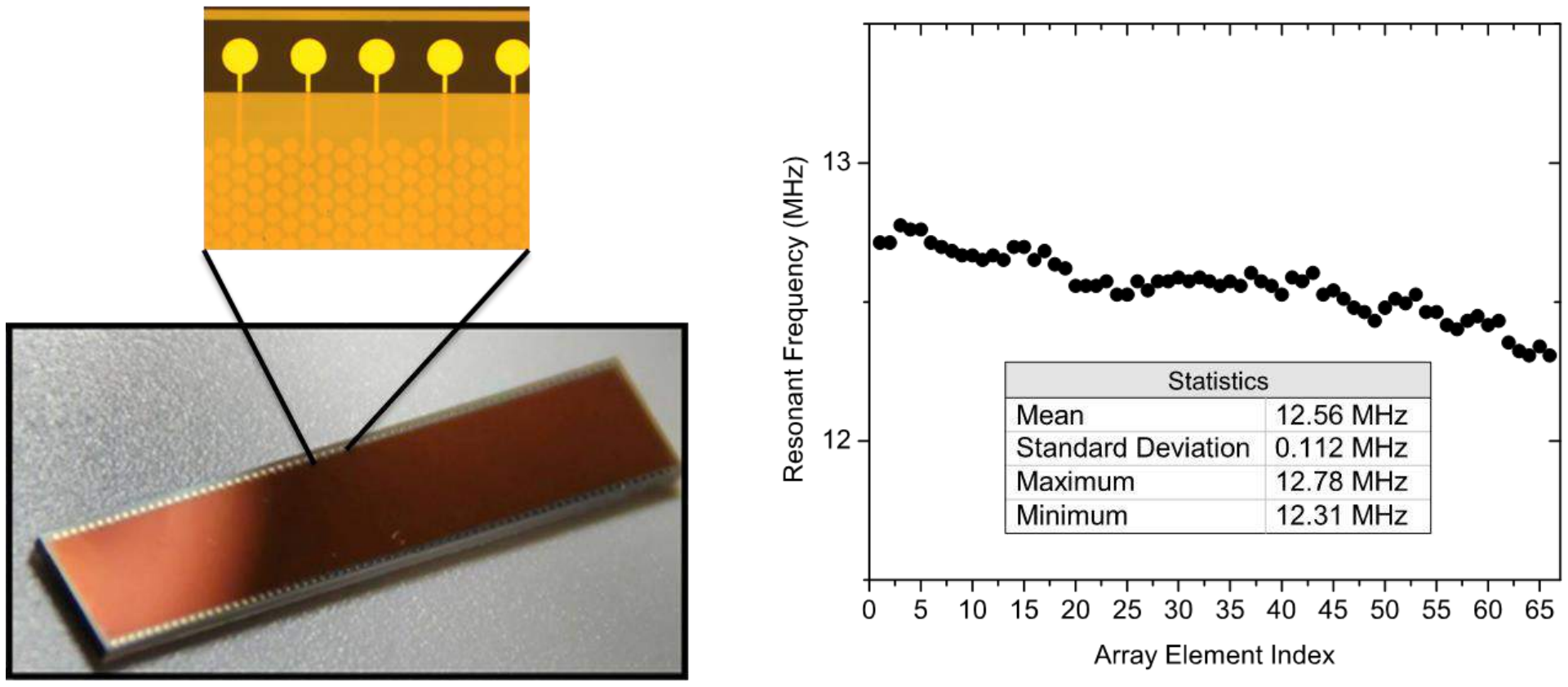

(a) (b)

**Figure 3-7: (a) A 66-element 1D CMUT array fabricated on a glass substrate using anodic bonding. (b) Array uniformity measurement.**

## 3.4 High-Frequency Broadband 1D CMUT array

As a special design for the platform process, we demonstrate a high-frequency (29-MHz) broadband (100% FBW) CMUT 1D array. The devices are fabricated using anodic bonding with only three photolithography steps. We also discuss the design guidelines for high-frequency broadband CMUTs using the simulations. A high fill factor and a thin plate are important for the broadband design. Small cell size is required for the increased center frequency. To improve the transducer sensitivity and to keep the collapse voltage low, the gap height should be small and a high-k dielectric insulation layer should be employed. The fabrication steps we report in this paper have good potential to meet the high-frequency broadband CMUT design requirements. So far we have demonstrated that we can define a 50-nm gap, bond to a post as narrow as 2 µm, and pattern a high-k dielectric layer on the bottom electrode.

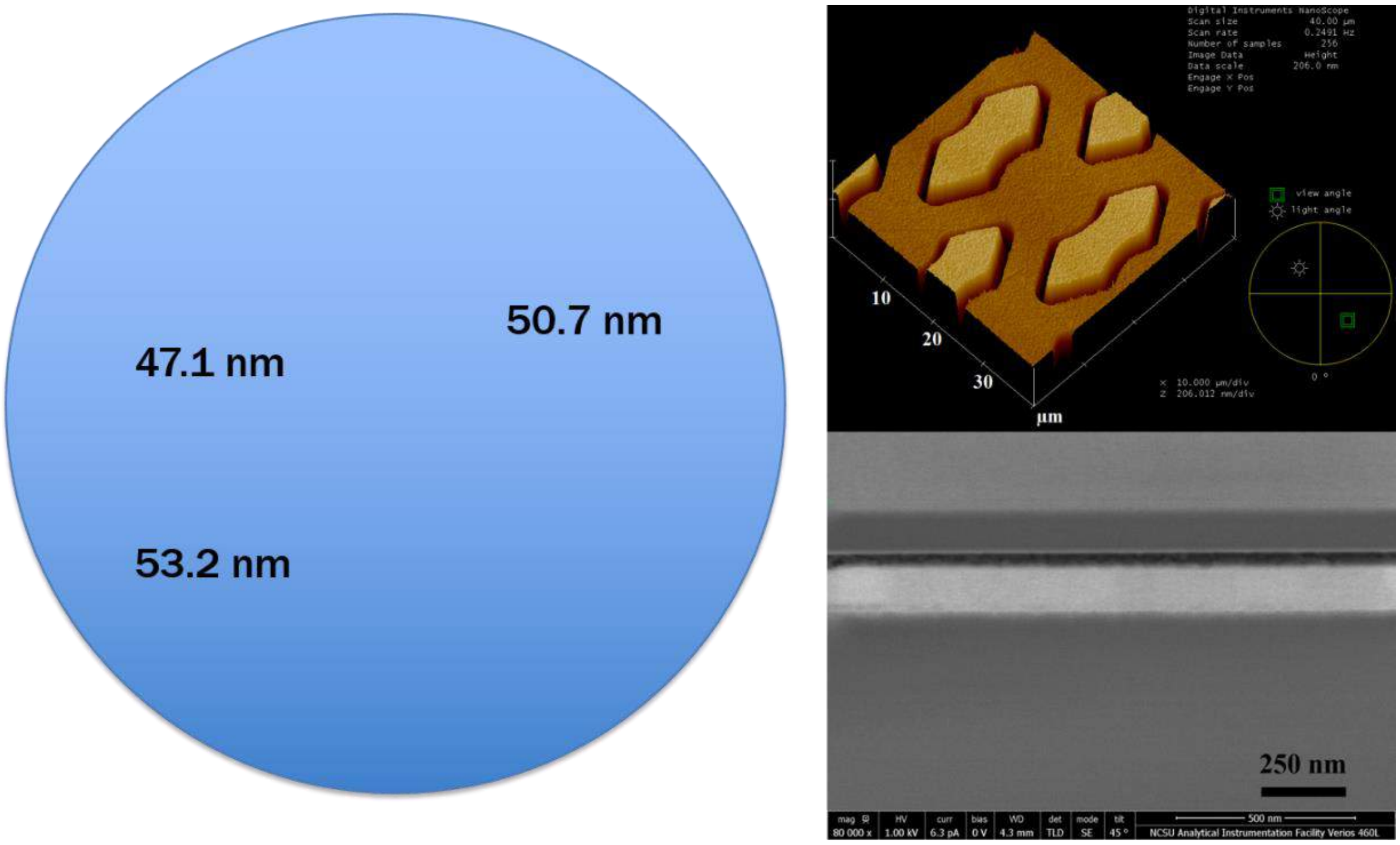

(a) (b)

**Figure 3-8: Gap height can be control as small as 50 nm through the fabrication process. (a) Different regions measured on the wafer. (b) AFM image showing 50-nm gap before bonding (top). SEM image showing 50-nm gap after bonding (bottom).**

The immersion test was performed in a tank filled with vegetable oil. A 1D array was diced and wire bonded to a chip carrier. The frequency response of the element in transmitting was measured using a hydrophone (Model HGL0200, Onda Corporation, Sunnyvale, CA) at a distance of 1.8 mm. The element was biased at 50-V DC and excited using a pulser/receiver (Model 5073PR, Olympus Corporation, Waltham, MA. Pulse repetition frequency (PRF): 200 Hz; Energy level: 2; Damping level: 1). The signal received by the hydrophone is shown in Figure 3-9(a) in the time domain. The corresponding frequency spectrum is shown in Figure 3-9(b). The frequency spectrum shows the transducer center frequency is 29 MHz and the 3-dB FBW is 100% after correcting for the pulse spectrum. The hydrophone calibration is available up to 40 MHz at the time. Therefore the spectrum is not corrected for the hydrophone response.

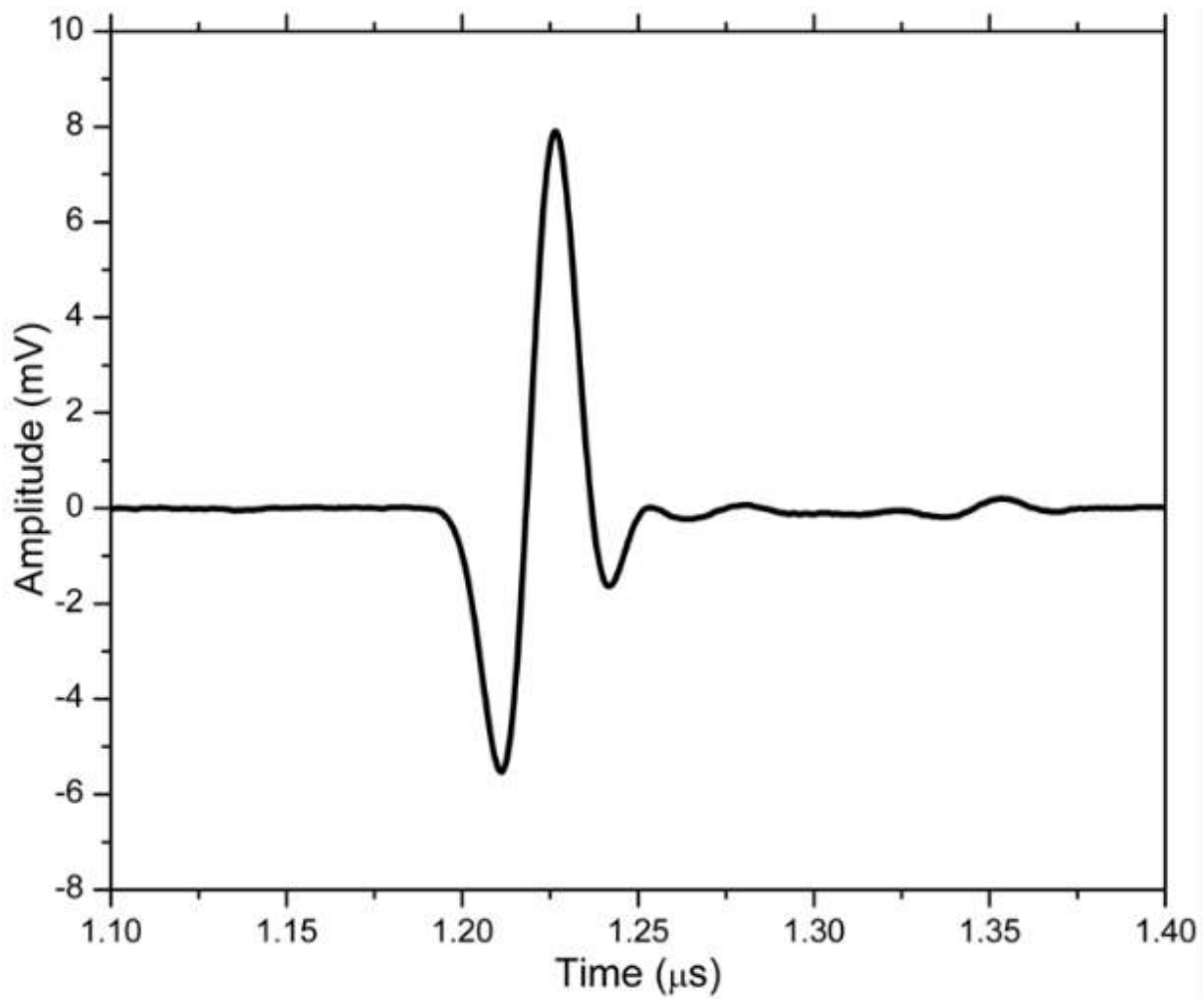

(a)

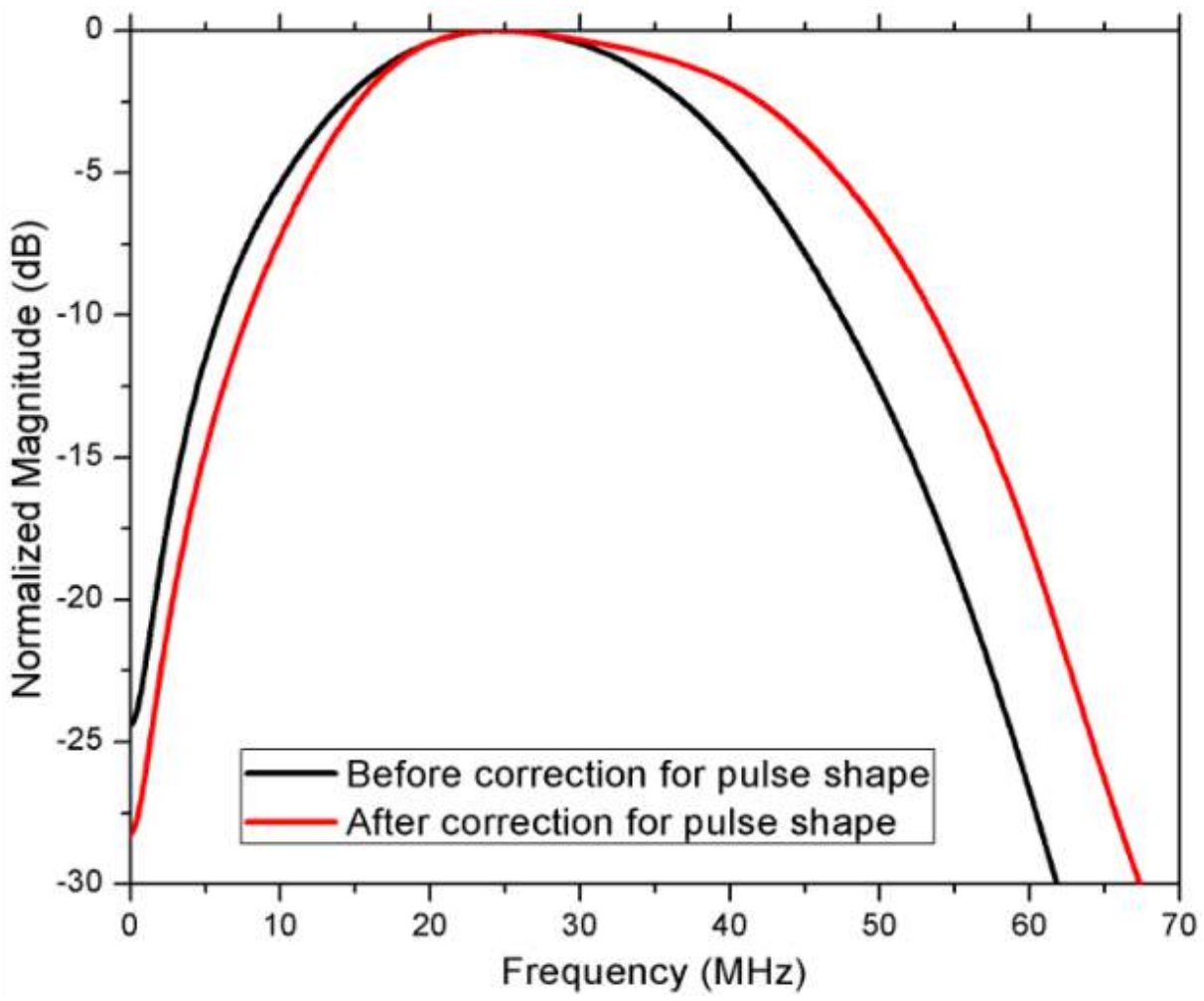

(b)

**Figure 3-9: (a) Experimental received signal by the hydrophone. (b) Corresponding frequency spectrum.**

## 3.5 Chapter Conclusions and Future Work

We have presented a fabrication process for CMUTs based on anodic bonding [54] [11]. This process offers the well-known advantages of wafer bonding such as good control over the thickness and mechanical properties of the plate and overall reduced process complexity. In addition to these general advantages of wafer bonding process, anodic bonding has the specific advantage of being more tolerant to roughness on the bonding surface. Furthermore, the narrower post structures are feasible with anodic bonding to maximize the fill factor, which is especially critical to achieve wide bandwidth at high frequencies. The maximum processing temperature in the presented approach is 350°C, which allows a patterned metal bottom electrode for reducing device series resistance. Use of the glass substrate helps reduce the parasitic capacitance and improves the dielectric reliability as the top plate and the bottom electrode mainly overlap on the active transducer area and not on the posts. Using the same process, we further demonstrated a high-frequency broadband 1D CMUT array using only three photolithography steps in the fabrication [55]. The fabricated CMUT was characterized in immersion. The CMUT shows 29-MHz center frequency and 100% FBW after correcting for pulse shape.

# Chapter 4 2D CMUT ARRAYS WITH TGV INTERCONNECTS

## 4.1 Introduction

Close integration of ultrasonic transducer arrays and front-end integrated circuits is critical for the overall ultrasound system efficiency and also a compact form factor. For 2D arrays and arrays used in ultrasound imaging catheters where the element area is small, the receiver electronics must be closely integrated with the transducer array to avoid additional parasitic capacitance introduced by the cables to preserve signal quality. CMUT technology has attracted a great deal of attention because of advantages such as ease of fabricating arrays and integrating them with supporting electronics, as well as wide bandwidth [14]. Two main methods have been developed for integrating CMUTs with supporting circuits: monolithic integration and hybrid integration [37]. For monolithic integration, one method is to co-process the CMUTs and the electronics side by side [56]. In this method, not only the substrate area is shared by CMUTs and electronic circuits, but also the properties and vertical dimensions of the layers used in the CMUT structure are limited by the CMOS process materials and film thicknesses. The other method is to fabricate the electronic circuit first and then build the CMUTs on top by post-processing [19], [44], [57], [58]. This method has a good area utilization and more dimensional control, but the CMUT process is still limited because of the temperature constraints set by the existing metal lines. The complexity of the overall process also increases.

## 4.2 Through-Wafer Interconnects

### 4.2.1 2D CMUT with TSV interconnects

Hybrid integration allows the optimization of CMUTs and electronics independently. In hybrid integration, through-wafer interconnects are established as part of CMUT arrays to enable a close connection with supporting electronics by chip-to-chip bonding or through an intermediate layer such as an interposer or a flex circuit [15], [59]. The current implementation of this hybrid integration approach is based on using through-silicon-via [46]. Since silicon is a semiconductor, a reverse biased PN junction or a metal-insulator-semiconductor (MIS) structure has to be used to isolate individual connections to each element in an array and to reduce the parasitic capacitance [60]. The implementation of an isolation structure complicates the fabrication process. Using polysilicon as a via and electrode material degrades the surface quality and wafer flatness, affecting subsequent processing steps. Last, TSVs can only be used for surface micromachining because the complicated process of making TSV degrades the wafer surface quality and make it not suitable for wafer bonding.

### 4.2.2 2D CMUT with Trench-Isolated Interconnects

An alternative interconnection method used for CMUTs is to create deep isolation trenches in a highly conducting silicon substrate, where the resulting pillar underneath each element serves as the interconnect from the front to the backside of the wafer [6]. This process is also complicated and creates reliability problems due to the exposed gaps between top and bottom electrodes of the CMUT array elements on the top surface.

### 4.2.3 2D CMUT with Through-Glass-Via Interconnects

Through-glass-vias (TGVs) have been used for advanced electronic packaging, especially for RF applications where parasitics are critical [61]. Anodic bonding is widely used for wafer-level hermetic MEMS packaging [62]. In recent years, fabrication of CMUTs on a glass substrate has aroused significant interest [10], [63], [64]. We have recently reported a process for fabricating vacuum-sealed CMUTs on a borosilicate glass substrate using anodic bonding [11]. This process benefits from all the advantages of wafer bonding, including process simplicity, control over plate thickness and properties, high fill factor, and ability to implement large vibrating cells. Additionally, it reduces the parasitic capacitance and series resistance benefiting from an insulating substrate and a metal bottom electrode, respectively.

The targeted implementation is illustrated in Figure 4-1, where the CMUTs are fabricated on a TGV substrate and directly flip-chip bonded to an integrated circuit (IC). The vibrating plate consists of a thin single-crystal silicon layer embedded between a silicon nitride insulation layer at the bottom and a metal electrode on top. The bottom electrodes are formed in the etched glass cavities and connected to the dedicated TGV interconnects giving each element electrical access from the backside. The metal layer deposited on top of the silicon plate is common to all elements in an array and is connected to a TGV for backside access. A stacked metal layer is deposited and patterned on the backside to form the bond pads.

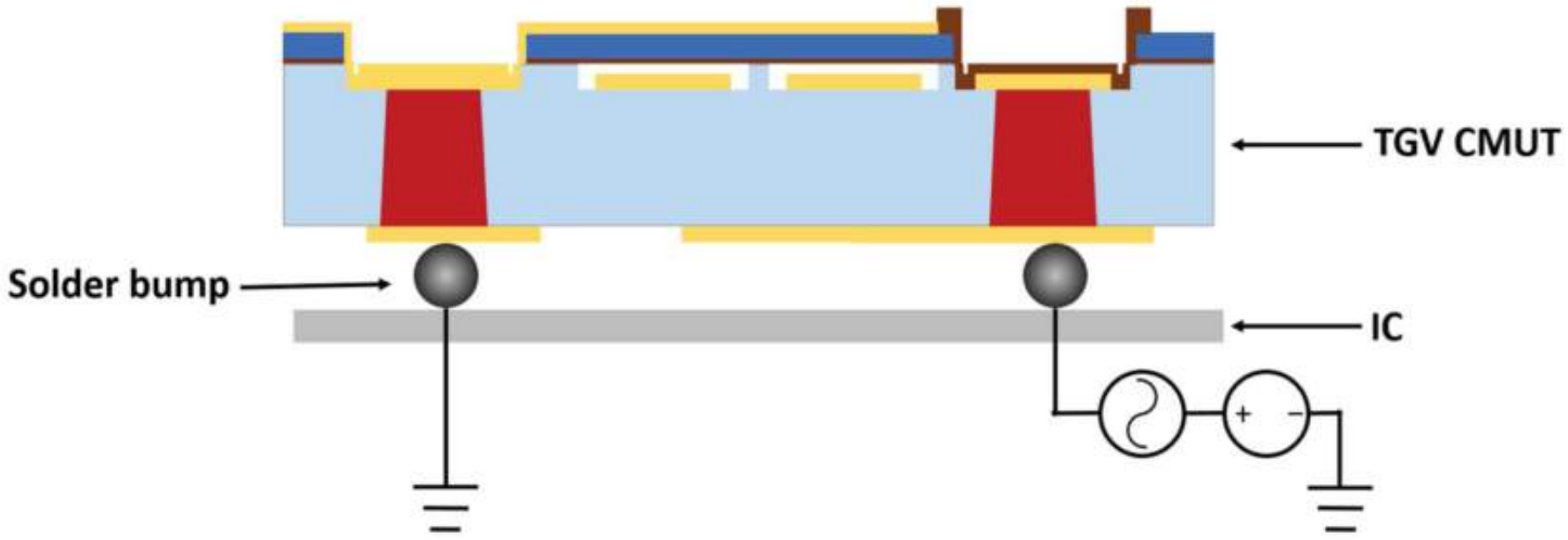

**Figure 4-1: A schematic cross-section of a completed CMUT element with TGV interconnects flip-chip bonded on an IC.**

# 4.3 Fabrication of 2D CMUT with TGV Interconnect

## 4.3.1 Formation of TGV Substrate

Five masks are used for the fabrication process presented in this section. The via on the left in the cross-sectional drawings and the via in the back in the 3D drawings are for the top electrode connection. The via on the right in the cross-sectional drawings and the via on the front in the 3D drawings are for the bottom electrode connection.

We use the same starting substrate and SOI wafer as in the platform process described in the last chapter. The vias are designed on the 100-mm wafer area for the both bottom and top electrode connections for different array geometries. The alignment marks are also formed by TGVs for aligning the subsequent layers to the via pattern. Through-wafer channels are created first in the 0.7-mm borosilicate glass substrate at the designed via locations by laser drilling (Figure 4-3). The through channels are then metalized by using the conductive copper paste technology (Triton Microtechnologies, Oro Valley, AZ) [65] (Figure 4-4). A benefit of this approach is that the paste has a thermal coefficient of expansion (TCE) matched to that of the glass substrate. Therefore, potential mechanical stress that can be caused by further heating steps and possible cracks around the vias can be minimized. Then, the wafer is sintered and polished. By selecting the appropriate coating material on the substrate before polishing and using a slurry loaded with composite particles as abrasive during the chemical-mechanical polishing (CMP) process, a smooth glass surface and a good copper-to-glass surface co-planarity are obtained [66], [67]. For a completed TGV wafer, the via location is in the range of ±5 µm from the designed location in both X and Y directions, which is the accuracy of the laser drilling. The diameter of the via is 70 µm at the laser entry side and 50 µm at the laser exit side. We build the CMUTs on the laseFr exit side and use the laser entry side for the backside pad formation. With 33207 vias formed on the 100-mm wafer area, the average wafer warp increased from ∼10 µm to ∼50 µm. Figure 4-5 shows the SEM cross-sectional image of a TGV embedded in the glass substrate and AFM image of the via surface on the laser exit side. At the via-glass boundary on the surface, copper and glass are at the same level in most regions around each via, which makes a reliable electrical connection between the CMUT electrodes and the TGVs possible.

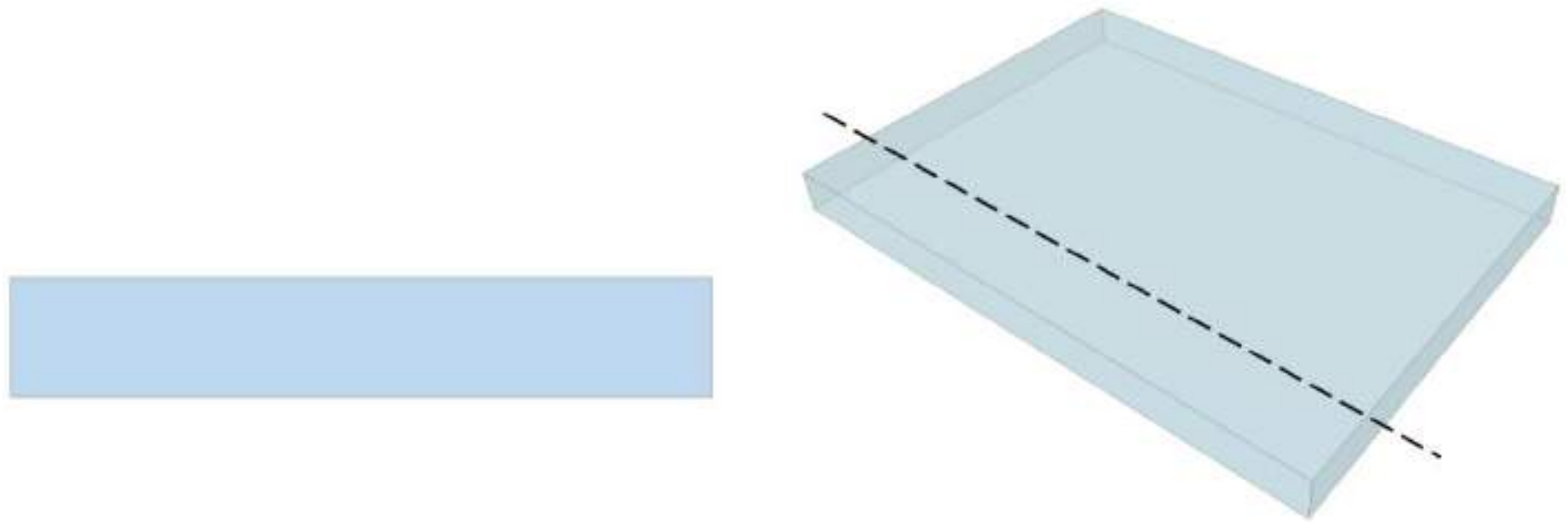

**Figure 4-2: Initial borosilicate substrate.**

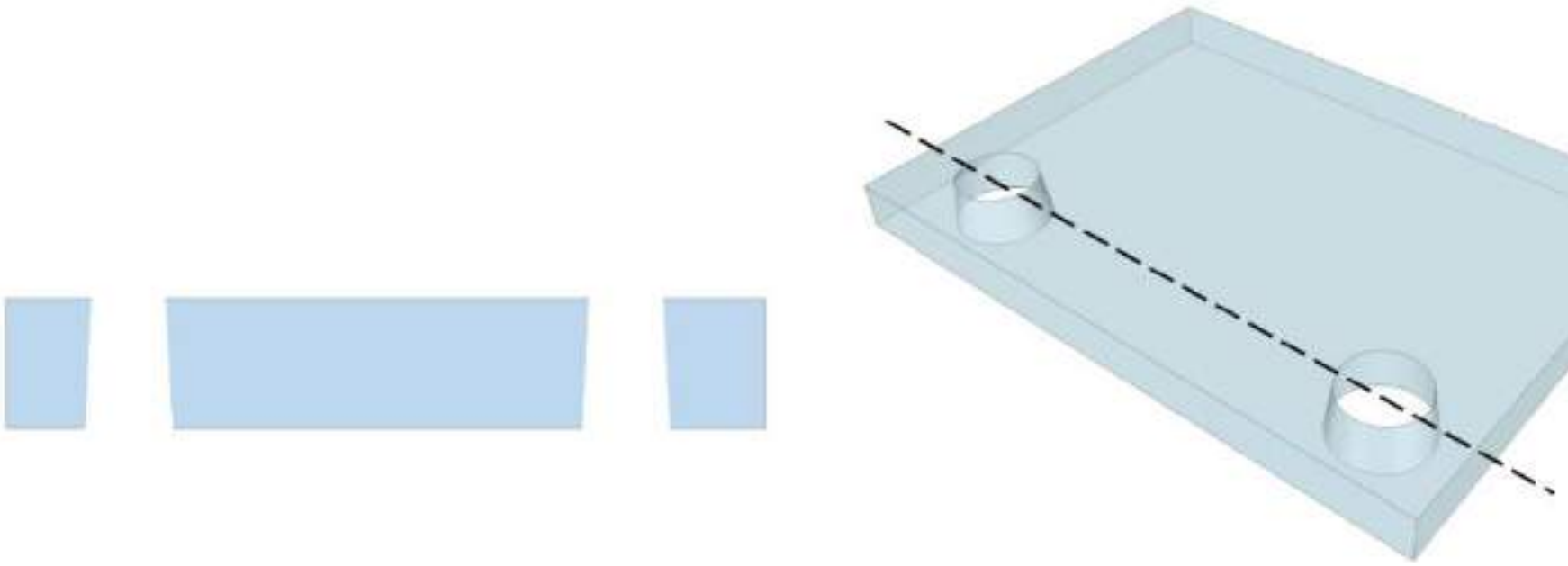

**Figure 4-3: Define patterned through-wafer holes by laser drilling.**

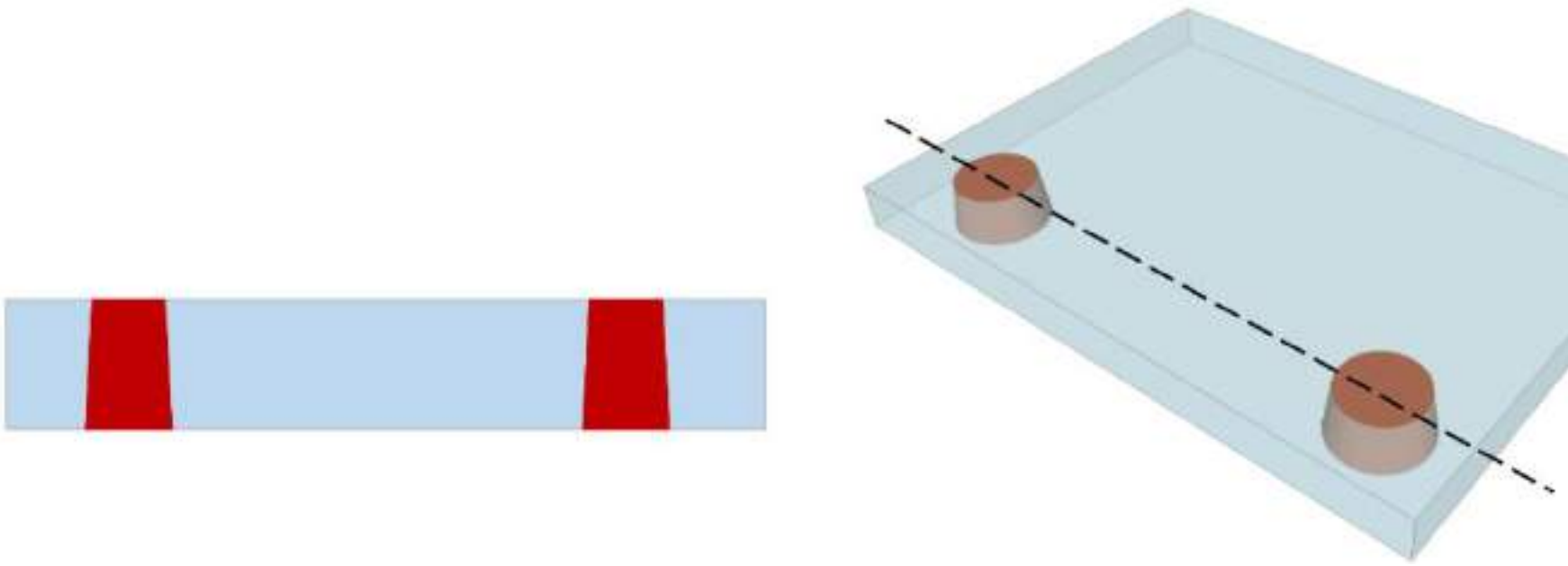

**Figure 4-4: Fill the through holes with copper paste.**

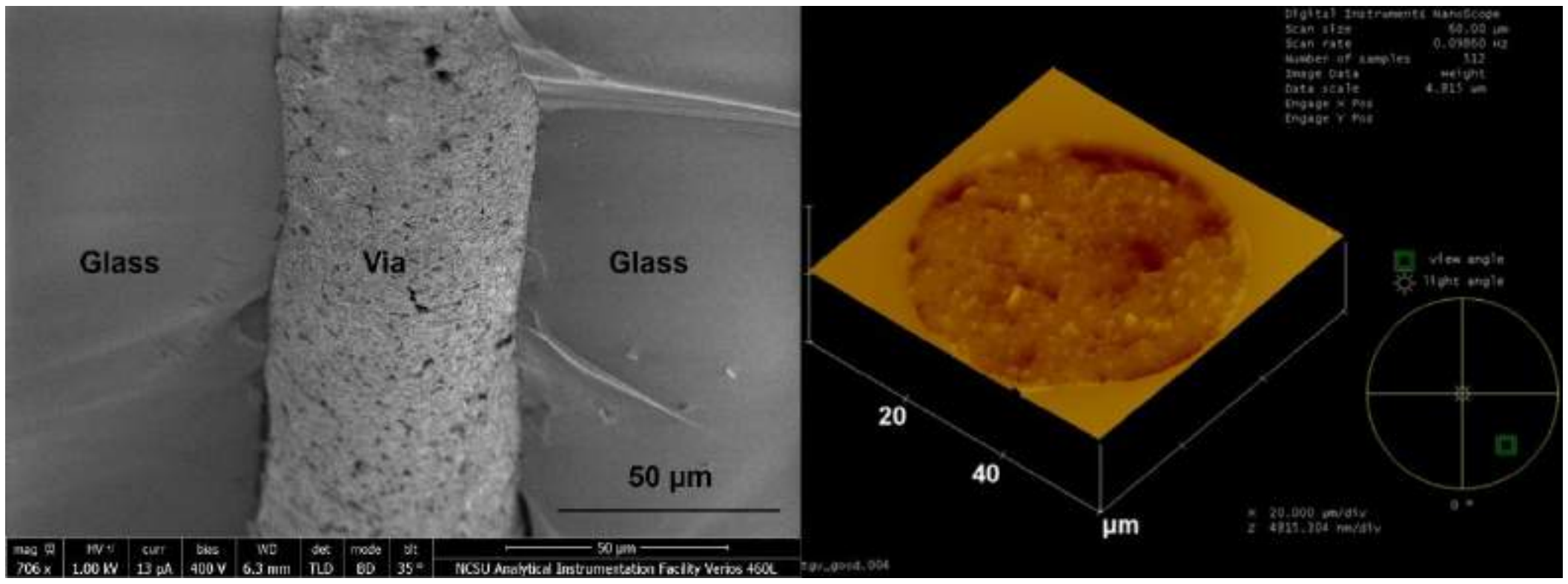

**Figure 4-5: SEM cross-section of the TGV embedded in the glass substrate (left). AFM image of the TGV surface profile (right).**

### 4.3.2 Forming Cavities and Bottom Electrodes Connected to the TGVs

The TGV substrate was cleaned using a heated NMP@70◦C solution to remove organics and contaminants from the surface. Acid-based cleaning solutions cannot be used because of the copper in the via. After cleaning, the cavities were patterned using a negative photoresist with the alignment of each element to the dedicated TGV. We used RIE to etch glass instead of buffered oxide etchant (BOE) in order to avoid peeling of the photoresist and not to damage copper vias. Dry etching also helps achieve a deeper cavity, which is desired for a thicker bottom electrode to build a reliable connection to the TGV. We performed RIE with SF6 gas to realize a glass cavity depth of 320 nm (Figure 4-6). After removing the photoresist, the bottom electrode was patterned by a second photolithography step using a 2-µm-thick negative photoresist (AZ-5214E IR, Clariant, Wiesbaden, Germany), which is suitable for lift-off. Prior to the bottom metal deposition, we wet etched the copper vias also 320 nm using a copper etchant (Copper Etchant 49-1, Transene Company, Inc., Danvers, MA) in order to keep the glass surface at the bottom of the cavity level with the copper via surface. This selective wet etchant does not attack the photoresist and helps remove the copper oxide on the via surface and consequently improve the electrical connectivity between the bottom electrode and the TGV. Then a stacked metal that consists of 20-nm chromium as an adhesion layer and 130-nm gold was deposited into the cavity by evaporation and defined by lift-off as the bottom electrode to obtain a 170-nm gap height and to build the electrical connection from the TGV to the bottom electrode (Figure 4-7). Figure 4-8 shows the top view of the processed substrate that is ready for anodic bonding.

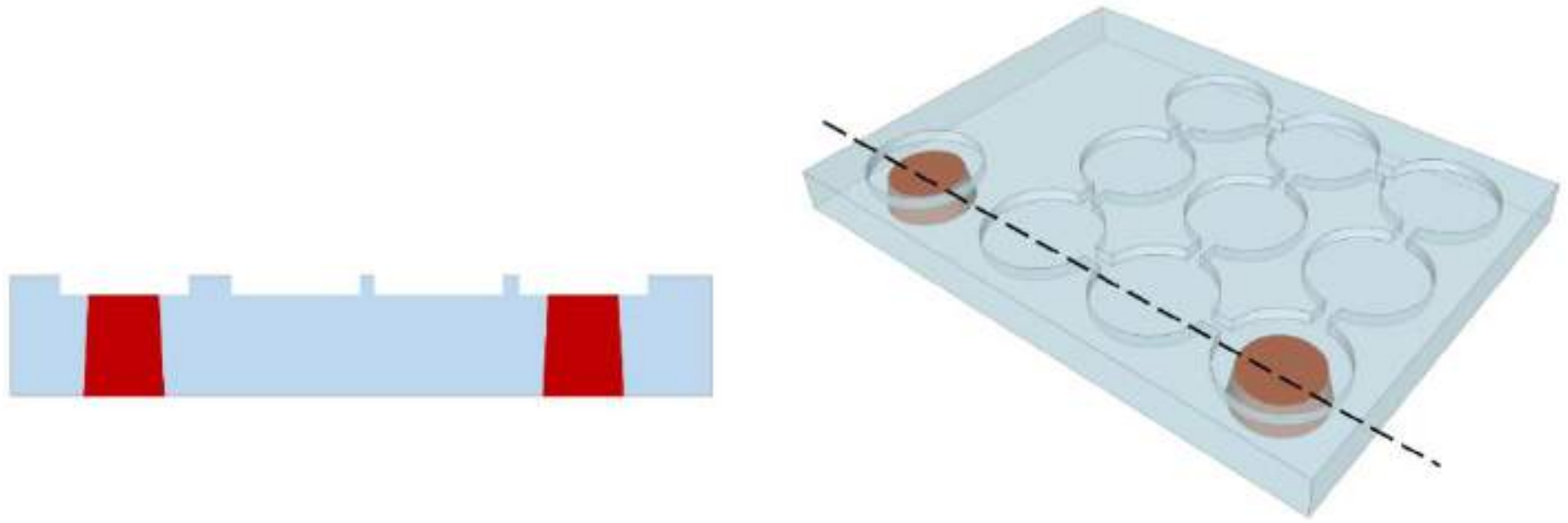

**Figure 4-6: Glass and copper etch to define CMUT cavities.**

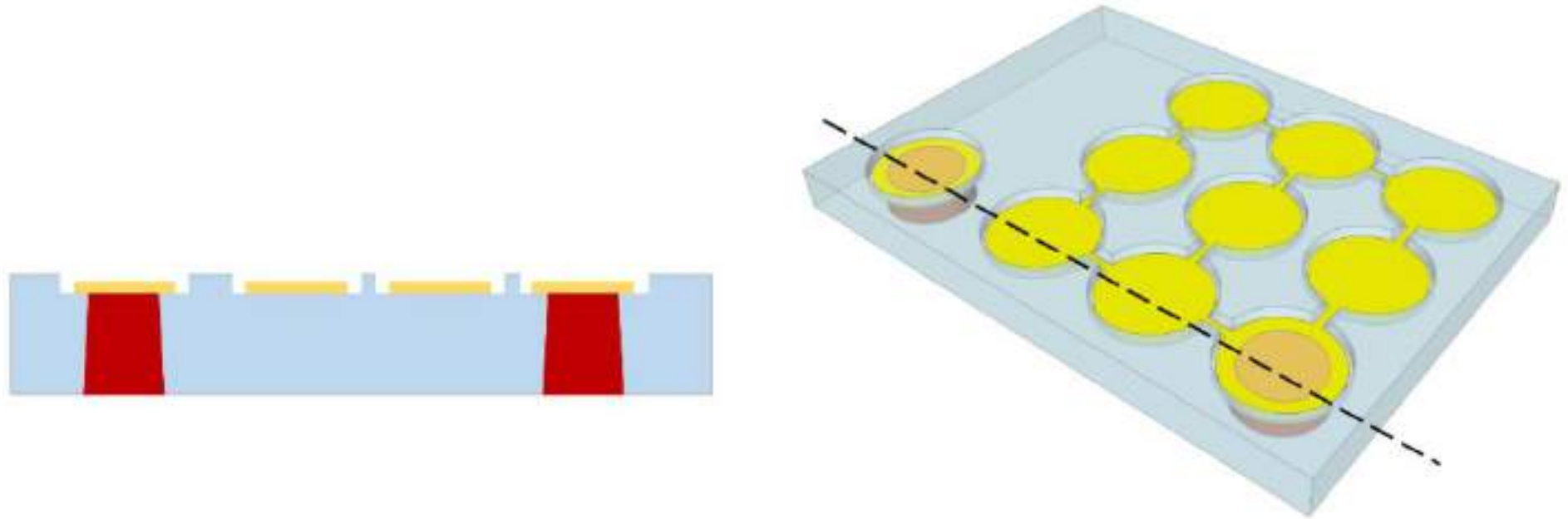

**Figure 4-7: Evaporation and lift-off the bottom electrode and forming an electrical connection to the dedicated TGV.**

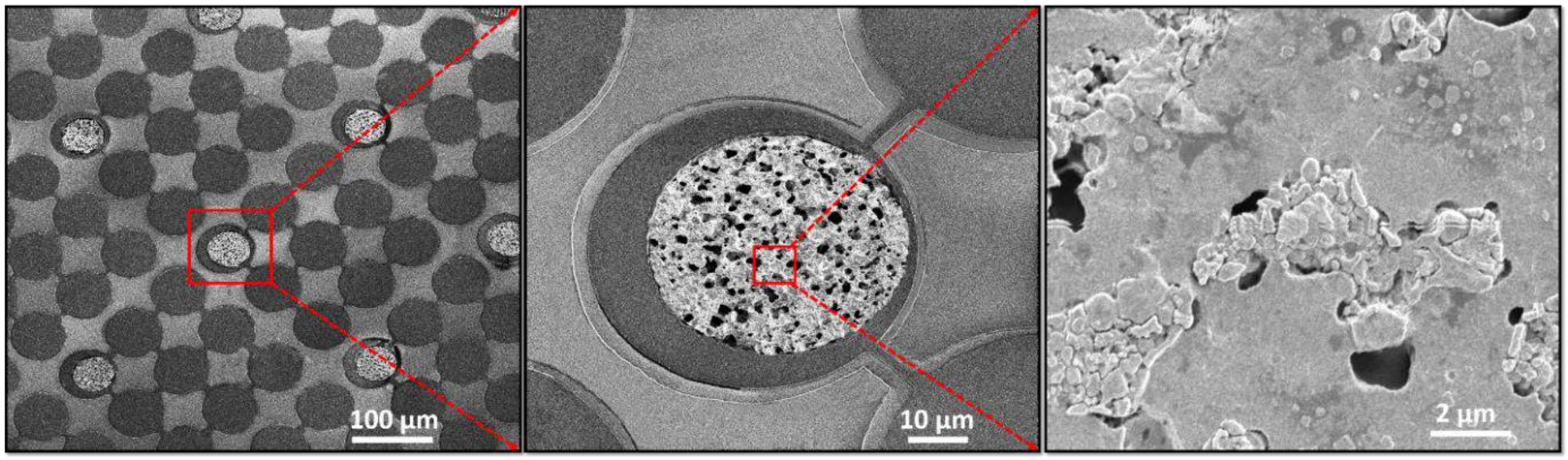

**Figure 4-8: SEM images of the completed glass substrate with TGVs that is ready for anodic bonding.**

### 4.3.3 Anodic Bonding

A 200-nm low-stress silicon nitride insulation layer was deposited on the device layer of the SOI wafer by plasma-enhanced chemical vapor deposition (PECVD) prior to bonding. The TGV wafer was cleaned with a solvent-based solution and the SOI wafer was cleaned using Piranha solution. The borosilicate glass surface and the nitride insulation layer surface were brought together in vacuum ($10^{-4}$ Torr) and then bonded at 350◦C under 2.5-kN down force in a semi-automatic bonding system (Model EVG510, EVG Group, St. Florian, Austria) (Figure 4-9). It is important to keep the wafer in vacuum before increasing the temperature to prevent copper oxidation. The handle layer of the SOI wafer was then ground down to 100 µm. At this step, a protection layer (ProTEK B3 alkaline protective coating, Brewer Science, Rolla, MO) was coated and cured on the backside in order to protect the vias during the handle wafer removal process. This protection layer also prevents any liquid from flowing into the bonded cavities through the via locations in case there is a leakage. A heated TMAH solution was used to selectively etch the remaining handle layer over the BOX layer. The silicon plate was released after removing the BOX layer in 10:1 BOE solution (Figure 4-10).

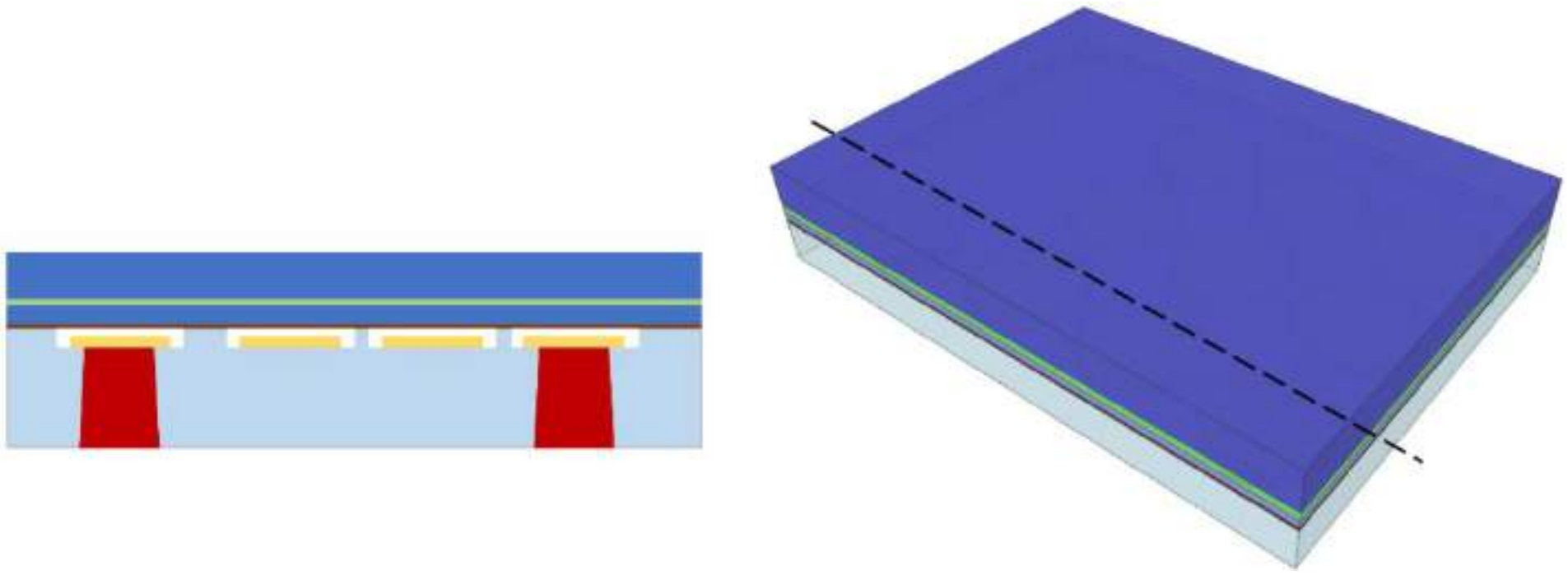

**Figure 4-9: Deposition of the silicon nitride insulation layer on SOI device layer. Anodic bonding in vacuum.**

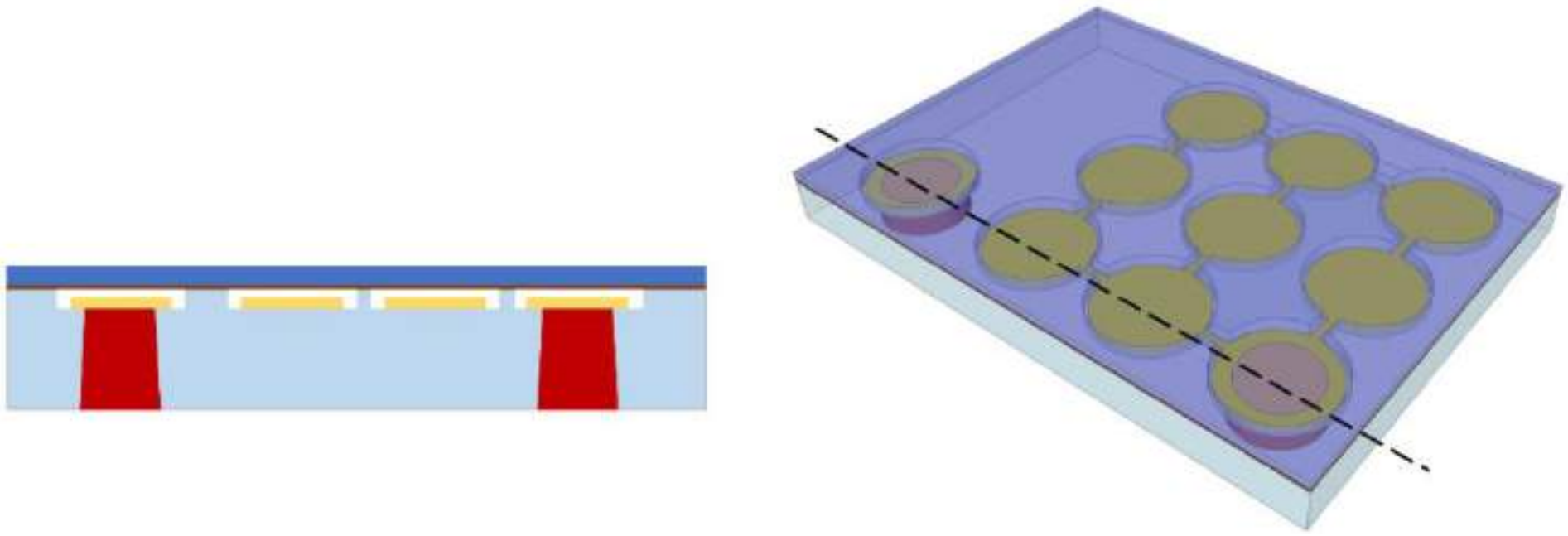

**Figure 4-10: Handle wafer and BOX layer removal.**

### 4.3.4 Reaching Vias and Sealing

We performed a third photolithography step in order to evacuate the gas generated during anodic bonding, and also to reach the vias for top electrode connection. The lines to guide the final dicing of the arrays were defined at the same time. For the gas evacuation, we chose to open the plate over the TGV for bottom metal connection (Figure 4-11). In this way, we could later seal this location so that the via is mechanically isolated from the active area but electrically connected to the bottom electrode. This improves the reliability of the process because CMUT operation will be independent of the TGV condition in case of failure of hermetic sealing around the via, which would break the CMUT vacuum and degrade the CMUT performance. This was observed in the first-generation devices, mainly because of some micro-cracking around the via boundary [68]. Prior to sealing, the photoresist on the front side was removed by oxygen plasma and the protection layer on the backside was removed by RIE using CF4 gas. The wafer was then sealed under vacuum with 1-μm conformal PECVD silicon nitride at 1000-mTorr chamber pressure and 350◦C temperature (Figure 4-12). In order to avoid copper oxidation, the wafer needs to stay in vacuum when chamber temperature is above 100◦C.

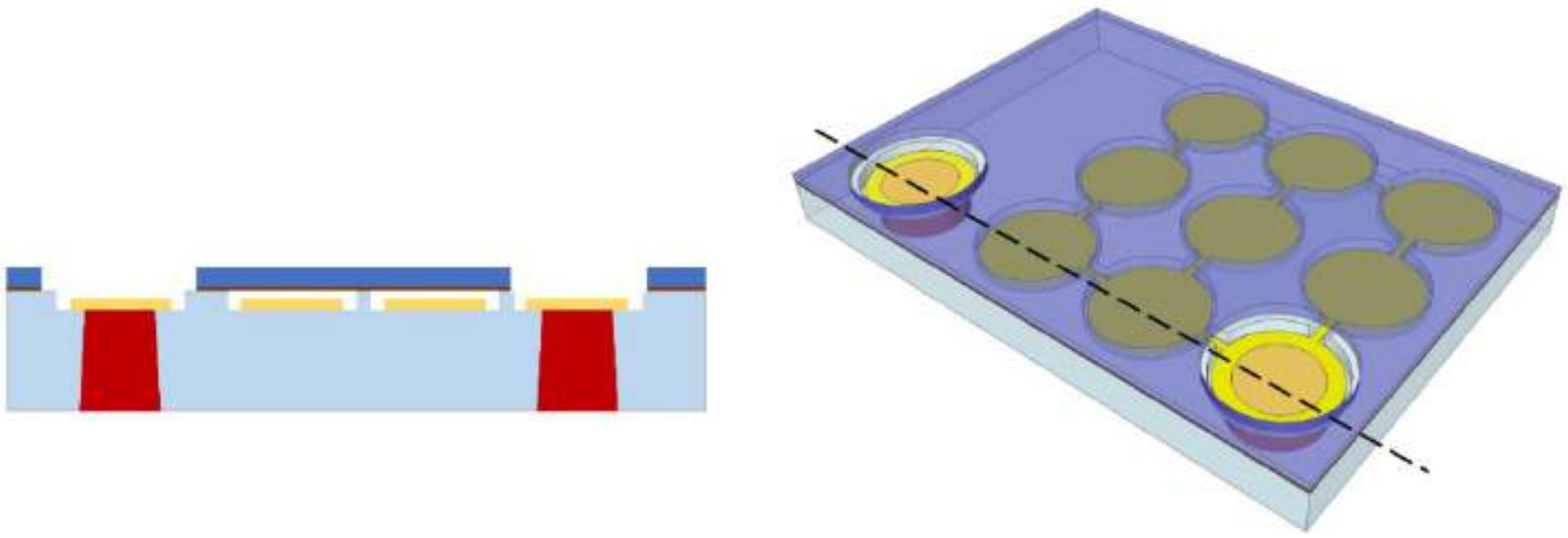

**Figure 4-11: Silicon/silicon nitride etch for gas evacuation, array separation, and reaching top electrode connection.**

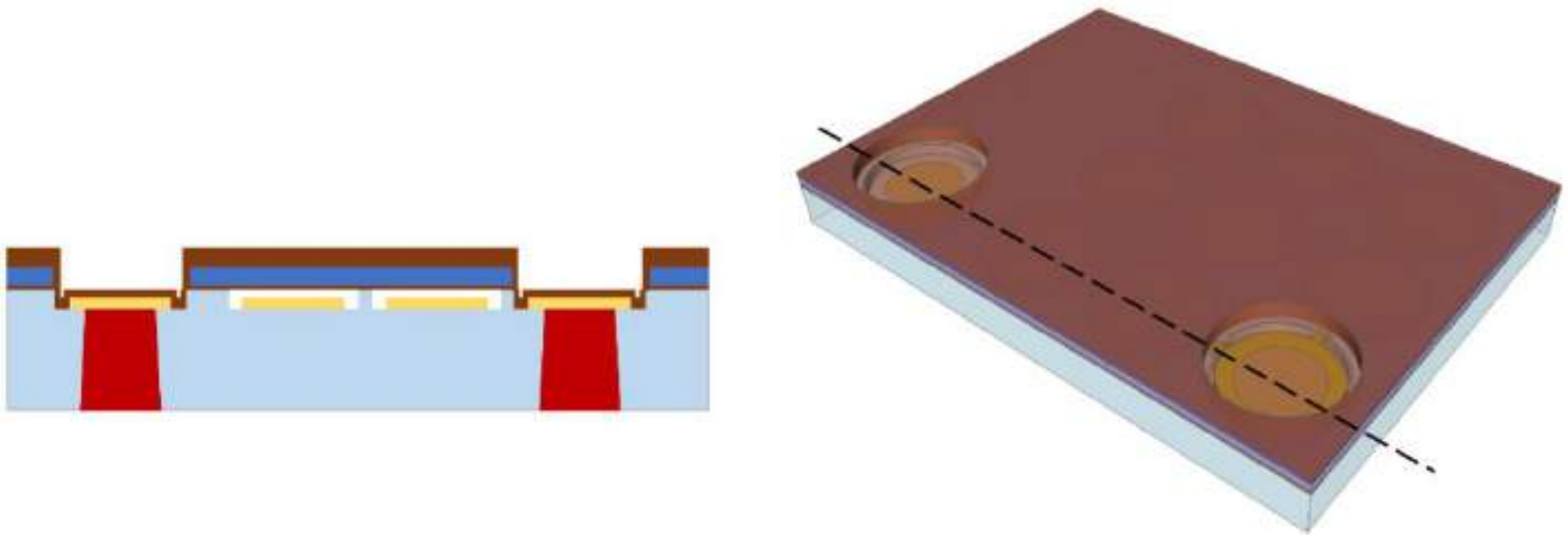

**Figure 4-12: Conformal PECVD silicon nitride deposition for sealing.**

### 4.3.5 Forming Electrical Contacts

To create the top electrode connections, the silicon nitride layer on the top electrode connection vias had to be removed. At this point, the silicon nitride deposited on the conductive silicon plate was also removed and the sealing layer was only left covering the cell that encloses bottom electrode TGV (Figure 4-13). After removing the negative photoresist using oxygen plasma, 20-nm chromium and 180-nm gold were deposited by dc sputtering over the entire wafer. The silicon nitride etching mask was used again, this time with positive photoresist to remove the metal on the sealing region in order to eliminate the parasitic capacitance at this location (Figure 4-14). The last photolithography was done on the backside to define the bottom electrode pads and a rectangular grid of top electrode connections to facilitate probing of elements. 20-nm chromium and 180-nm gold were deposited and then lifted off on the backside (Figure 4-15). At this step, the device fabrication was completed. Figure 4-16 shows completed device. On the left is the SEM cross-section of a completed CMUT cell with the dedicated TGV for bottom electrode connection. On the right are the optical image of the front side and backside of the fabricated 16×16-element 2D CMUT array. The physical parameters of the fabricated transducer elements are shown in Table 4-1

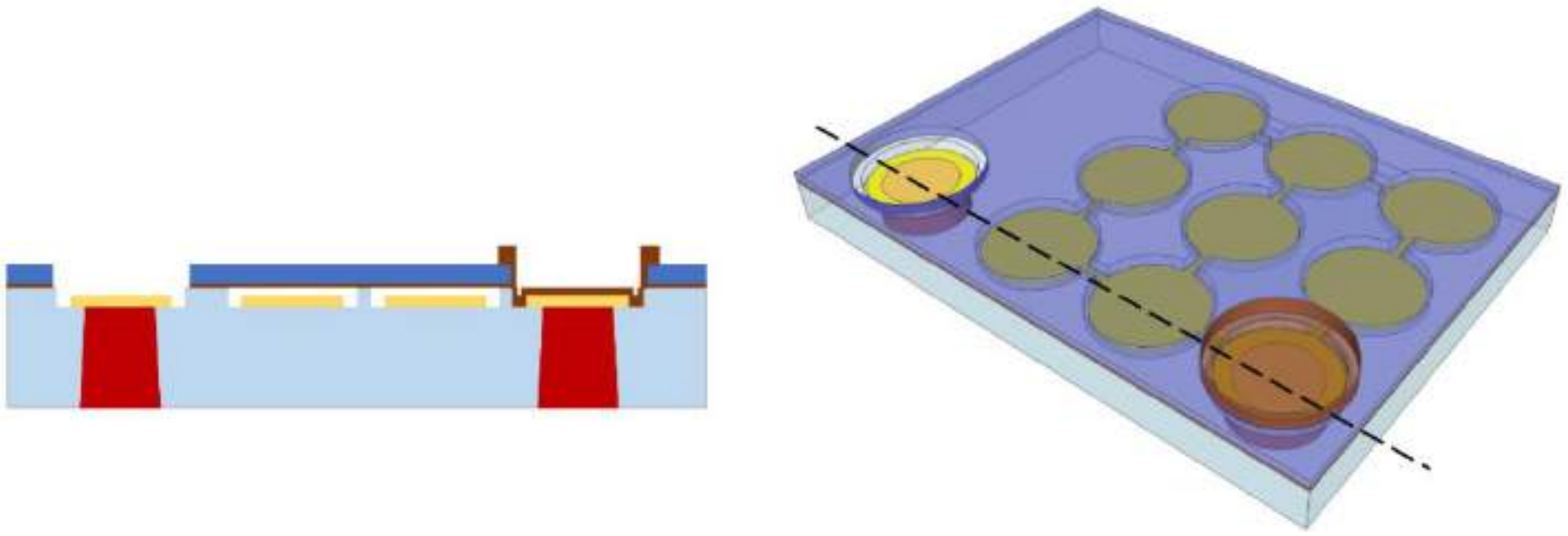

**Figure 4-13: Silicon nitride etch to reach the plate and top electrode TGV.**

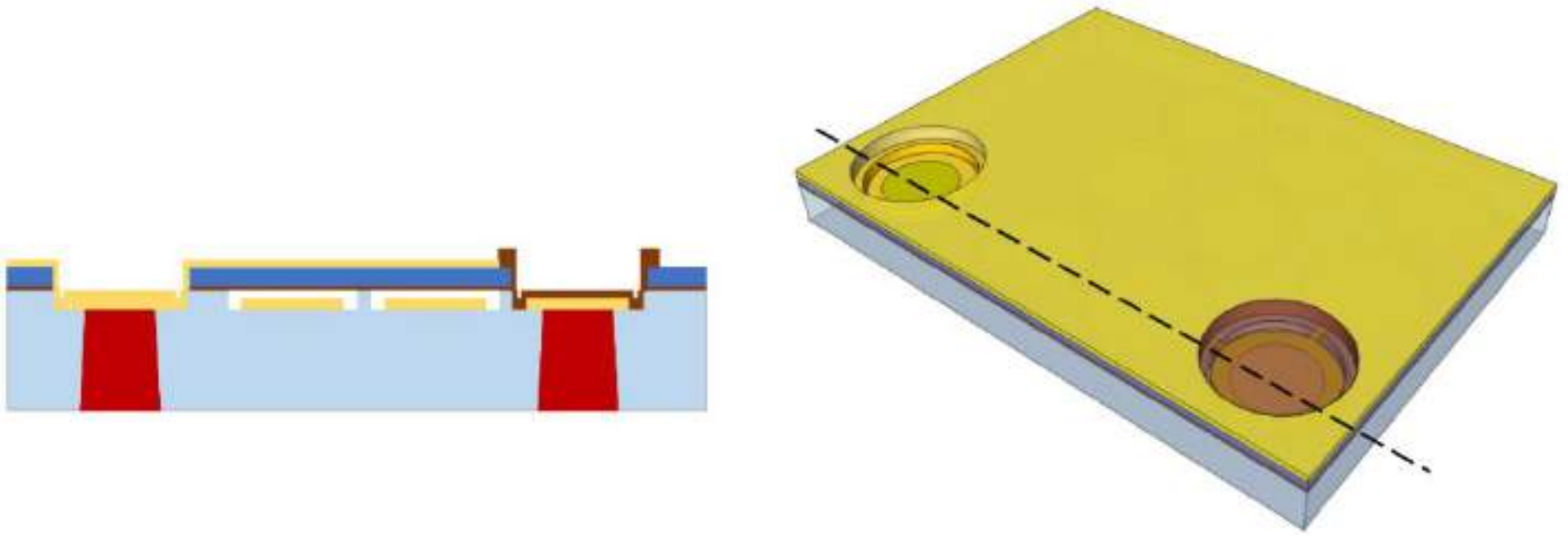

**Figure 4-14: Sputter top electrode and build a connection to dedicated TGV.**

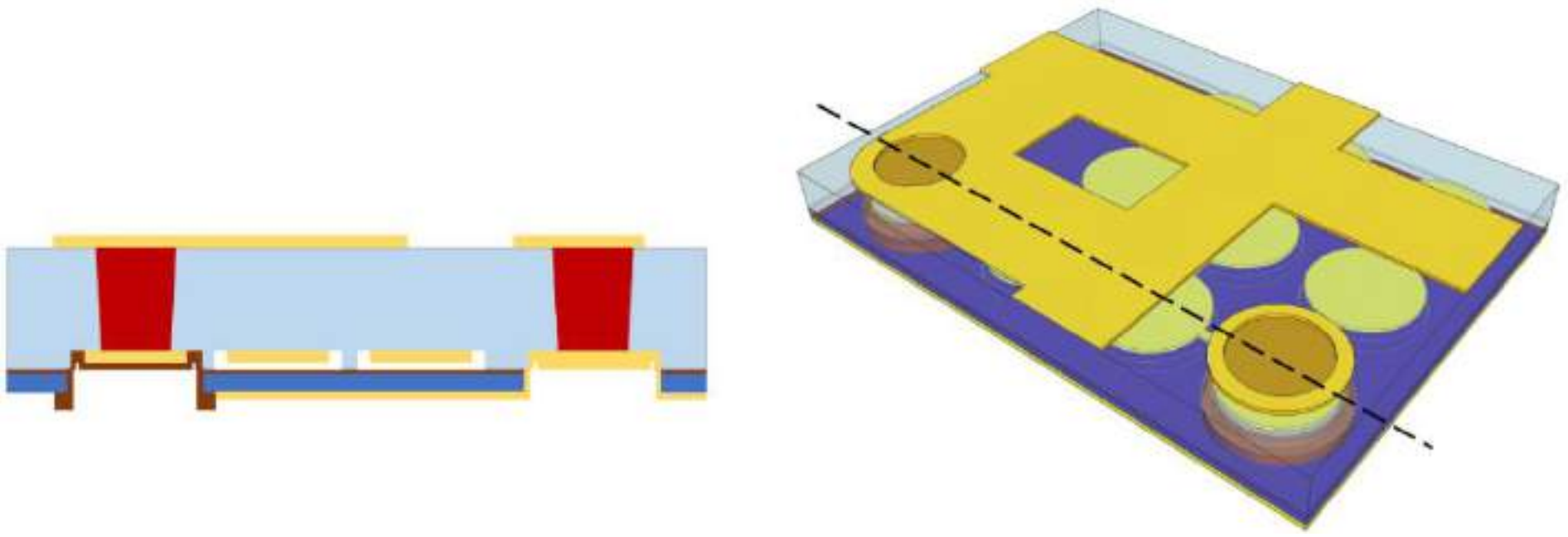

**Figure 4-15: Evaporation and lift-off for backside contact pad and testing grid.**

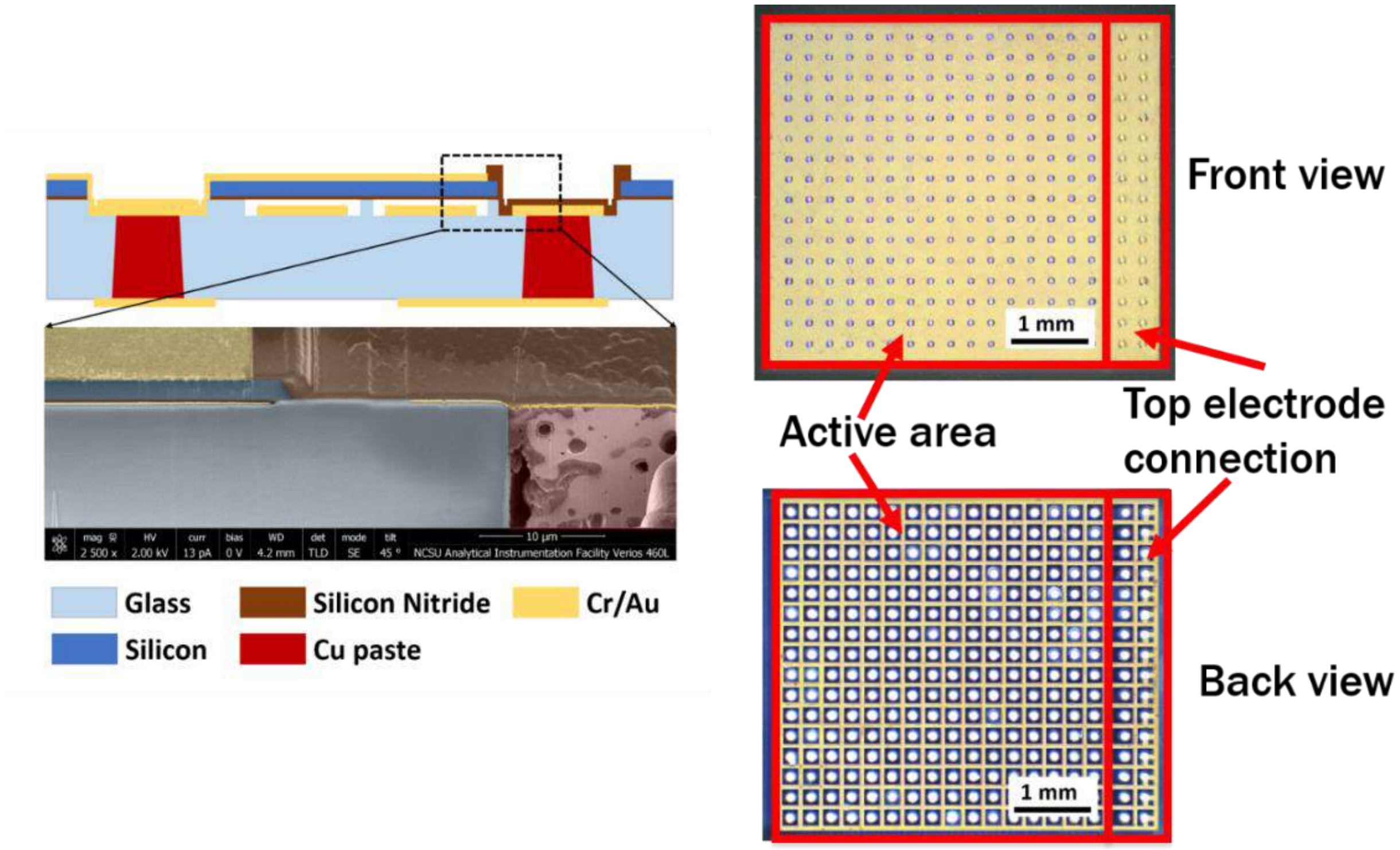

**Figure 4-16: Completed device. SEM (Left). Optical photos (Right)**

**Table 4-1: The physical dimensions of the fabricated 2D CMUT array element.**

| | |
|---|---|
| Shape of the cell | Circular |
| Cell width | 78 µm |
| Cell-to-cell distance | 4 µm |
| Top metal thickness | 0.2 µm |
| Silicon layer thickness | 1.5 µm |
| Insulating layer thickness | 0.2 µm |
| Gap height | 0.17 µm |
| Bottom metal thickness | 0.15 µm |
| Substrate thickness | 700 µm |
| Number of cells per element | 8 |
| Length of an element | 243 |
| Width of an element | 243 |
| Element pitch | 250 |

## 4.4 Device Characterization

### 4.4.1 Static Deflection Under Atmospheric Pressure

One of the main features of the 2D CMUT with TGVs is that a reliable vacuum sealing of the active CMUT cells could be achieved by mechanically isolating the through-glass via from the active CMUT cells. The achievement of vacuum sealing can be confirmed by measuring the plate deflection under atmospheric pressure. We used a stylus surface profilometer (Dektak 150, Veeco Instruments Inc, Plainview, NY) and measured a maximum deflection of 80 nm in the center of a circular CMUT cell (Figure 4-17). The finite element model (ANSYS v.15, ANSYS, Inc., Canonsburg, PA) predicts that the atmospheric deflection is 78 nm, confirms the sealing, and also proves that there is no significant stress generated on the plate during the fabrication.

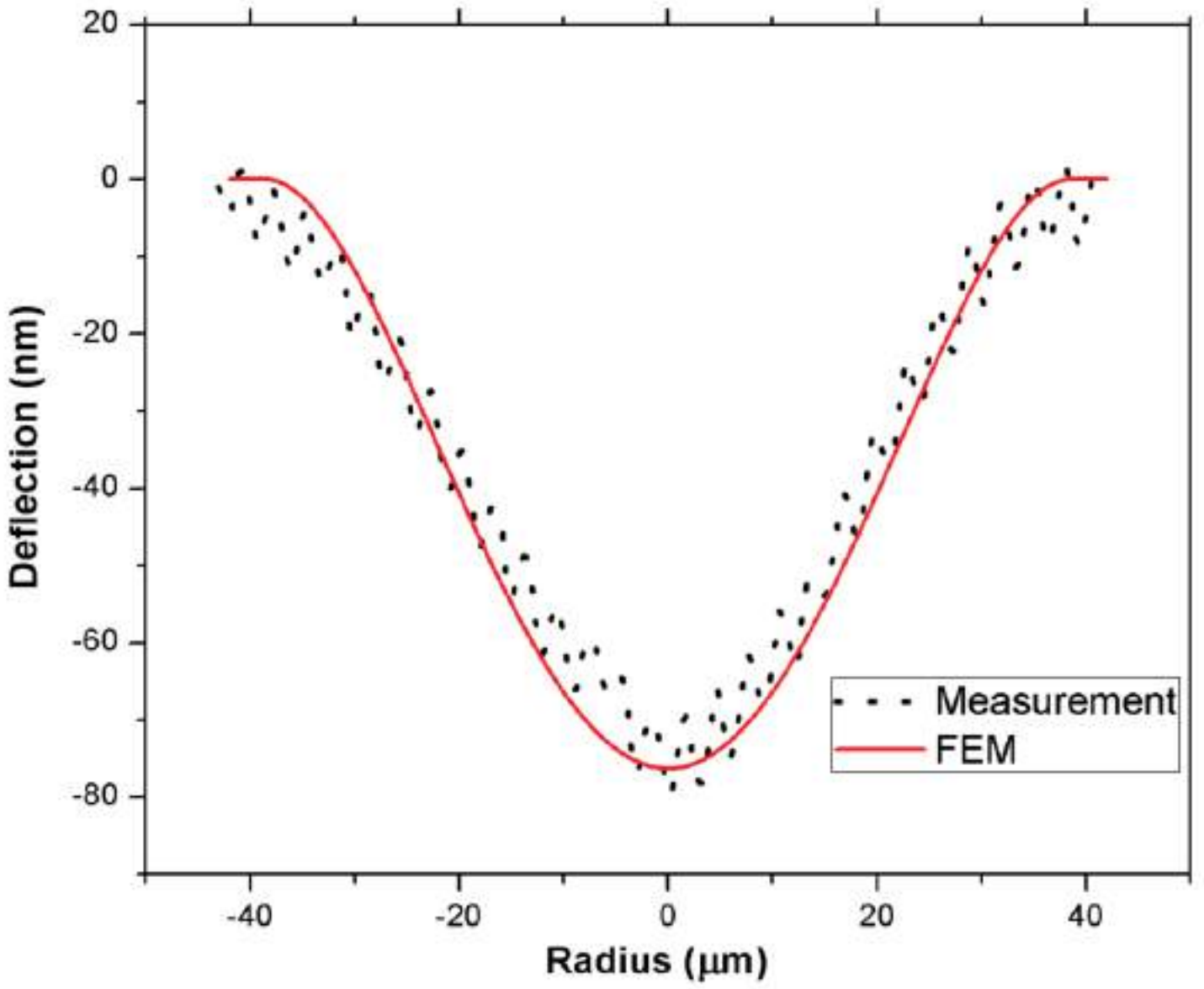

**Figure 4-17: Measured and simulated deflection profile of the plate after the device fabrication was completed.**

### 4.4.2 Parasitic Resistance and Capacitance of the TGVs

One of the main motivations for using copper through-glass vias for interconnection is the reduced parasitic resistance and parasitic capacitance. We designed via test structures with different pitch (125 µm, 190 µm, and 250 µm) to characterize the via resistance and via-to-via capacitance as shown in the left panel of Figure 4-18. Both resistance and capacitance measurements were performed by accessing the vias from the backside of the wafer after the backside metal pad formation. The resistance test vias are connected on the front side by a metal layer that is formed at the step of bottom metal deposition. The measurement setups are shown in the right panel of Figure 4-18. The via resistance test structures include two vias connected with a metal line. Therefore the resistance includes two TGV resistances in series with the resistance of the 20-µm-wide metal line between the vias. We measured 10 test structures for each pitch using a multimeter (U1272A Handheld Digital Multimeter, Agilent, Santa Clara, CA) connected to two needle probes in a probe station. The resistance distribution is shown in Figure 4-19 and the average resistance of structures with 125-µm, 190-µm, and 250-µm pitch are 6.8 Ω, 9.7 Ω, and 13.1 Ω, respectively. As a result, the resistance of a single via including the contact resistance is calculated as approximately 2 Ω. The bottom gold sheet resistance is approximately 1 Ω /sq, which matches the sheet resistance of 130-nm-thick gold that has been reported in the literature [69].

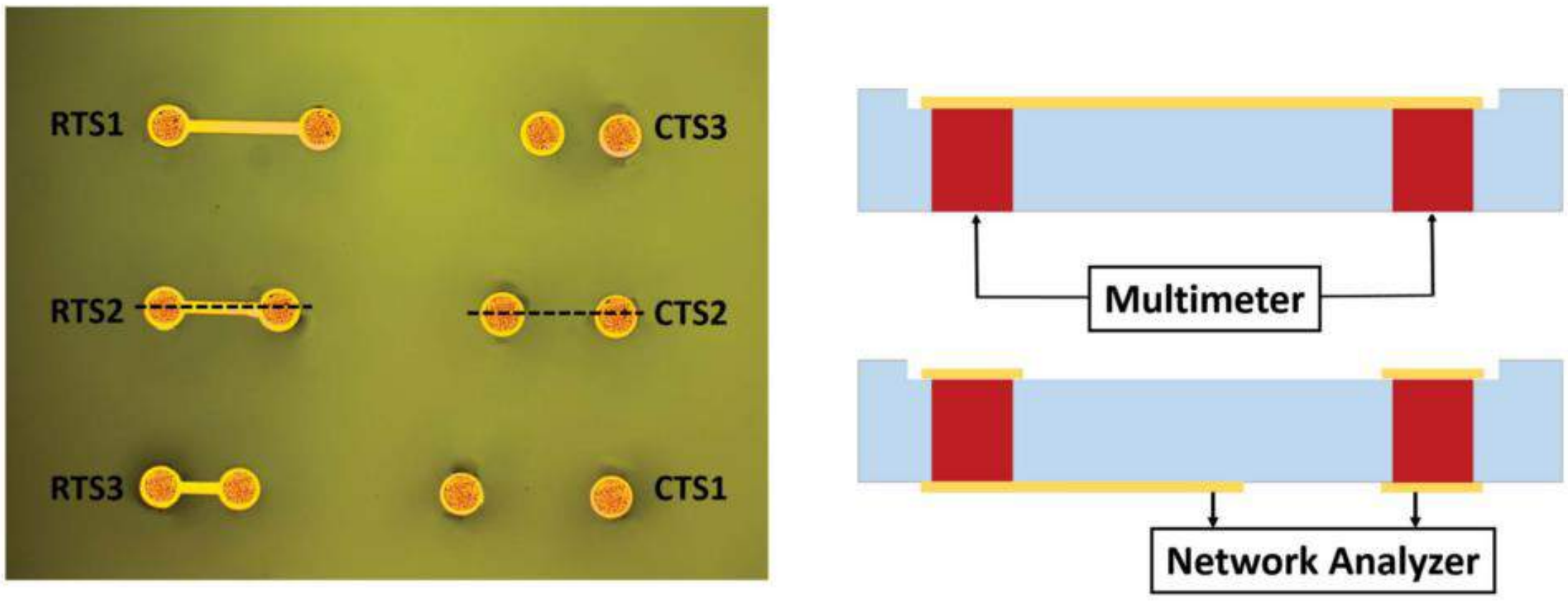

**Figure 4-18: Via test structures (left panel), via resistance measurement setup (right panel top), and via-to-via capacitance measurement setup (right panel bottom).**

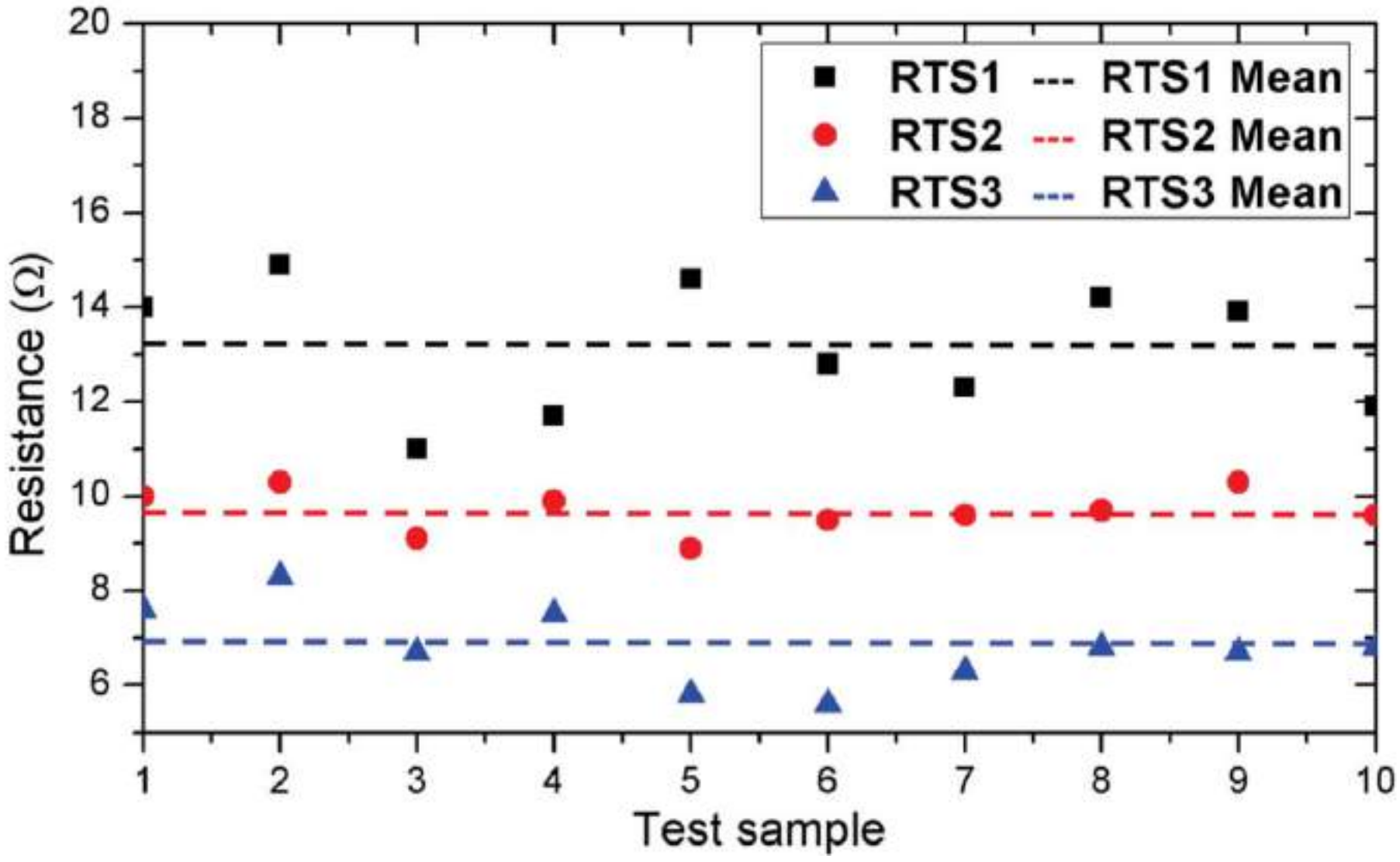

**Figure 4-19: Measured results for resistance test structures.**

The via-to-via capacitance is mainly contributed by the two vias as electrodes with the glass as a dielectric between them. The metal pads on the via will only contribute a negligible amount of capacitance. The capacitance measurement is performed by using a 125-µm pitch coplanar microwave probe (ModelACP40-GSG-125, CascadeMicrotech, Beaverton, OR) connected to a network analyzer (Model E5061B, Agilent Technologies, Inc., Santa Clara, CA). The calibration was carefully done in a tight frequency range from 10 MHz to 10.000005 MHz in order to measure the femtofarad-level capacitance. The measured capacitance distribution is shown in Figure 4-20 and the average capacitance of structures with 125-µm, 190-µm, and 250-µm pitch are 47.8 fF, 34.8 fF, and 21.0 fF, respectively. The measured values are confirmed by the finite element model and analytical calculation using Eq.1. In the equation, 2d is the center-to-center pitch of two vias; R is the via radius; l is the substrate thickness, i.e, via length. Some variance between the model and the measurement could be due to the fact that the actual via is tapered while the model assumes the via to be cylindrical. The comparison of the measurements and models is summarized in Table 4-2. One should also note that the parasitic capacitance is between the signal electrode (bottom electrode) and ground (top electrode) for each element. Hence the measured via parasitic capacitance represents the worst case. To that end, the measured via-to-via capacitance is a more accurate representation for element-to-element electrical coupling. In any case, given the CMUT device capacitance is usually on the order of picofarads, the parasitic capacitance introduced by the TGV interconnects is negligible.

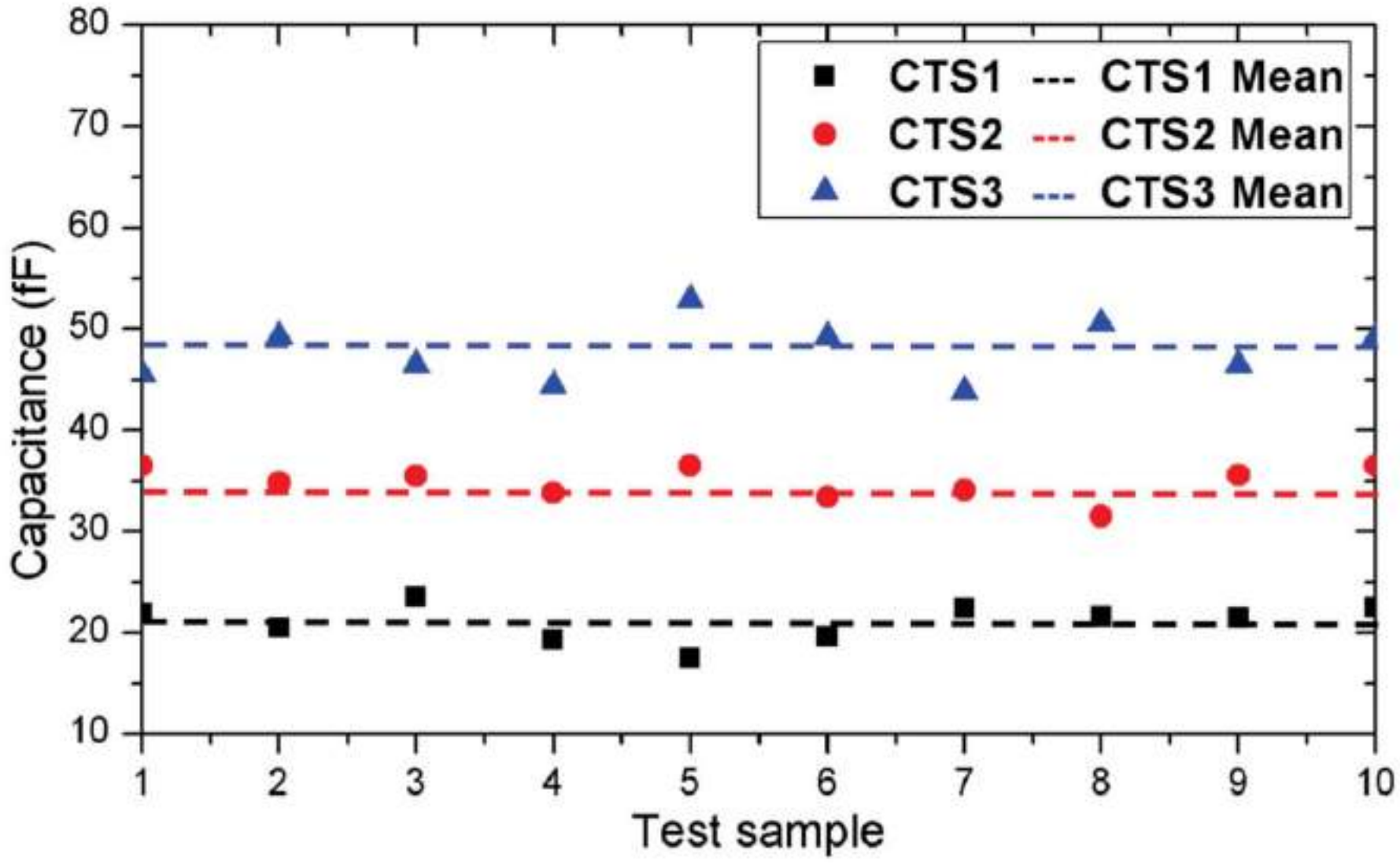

**Figure 4-20: Measured results for capacitance test structures.**

$$C = \frac{\pi \varepsilon_0 \varepsilon_r l}{\ln[\frac{d}{R} + \sqrt{(\frac{d}{R})^2 - 1}]} \qquad (2.31)$$

**Table 4-2: Average via-to-via capacitance: Measurement and Simulation Results**

| Via pitch | Measured | FEM | Analytical |
|---|---|---|---|
| 125 µm | 47.8 fF | 51.7 fF | 54.1 fF |
| 190 µm | 34.8 fF | 37.9 fF | 38.6 fF |
| 250 µm | 21 fF | 30.5 fF | 32.8 fF |

### 4.4.3 Electrical Input Impedance in Air

We tested the fabricated elements in air using a network analyzer (Model E5061B, Agilent Technologies, Inc., Santa Clara, CA) with an internal dc voltage source available up to 40 V. We probed the elements from the backside of the wafer. The input electrical input impedance are measured in air (Figure 4-21). The open-circuit resonance frequency of a 2D CMUT array element was measured as 3.32 MHz at 15-V dc voltage, which is approximately 80% of the pull-in voltage. The baseline in the real part corresponds to the series resistance of the device that includes via resistance as well as the resistance of the bottom electrode, which is measured as 21 Ω at 5 MHz from Figure 4-21(a). The device capacitance is calculated as 2274 fF at 5 MHz from Figure 4-21(b). The resonance frequency, collapse voltage, and device capacitance were simulated using the equivalent circuit model [70] and by finite element model using TRANS 126 electromechanical transducer elements for direct coupling of electrostatic and structural domains. The material properties used in the simulations are listed in Table 4-3. The simulations and the measurements are compared in Table 4-4. The results show that the fabricated CMUTs with TGV interconnects have a small parasitic capacitance (approximately 200 fF) and operate as predicted by the models.

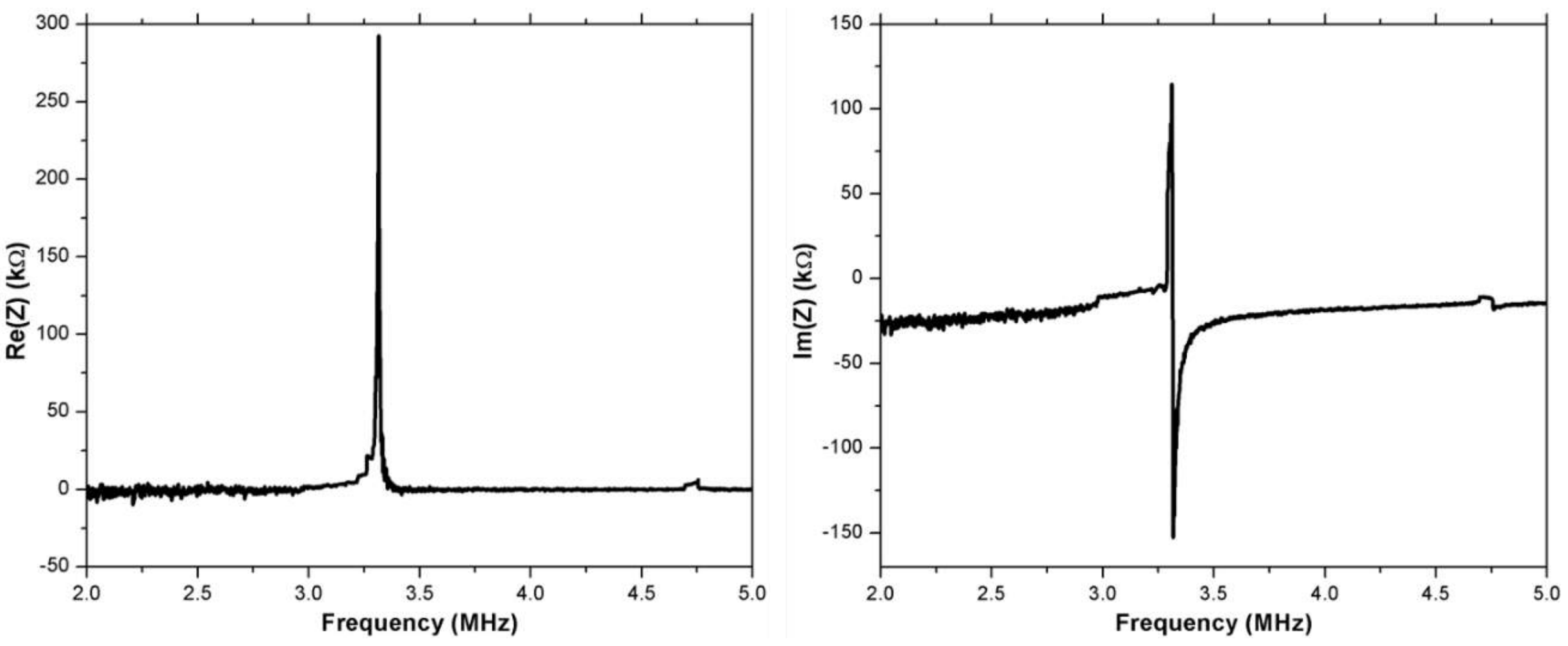

(a) (b)

**Figure 4-21: Impedance measurements (Vdc = 15 V) (a) Real part of the electrical input impedance; (b) Imaginary part of the electrical input impedance.**

**Table 4-3: Material Properties Used in Simulations.**

| | Glass | Silicon | Silicon Nitride | Gold |
|---|---|---|---|---|
| Young's Modulus, GPa | | 148 | 260 | 70 |
| Density, $kg/m^3$ | | 2328 | 3100 | 3300 |
| Poisson ratio | | 0.17 | 0.27 | 0.33 |
| Relative permittivity | 4.6 | 11.7 | 5.7 | |

**Table 4-4: Simulation Versus measured device performance in air**

| | **Measured** | **FEM** | **Analytical** |
|---|---|---|---|
| Resonance Frequency | 3.32 MHz | 3.47 MHz | 3.34 MHz |
| Collapse Voltage | 17.5 V | 15.4 V | 17 V |
| Device Capacitance | 2274 fF | 2001 fF | 2054 fF |

## 4.5 Chapter Conclusions and Future Work

We presented a CMUT fabrication process that integrates TGV interconnects and anodic bonding [68], [71]. The process is low-temperature and eliminates the need for an insulating lining for making through-wafer interconnects. Anodic bonding has the specific advantage of being more tolerant to roughness on the bonding surface compared to commonly used fusion bonding technique. By opening the plate over the bottom electrode TGV and then re-sealing, a reliable vacuum seal can be achieved for all the elements. The use of glass as substrate and metal for interconnects and electrodes reduce the parasitic capacitance and the series resistance of the CMUTs. The resistance of a single via is measured as 2 Ω. The via-to-via capacitance of a 250-µm-pitch via pair is measured as 21 fF. The impedance measurements demonstrate the fabricated device has low parasitics and operates as the models predict.

In the presented approach, by decreasing the substrate thickness, a smaller via diameter could be realized and thus more space and flexibility in array design can be achieved. 40-µm vias with 80-µm pitch have already been demonstrated in a 0.65-mm-thick borosilicate glass wafer [22]. 20-µm vias are possible in 0.3-mm-thick borosilicate glass. Several aspects of the process flow could be further improved. First, the insulation silicon nitride could be replaced by an ALD $HfO_2$ layer defined in the CMUT cavities by a lift-off process. We demonstrated the feasibility of this approach in [55]. Also, for a better plate thickness control, an etch-stop could be added after the handle removal process to avoid the silicon plate over-etch in sealing nitride etching step. First and foremost, this paper demonstrates the process feasibility for incorporating TGVs in 2D CMUT array fabrication using anodic bonding. The fabricated arrays need to be flip-chip bonded to the electronics for immersion measurements, which is outside the scope of the presented work but will be investigated in the future.

# Chapter 5 A MEMS T/R SWITCH EMBEDDED IN CMUT STRUCTURE

Ultrasound imaging systems require high-voltage transmit pulses (>100 $V_{pp}$) in the transmit mode and require low-noise amplification of the received echo signals in the receive mode. Transmit/receive (T/R) switches are critical components in an ultrasound imaging system as the system needs to frequently switch back and forth between the transmit mode and receive mode during imaging. An ideal T/R switch should act like an open circuit in the off-state and short circuit in on-state.

Traditionally, the T/R switches are implemented based on electronic components such as diodes or field-effect transistors (FETs) [72][73], which is not ideal as they introduce noise and distortion to the received signals and fail to completely isolate the receiving paths from the transmit pulse causing some extended dead zone in the near field. A micro-electromechanical (MEMS) switch can significantly help as it provides low insertion loss in closed state and a high-isolation in open state. In this chapter, we present a fast-switching (switching on in 1.25 μs, switching off in 80 ns), low control voltage (2.5 V), high isolation (<-50 dB), DC-contact mode MEMS switch that could be integrated and co-processed into a CMUT element. The significance of this work is to improve the CMUT based ultrasound imaging system efficiency and ease the high-voltage requirements of the front-end circuits.

## 5.1 Transmit/Receive Switching in an Ultrasound Imaging System

In many current ultrasonic imaging systems, the transducer array located in a hand-held probe is connected to the main processing unit via a bundle of cables. The transmit circuitry and receive circuitry are all located in the main processing unit. A close integration of the transducer array and the electronics is highly desired to minimize the effect of cable losses and parasitic capacitance, which could degrade the noise performance of the system, and limit the achievable bandwidth. Furthermore, integration helps to decrease the number of connections needed to the main processing unit by multiplexing several channels. For some applications such as intraoperative navigation and intravascular diagnostic imaging, the size of the probe and the number of the cables connected to the probe are even more important.

A significant challenge of the ultrasound analog front-end electronics is the requirement for high-voltage excitation signals. The front-end integrated circuit (IC) needs to be capable of passing through the high-voltage excitation pulse and also amplifying the received small echo signal. Therefore, an important task is to facilitate the use of high-voltage excitation pulse while protecting the low-voltage amplifiers.

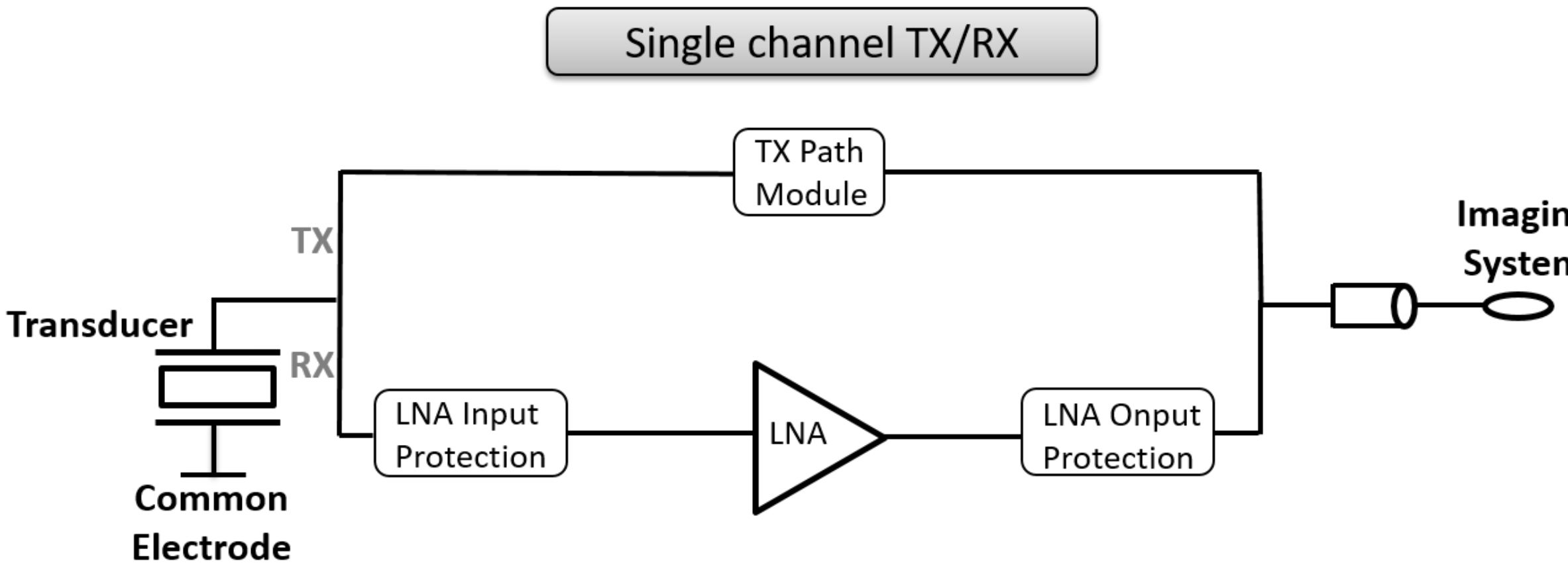

**Figure 5-1: Top-level block diagram of a single TX/RX channel for ultrasound frontends**

The common implementations of the T/R switch in ultrasound frontends are based on using a diode bridge or cross-coupled diode pair [59]. The top-level block diagram of a single T/R channel is shown in Figure 5-2.

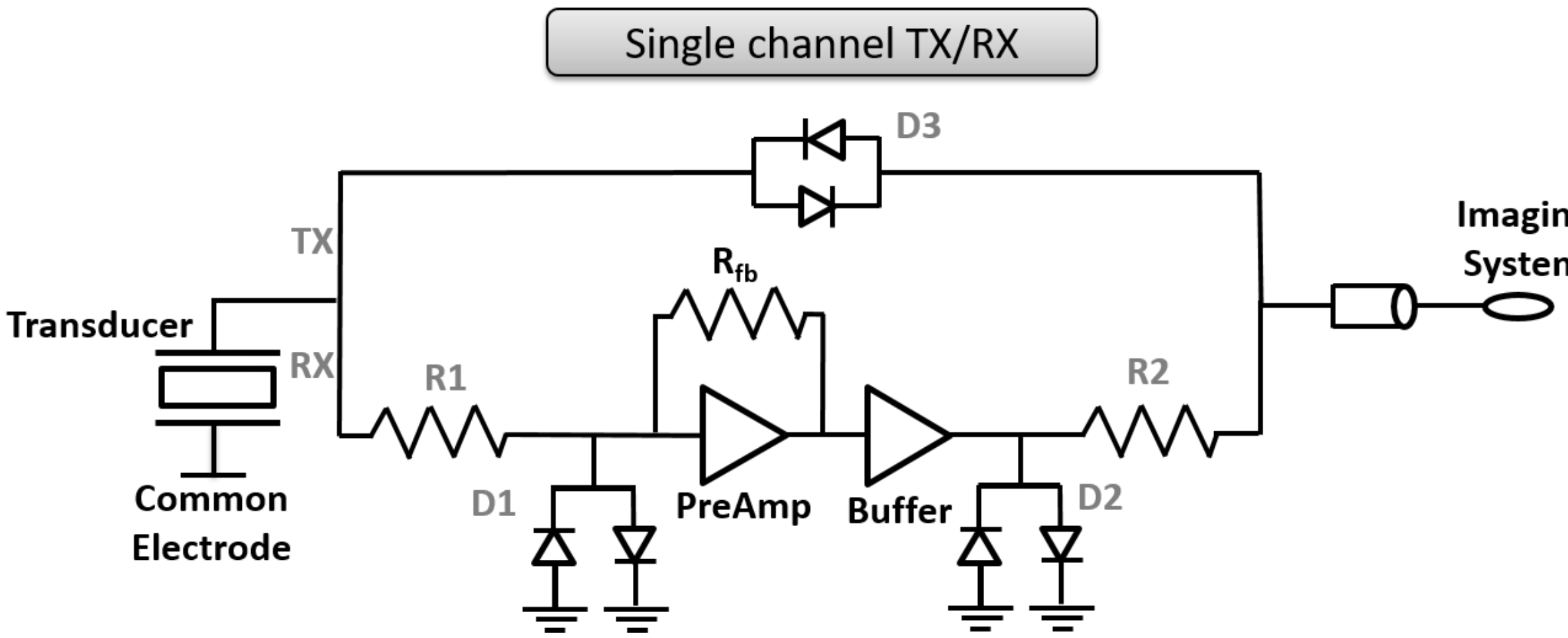

**Figure 5-2: Current implementation of T/R switches using diodes pairs (single channel)**

The circuit architecture can be divided into two paths. In the transmit paths on top, the diode expander circuitry *D3* allows the transmit pulse to pass through during transmit cycle, while it presents an open circuit to the transducer during receive cycle. In the IC implementation, each diode in the expander circuitry was implemented in an electrically isolated region of the n-type epi-layer. To make sure that the parasitic bipolar transistors do not adversely affect the operation of the diodes, the n-type epi region was connected to the p-side of the diode. In the receive path at the bottom, a trans-impedance preamplifier (TIA) is followed by a buffer. Both the input and the output of the receiving path is protected from the high-voltage pulses using diode limiters, which composed of a resistor and two back-to-back diodes (*D1* and *D2)*.

Although this circuit architecture provides an automatic T/R switching mechanism, it suffers from several disadvantages [72]. First, the resistor *R1* and *R2* plays an important role of the trade-off between power losses and thermal noise. During the pulsed excitation, both the transducer and the two resistors load the driving circuit. To limit the power losses, high resistor values are desired. However, this results in additional noise to the received signals. Also, the diode-based switches could introduce some distortion to the signals due to nonlinearity. Third, this implementation fails to completely isolate the receiver from the transmit pulse and therefore the input of the TIA on the receiving paths is charged by the transmit pulse. During the receiving cycle, the discharge has to occur first in order to bring the TIA back to its operating point, which causes some extended dead zone in the near field on the reconstructed image.

To overcome the above shortcomings, the development for a high isolation, low insertion loss, fast-speed T/R switch that could be integrated with an ultrasound transducer array is highly needed. In addition to excellent T/R isolation, a MEMS switch implementation could also benefit the system noise performance in two ways: first, a low-voltage process can be used for the amplifier, which is optimized for low-noise performance. Second, it will eliminate the high-voltage protection circuitry and reduce the overall electronic noise. Besides, a MEMS switch will also help improve the linearity of the front end.

## 5.2 MEMS Switches

MEMS switches have been widely studied for decades, mainly for RF applications, such as radar systems, satellite communication systems, and wireless communication systems. MEMS switches have some advantages that make it closer to an ideal T/R switch required in an ultrasound imaging system front-end:

- Very high isolation in the off-state and low insertion loss in the on-state. MEMS switches are fabricated with air gaps, therefore have a very low off-state capacitance (2-4 fF) and result in excellent isolation.
- Near-zero power consumption. Although tens of volts electrostatic actuation voltage is needed, MEMS switches have near-zero power dissipation because they do not consume any current.
- Good linearity. MEMS switches are very linear devices and therefore can minimize signal distortion or intermodulation.
- Low cost and ease of fabrication. MEMS devices are fabricated by micromachining techniques on substrates such as silicon, glass, GaAs, or low-temperature cofired ceramics (LTCC).

On the other hand, the currently available MEMS switches have some well-known problems:

- Relatively low-speed. The switching speed of most RF MEMS switches that has been reported is around 2-40 µs.
- Power handling. Most RF MEMS switches cannot handle power more than 20-50 mW with good reliability.
- High-voltage drive. Electrostatic MEMS switches require 20-80 V actuation voltage for reliable operation.
- Reliability. Many systems require switches with 20-200 billion cycles. However, the current mature RF MEMS switches can switch 0.1-10 billion cycles.
- Packaging. Packaging costs are high as MEMS switches need to be packaged in inert atmospheres such as nitrogen and argon.

An example of an RF MEMS switch is shown in Figure 5-3. Typically, RF MEMS switches have a lateral dimension of hundreds of microns, and an air gap of several microns.

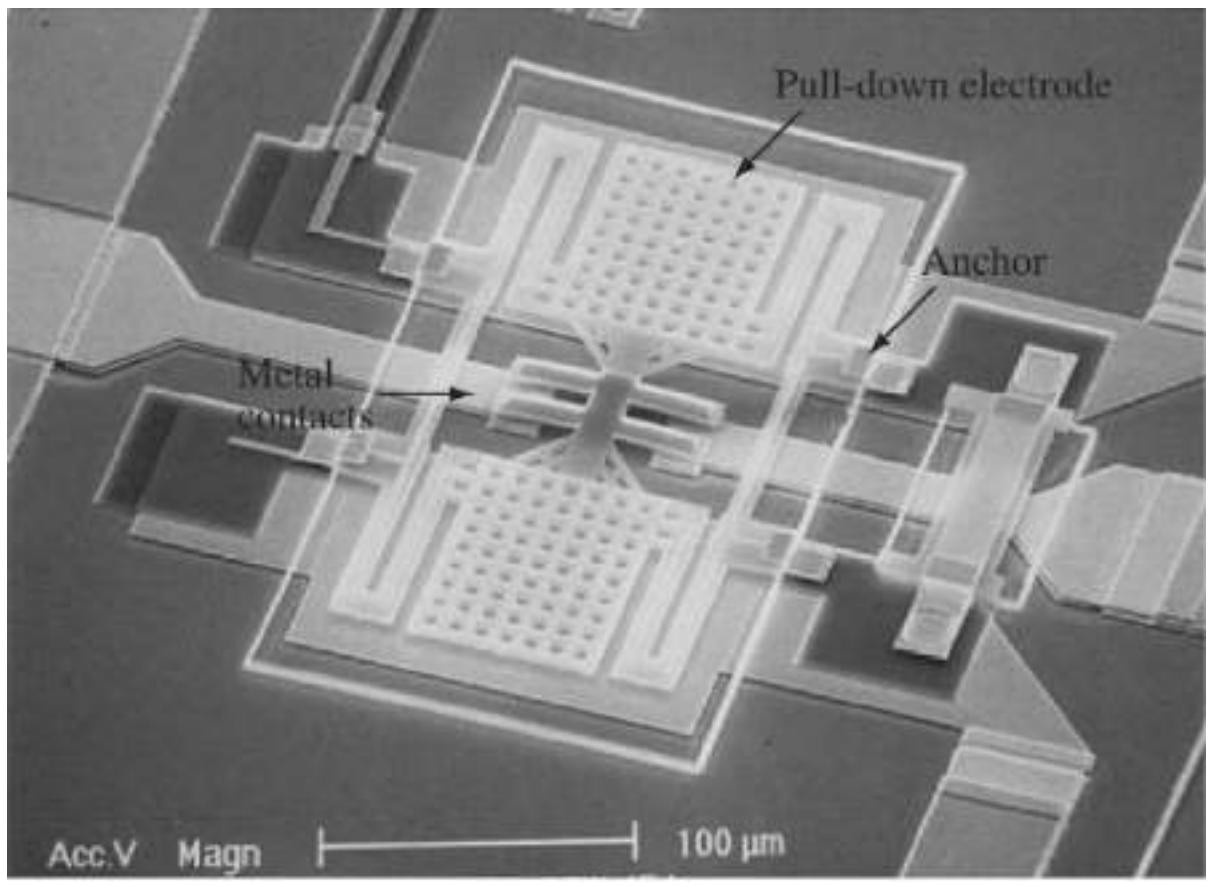

**Figure 5-3: A RF MEMS switch example: Rockwell Scientific MEMS series switch** [74]

**Table 5-1: Important dimensional parameters for the Rockwell MEMS switch** [75]**.**

| ***Horizontal dimensions*** | |
|---|---|
| Total Length | 250 µm |
| Total Width | 150 µm |
| Control electrode length | 75 µm |
| Control electrode width | 75 µm |
| RF line length | 200 µm |
| RF line width | 20-40 µm |
| ***Vertical dimensions*** | |
| Gap height | 2-2.5 µm |
| Bottom metal thickness | 2 µm |
| Bump thickness | 1 µm |
| Contact area | 400 $\mu m^2$ |
| Cantilever thickness | 2 µm |

However, the MEMS switches designed for RF applications usually cannot be directly applied to an ultrasound imaging system mainly due to the following two reasons. First, most of the RF MEMS switches do not meet the switching speed requirement for an ultrasound imaging system front-end [76]. Also, the previously reported MEMS switches do not have a compatible process flow and cannot be conveniently integrated with ultrasound transducers.

CMUTs and MEMS switches are both electrostatically actuated capacitors with a moving plate and share similar fabrication technologies. Although the integration of these two components seems a natural decision, there has been no demonstration of CMUTs with embedded MEMS T/R switches. Integrating a MEMS switch on the same substrate with the transducers will not only improve the system performance contributed by its high isolation and low insertion loss but also it can enable implementation of the LNA circuitry in a standard low-voltage CMOS process. This front-end amplification is essential to preserve the bandwidth and signal integrity, especially when small transducer elements are loaded by a long cable.

We propose to design and fabricate a fast-switching, low-control-voltage, DC-contact-mode series MEMS switch that can be integrated and co-fabricated with CMUTs on the same substrate using our previously reported anodic-bonding-based process. A highly integrated CMUT based ultrasound system could improve the system performance, ease the high-voltage requirement of the front-end electronics and therefore enable new imaging, diagnostic, and therapeutic applications.

## 5.3 Design of a MEMS Switch Embedded in the CMUT Structure

Since the basic building block of a CMUT is a parallel plate capacitor that pulls in due to mechanical instability at a certain electrical field, an electromechanical switch can be easily embedded in the CMUT structure. The electrostatic force that pulls down CMUT plate is also what makes a MEMS switch work. Therefore integration of these two components on the same substrate is highly feasible.

### 5.3.1 Switch Structure Design

We need to make some modifications to a single CMUT cell in order to build a switch as shown in Figure 5-4. In the proposed MEMS switch structure, we have an interrupted transmission line crossing the midline of the device, so the switch is normally in OFF-state. On the two sides of this transmission line, there are control electrodes to apply the electrostatic force to pull the plate down. A metal contact is defined under the silicon nitride insulation layer at the bottom side of the plate so that it closes the circuit in the interrupted transmission line when the plate is pulled down, which is the switch ON-state. The insulation layer is needed so that the RF signal does not interfere with the control signal. This approach will only require the distribution of digital control signals for the switches and analog signals lines.

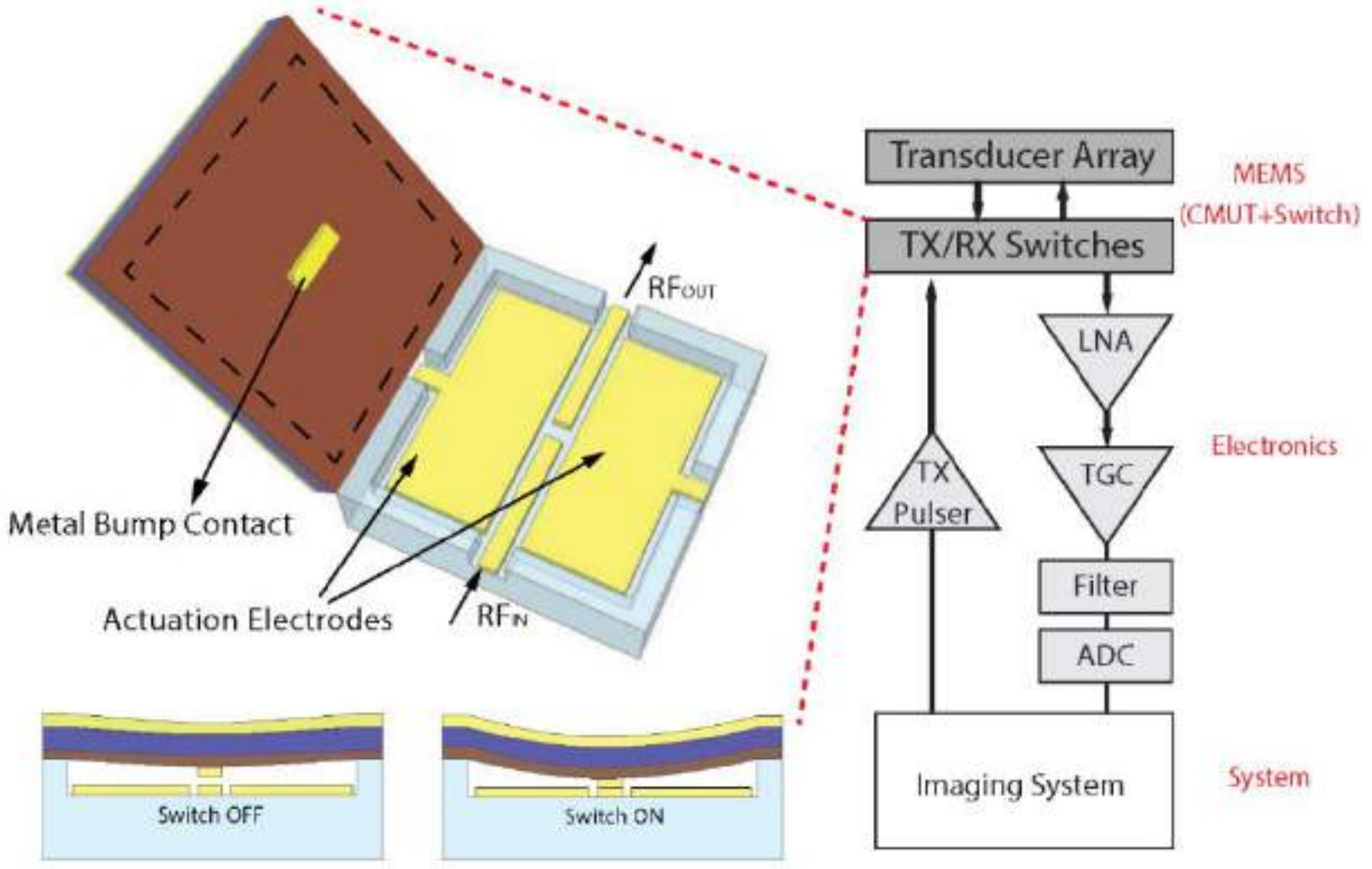

(a) (b)

**Figure 5-4: (a) MEMS switch structure modified from a single CMUT cell. (b) T/R switch in an ultrasound imaging system.**

The designed MEMS switches could be integrated into a CMUT 1D array in the following two ways as shown in Figure 5-5. In one design [Figure 5-5(a)], one switch will be placed on both sides of a 1D CMUT array element. The control signal for the switch on each side will be 180º out of phase. Therefore, when the top switch is on and bottom switch is off, the element will be connected to the high-voltage transmitting circuitry and disconnected from the receivers. On the contrary, when the control signal is reversed, the transmitter will be disconnected and the receiver will be connected to the element. All the T/R switching will take place at the same time, therefore only one control signal and its complement are required for the whole array. In the actual imaging case, the "on time" for TX switches can be set according to the maximum delay required for transmit beamforming. The second configuration [Figure 5-5(b)] is more suitable for a direct connection to a standard imaging system. As the additional switch provides protection for the output of the LNA, and hence only one cable is needed to pass through the high-voltage transmit pulse and received the amplified echo signal to the system.

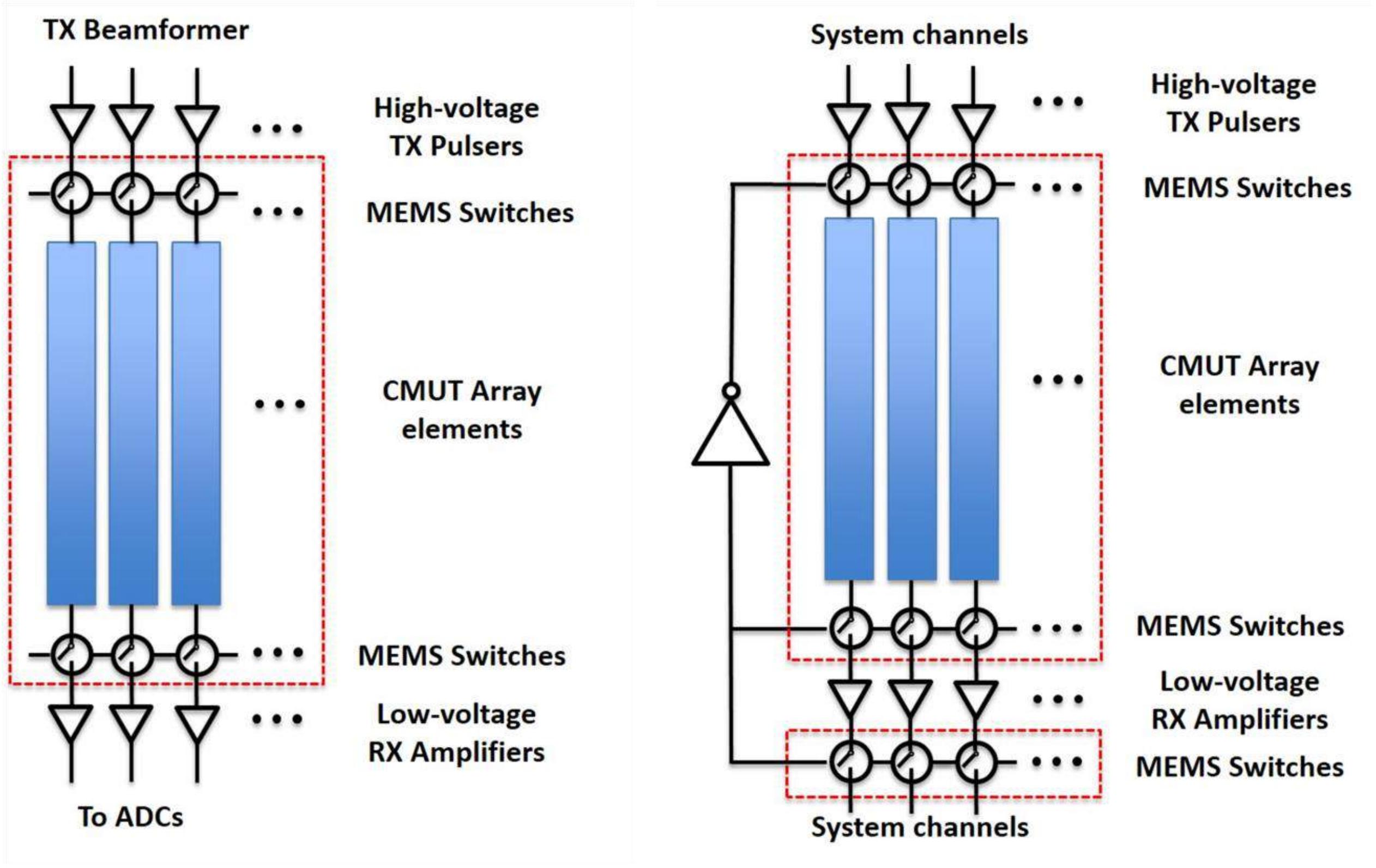

**(a)** **(b)**

**Figure 5-5: (a) Two-switch configuration where TX and RX paths are totally isolated. (b) Three-switch configuration for seamless integration of low-voltage frontend IC with a standard imaging system.**

### 5.3.2 Analytical Calculation of Switching Time:

It has been reported that the switching time is mainly determined by the device mechanical resonant frequency $f_0$. For an inertia-limited system (with a small damping coefficient and Q≥2), the switching time $t_s$ depends on the device resonant frequency $f_0$, pull-in voltage $V_p$, and actuation voltage $V_s$, as shown in Eq. (5.1)

$$t_s \approx 3.67 \frac{V_p}{V_s 2\pi f_0} \tag{5.1}$$

The vacuum-backed membrane structure with small dimensions gives CMUTs the ability to make devices with MHz resonant frequencies which translate to fast switching. Considering that the CMUT resonant frequency for medical applications is usually in the range of several MHz, pull-in voltage is in the range of tens of volts, Table I presents several examples of fast switching time assuming typical CMUT operation parameters.

**Table 5-2: Examples of the calculated switching time**

| $f_0$ | $V_p$ | $V_s$ | $t_s$ |
|---|---|---|---|
| 1 MHz | 10 V | 13 V | 450 ns |
| 3 MHz | 15 V | 20 V | 146 ns |
| 5 MHz | 20 V | 26 V | 90 ns |

### 5.3.3 FEM Modeling:

In order to get the device dimensions and verify the analytic calculation results, a finite-element model (FEM) was developed in ANSYS APDL (ANSYS v.17.2, ANSYS Inc., Canonsburg, PA, USA). CMUT structure has been widely studied using FEM models [77][35]. Although the switch structure is similar to a CMUT, the operating principle and the performance specifications are different. Also, the metal contact, RF line, and the control electrodes need to be defined in the switch structure.

A full 3D model is developed in order to get accurate results. We start the modeling with the plate geometry definition. The plate is composed of a silicon plate layer, a stacked Cr/Au layer top electrode on the plate, a silicon nitride insulation layer underneath, and a Cr/Au metal contact at the bottom of the insulation layer. The material properties of each layer are listed in Table 5-3. SOLID45 elements are suitable for the plate modeling because it has plasticity, creep, swelling, stress stiffening, large deflection, and large strain capabilities. The plate is fixed by the 10-μm outer boundary at the bottom of the silicon nitride layer, which represents the post region formed by anodic bonding.

**Table 5-3: Material properties used in FEM simulations**

| | Silicon | Silicon nitride | Gold | Immersion medium |
|---|---|---|---|---|
| Young's modulus (GPa) | 148 | 296 | 79 | |
| Density (kg/m$^3$) | 2328 | 3187 | 19300 | 1000 |
| Poisson ratio | 0.17 | 0.27 | 0.44 | |
| Relative permittivity | | 5.7 | | |
| Speed of sound (m/s) | | | | 1500 |

The control electrodes and the metal contact are differentiated using two types of TRANS126 electromechanical transducer elements according to their gap distances. For the control electrodes, the gap height is the effective gap height

$$T_{eff} = T_{gap} + \frac{T_{SiN}}{\varepsilon_r} \ , \tag{5.2}$$

and the close gap height is 0. For the metal contact, the gap height is $T_{gap}$, and the close gap height is the metal contact height $T_{contact}$. We first perform the static simulation in air to model device steady-state behavior, including the atmospheric deflection and the plate deflection with DC bias.

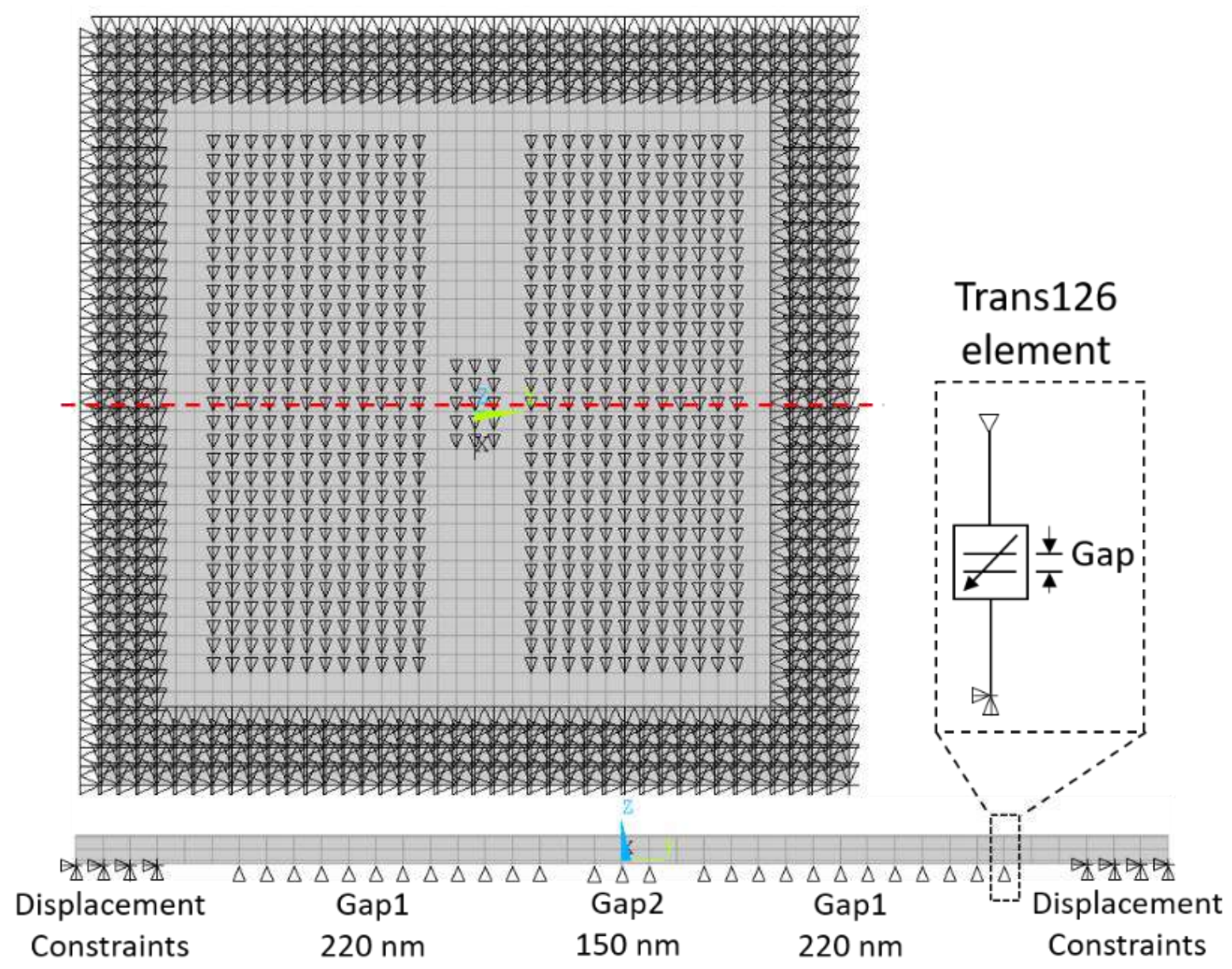

**Figure 5-6: Plate structure model and created elements for static simulation.**

A more important aspect of the switch is the dynamic behavior, which is the switching speed. The switch is next to the CMUTs and will operate in immersion, which is beneficial to the switching speed because the medium will increase the damping and reduce the plate oscillation. We model the immersion medium using a sphere medium using FLUID30 element. The fluid-structure interface is specified using proper fluid-structure interaction flags. FLUID130 element is used for an absorbing boundary to extend the fluid domain to infinity. Transient simulation with the full method is used to including the all the non-linearities. The time step is set to be smaller than $\frac{1}{20f}$ to capture the mode frequency at $f$, which is first determined by performing a harmonic analysis. The simulated results are compared and discussed with the actual device measurement results in section 5.5.

## 5.4 Fabrication Process

The fabrication process flow is shown in Figure 5-7. The first cell on the left of the cross-sectional drawing and in the front of 3D drawing represent the switch. The other cells represent the co-fabricated CMUT element. The initial glass substrate was the same wafer that was used in the platform process [Figure 5-7(a)]. First, the substrate was cleaned using a piranha solution for removal of organics and other contaminants. Then the cavities of the switch and the CMUT were patterned and etched to 350-nm deep simultaneously [Figure 5-7(b)]. The photoresist was then removed and the bottom metal pattern was defined by a second photolithography step using the same photoresist. After hard baking the photoresist for 5 minutes at 125ºC, a 1-min BOE etch was performed to smooth the bottom of the cavity and also to enlarge the photoresist undercut to facilitate lift-off. A stacked metal layer that consists of 20-nm chromium and 100-nm gold was formed in the etched cavities by evaporation and liftoff [Figure 5-7(c)]. The metal layer serves as the bottom electrode for the CMUT and also control electrodes and RF line for the switch. Figure 5-10(a) shows the optical image of the switch and the CMUT after the completion of process on the glass substrate. Figure 5-8(a) shows the AFM image of the center region of the switch cell.

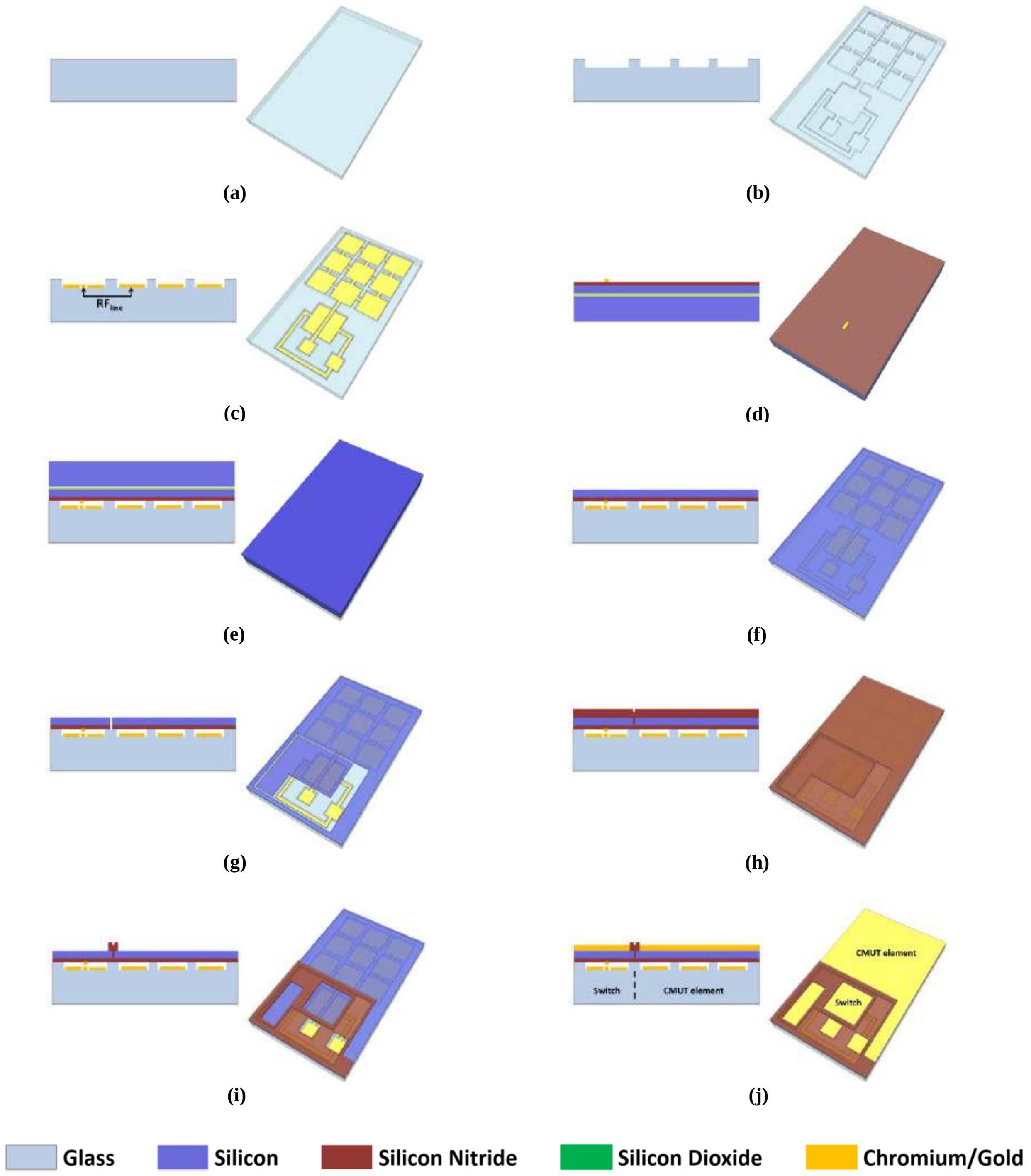

**Figure 5-7: Fabrication process flow**

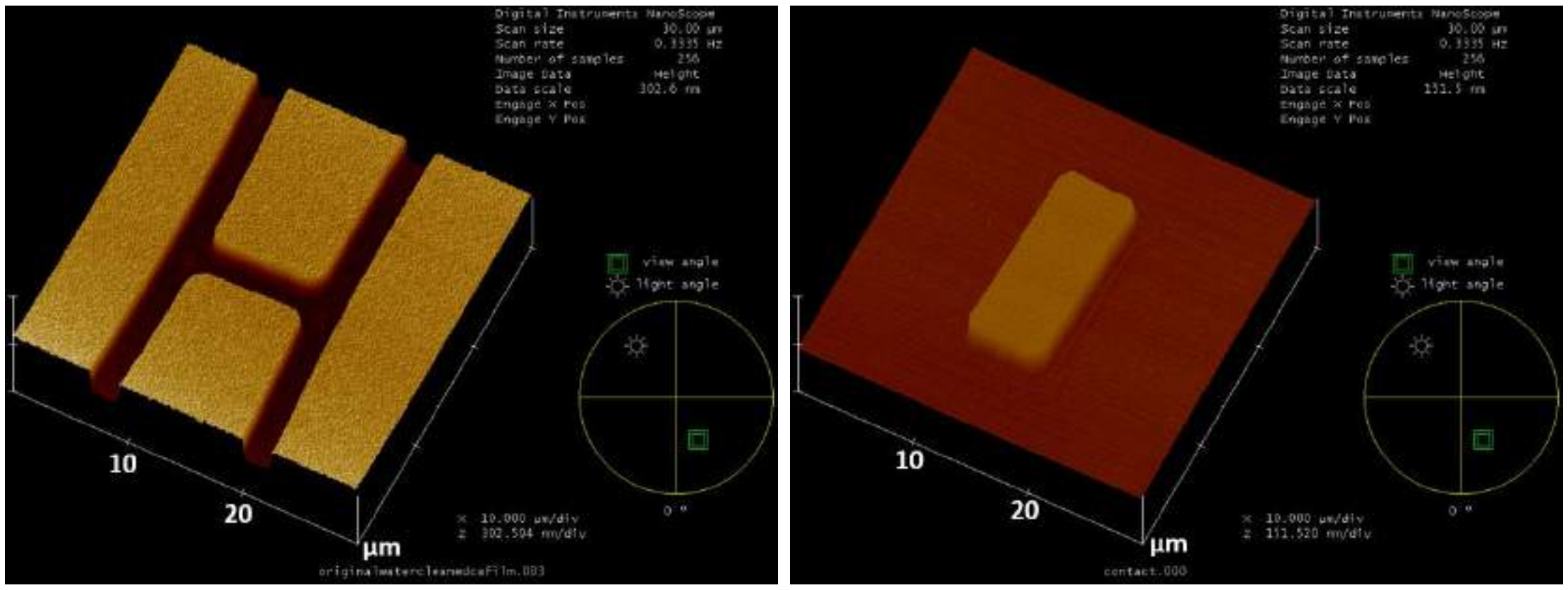

**Figure 5-8: AFM image of the center region of the switch cell (left), and the corresponding metal contact.**

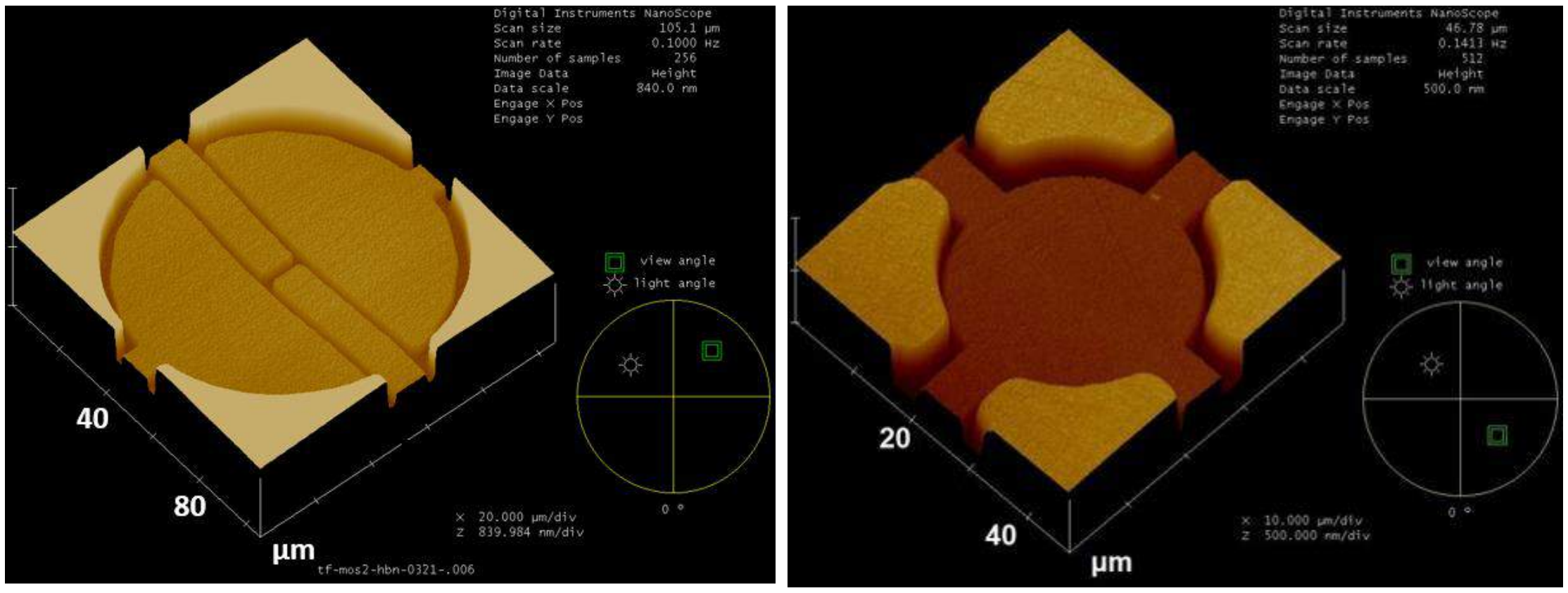

**Figure 5-9: AFM images of the switch structure and the co-fabricated CMUT structure**

Next, we deposited a 200-nm PECVD silicon nitride insulation layer on the device layer of the SOI wafer. The bump can then be formed on the insulation layer by evaporating and lifting off a stacked metal layer of 20-nm chromium and 50-nm gold [Figure 5-7(d)]. The insulation layer is needed for both the switch and the CMUT: it prevents the short circuit when the CMUT pulls in, and isolate the metal contact from the top plate electrode for the switch.

The processed glass and SOI wafers were first aligned and then anodically bonded at 350◦C under 2.5-kN down force in vacuum ($10^{-4}$ Torr) using an in-situ aligned wafer bonding system (Model AML-AWB-04, Applied Microengineering Ltd, Oxfordshire, United Kingdom) [Figure 5-7(e)]. To release the plate, the handle layer of the SOI wafer was ground down to 100 μm and the remaining handle layer and the BOX layer were removed using a 10% TMAH at 80ºC and a 10:1 BOE solution, respectively [Figure 5-7(f)]. Figure 5-10(b) displays the optical image that shows the bonded SOI device layer as the plate and a close-up view of the aligned metal bump over the RF line.

In the next step, we dry-etched the silicon/silicon nitride plate [Figure 5-7(g)]. This step serves the following purposes. First, it evacuates the generated gas during anodic bonding in the switch and CMUT cells, so that the cells could be later sealed in vacuum. Second, it isolates the top electrode plate of the switch and the CMUTs to enable independent biasing of the CMUT and the switch cell. Third, it allows accessing the bottom pads of the switch and CMUTs. Additionally, this step defines the dicing lines for device separation. After the etching, the photoresist was removed using oxygen plasma. Figure 5-10c shows the optical image after this step.

We deposited 800-nm conformal PECVD silicon nitride on the entire wafer surface to seal the cavities [Figure 5-7(h)]. After that, the silicon nitride needs to be etched in order to access the top plates and also the bottom electrodes for both the switch and the CMUT. The sealing silicon nitride was only left at the places where sealing is needed [Figure 5-7(i)]. Then we transferred the wafer to the evaporation chamber without removing the photoresist. Finally, a stacked layer of 20-nm chromium and 180-nm gold was evaporated and lifted off to form electrical contacts [Figure 5-7(j)]. Figure 5-10(d) shows the optical image of the final device.

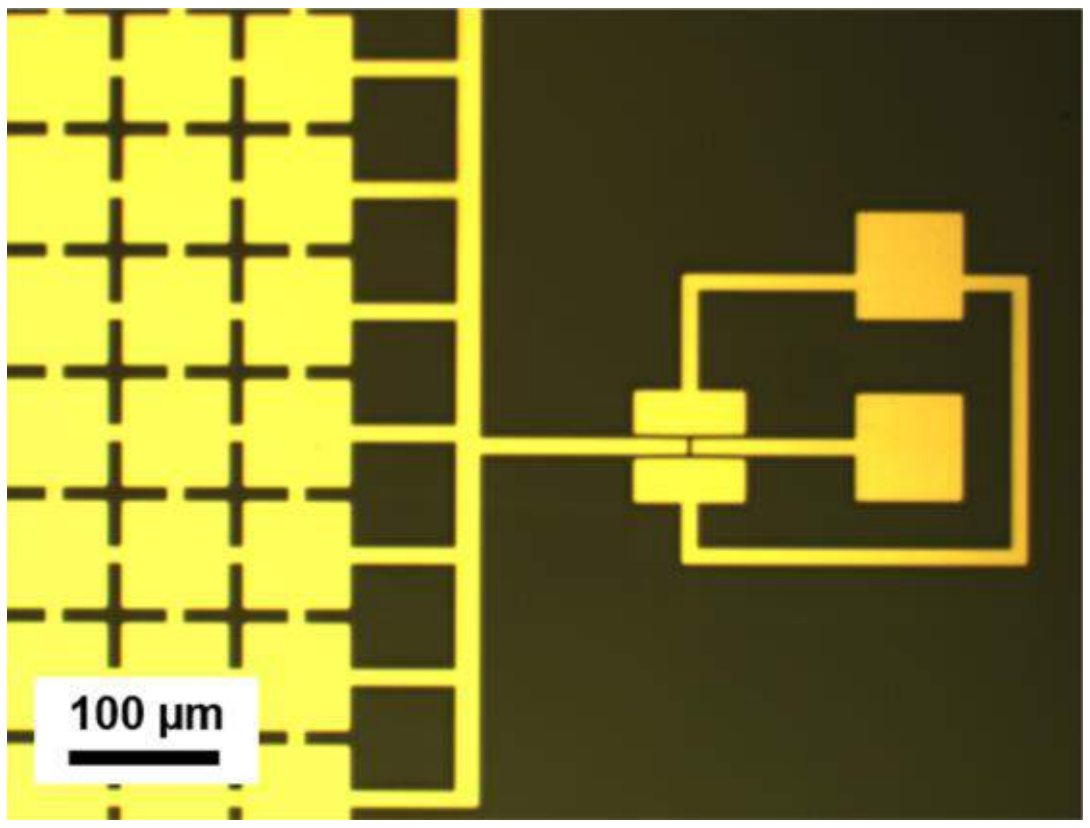

**(a) Bottom metal formation**

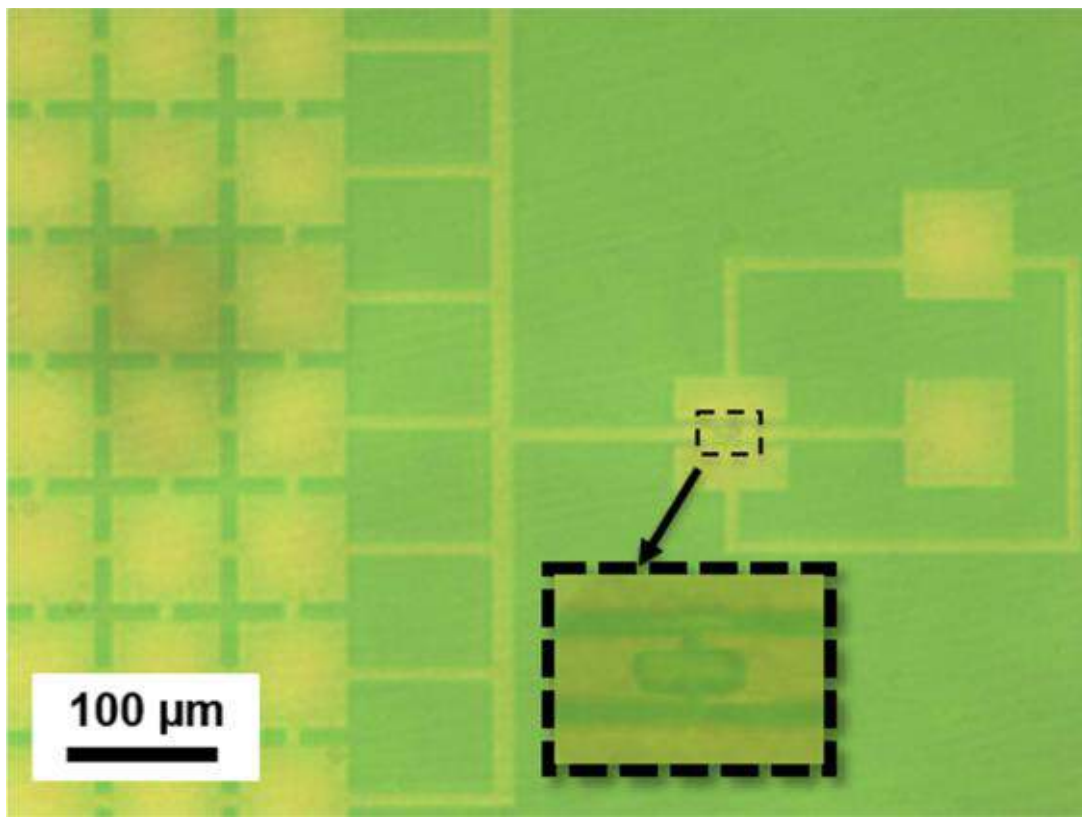

**(b) Aligned anodic bonding**

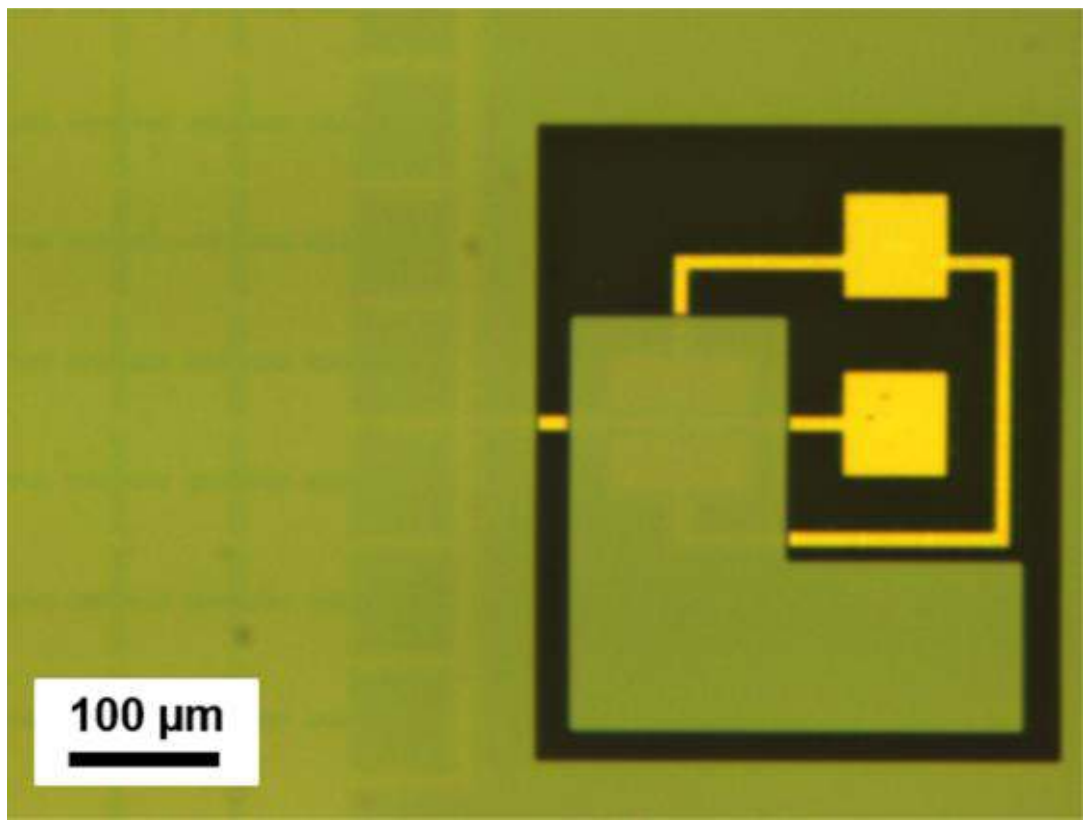

**(c) Top plate etch**

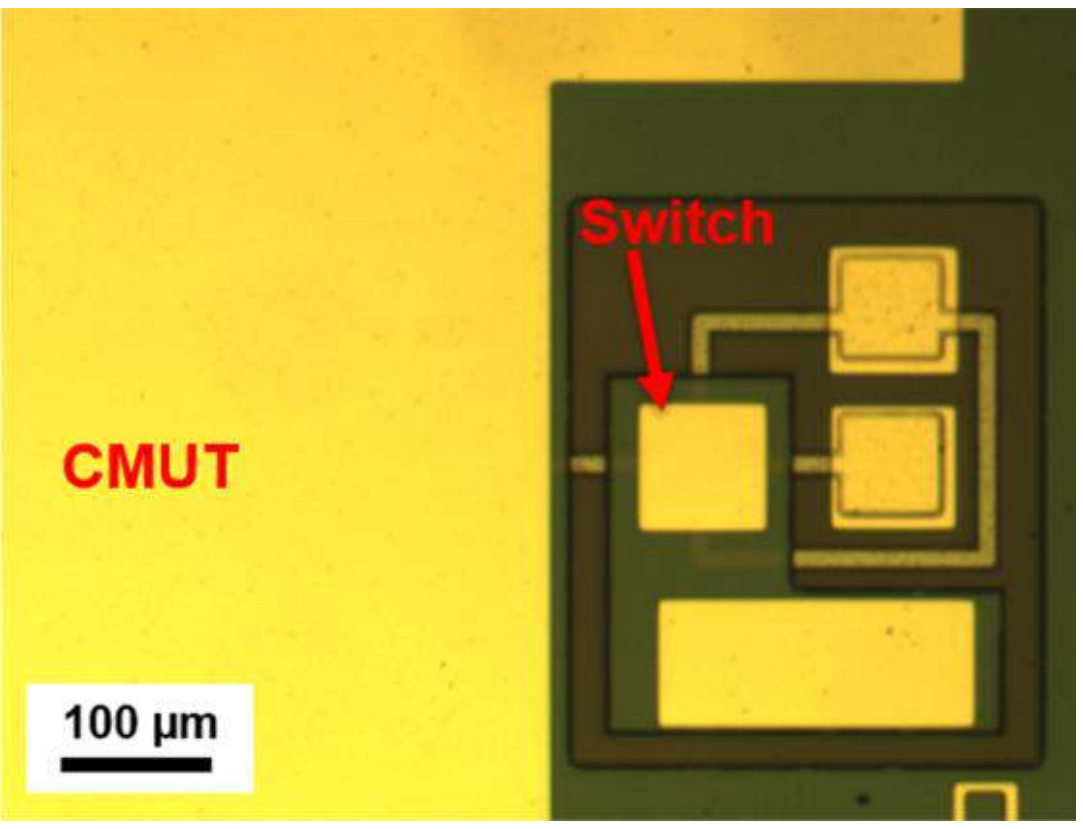

**(d) Completed devices**

**Figure 5-10: Optical images of the critical processing steps**

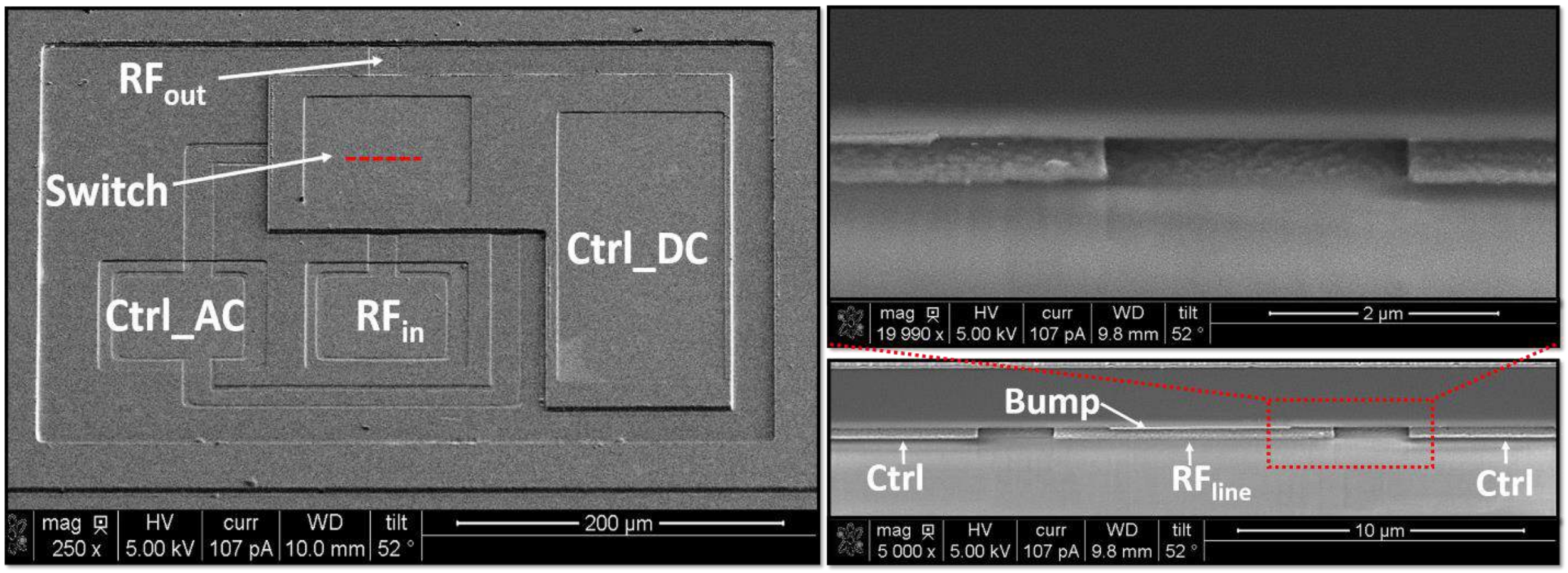

**Figure 5-11: Cross-sectional SEM image of the finished switch**

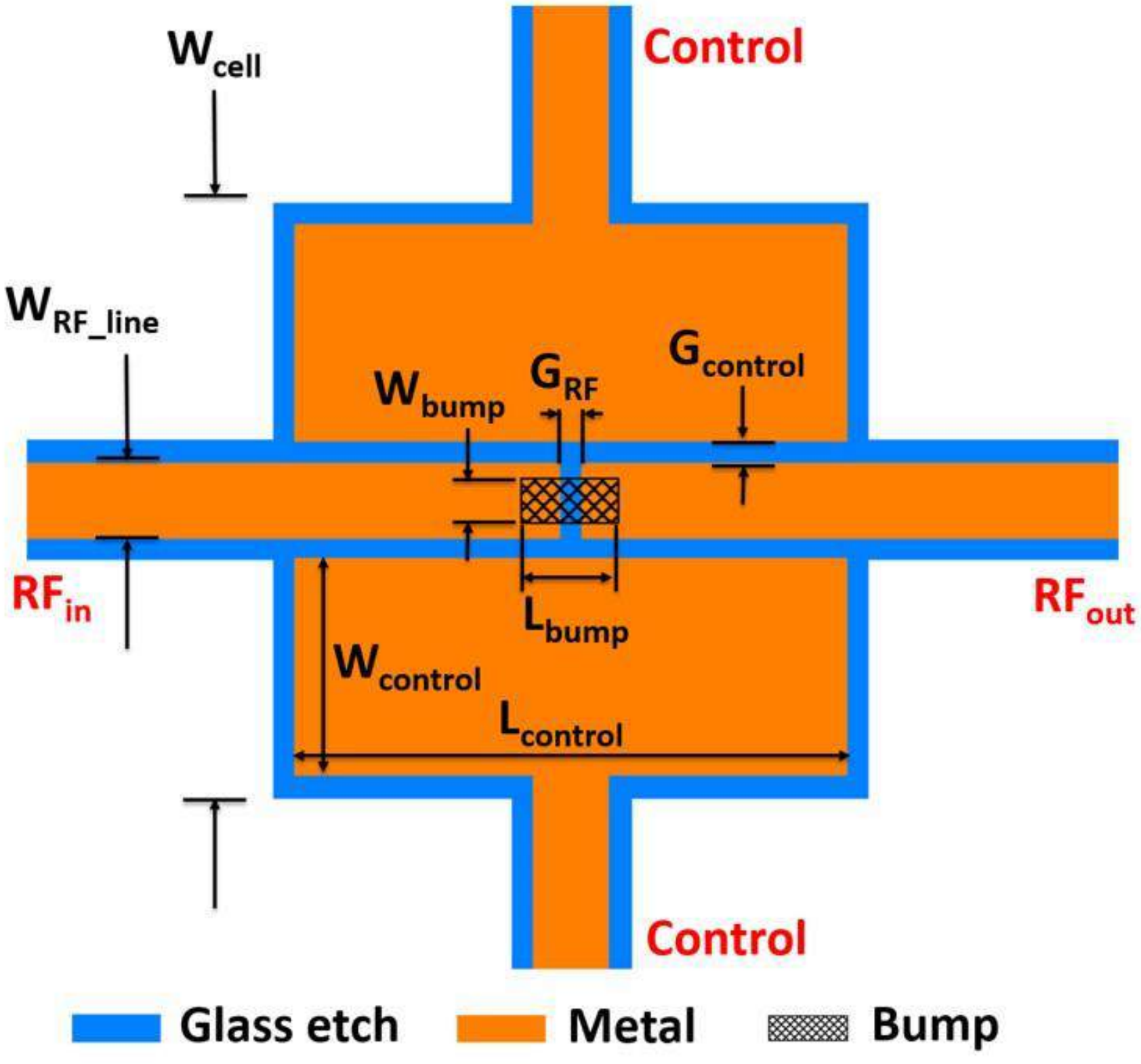

**Table 5-4: Physical parameters of the tested switch**

| ***Horizontal dimensions*** | |
|---|---|
| Shape of the cell | Square |
| Cell width, $W_{cell}$ | 80 µm |
| Control electrode width, $W_{control}$ | 29 µm |
| Control electrode length, $L_{control}$ | 74 µm |
| Bump width, $W_{bump}$ | 6 µm |
| Bump length, $L_{bump}$ | 13 µm |
| RF line width, $W_{RF_line}$ | 10 µm |
| Ctrl electrode to $RF_{line}$ gap,$G_{control}$ | 3 µm |
| RF line gap, $G_{RF}$ | 3 µm |
| ***Vertical dimensions*** | |
| Substrate thickness, | 700 µm |
| Glass etching depth, | 0.35 µm |
| Bottom metal thickness, | 0.15 µm |
| Bump thickness, | 0.07 µm |
| Insulation layer thickness, | 0.2 µm |
| Silicon plate thickness, | 2 µm |
| Top metal thickness, | 0.2 µm |

## 5.5 Device Characterization

### 5.5.1 Experimental Setup

The switch die was diced from the wafer and the switch pads were wire bonded to a chip carrier. During the wire bonding and test, electrostatic discharge (ESD) should be prevented to avoid damaging the switches due to stiction or dielectric breakdown [78]. A printed circuit board (PCB) was designed for the test to reduce the cable parasitics.

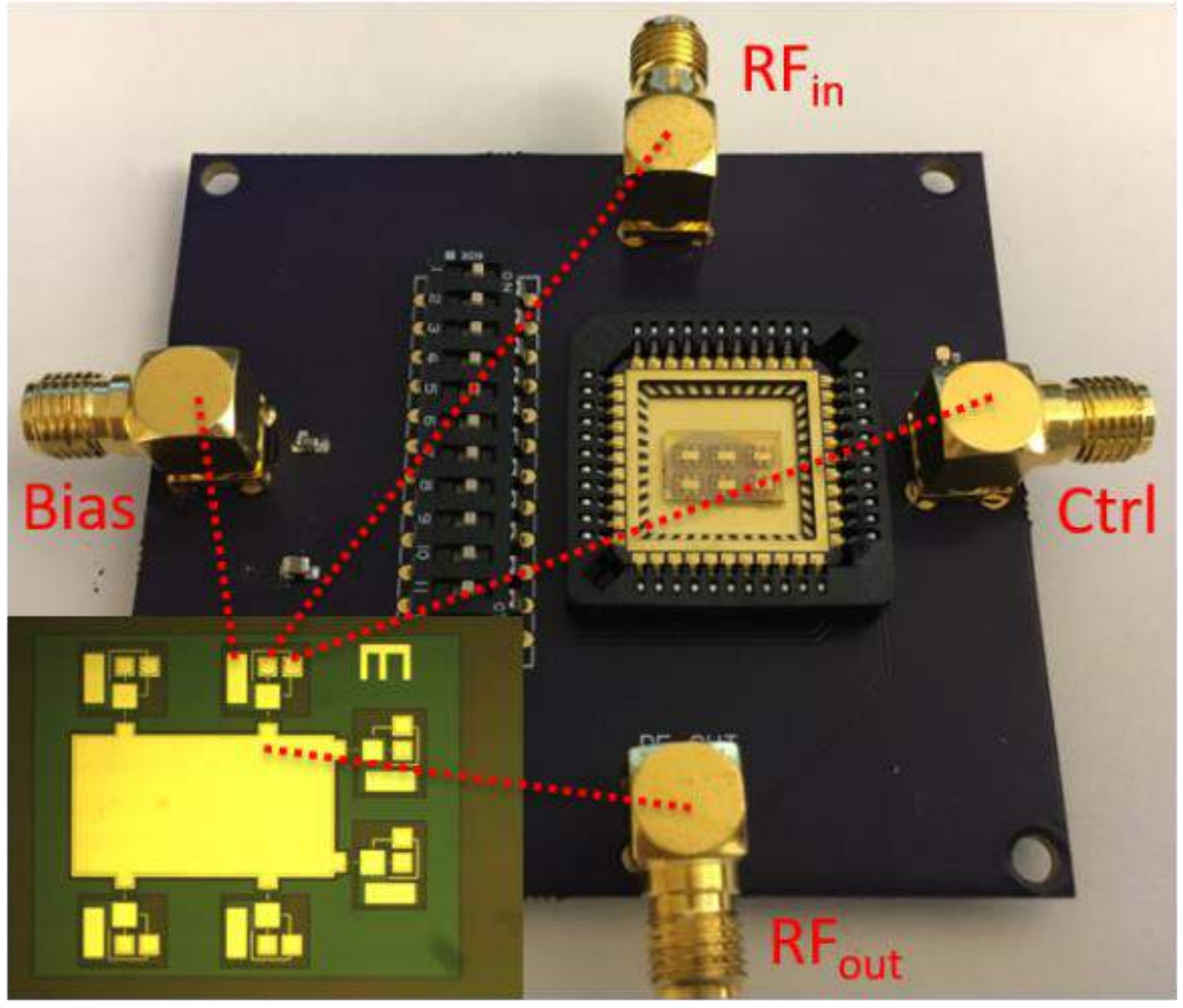

**Figure 5-12: A switch die wire bonded on the PCB.**

### 5.5.2 Static Characterization

The PCB with the wire bonded MEMS switch was first characterized in air to study the steady-state behavior of the MEMS switch. A dc power source (Model PS310, Stanford Research Systems, Sunnyvale, CA) was connected through the PCB to the top plate of the switch and the control electrodes were terminated to ground. A multimeter (Model U1272A, Agilent Technologies, Santa Clara, CA) was connected between the $RF_{in}$ and $RF_{out}$ to measure the resistance of the switch. In the meanwhile, a Wyko surface optical profiler (Wyko NT1100, Veeco Instruments Inc., Plainview, NY) was used to measure the plate surface profile. The dc bias was increased from 0 V to 90 V with a step of 1 V and a dwell time of 10 s to let the switch reach the steady state. The measured switch surface center deflection at different dc bias voltages are shown in Figure 5-13(a). The FEM static simulation results are shown in Figure 5-13(b). A good agreement between the model and the measurement was observed, confirming that the MEMS switch is vacuum sealed and a good dimensional control and stress control was achieved. From the steady-state measurement, it was observed that the switch turned on and off both at 68 V. The on-resistance was measured as approximately 75 Ω. Given the $RF_{in}$ to $RF_{out}$ metal path resistance is 25 Ω, the switch contact resistance is approximately 50 Ω, which could be reduced by optimizing the switch geometry. Figure 5-14 shows the conductivity between the $RF_{in}$ and $RF_{out}$ from 40 V bias voltage to 90 V bias voltage. The inset graphs are the switch surface profile for the off-state (65 V) and on-state (70 V).

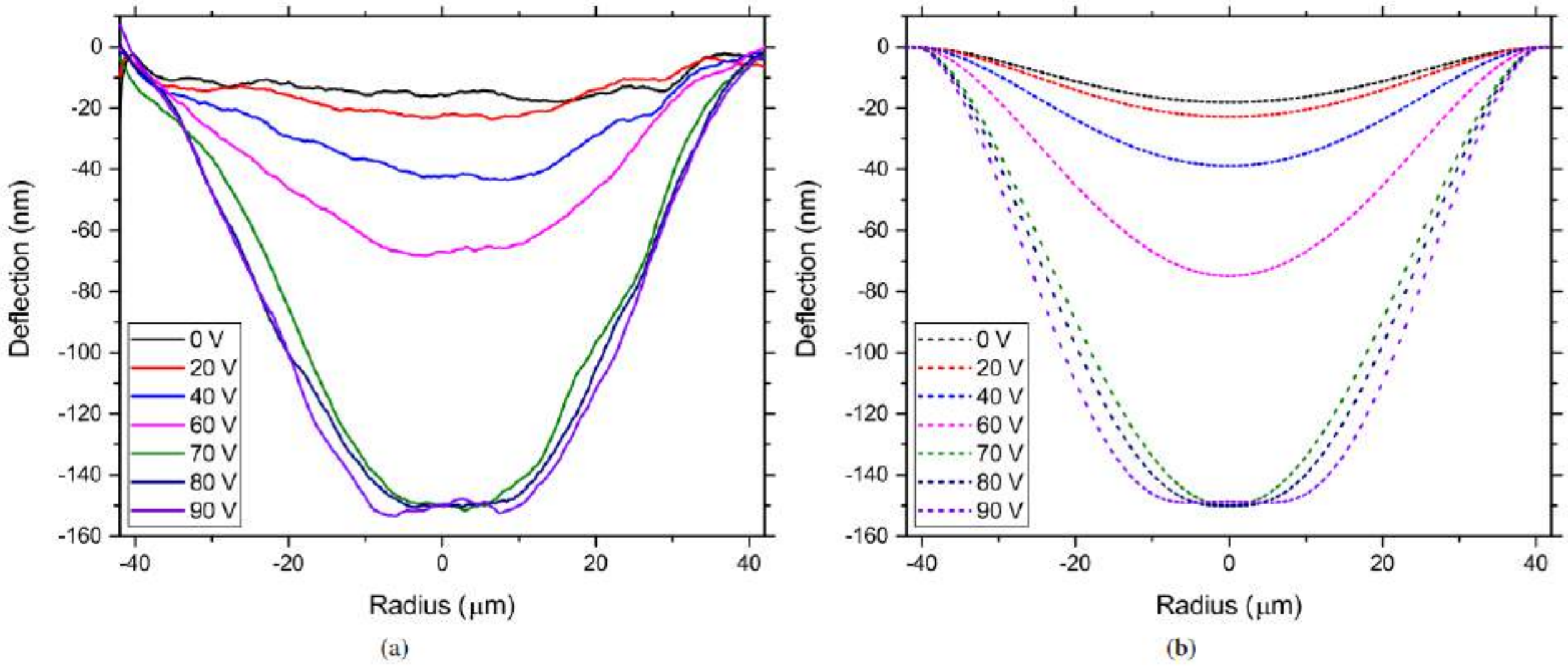

**Figure 5-13: Static deflection at different bias voltages. (a) Wyko measured results. (b) FEM static simulation results.**

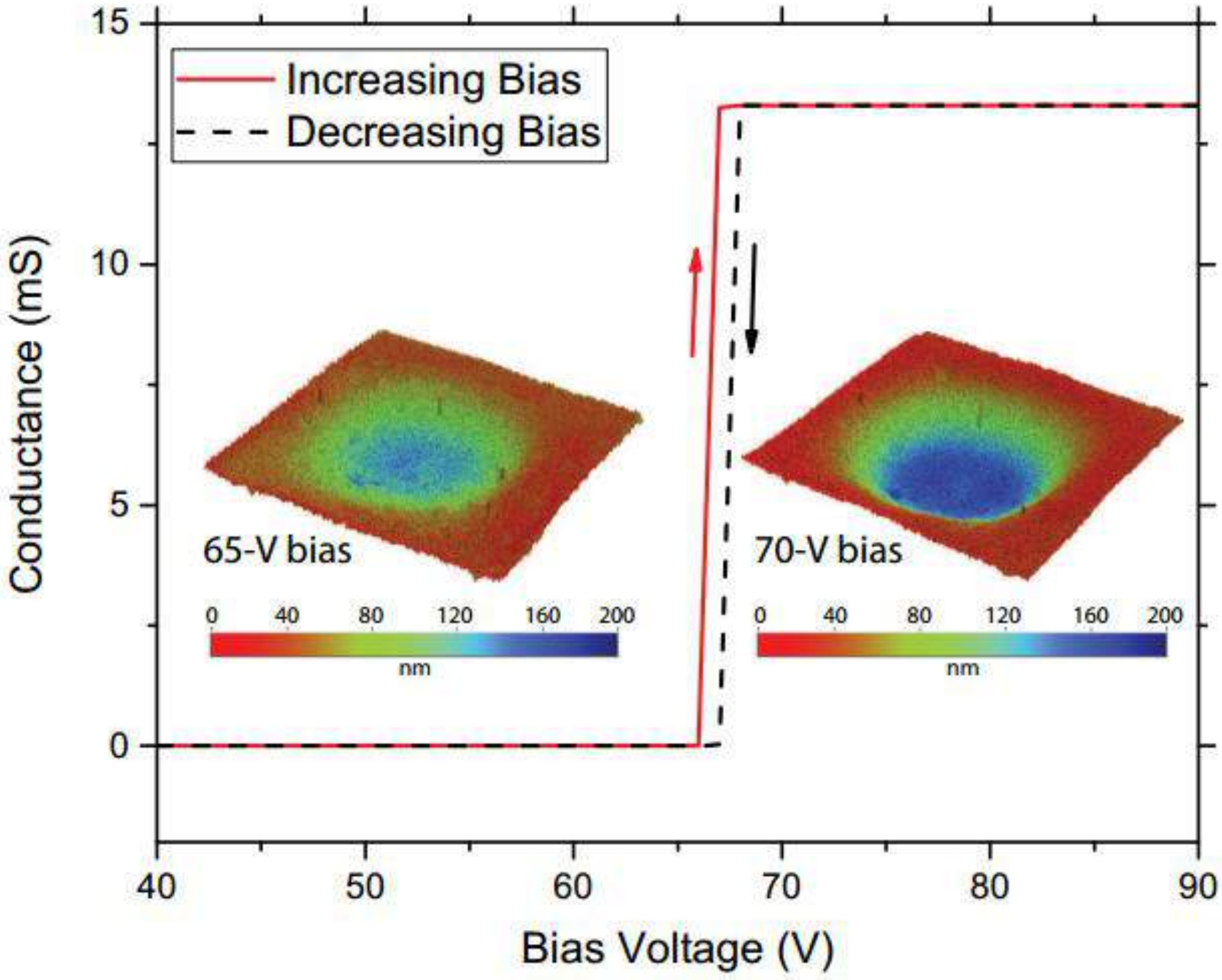

**Figure 5-14: Steady-state dc measurement.**

### 5.5.3 Dynamic Characterization

For medical ultrasound imaging applications, the MEMS T/R switches integrated next to the CMUT elements need to work in immersion, which is also beneficial for the switching speed of the MEMS switch because the medium damping will reduce the plate oscillation. A low-voltage control signal is desired for an ultrasound imaging system front end to ease the requirement of circuit design. It is also required for switch operation in order to minimize the oscillation amplitude during the switching. Therefore, we biased the switch at 67 V, which is close enough to its switching voltage (68 V). Then we increased the control signal amplitude from 0-V with a 0.1-V step. We found that the switch could be operated at a control signal amplitude of 2.5 V. Another important factor related to the switching speed is the rise time and the fall time of the control signal. We investigated this behavior using a single-point laser Doppler vibrometer (OFV-534 sensor head, OFV-5000 controller, Polytec GmbH, Waldbronn, Germany). A 10× microscope lens was used to concentrate the laser beam to a 3-µm diameter spot on the top plate over the metal bump. The DD-300 decoder with a frequency range from 30 kHz to 24 MHz was used to filter out the low-frequency noise. We used vegetable oil as the immersion medium due to the exposed pads and bonding wires. The laser was focused through the oil on the plate surface. A dual-channel waveform generator (33522A, Agilent Technologies, Santa Clara, CA) was used to deliver the control signal and also the RF signal to the MEMS switch. The control signal was generated using the pulse mode in order to enable modification of the edge speed. For the rising edge, the rise time was changed from 20 ns to 1 µs. The control signal and the $RF_{out}$ signal were recorded using an oscilloscope (MSO-X 3024A, Agilent

Technologies, Santa Clara, CA). In the meanwhile, the vibration of the plate was measured using the vibrometer. Figure 5-15 shows the recorded RF signal and the corresponding plate displacement for a rise times of 100-ns, 200-ns, and 300-ns. The measured displacement results were compared to FEM simulation result.

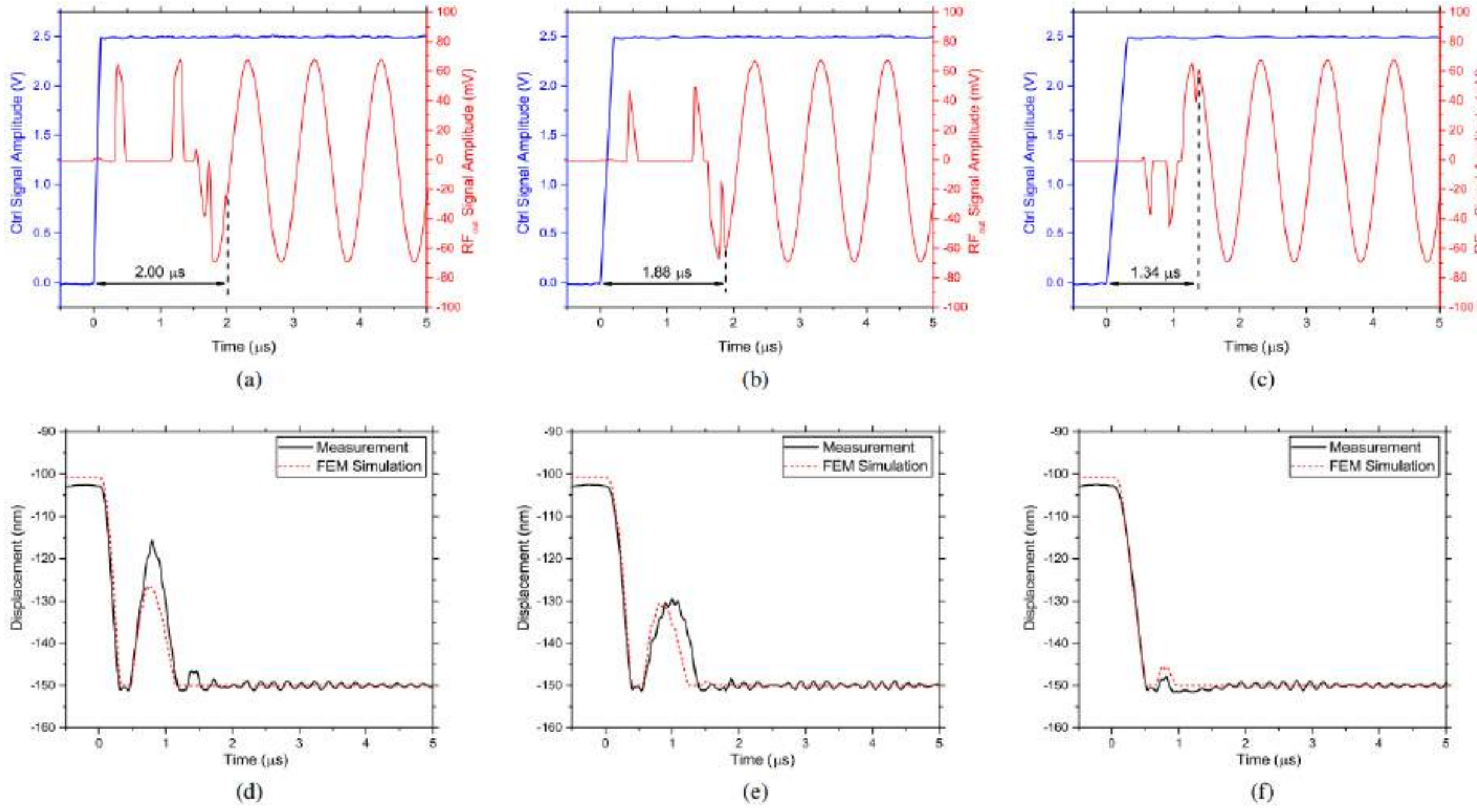

**Figure 5-15: Dynamic characterization with a control signal rise time of (a) 100 ns; (b) 200 ns; and (c) 300 ns. Figure (d)-(f) shows the corresponding plate center displacement.**

It can be observed that the switching time is first dominated by the metal bump rebound after the first contact. As the rise time increases, the rebound becomes less significant and reduces the switching time, which is due to the center frequency of the rising edge deviate from the plate mechanical resonant frequency. However, as the rise time further increases, the switching time is dominated by the prolonged rise time and therefore increase again as shown in Figure 5-16. For the switch under measurement, the optimum switching time is 1.34 μs and the corresponding rise time is 300 ns.

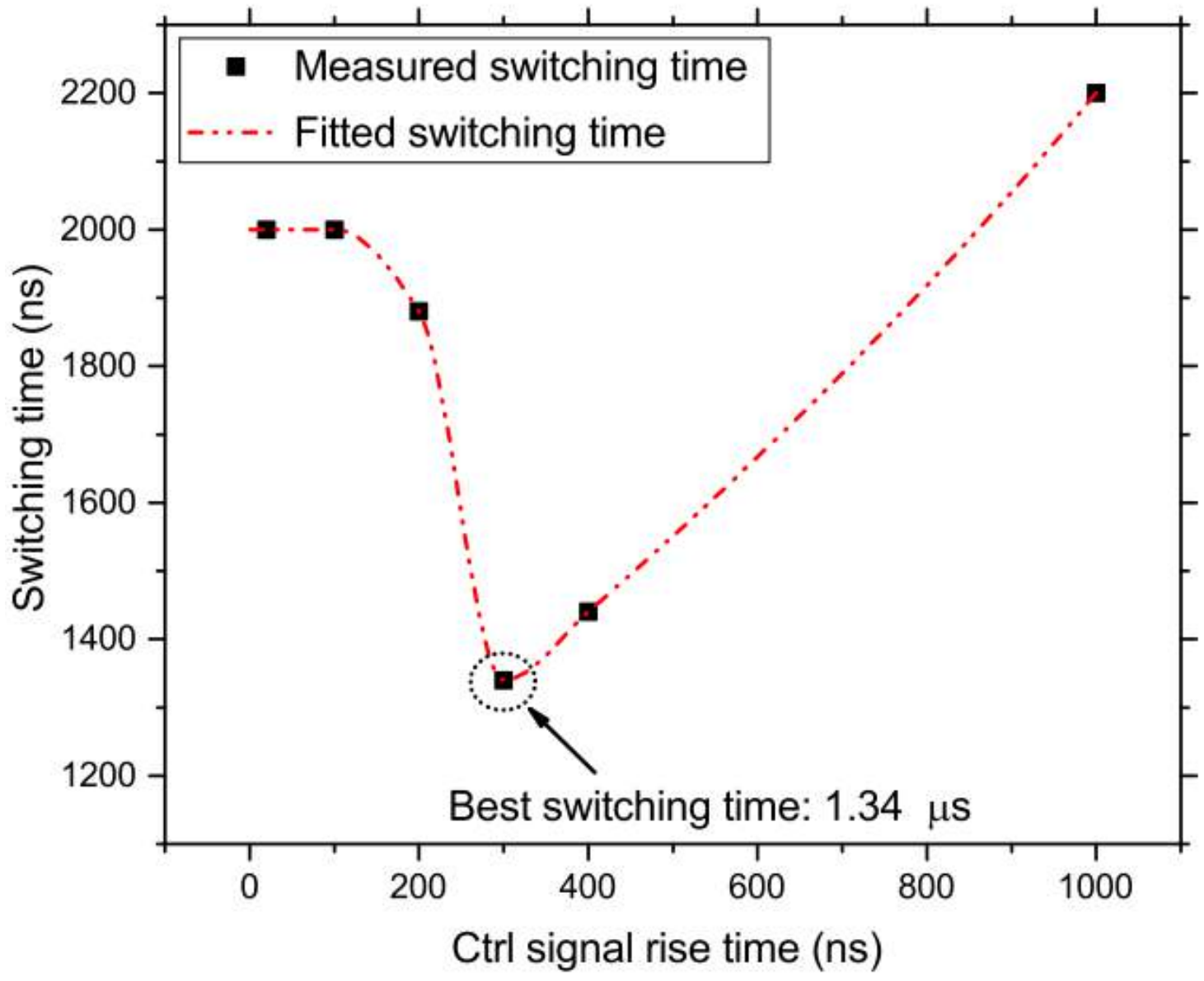

**Figure 5-16: Switch switching time versus control signal rise time.**

For the release time, the speed is only determined by the fall time of the control signal. Figure 5-17 shows the recorded RF signal and its corresponded displacement compared to the FEM simulation for the fall time of 100 ns, 200 ns, and 300 ns, respectively. Figure 5-18 shows the relationship between the fall time and the release time. It can be observed that the fall time and the release time have a linear relationship. Therefore, a fast fall time is desired for a fast switching off. For the switch under test, the optimum release time is 80 ns while the fall time is 20 ns.

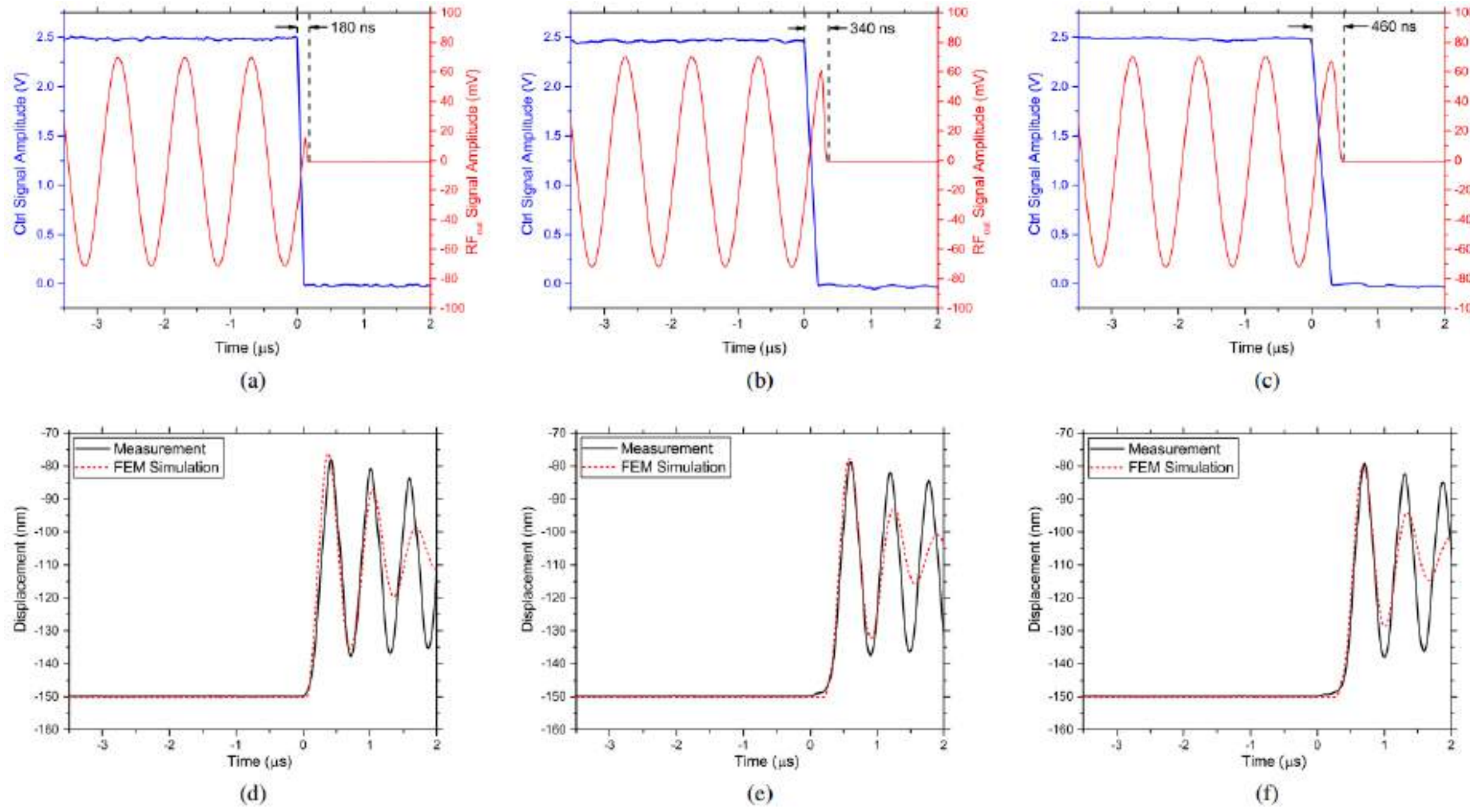

**Figure 5-17: Dynamic characterization with a control signal fall time of (a) 100 ns; (b) 200 ns; and (c) 300 ns, respectively. Figure (d)-(f) shows the corresponding center displacement.**

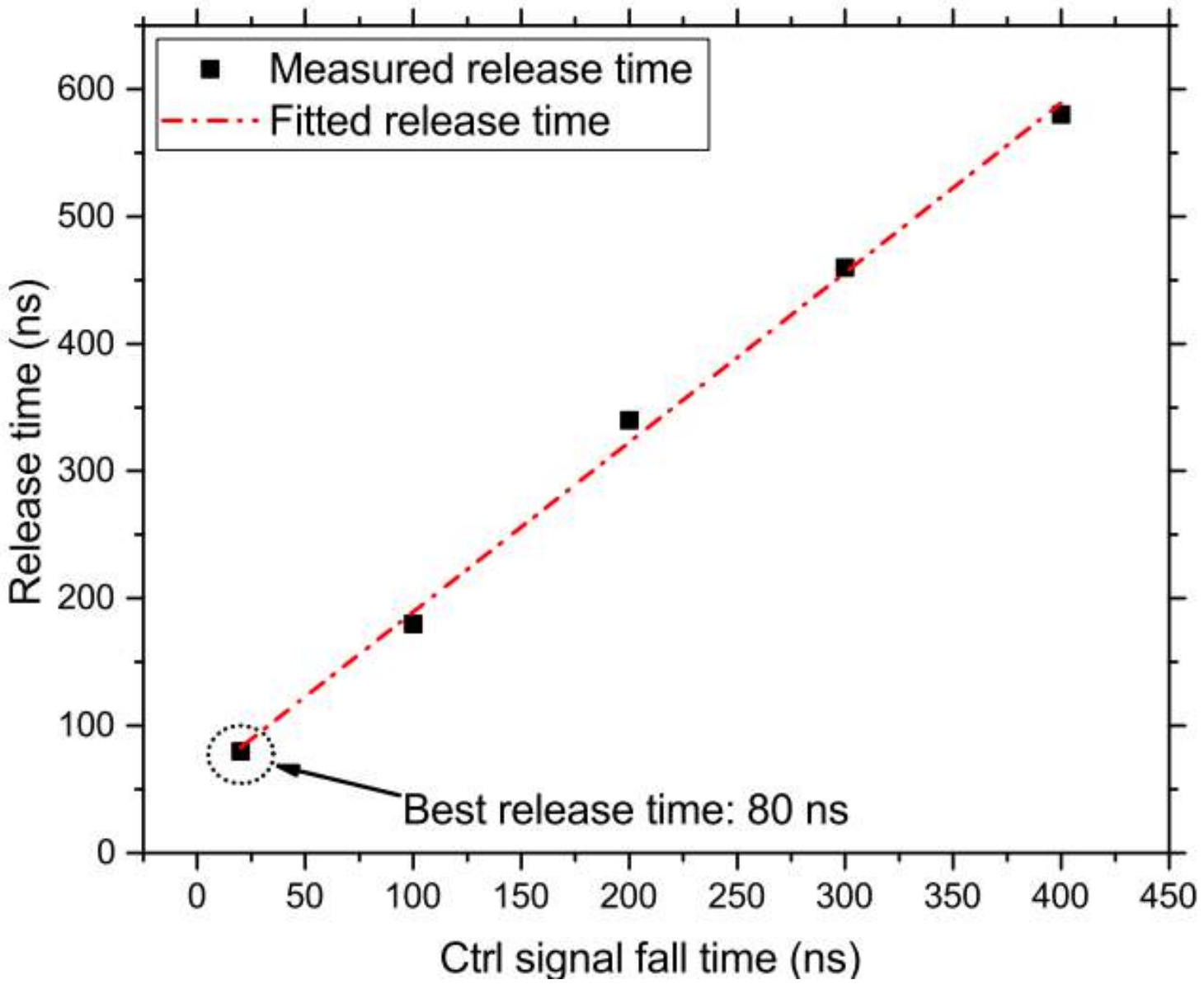

**Figure 5-18: Switch release time versus control signal fall time.**

Finally, we used a 2.5-Vpp, 1-kHz square wave as the control signal with the optimum rise time and fall time. A 1-MHz, 300-mVpp continuous sinusoidal wave was first used as the RF input. Figure 5-19(a) shows the measured output signal. Then we used a 1-MHz, 5-Vpp unipolar pulse wave as the input and the output is shown in Figure 5-19(b). The results demonstrate that the fabricated MEMS switch is suitable for ultrasound imaging system operating frequency.

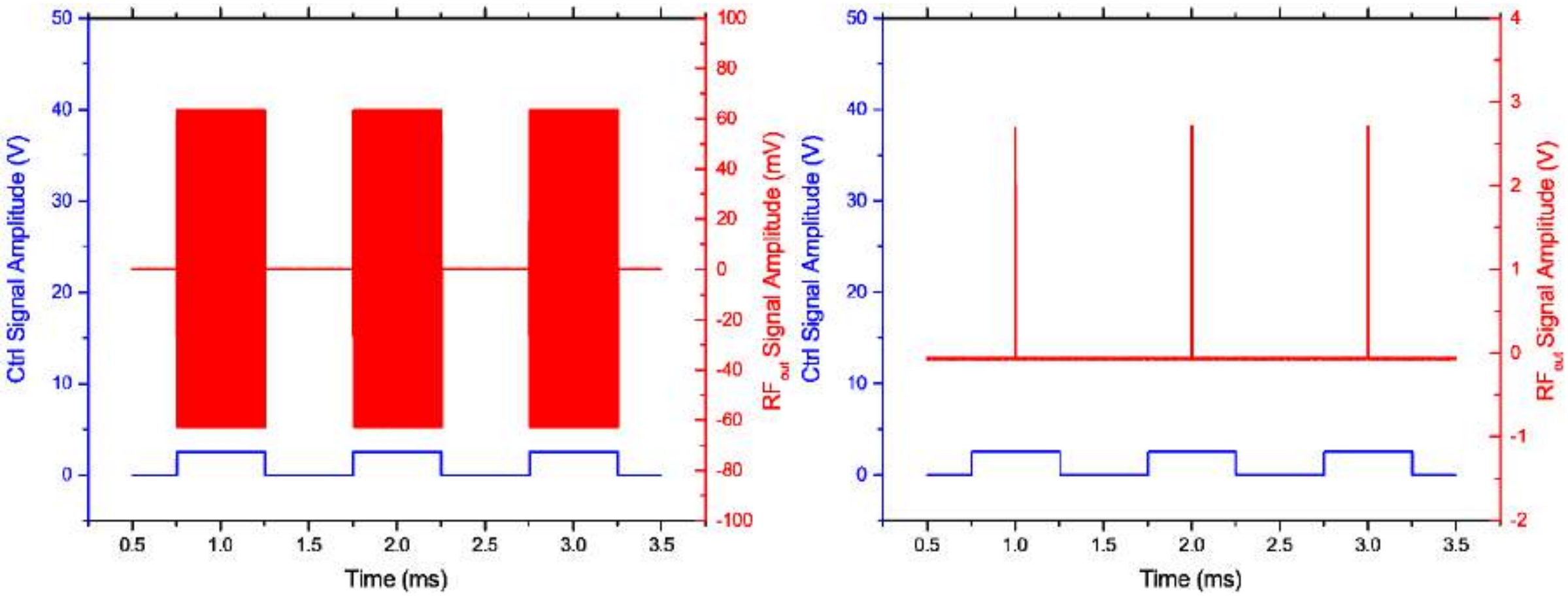

(a) (b)

**Figure 5-19: (a) 1-MHz, 300-mV$_{pp}$ CW input signal turned on and off by a 1-kHz control signal. (b) 500-ns, 5-V$_{pp}$ pulse signal turned on and off by a 1-kHz control signal.**

## 5.6 Chapter Conclusions and Future Work

We designed and fabricated a MEMS T/R switch that could be co-fabricated and closely integrated with a CMUT element on the same glass substrate using anodic bonding [79], [80]. A switch test structure was characterized. The static deflection measurement agrees well with the FEM simulation results, confirmed that fabricated switch is vacuum-sealed and has a good dimension and stress control. The static measurement demonstrates the switch has a switching voltage of 68 V and a contact resistance of 50 Ω, which could be reduced by optimizing the switch geometry. We further investigated into the switching mechanism using a dynamic characterization. The result shows the switching time is first dominated by metal bump rebound and then dominated by the control signal rise time. An optimum rise time exists for the fastest on-switching. For the switch under test, the best switching time is 1.34 µs with 300-ns control signal rise time. The release time is only dominated by the fall time. A shorter fall time is desired for a fast release. A release time of 80 ns has been demonstrated with 20-ns control signal fall time. We also demonstrated that the MEMS switch could be operated by a 2.5-V, 1-kHz control signal to conduct and block a 1-MHz, 300-m$V_{pp}$ continuous sinusoidal wave as an input signal, and 500-ns , 5-$V_{pp}$ unipolar pulse wave as an input signal. This work demonstrates that the MEMS T/R switch can meet the operating frequency requirement in an ultrasound imaging system.

The fabricated switch needs to be tested with the co-fabricated CMUTs to demonstrate its capability in an ultrasound imaging system, which will be investigated in the future. There are several aspects of the MEMS switch that could be further improved. First, a thicker metal bump could be used to reduce the series resistance of the switch, and thus improve the power handling capability, which is especially important for transmitting high-pressure ultrasound and also applications where a high-power continuous wave is needed such as HIFU. Also, some other metal, such as Pt and Pd, could be employed to improve the power handling and reliability of a MEMS switch [81]–[84]. Second, the gap height should be reduced so that the dc bias voltage can be minimized, which will also reduce the plate oscillation amplitude and make the switch faster. Besides, the switch cell could be designed at a higher frequency for an even faster-switching speed and integrate with high-frequency CMUT arrays as reported in Chapter 3. Furthermore, the proposed MEMS switch structure is compatible with 2D CMUT array fabrication process that we reported in Chapter 4 and could enable reconfigurable transducer array [85].

# Chapter 6 TRANSPARENT CMUTS ON GLASS SUBSTRATES

## 6.1 Applications of Transparent Transducers

The combination of ultrasonics and optics is desired in many applications such as integrating ultrasound sensing with flat panel displays, embedded optical vibrometry, and photoacoustic imaging [86][32]. The opacity of materials used in conventional piezoelectric transducer constructs has severely restricted such applications. Lithium niobate with indium tin oxide (ITO) electrodes has been investigated to build a transparent piezoelectric transducer [87]. However, transparent CMUTs are desired because of the benefits such as ease of fabrication and integration, and broad bandwidth. Our platform process enables fabrication of CMUT on a glass substrate that has a high optical transmittance. A transparent CMUT structure can be realized by replacing the metal electrode with transparent electrodes.

Indium tin oxide (ITO) has high optical transparency and electrical conductivity and hence has been used as a transparent conductor in devices such as light emitting diodes (LED) and liquid crystal displays (LCD) [88]. In this work, we use ITO as the bottom electrode of CMUTs to improve the optical transparency.

The CMUTs with ITO bottom electrodes are fabricated based on our anodic bonding process. We had also applied the process to make CMUTs that has silicon nitride instead of metal on the silicon plate [55]. Glass substrate and silicon nitride have good transparency. The thin silicon plate has some limited transparency in the red to NIR wavelength range. The major limitation in the total transmission through the CMUT structure we have reported in our previous work was the metal (Cr/Au) bottom electrodes. In this work, we use ITO as bottom electrode material instead of Cr/Au. In the following section, the fabrication process and the optical and acoustical characterization results are first presented. Then, we use the device for experiment backward-mode photoacoustic imaging. In section 6.4, we further improved the transparency by replacing the silicon plate with a glass plate. The device is designed for applications where optics and acoustics need to be combined, such as integrating an ultrasound sensor under a flat panel display.

## 6.2 CMUT with ITO Bottom Electrodes for Improved Transparency

### 6.2.1 Device Fabrication:

The simplified fabrication process flow is shown in Figure 6-1, which is similar to the platform process reported in Chapter 3 with replacing the metal bottom electrode to ITO electrode. CMUT cavities are patterned after cleaning the wafer. The wafer was baked at 125ºC between the wet etching cycles to prevent photoresist from peeling off. The BOE etching process has a lateral-to-vertical etch rate ratio of 10:1. Therefore a 3.5-µm undercut can be achieved during the cavity formation, which is beneficial for the later ITO lift-off step Figure 6-1(a). The wafer was then transferred into an RF sputtering system without removing the photoresist. An ITO film with a thickness of 170 nm was sputtered over the wafer in ambient temperature and then lifted off in a heated NMP solution. Then the wafer was annealed at 450ºC for 5 minutes in a rapid thermal annealing (RTA) system to improve the ITO conductivity and transparency [Figure 6-1(c)]. Figure 6-2 shows optical image and AFM image of the ITO bottom electrodes defined in the glass cavities.

The SOI wafer and the processed glass substrate were bonded together by anodic bonding. The top plate was formed after the handle wafer and BOX layer removal [Figure 6-1(d)]. The silicon plate was etched at the bottom pad location to evacuate the gas generated during anodic bonding [Figure 6-1(e)]. The device was sealed using a 1-µm conformal PECVD silicon nitride [Figure 6-1(f)]. The sealing nitride was etched to reach the conductive top plate silicon and the bottom electrode to form electrical contacts Figure 6-1(g). The device fabrication was completed after evaporating and lifting off a stacked layer of 20-nm chromium and 180-nm gold as the bond pads [Figure 6-1(h)].

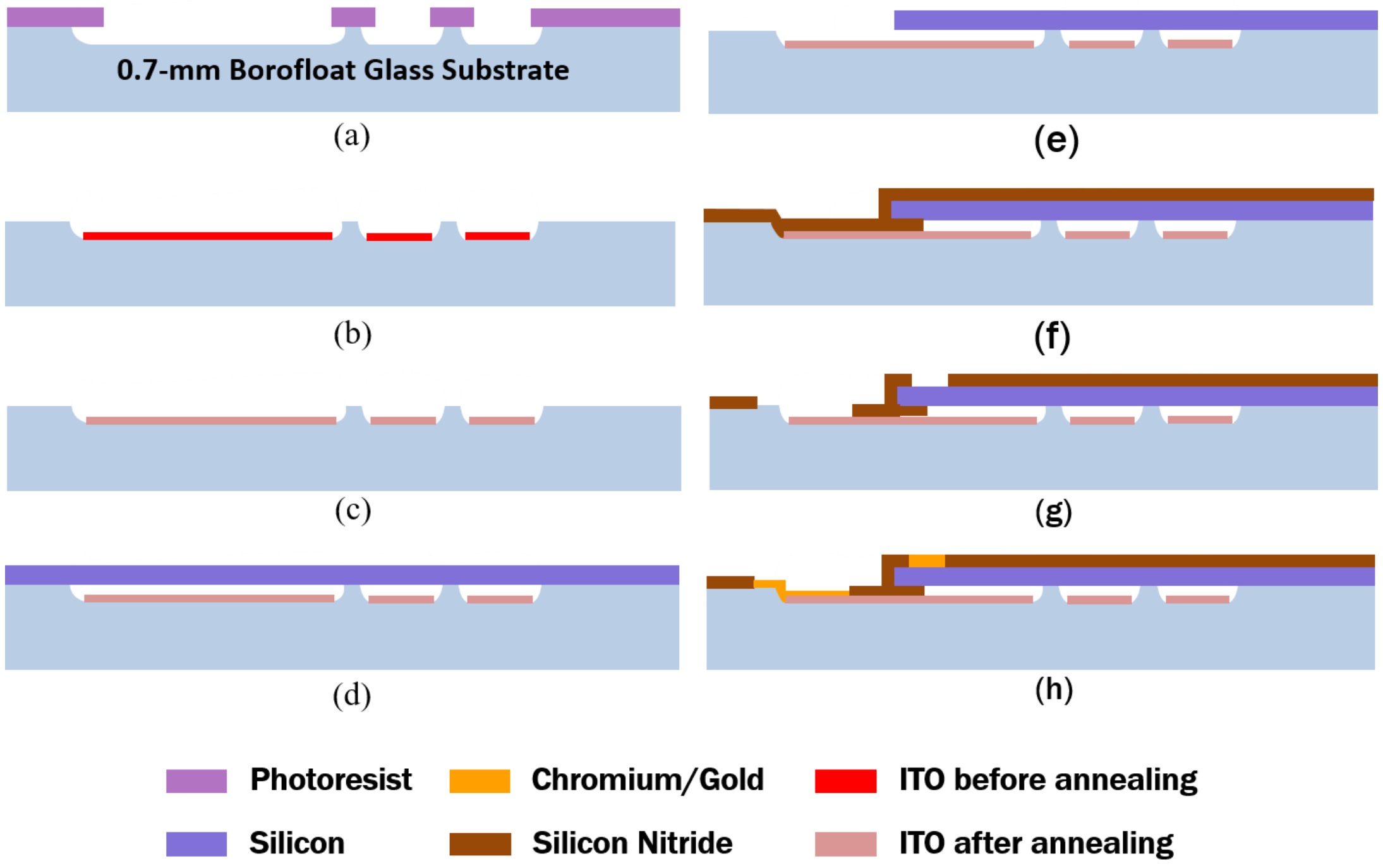

**Figure 6-1: Simplified fabrication process flow.**

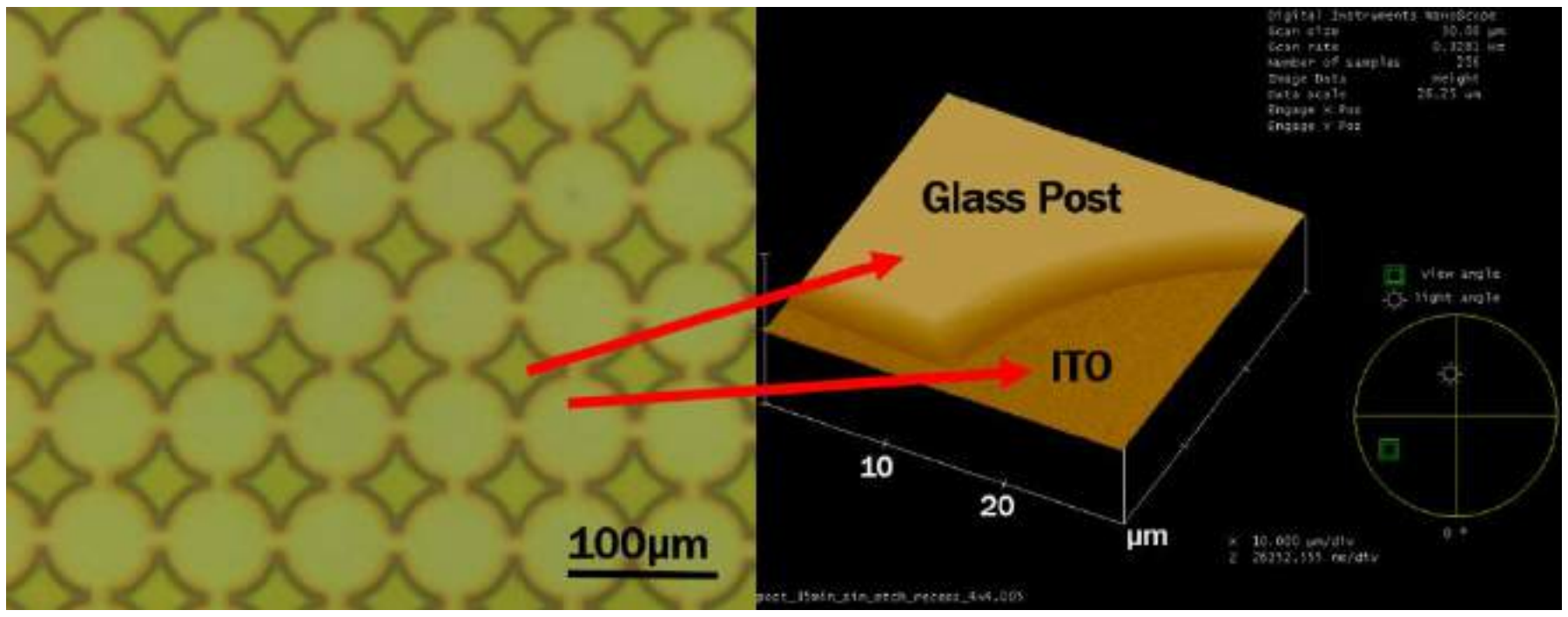

**Figure 6-2: Optical image after the ITO bottom electrode definition (left). AFM image showing the profile of the glass post and ITO bottom electrode (right).**

Figure 6-3 shows the optical image of a finished CMUT element with ITO bottom electrode and circular cells with a diameter of 78 µm and a plate thickness of 2.5 µm. The atmospheric deflection measurement in the center of a cell confirmed the devices are vacuum-sealed.

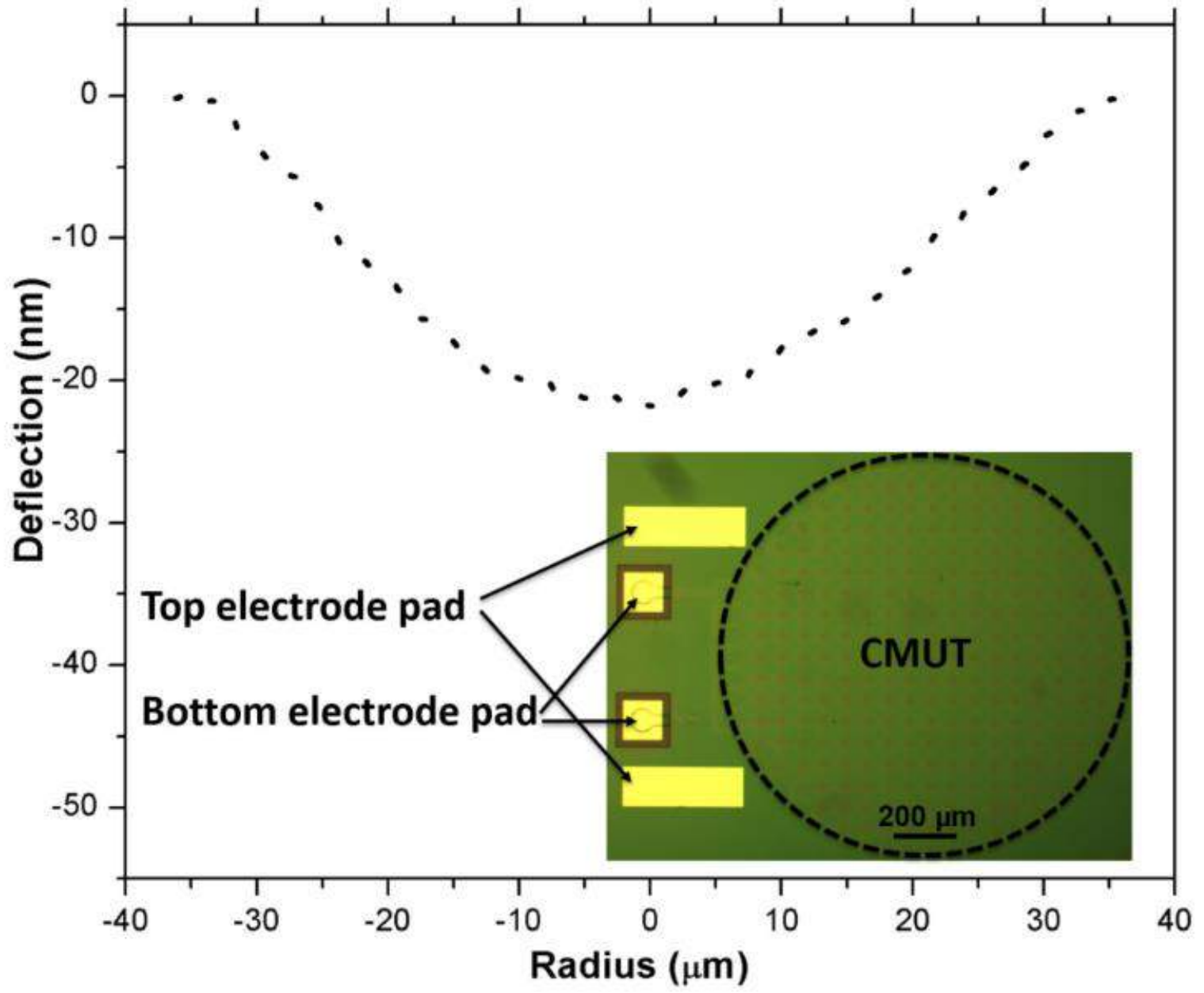

**Figure 6-3: Atmospheric deflection measurement of a finished CMUT element. The inset shows a finished CMUT element.**

### 6.2.2 Device Characterization:

Figure 6-4 is the optical images of the CMUTs fabricated with ITO bottom electrodes (left) and metal electrodes (right) under a microscope with backside illumination. The die with six CMUT elements was placed above an “NC STATE UNIVERSITY” logo. From the optical image, it is clear that the device with ITO bottom electrode has an improved transparency in visible light range.

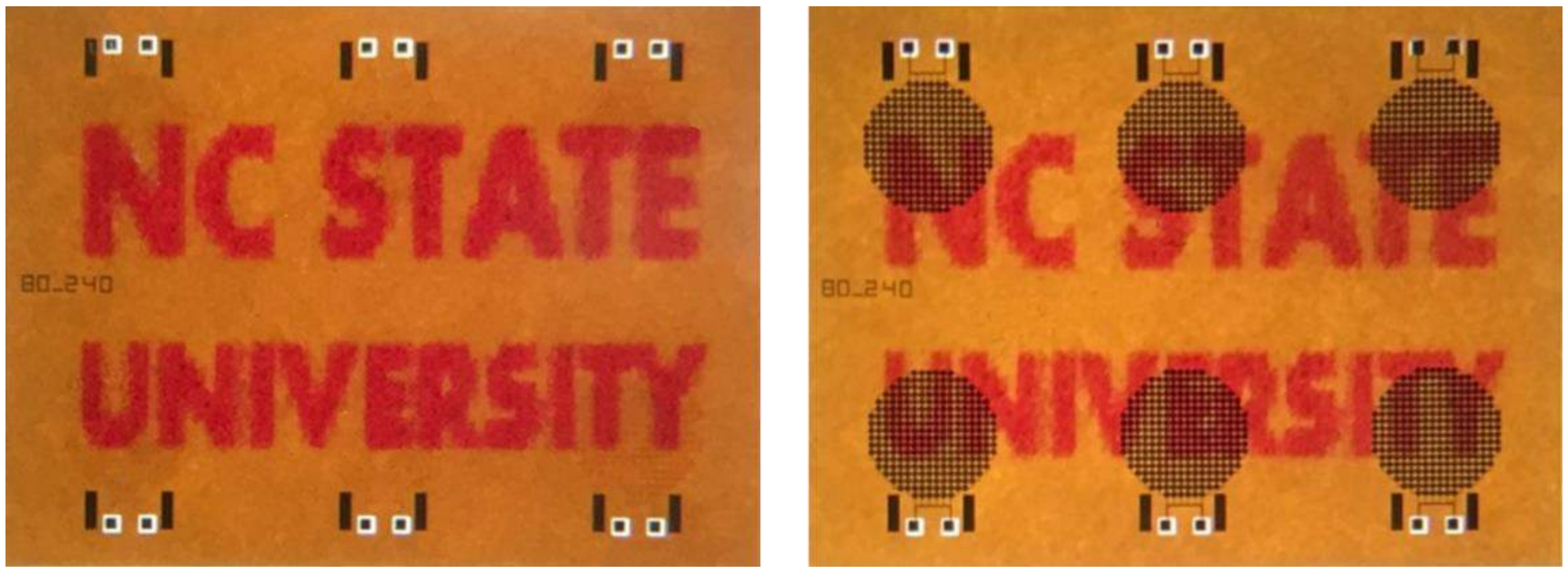

**Figure 6-4: (a) Six CMUT elements with ITO bottom electrodes show the improved transparency of the bottom electrodes. (b) Six CMUT elements with Cr/Au bottom electrodes presented for comparison.**

To further characterize the transmission coefficient, the ITO and Cr/Au bottom electrodes and the finished devices were measured using a spectrophotometer (Cary 60 UV-Vis, Agilent Technologies, Santa Clara, CA) from 400 nm to 1000 nm wavelength range. Figure 6-5(a) shows the transmission through the 150-nm ITO bottom electrode on the glass before and after annealing, in comparison to 150-nm Cr/Au bottom electrode on the glass. Figure 6-5(b) shows the optical transmission through the final device with ITO bottom electrodes and Cr/Au electrodes. It is clear that the CMUTs with ITO bottom electrodes have a significant transmission improvement in the measured wavelength range. However, the 2-µm silicon plate is the main hurdle for the transmission in the shorter wavelength range.

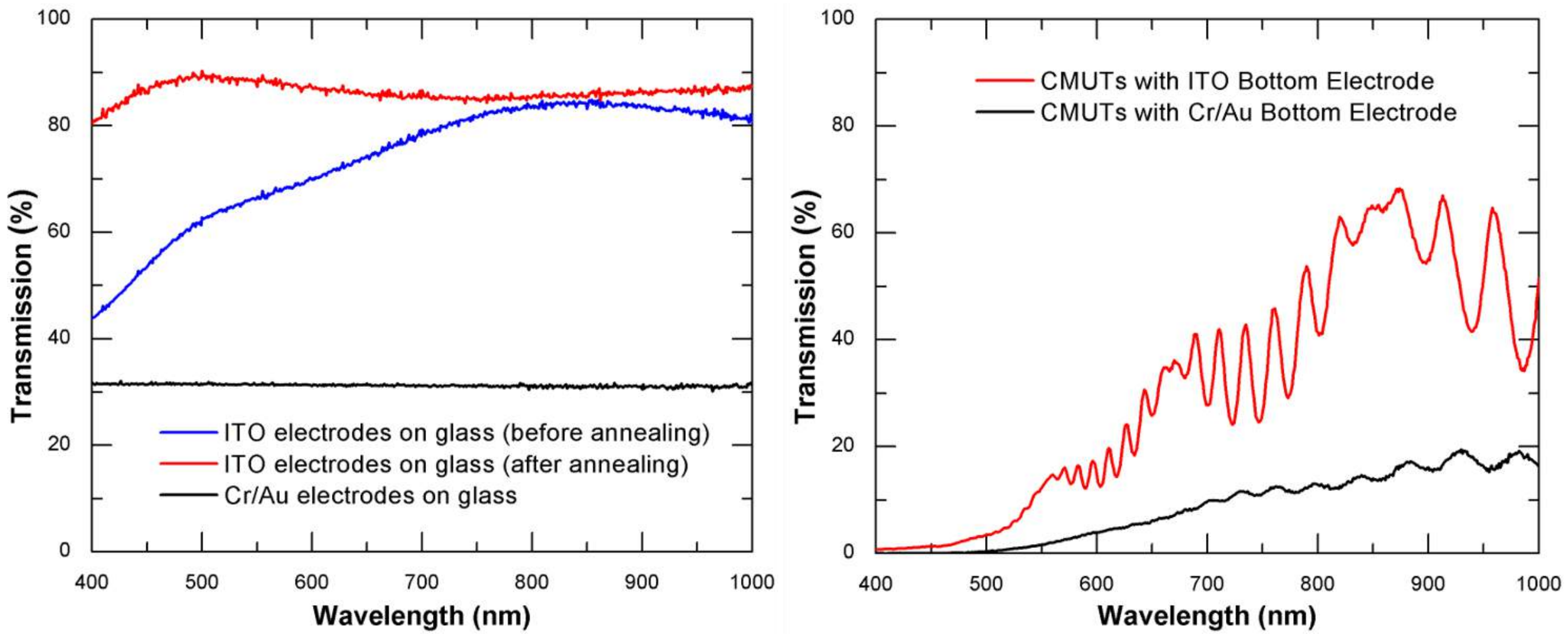

(a) (b)

**Figure 6-5: (a) Transmission measurement. Transmission of the bottom. (b) Transmission of the final devices.**

The real and imaginary parts of the electrical input impedance of the fabricated CMUT was measured in air using a network analyzer (Model E5061B, Agilent Technologies, Inc., Santa Clara, CA) [Figure 6-6(a)] . The open circuit resonance frequency of the CMUT element was measured as 3.62 MHz at 18-V dc voltage, which is approximately 75% of the pull-in voltage. The 1-kΩ baseline in the real part corresponds to the series resistance of the device, which is mainly contributed by the resistance of the patterned ITO bottom electrode.

This resistance could be reduced by depositing a thicker ITO layer or using parallel connections to the pads. For the PAI experiment, we wire-bonded two connections to two pads reaching the ITO bottom electrode to reduce the series resistance.

A pulse-echo measurement was performed in vegetable oil to characterize the small-signal bandwidth in immersion and also to help quantify the effects of optical absorption in the silicon plate on the generation of spurious transmit signals. The CMUT was placed 1.2 cm away from a plane reflector and was biased at 18-V dc voltage (75% $V_{pull\text{-}in}$). A 1-V, 250-ns pulse was used to excite the device. The received echo signal and its Fourier transform are shown in Figure 6-6(b). The center frequency of the CMUT is 1.4 MHz with a 6-dB fractional bandwidth of 105%.

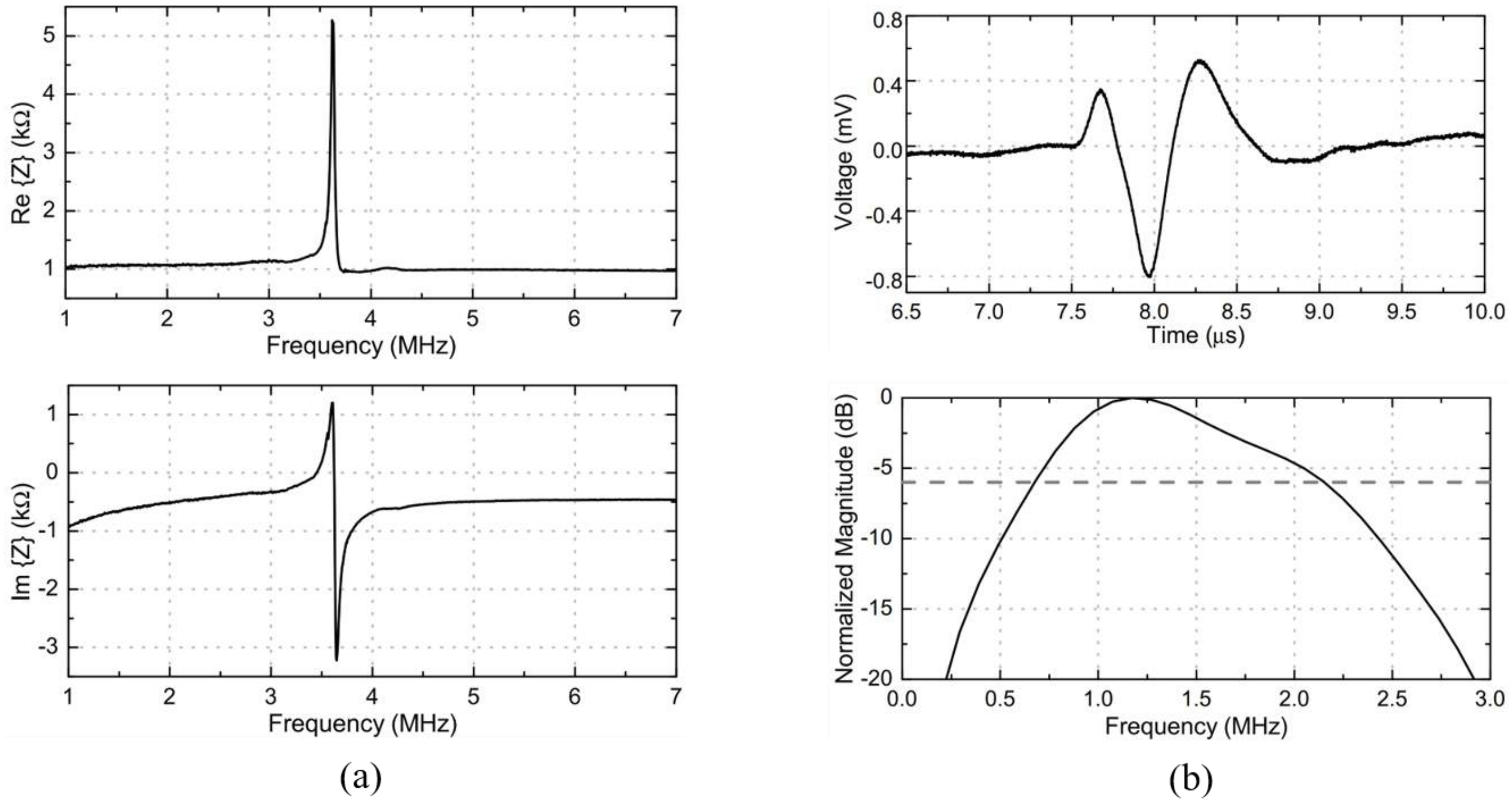

**Figure 6-6: (a) Electrical input impedance measurement in air. (b) Pulse-echo measurement in vegetable oil.**

## 6.3 Backward-mode Photoacoustic Imaging

### 6.3.1 Background of Photoacoustic Imaging:

Photoacoustic imaging (PAI) is emerging as an attractive hybrid imaging modality that combines the physics of optical imaging and ultrasound imaging [89]. It is a complement to the traditional pulse-echo ultrasound imaging since it can provide optical contrast information. Compared to optical imaging techniques, PAI provides excellent spatial resolution with deeper penetration primarily determined by ultrasound wavelength [21].

An important challenge in PAI is the arrangement of the laser source and the ultrasound transducer. Various approaches have been demonstrated for different applications [90]. One of the commonly used arrangement is to have the light source illuminate the target from the single side or two sides at a right angle to the acoustic path [91]. Another implementation is to have the light source integrated as two fiber bundles along the length direction of a 1D transducer array [92], [93].

However, this method does not illuminate the surface area under the transducer array and results in a blind spot in front of the transducer. Also, it is difficult to use this approach with 2D arrays, which would result in a larger area under the 2D transducer array not being illuminated. This approach also limits the footprint of the imaging probe as the fiber bundle and the closure required to place it next to the ultrasound array occupy extra space. The small footprint is especially important for intracavital probes. The light illumination has also been distributed using a spherical mirror [94]. Another approach a is a face-to-face arrangement which is also referred to as forward-mode [95]–[97], which is not practical for many clinical applications.

In some diagnostic imaging and image-guided therapeutic applications, it is desirable to have the light source integrated behind the ultrasound transducer for implementing the so-called backward-mode PAI. First of all, this approach would allow a more uniform illumination in the near field. Also, it could lead to a compact form factor, which is important for therapy monitoring applications using a catheter-based imaging system. One approach to implement the backside illumination is to use a ring array or annular array and introduce an optical fiber bundle into the central lumen [93], [98], or place a forward-viewing ultrasound catheter in the central lumen of an annular light ring [99]. However, the former arrangement limits array geometry and illuminated area and the latter results in a larger catheter size. The ideal implementation is to have a transparent transducer which allows the laser illumination to go through with minimal self-absorption. Some transparent piezoelectric materials such as polyvinylidene fluoride (PVDF) and lead lanthanum zirconate titanate (PLZT) have been proposed for such purpose [100], [101]. A PVDF annular array has been adopted with through-hole illumination for PAI [102]. Also, transparent Fabry-Perot polymer film optical ultrasound sensors and optical micro-ring resonator (MRR) have been investigated for implementation of backward-mode PAI [103], [104]. However, there are only limited reports on using CMUTs for such an operation. An earlier attempt was based on CMUTs fabricated on a silicon substrate [86], where transmission is limited and the wavelength of the light source has to be above 1000 nm to reduce the silicon substrate absorption.

The above-described CMUT on a glass substrate with indium tin oxide (ITO) bottom electrode improved the device transmission to 30% to 70% in the wavelength range from 700 nm to 900 nm, which is commonly used for in-vivo photoacoustic imaging. We will use this device in immersion and have laser illumination through the fabricated CMUT to demonstrate preliminary photoacoustic imaging results.

### 6.3.2 Experimental Setup:

The diagram of the experimental setup for photoacoustic imaging is shown in Figure 6-7. The CMUT die was mounted on a PCB that has a bias-T circuit and switches to select individual CMUT elements for testing. The PCB was designed with a rectangular cutout in the center to allow light to pass through the CMUT die, as shown in the inset of Figure 6-7(b). After wire bonding, the CMUT and the laser output were fixed using a 3D-printed holder and mounted on a 3-axis linear stage (model PRO165, Aerotech Inc., Pittsburgh, PA, USA) to enable mechanical scanning. The 3D-printed holder was used to ensure the relative position of the CMUT and the laser output does not change during the experiment. A dc power supply (model E3631A, Agilent Technologies, Santa Clara, CA) was connected to the PCB and the signal received by the CMUT was filtered and amplified by a receiver (model 5072PR, Olympus Corporation, Tokyo, Japan). The filtered and amplified signal was recorded by a PC-controlled digitizer (model NI PCI-5124, National Instruments, Austin, TX). The excitation laser source is a fiber-coupled optical parametric oscillator (OPO) pumped by a Q-Switched Nd: YAG laser (model Phocus Mobile, Opotek Inc., Carlsbad, CA) with a wavelength range from 690 nm to 950 nm. The laser pulses had a pulse-width of 4.5 ns and a repetition rate of 20 Hz. The output energy of the laser was calibrated using a pyroelectric energy detector (model: QE25, Gentec Inc., Quebec City, Canada). The output of the laser was coupled to the backside of the CMUT die using a fiber bundle with a diameter of 5 mm. The target and the coupling medium were placed in a glass container under the PCB. The bottom view of the PCB is shown in Figure 6-8(a) with the inset graph indicating the relative location of the laser beam and the CMUT elements. We chose to use the number 2 CMUT element on the die because the light illuminated through this entire device. The side view of the fiber bundle, the PCB, and the holder is shown in Figure 6-8(b).

Two different targets were imaged. The first imaging target was a 0.7-mm diameter pencil lead. The pencil lead was suspended 2 cm above the bottom surface of the glass container. Vegetable oil was used as the medium instead of water as the transducer surface and bond wires were not electrically insulated. We filled the vegetable oil up to approximately 1.5 cm above the pencil lead. Then we lowered the holder until the CMUT surface touched the oil. The laser beam output from the fiber bundle into the CMUT chip had a wavelength of 830 nm and a fluence of 12 $mJ/cm^2$. The CMUT was biased at 18-V dc voltage. We set the receiver gain at 20 dB and the cutoff frequency of the low-pass filter at 10 MHz. The transducer was scanned across the pencil lead and the received signals at each location were sampled at a rate of 200 MSa/s, digitized, and averaged 16 times to improve the SNR before recording.

The second target was designed to better mimic the condition of biological tissues. We looped a polyethylene tube with an inner diameter of 2.3 mm and an outer diameter of 3.6 mm and filled it with an indocyanine green (ICG) solution (50-µM), which is commonly used as a contrast enhancement agent in PAI. The tube was then suspended using fishing lines inside the glass container. Then we filled up the container with a mixed solution of 1% Agar and 1% Intralipid (20% intravenous fat emulsion) in DI water to build the photoacoustic imaging phantom. After the phantom was solidified, we added a 5-mm oil layer on top of the solid phantom for acoustic coupling. The CMUT was again biased at 18-V dc voltage. This time the received signals were amplified with 40-dB gain. The laser wavelength was chosen as 790 nm to match the maximum absorption of the ICG solution. In order to get a stronger PA signal, this time we used laser output fluence of 20 mJ/cm2 into the CMUT chip. By mechanically scanning in x and y directions, volumetric data was recorded at a sampling rate of 200 MSa/s and by averaging of each scan line 16 times.

Photoacoustic images were reconstructed using the standard delay-and-sum (DAS) beamforming algorithms [105] along with a coherence factor (CF) weighting [106]. Prior to image reconstruction, every A-scan S(t) was processed as in [107]:

$$S_{processed}(t) = S(t) - t\frac{\partial S(t)}{\partial t}, \tag{6.1}$$

$$S_{processed}(i) = S(i) - i(S(i) - S(i-1)), \tag{6.2}$$

This preprocessing suppresses the low-frequency component in the signal. Then, the A-scans were filtered by a 0.15-MHz 4.5-MHz band-pass filter to eliminate out-of-band noise. After that, DAS receive-only beamforming was applied to form the PA image. Considering the radiation pattern of the CMUT and to maximize the image SNR, a threshold value of 14° was chosen and the contribution from an element was not taken into consideration if the angle from its normal to the pixel location was larger than the threshold. Finally, envelope detection was performed and the image was multiplied by the coherence factor map. Logarithmic compression was performed before displaying the PA image.

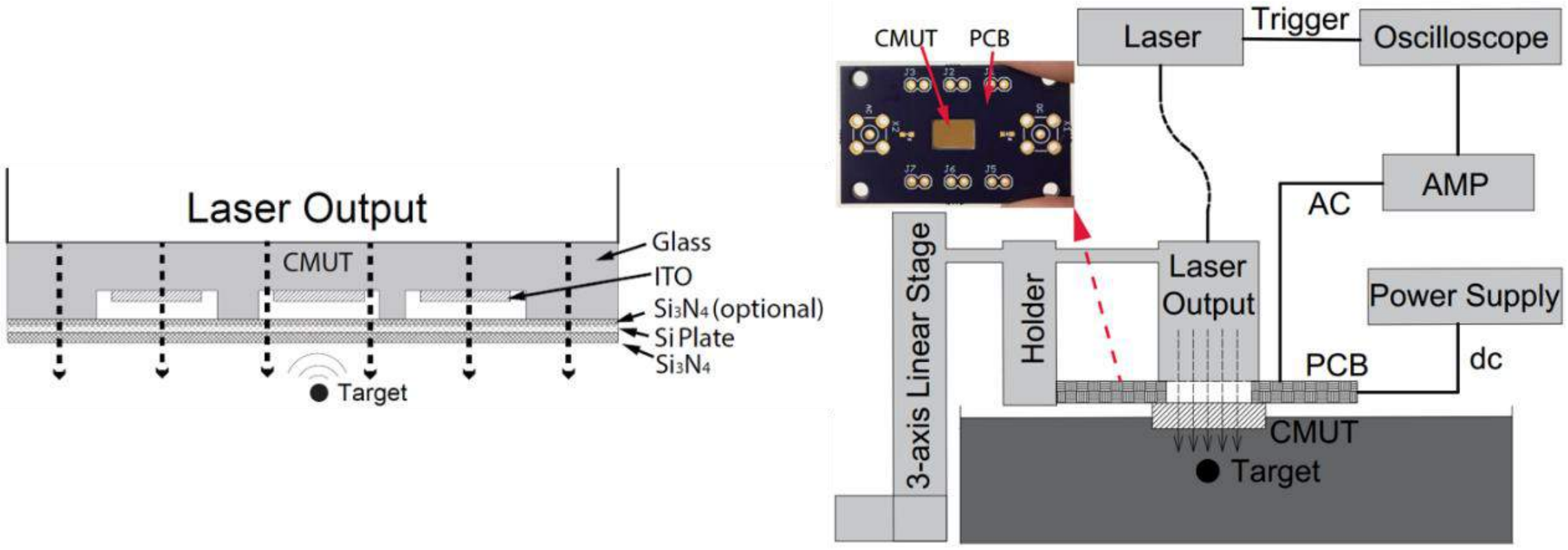

(a) (b)

**Figure 6-7: A schematic diagram of the CMUT with improved transparency**

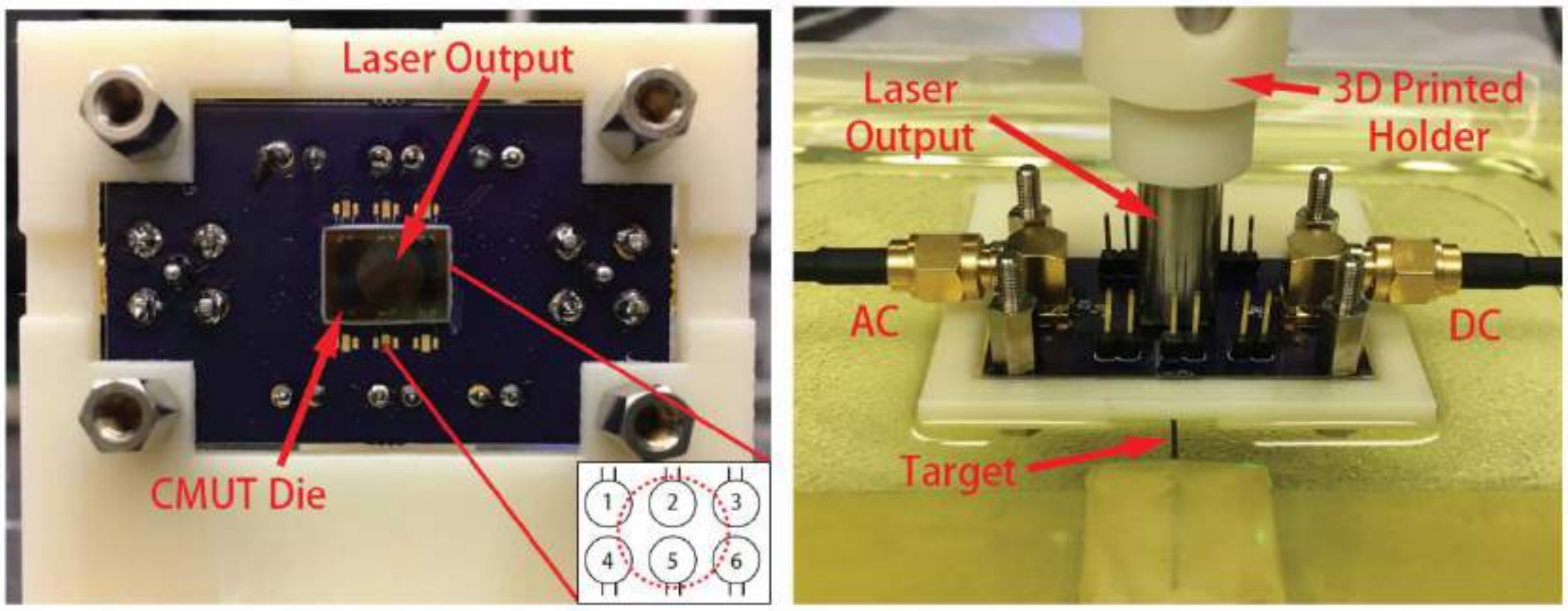

(a) (b)

**Figure 6-8: (a) Bottom view of the PCB with CMUT (inset graph indicates the relative location of the laser output and CMUT elements). (b) Side view of PCB attached on the holder with the optical fiber bundle in the back.**

### 6.3.3 Backward-mode PAI Results:

#### 6.3.3.1 Graphite Target

The experimental results of imaging of the pencil lead in oil are shown in Figure 6-9. A sample A-scan at X=0 mm is shown in Figure 6-9(a). Four signals (S1, S2, S3, S4) are marked on the A-scan with the signal paths shown in Figure 6-9(b). As the 4.5-ns wide laser pulse shines through the CMUT, some of the optical energy is absorbed in the silicon plate and converted to heat. Thermoelastic expansion of the silicon plate caused by rapid heating and cooling will set the plate into vibration at its natural frequency in oil, which results in the Sl on the A-scan. S2 is the received PA signal generated by the pencil lead. The tail signal after S2 is because of the substrate ringing of the device and the reverberations in the pencil lead. S3 is the pulse-echo signal transmitted due to S1 and reflected by the pencil lead, and therefore occurred at double the distance compared to S2. S4 is the PA signal generated by the pencil lead reflected by the silicon plate and then reflected back by the pencil lead. Therefore it appeared at three times the target depth. The reconstructed B-scan image of the cross section of the pencil lead is shown in Figure 6-9(c) with 40-dB dynamic range. The pencil lead was seen at the depth of approximately 12 mm. Substrate ringing and the reverberations in pencil lead can be observed after the main signal. At the distance of 24 mm, a weaker signal (34 dB lower than the pencil lead) was detected, which is due to the pulse-echo signal generated by the silicon plate absorption and corresponds to S3 on the A-scan.

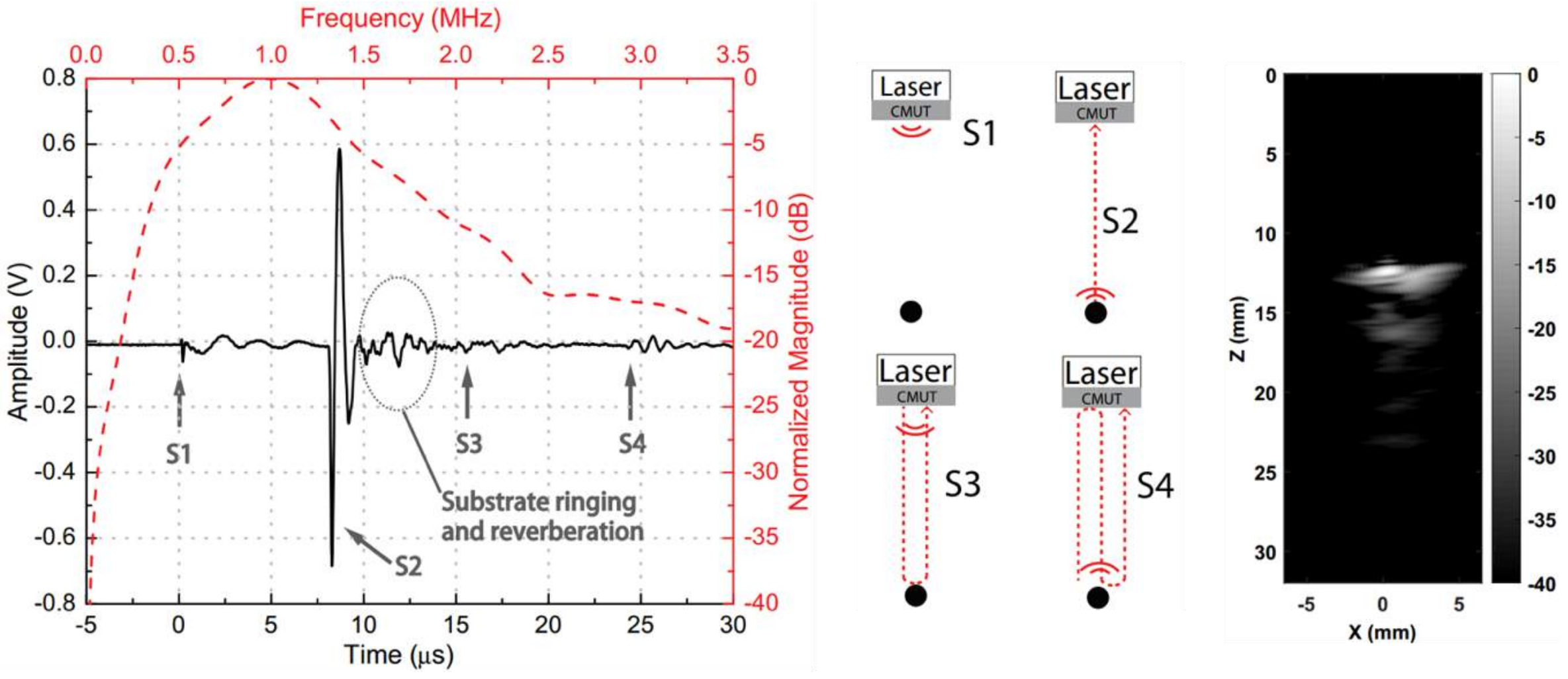

(a) (b) (c)

**Figure 6-9: Pencil lead cross-sectional PA imaging results: (a) A sample A-scan at X=0 mm and its Fourier transform after applying a Gaussian window; (b) Signal paths of the four signals on the A-scan; (c) Reconstructed image.**

6.3.3.2 Silicon Plate Absorption Spurious Signal

In order to further evaluate the effect of the silicon plate absorption, we designed the following two experiments. First, we compared the photoacoustic signals generated by the light absorption in the pencil lead (referred as $PA_{target}$ in the following context) and the pulse-echo signals generated by photoacoustically induced vibration of the silicon plate (referred as $PA_{cmut}$ in the following context) for the same travel distance in oil across the 690 nm to 950 nm wavelength range. $PA_{target}$ was first recorded as described for the imaging experiments. To record $PA_{cmut}$ also at 12-mm travel distance, we placed the CMUT 6-mm away from the glass container bottom, which served as a plane reflector without generating interfering PA signals. The laser wavelength was scanned from 690 nm to 950 nm wavelength range with a step of 10 nm. The received signals were normalized to 1-mJ/cm$^2$ laser energy through the CMUT. The results are plotted in Figure 6-10(a) with curve fitting. It can be seen that for the same travel distance, $PA_{cmut}$ is much smaller than $PA_{target}$ (approximately 30 dB lower at the wavelength of 830 nm).

In the second experiment, we compare the $PA_{cmut}$ to a pulse-echo signal generated by the electrical excitation ($PE_{cmut}$) for the same travel distance. The aim of this experiment is to find an equivalent excitation voltage for the CMUT that would generate a $PE_{cmut}$ equals to $PA_{cmut}$. We use a 250-ns, 1-V unipolar pulse to perform a regular pulse-echo test. The received echo signal amplitude was 1.5 $mV_{pp}$. Thus the equivalent electrical excitation amplitude can be calculated for 1-mJ/cm$^2$ laser excitation through the CMUT [Figure 6-10(b)]. At the wavelength of 830 nm, the pulse-echo signal generated by photoacoustically induced vibration of the silicon plate using 1-mJ laser power is equivalent to that generated by the CMUT using 0.29-V electrical excitation.

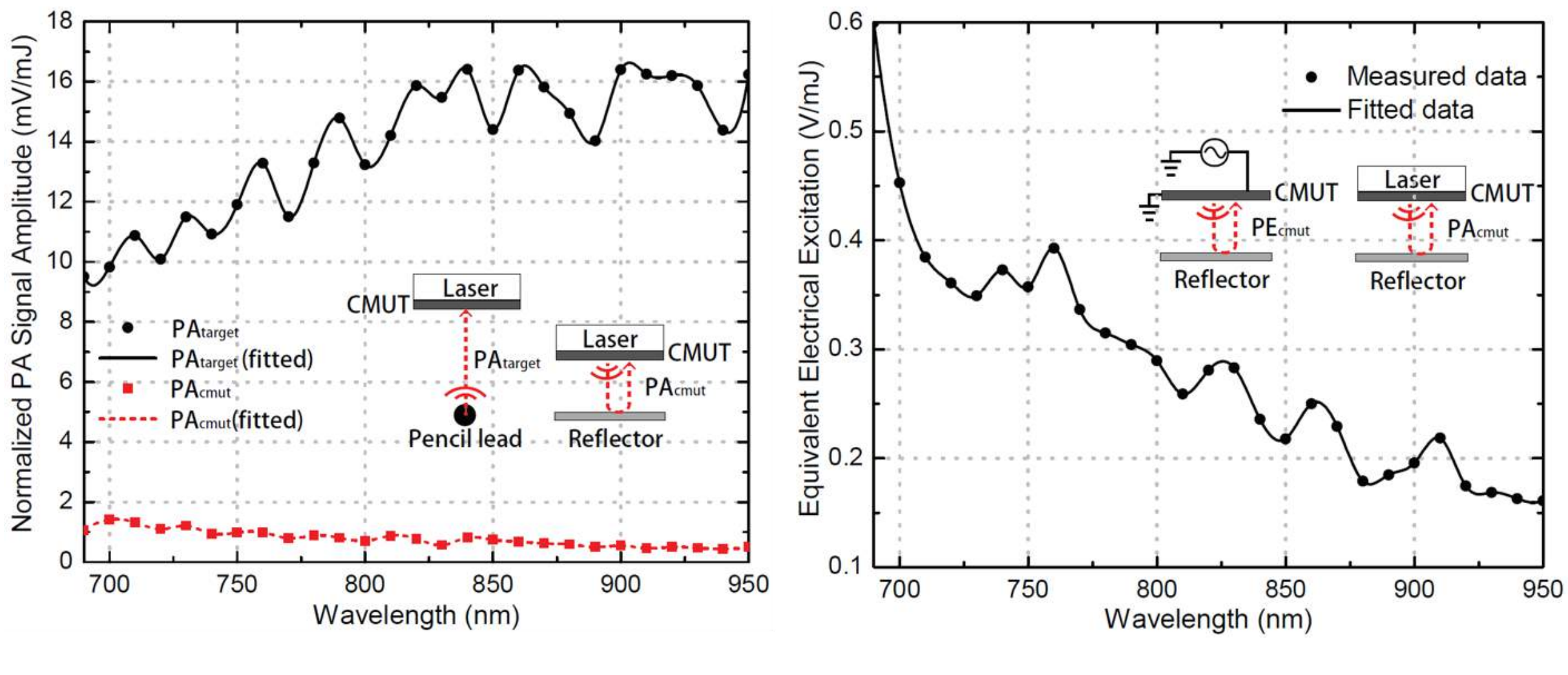

(a) (b)

**Figure 6-10: (a) Received PA signal using 1-mJ laser power at 12-mm travel distance: one-way pencil lead signal ($PA_{target}$) and two-way silicon plate signal ($PA_{cmut}$). (b) Equivalent electrical excitation amplitude for 1 mJ laser excitation.**

#### 6.3.3.3 ICG Solution Target

The photograph of the phantom is shown in Figure 6-11(a), where a looped polyethylene tube filled with the ICG solution was embedded in the tissue mimicking material and suspended using fishing lines. 3D image reconstruction was performed and then the volumetric data was rendered using a medical image viewing software (Osirix, Pixmeo SARL, Bernex, Switzerland) [108] and displayed by using maximum intensity projection (MIP) [Figure 6-11(b)]. The ICG tube and the fishing line node could be seen in the rendered 3D image.

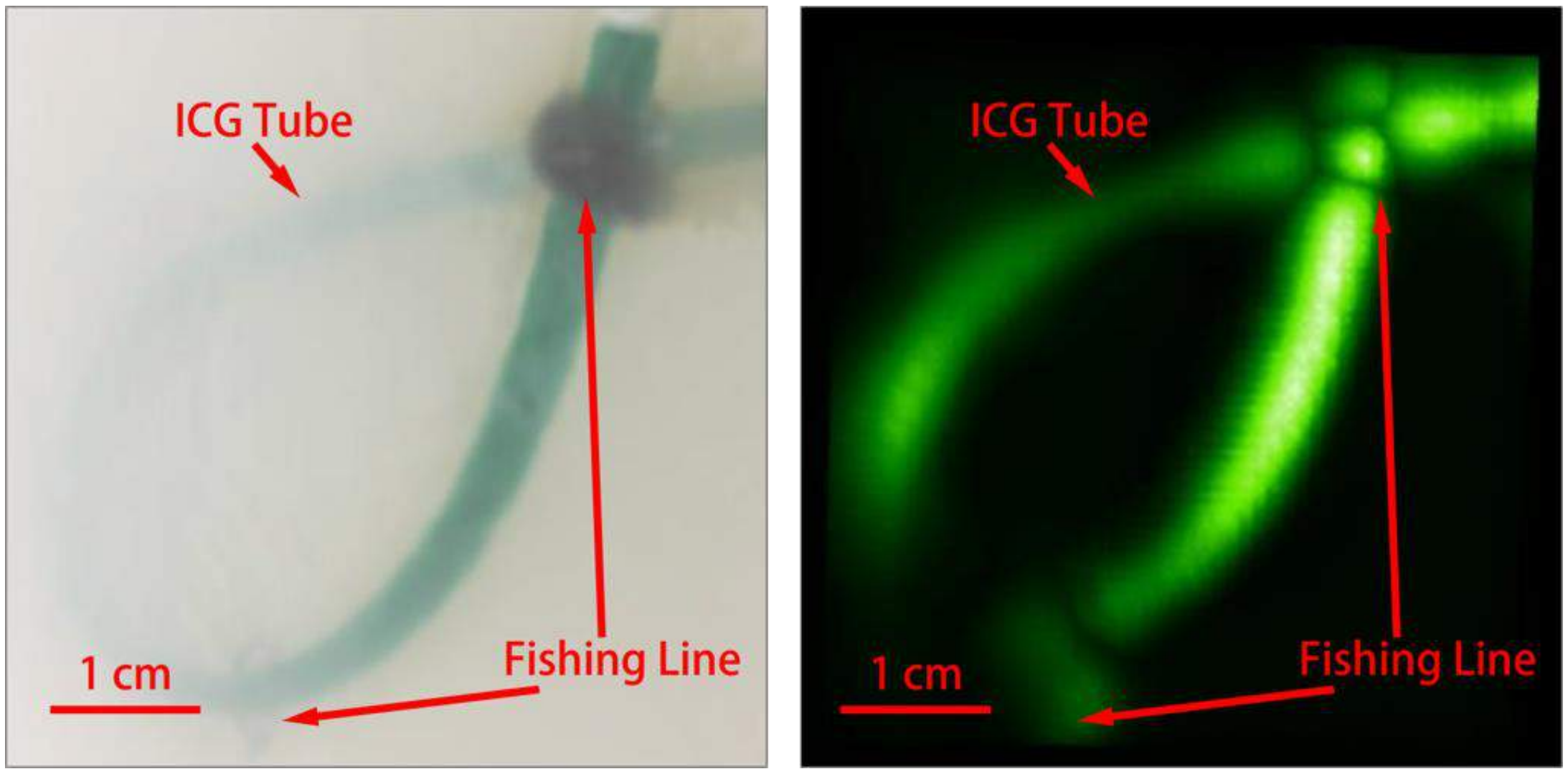

(a) (b)

**Figure 6-11: ICG tube phantom (a) Photograph of the ICG-filled polyethylene tube. (b) 3D rendered an image of the ICG-filled polyethylene tube.**

## 6.4 An Optically Transparent CMUT Fabricated Using SU-8 or BCB Adhesive Wafer Bonding

Fully transparent ultrasonic transducers are desired in consumer electronics such as integrating large-area ultrasound sensing or directed sound generation integrated with an electronic display. Transparent piezoelectric transducers have been reported using transparent piezoelectric materials, such as PVDF [100], LNO [87], and transparent PZT thin film [109]. Also, AlN-based PMUTs have been fabricated on a transparent substrate [110]. Besides, optical ultrasound sensors can be used as a transparent ultrasound receiver [111], [112].

Transparent capacitive micromachined ultrasonic transducers (CMUTs) are desired to take advantage of the wide bandwidth, ease of fabrication, and broad selection of processing materials. Transparent CMUTs with polymer-based vibrating plates have been fabricated using a sacrificial release process [32] and a roll-lamination process [113]. Wafer bonding technique has various benefits for fabrication of CMUTs. We have fabricated CMUTs on glass substrates using anodic bonding. We have also demonstrated a CMUT with indium tin oxide (ITO) bottom electrode for improved transparency for the backward-mode photoacoustic imaging application.

Here we have designed and fabricated an air-coupled CMUT to achieve high optical transparency in the full visible wavelength range by replacing the silicon plate with a glass plate. The transparent CMUT structure was realized with two ITO coated glass wafers bonded together using adhesive wafer bonding. Adhesive wafer bonding with photosensitive benzocyclobutene (photo-BCB) has been demonstrated for CMUT fabrication on a silicon substrate [114]. Here we investigate both SU-8 photoresist and photo-BCB as the adhesive bonding layer, which are also used to form the cavities and the insulation layer in the CMUT structure. SU-8 has the benefits of low cost and ease of processing. BCB has the benefits of high dielectric breakdown voltage and high bonding strength.

The transparent CMUT was designed as a single transducer on a 100-mm-diameter wafer, specifically for display-based air-coupled applications. The fabrication process flow is depicted in Figure 6-12 and described below.

The starting substrate was a 1.1-mm-thick, 100-mm-diameter glass wafer (HQ-Float Glass, Präzisions Glas & Optik GmbH) with a flat cut, coated with a 200-nm ITO film on top of a thin $SiO_2$ barrier passivation layer, which is suitable for display technology [Figure 6-12(a0)]. A patterned photoresist was used as a mask for etching the ITO bottom electrode in a heated mixture of hydrochloric acid (35% HCl), nitric acid (65% HNO3), and deionized (DI) water with a mixing ratio of 1:0.08:1 [Figure 6-12(a1)].

To form the cavities, on one wafer SU-8-5 was spun at a rate of 3000 rpm to achieve a uniform layer and sequentially patterned to create cavities aligned to the bottom ITO electrodes. On another wafer, we used a commercially available bisbenzocyclobutene (BCB) electronic resin (CYCLOTENE 4022-35, Dow Chemical Company, Midland, MI) to form cavities. An adhesion promoter (AP3000, Dow Chemical Company, Midland, MI) was first applied and the photo-BCB was spun at 2500 rpm. After exposure and puddle development, cavities aligned to the bottom ITO electrode were realized Figure 6-13(a). The cavities are electrically connected to the bottom pad but physically isolated from the atmosphere so that vacuum sealing of the device could be achieved after the wafer bonding [Figure 6-12(a2)].

The plate wafer was implemented using a 175-µm-thick, 100-mm-diameter glass (D263-T Glass, Präzisions Glas & Optik GmbH) wafer with a flat cut, coated with a 200-nm ITO film on top of a thin $SiO_2$ barrier passivation layer [Figure 6-12(b0)]. ITO etching was performed to remove part of the top ITO electrode to avoid electrical field being applied to the polymer [Figure 6-12(b1)]. For SU-8 bonding process, SU-8 2 was span at 3000 rpm to form a 1.5-µm layer on the ITO film. For BCB bonding process, BCB resin (CYCLOTENE 4022-35, Dow Chemical Company, Midland, MI) was spun at a rate of 6000 rpm to form a 2-µm layer on the ITO film. The polymer layer serves as the insulation layer for the CMUT and also the bonding material. We patterned the polymer layer so that the ITO top electrode could be exposed through the flat cut on the substrate for electrical contact. At this step, the top plate wafer was also ready for wafer bonding [Figure 6-12(b2)].

Adhesive bonding was performed in a wafer bonder (AML-AWB-04, Applied Microengineering Ltd, Oxfordshire, United Kingdom). The top and bottom wafers were manually aligned so that the ITO contact pads could be accessed through the wafer flat cuts on both wafers. The wafer pair was then brought into contact with 0.3-MPa bonding pressure applied over the 100-mm wafer area. For SU-8 bonding, the wafer pair under pressure was heated up to 120°C and maintained for 1 h. For BCB bonding, the wafer pair under pressure was first heated up to 150°C, maintained for 15 minutes, and then heated up 250°C and maintained for 1 h, which finished the fabrication process. The physical dimensions of the fabricated CMUTs are shown in Table 6-1.

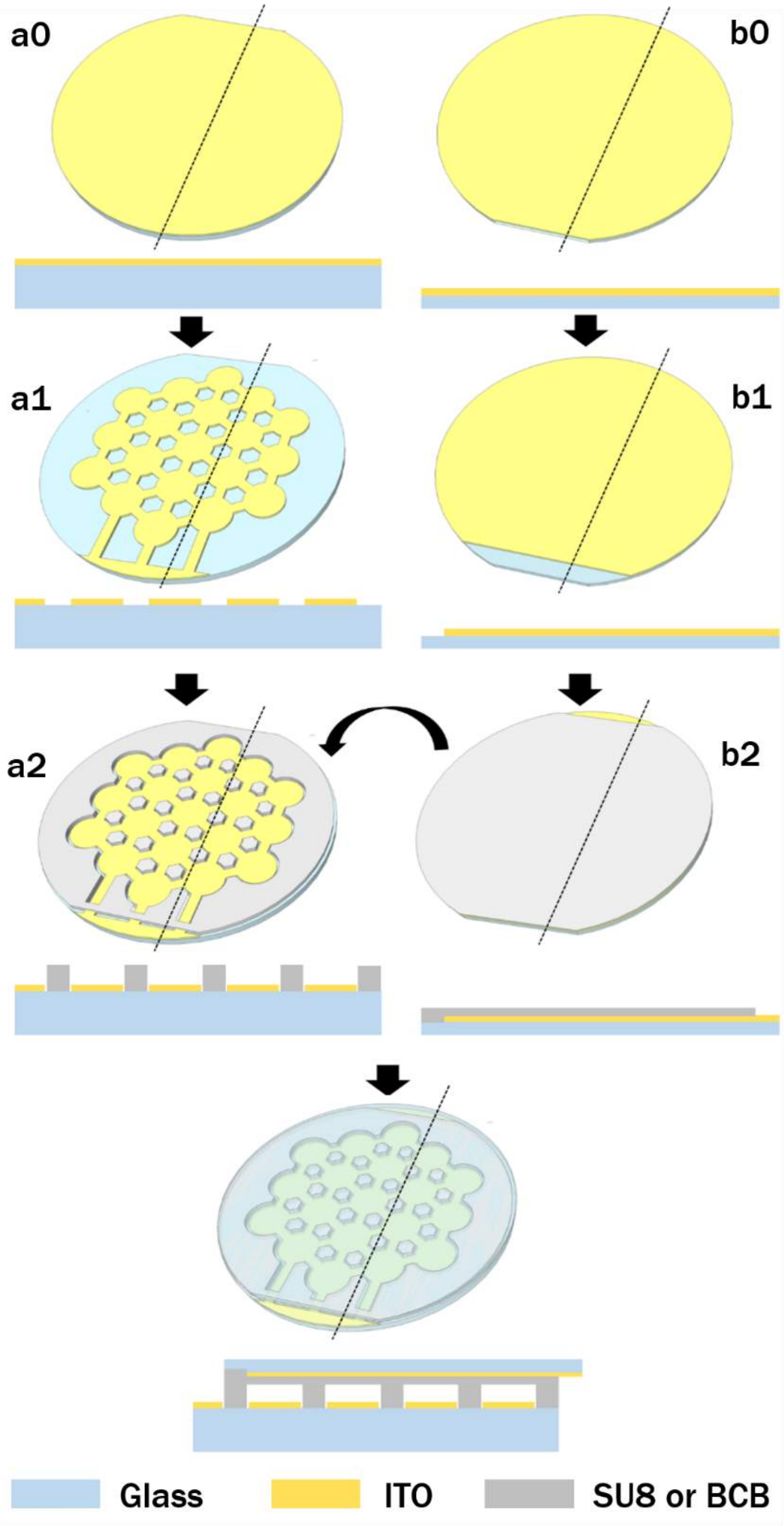

**Figure 6-12: Fabrication of fully transparent CMUT using adhesive bonding.**

**Table 6-1: Physical Dimensions of the Fabricated Transparent CMUTs**

| | **SU-8 bonding** | **BCB bonding** |
|---|---|---|
| **Shape of cells** | Circular | Circular |
| **Cell radius** | 2.8 mm | 2.8 mm |
| **Cell-to-cell distance** | 400 µm | 400 µm |
| **Cavity depth** | 4.4 µm | 3.6 µm |
| **Bottom ITO thickness** | 200 nm | 200 nm |
| **Insulation layer thickness** | 1 µm | 2 µm |
| **Top ITO thickness** | 200 nm | 200 nm |
| **Plate thickness** | 175 µm | 175 µm |

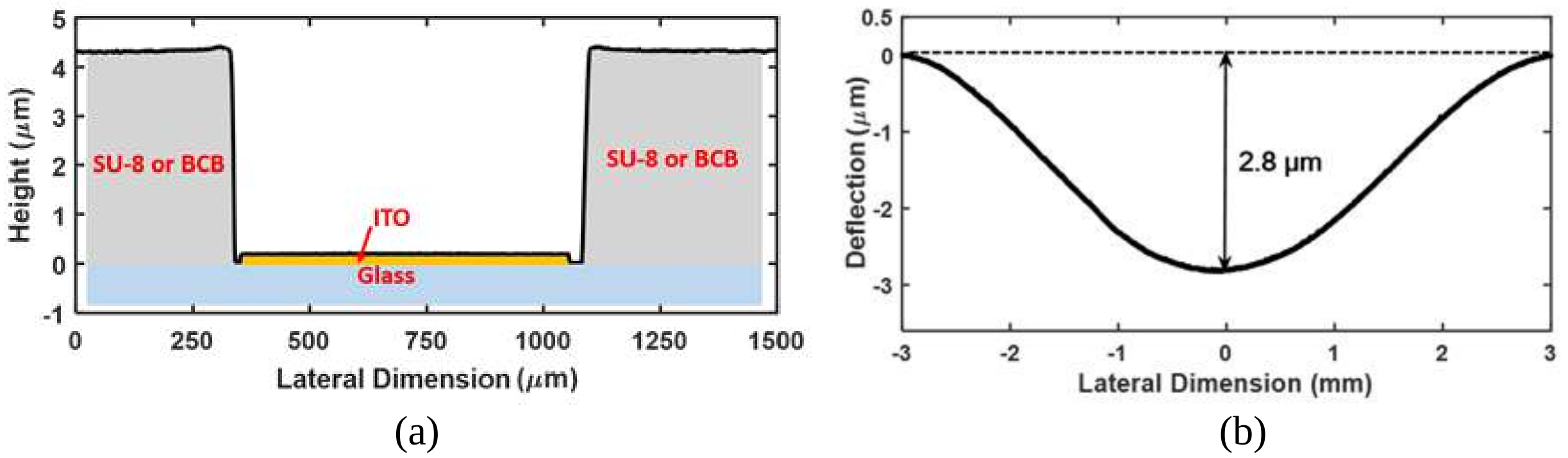

**Figure 6-13: (a) Profilometer measurement of the formed cavities with ITO bottom electrodes on a glass substrate. (b) Atmospheric deflection of a fabricated CMUT cell.**

Figure 6-14(a) shows the optical image of the single CMUT element fabricated using the SU-8 bonding process. The wafer was placed on an “NC STATE UNIVERSITY” logo to show the transparency. In comparison, Figure 6-14(b) shows the CMUT element fabricated using the BCB bonding process. The optical picture is shown on the top and the measured transmittance was shown at the bottom. The average transmittance was then calculated according to the fill factor of the CMUT active area, which is shown as the red solid curves on the plots, indicating that both devices have 70%-80% optical transmittance in the full visible wavelength range.

The electrical connections to an SMA connector were established using silver epoxy on ITO contact pads. The electrical input impedance of the fabricated CMUT elements was then measured in air using a calibrated network analyzer (Model: E5061B, Agilent Technologies, Santa Clara, CA). Both CMUT elements showed a resonance frequency of approximately 62 kHz and a series resistance of approximately 30 Ω. The real and imaginary parts of the electrical input impedance are shown in Figure 6-15(a) for the SU-8 process and in Figure 6-15(b) for the BCB process.

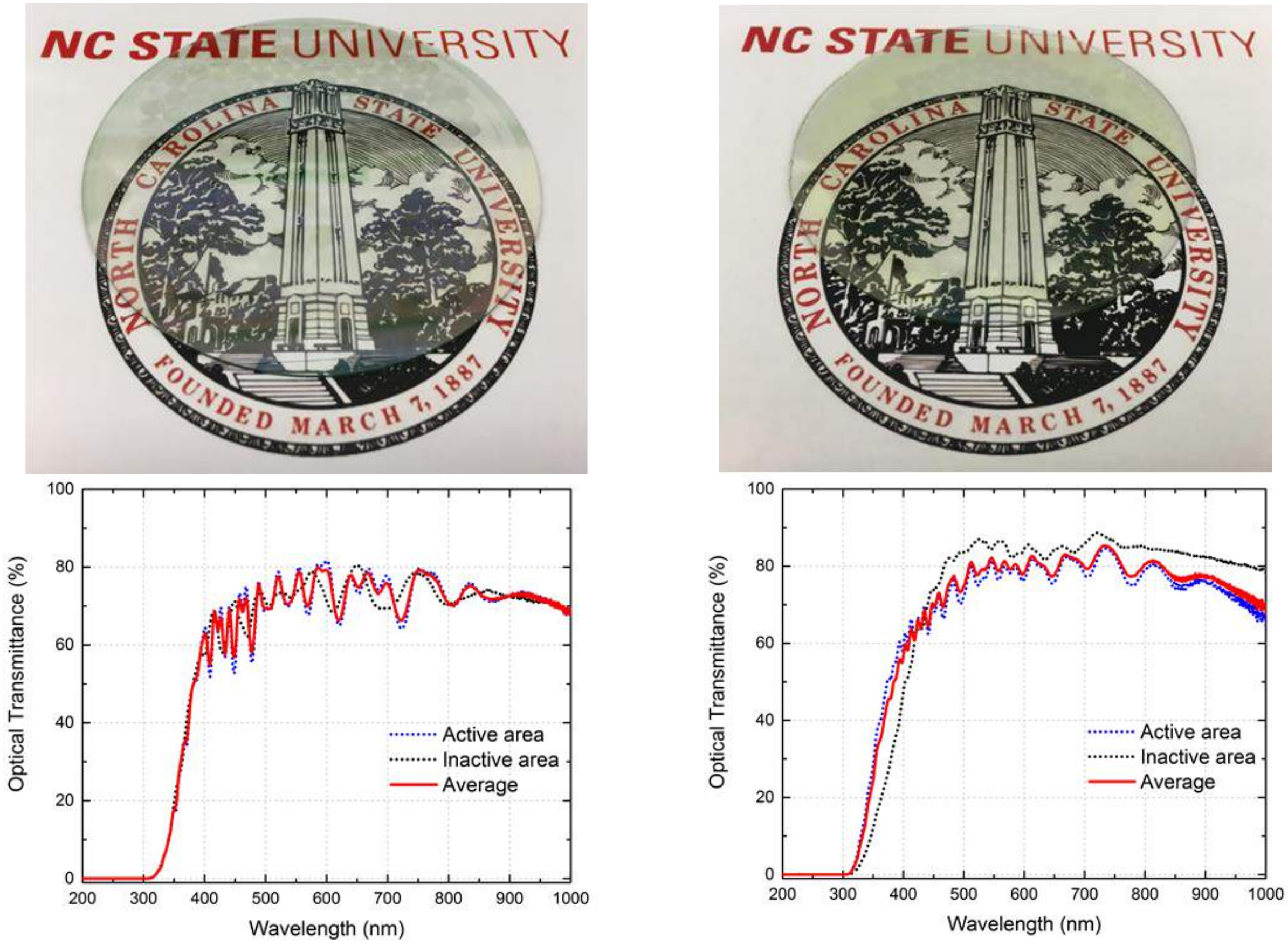

(a) (b)

**Figure 6-14: Fabricated fully transparent CMUT. Optical photos (top), transmittance measurement (bottom). (a) SU-8 bonded wafer pair. (b) BCB bonded wafer pair.**

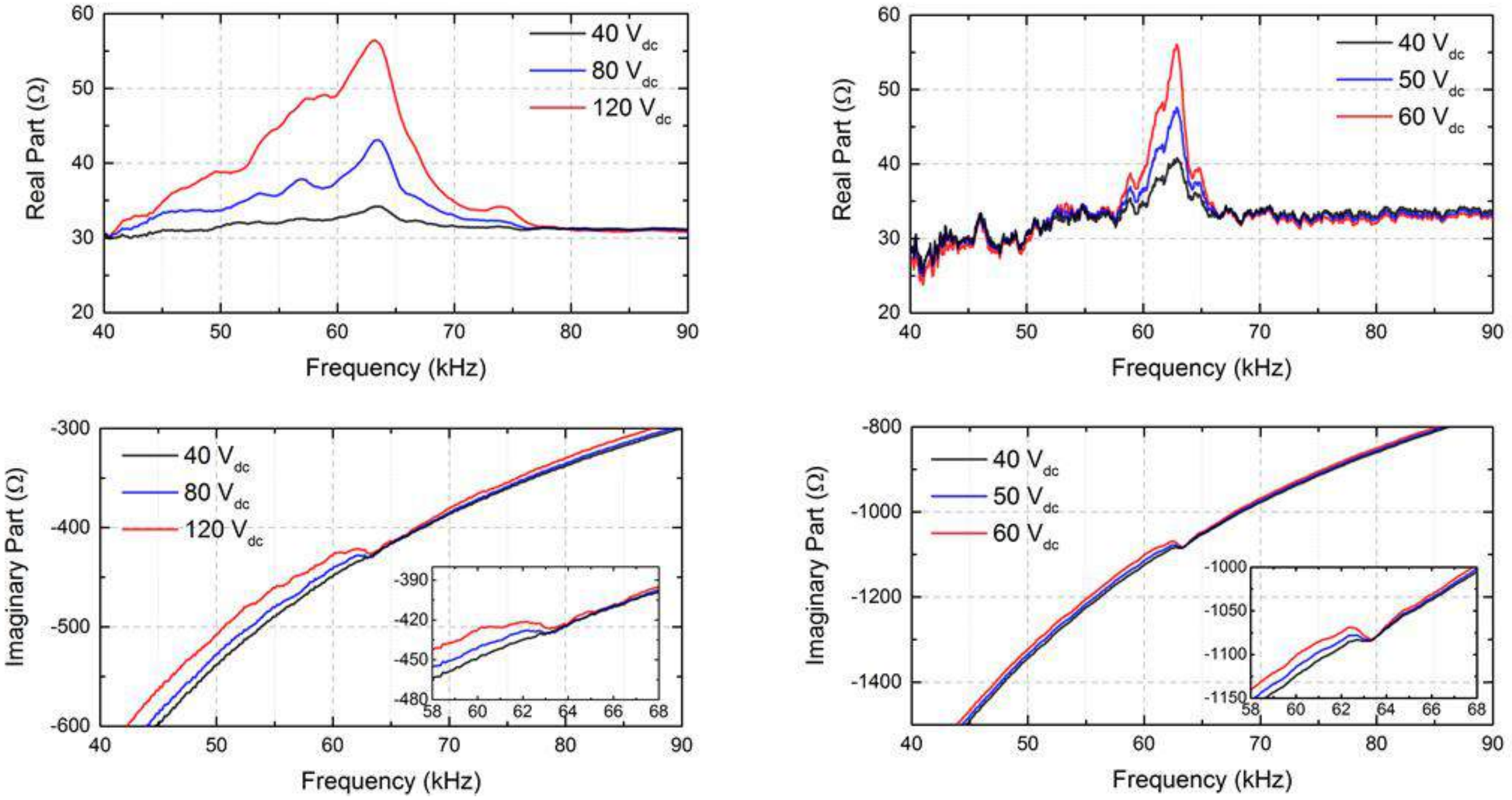

(a) (b)

**Figure 6-15: Input electrical impedance measurement. Real part (top). Imaginary part (bottom). (a) SU-8 bonded wafer pair. (b) BCB bonded wafer pair.**

Although the CMUTs reported here were designed with cavity depth of several microns, the thinner gap can be realized by modifying the polymer coating process. A CMUT fabricated using photo-BCB with several hundreds of nanometers gap has been reported before and SU-8 as thin as 0.5 μm is commercially available. Besides, the plate could be implemented using a thinner glass wafer to extend the design flexibility. The electrical contacts in the demonstrated CMUT were realized by exposing the contact pad using the wafer flat cut. In order to facilitate integration with electronics, through-glass-vias could be incorporated to access the ITO electrodes from the backside of the wafer. Furthermore, CMUTs can be implemented with a vacuum-sealed gap or with open cavities.

## 6.5 Chapter Conclusions and Future Work

In this chapter, we reported a CMUT fabricated on a glass substrate with ITO bottom electrodes for improved transparency [13]. The CMUT plate was formed by anodic bonding a 2-µm thick SOI silicon device layer on to borosilicate glass. The completed device has 40% to 70% transmission in the wavelength range from 700 nm to 900 nm. The resonant frequency was 3.62 MHz in air. Center frequency in immersion was 1.4 MHz and the fractional bandwidth was 105%. The CMUT was mounted on a custom-designed PCB with a cutout in the center to allow the laser illumination from the backside. A pencil lead cross section was first imaged and the effect of the silicon plate absorption was investigated. The artifact introduced by the spurious transmit signal caused by the light absorption in the silicon plate is 34 dB lower compared to the photoacoustic image intensity of the pencil lead at a wavelength of 830 nm. The second target, a polyethylene tube filled with 50-µM ICG solution was embedded in a tissue-mimicking material. The volumetric image was reconstructed by mechanically scanning the CMUT element with the fiber bundle on the backside in x and y directions. The two preliminary PA experiments demonstrated that CMUTs with the proposed fabrication approach is promising for backward-mode photoacoustic imaging [115].

We are currently working on designing and fabricating 1D and 2D CMUT arrays with ITO bottom electrodes as well as further improving the plate transparency to reduce them light absorption in the device. By doing that a faster data acquisition rate and a better image quality can be realized. In the long term, we will implement the through-illumination approach in intravascular/intracardiac imaging, endoscopy, laparoscopy, and other intracavital applications that require compact integration of optical and acoustic components for photoacoustic imaging.

We also demonstrated a fully transparent CMUT with glass plate, glass substrate, and ITO electrodes fabricated using SU-8/BCB-based adhesive bonding [116]. The preliminary measurements show the functionality of the device and indicated the transmittance of ~80% in the visible wavelength range.

# Chapter 7 Summary

The realization of densely populated CMUT arrays with TGV interconnects and MEMS T/R switches on glass substrates are important for simplified transducer fabrication and the front-end electronics integration. Also, parasitics are significantly reduced benefiting from the insulating glass substrate, which could significantly improve the efficiency of the transducer. Furthermore, a transparent CMUT was fabricated by taking the advantage of glass substrate transparency, which can find its application in both medical applications and consumer electronics. The above components were described in detail in this dissertation to establish a tool kit for the development of the next-generation ultrasound transducers.

We use borosilicate glass as an initial substrate and use anodic bonding to implement the movable plate of CMUTs. We first developed a platform process to fabricate vacuum-sealed CMUTs on a glass substrate using a three-mask process. The fabricated CMUT was characterized in air and also in immersion. The measured performance agrees well with finite-element modeling results (FEM), which indicates low parasitics and good dimensional control was achieved. A high-frequency (30-MHz), broadband (100%) CMUT array was further demonstrated using the same process flow.

2D CMUT array was then developed on a glass substrate by incorporating through-glass-via (TGV) interconnects. The fabrication flow to fabricate 2D CMUT array with vacuum-backing is presented. The parasitic capacitance of a via pair of a 250-µm pitch was measured as 21 fF. The resistance of a single via plus via-to-electrode contact resistance was 2 $\Omega$. A CMUT element from a $16\times16$ 2D array was characterized by measuring its input electrical impedance.

A MEMS switch is also enabled by the platform process. We designed the MEMS switch using FEM and fabricated the MEMS switch and CMUT side by side. Static characterization was performed in air and dynamic characterization was performed in immersion. The MEMS switch has a dc switching voltage of 68 V and could be operated using a control voltage of 2.5 V. The optimum switching time is 1.34 μs and release time is 80 ns.

Glass substrate transparency enables applications where acoustics and optics are combined. A CMUT element with improved transparency was first fabricated and used for backward-mode photoacoustic imaging application. A graphite target and a tube filled with ICG solution were imaged and the effect of silicon plate absorption was discussed. Besides, CMUTs for consumer electronics applications is appearing as another important aspect for the CMUT commercialization. We developed a fully transparent CMUT by adhesively bonding two ITO coated glass substrates. The device shows 60%-80% optical transmission across the visible wavelength range and the input electrical impedance measurements demonstrated the functionality of the fabricated device. Such a device can be used for integrating ultrasound with flat panel display, including embedding parametric speakers under a TV screen, and integrating finger printer sensors under a smartphone display panel.

## BIBILOGRAPHY

[1] K. F. Graff, “Ultrasonics: historical aspects,” *Proc. IEEE Ultrason. Symp.*, 1977.

[2] M. I. Haller and B. T. Khuri-Yakub, “A surface micromachined electrostatic ultrasonic air transducer,” *IEEE Trans. Ultrason. Ferroelectr. Freq. Control*, vol. 43, no. 1, pp. 1–6, 1996.

[3] Frederick V. Hunt, *Electroacoustics: The Analysis of Transduction, and Its Historical Background*. 1954.

[4] Y. Huang, A. S. Ergun, E. Hæggström, and M. H. Badi, “Fabricating Capacitive Micromachined Ultrasonic Transducers With Wafer-Bonding Technology,” *J. Microelectromech. Syst.*, vol. 12, no. 2, pp. 128–137, 2003.

[5] C. H. Cheng, E. M. Chow, X. Jin, S. Ergun, and B. T. Khuri-Yakub, “An efficient electrical addressing method using through-wafer vias for two-dimensional ultrasonic arrays,” in *Proc. IEEE Ultrason. Symp.*, 2000, pp. 1179–1182.

[6] X. Zhuang, A. S. Ergun, Y. Huang, I. O. Wygant, Ö. Oralkan, and B. T. Khuri-Yakub, “Integration of trench-isolated through-wafer interconnects with 2D capacitive micromachined ultrasonic transducer arrays.,” *Sens. Actuators. A, Phys.*, vol. 138, no. 1, pp. 221–229, 2007.

[7] D. T. Yeh, Ö. Oralkan, I. O. Wygant, A. Sanli Ergun, J. H. Wong, and B. T. Khuri-Yakub,

"High-resolution imaging with high-frequency 1-D linear CMUT arrays," in *Proc. IEEE Ultrason. Symp.*, 2005, vol. 1, no. c, pp. 665–668.

[8] A. S. Logan, L. L. Wong, and J. T. W. Yeow, "2-D CMUT Wafer Bonded Imaging Arrays with a Row-Column Addressing Scheme," in *Proc. IEEE Ultrason. Symp.*, 2009, pp. 984–987.

[9] J. Knight, J. McLean, and F. L. Degertekin, "Low temperature fabrication of immersion capacitive micromachined ultrasonic transducers on silicon and dielectric substrates," *IEEE Trans. Ultrason. Ferroelectr. Freq. Control*, vol. 51, no. 10, pp. 1324–1333, Oct. 2004.

[10] M. Bellaredj, G. Bourbon, V. Walter, P. Le Moal, and M. Berthillier, "Anodic bonding using SOI wafer for fabrication of capacitive micromachined ultrasonic transducers," *J. Micromech. Microeng.*, vol. 24, no. 2, p. 25009, 2014.

[11] F. Y. Yamaner, X. Zhang, and Ö. Oralkan, "A three-mask process for fabricating vacuum-sealed capacitive micromachined ultrasonic transducers using anodic bonding," *IEEE Trans. Ultrason. Ferroelect., Freq. Contr.*, vol. 62, no. 5, pp. 972–982, 2015.

[12] Z. Li, L. L. P. Wong, A. I. H. Chen, S. Na, J. Sun, and J. T. W. Yeow, "Fabrication of capacitive micromachined ultrasonic transducers based on adhesive wafer bonding technique," *J. Micromech. Microeng.*, vol. 26, no. 11, p. 115019, 2016.

[13] X. Zhang, O. Adelegan, F. Y. Yamaner, and Ö. Oralkan, "CMUTs on glass with ITO

bottom electrodes for improved transparency," in *Proc. IEEE Ultrason. Symp.*, 2016, pp. 1–4.

[14] Ö. Oralkan, M. Karaman, U. Demirci, K. Kaviani, T. H. Lee, and B. T. Khuri-yakub, "Capacitive micromachined ultrasonic transducers: Next-generation arrays for acoustic imaging?," *IEEE Trans. Ultrason. Ferroelect., Freq. Contr.*, vol. 49, pp. 1596–1610, 2002.

[15] A. Bhuyan *et al.*, "Integrated circuits for volumetric ultrasound imaging with 2-D CMUT arrays.," *IEEE Trans. Biomed. Circuits Syst.*, vol. 7, no. 6, pp. 796–804, Dec. 2013.

[16] I. O. Wygant *et al.*, "Integration of 2D CMUT arrays with front-end electronics for volumetric ultrasound imaging.," *IEEE Trans. Ultrason. Ferroelectr. Freq. Control*, vol. 55, no. 2, pp. 327–42, Feb. 2008.

[17] D. T. Yeh, O. Oralkan, I. O. Wygant, M. O'Donnell, and B. T. Khuri-Yakub, "3-D ultrasound imaging using a forward-looking CMUT ring array for intravascular/intracardiac applications," *IEEE Trans. Ultrason. Ferroelectr. Freq. Control*, vol. 53, no. 6, pp. 1202–1211, Jun. 2006.

[18] G. Gurun *et al.*, "Single-chip CMUT-on-CMOS front-end system for real-time volumetric IVUS and ICE imaging.," *IEEE Trans. Ultrason. Ferroelectr. Freq. Control*, vol. 61, no. 2, pp. 239–50, Feb. 2014.

[19] J. Zahorian *et al.*, "Monolithic CMUT-on-CMOS integration for intravascular ultrasound

applications," *IEEE Trans. Ultrason. Ferroelect., Freq. Contr.*, vol. 58, no. 12, pp. 2659–2667, 2011.

[20] S. Vaithilingam *et al.*, "Three-dimensional photoacoustic imaging using a two-dimensional CMUT array.," *IEEE Trans. Ultrason. Ferroelectr. Freq. Control*, vol. 56, no. 11, pp. 2411–9, Nov. 2009.

[21] S. Kothapalli, T.-J. Ma, S. Vaithilingam, Ö. Oralkan, B. T. Khuri-Yakub, and S. S. Gambhir, "Deep Tissue Photoacoustic Imaging Using a Miniaturized 2-D Capacitive Micromachined Ultrasonic Transducer Array," *IEEE Tran. Biomed. Eng.*, vol. 59, no. 5, pp. 1199–1204, 2012.

[22] F. O. R. I. Release, "www.hitachi.com/New/cnews/month/2017/03/170302a.pdf," 2017.

[23] D. Zhao, S. Zhuang, and R. Daigle, "A Commercialized High Frequency CMUT Probe for Medical Ultrasound Imaging," in *Proc. IEEE Ultrason. Symp.*, 2015, pp. 3–6.

[24] S. H. Wong, M. Kupnik, R. D. Watkins, K. Butts-pauly, and B. T. P. Khuri-yakub, "Capacitive Micromachined Ultrasonic Transducers for Therapeutic Ultrasound Applications," *IEEE Trans Biomed Eng.*, vol. 57, no. 1, pp. 114–123, 2010.

[25] J. H. Jang *et al.*, "Dual-Mode Integrated Circuit for Imaging and HIFU With 2-D CMUT Arrays," in *Proc. IEEE Ultrason. Symp.*, 2015, pp. 3–6.

[26] H. J. Lee, K. K. Park, and M. Kupnik, "Chemical Vapor Detection Using a Capacitive

Micromachined," *Anal. Chem.*, pp. 9314–9320, 2011.

[27] Q. Stedman, K. Park, and B. T. Khuri-yakub, "Distinguishing Chemicals Using CMUT Chemical Sensor Array and Artificial Neural Networks," in *Proc. IEEE Ultrason. Symp.*, 2014, pp. 162–165.

[28] M. M. Mahmud *et al.*, "A low-power gas sensor for environmental monitoring using a capacitive micromachined ultrasonic transducer," in *Proc. IEEE Sensors Conf.*, 2014, pp. 677–680.

[29] I. O. Wygant *et al.*, "50 kHz capacitive micromachined ultrasonic transducers for generation of highly directional sound with parametric arrays," *IEEE Trans. Ultrason. Ferroelectr. Freq. Control*, vol. 56, no. 1, pp. 193–203, 2009.

[30] N. Lamberti, G. Caliano, A. Iula, and A. Stuart, "Sensors and Actuators A : Physical A high frequency cMUT probe for ultrasound imaging of fingerprints," *Sensors Actuators A. Phys.*, vol. 172, no. 2, pp. 561–569, 2011.

[31] P. Fesenko, "Capacitive micromachined ultrasonic transducer ( cMUT ) for biometric applications Capacitive micromachined ultrasonic transducer ( cMUT ) for biometric applications," *Master Thesis*, 2012.

[32] T. I. Chiu, H. C. Deng, S. Y. Chang, and S. B. Luo, "Implementation of ultrasonic touchless interactive panel using the polymer-based CMUT array," in *Proc. IEEE Sensors Conf.*, 2009, pp. 625–630.

[33] W. H. Peake, “The lowest resonant frequency of a water-loaded circular plate,” *J. Acoust. Soc. Am.*, vol. 26, pp. 166–168, 1954.

[34] J. D. Irwin, *Basic Engineering Circuit Analysis*. 1996.

[35] G. G. Yaralioglu, A. S. Ergun, B. Bayram, E. Haeggstrom, and B. T. Khuri-Yakub, “Calculation and measurement of electromechanical coupling coefficient of capacitive micromachined ultrasonic transducers,” *IEEE Trans. Ultrason. Ferroelect., Freq. Contr.*, vol. 50, no. 4, pp. 449–456, Apr. 2003.

[36] K. K. Park, H. Lee, M. Kupnik, and B. T. Khuri-yakub, “Fabrication of Capacitive Micromachined Ultrasonic Transducers via Local Oxidation and Direct Wafer Bonding,” *Microelectromechanical Syst. J.*, vol. 20, no. 1, pp. 95–103, 2011.

[37] B. T. Khuri-Yakub and Ö. Oralkan, “Capacitive micromachined ultrasonic transducers for medical imaging and therapy.,” *J. Micromech. Microeng.*, vol. 21, no. 5, pp. 54004–54014, 2011.

[38] M. Kupnik, S. Vaithilingam, K. Torashima, I. O. Wygant, and B. T. Khuri-yakub, “CMUT Fabrication Based on a Thick Buried Oxide Layer,” in *Proc. IEEE Ultrason. Symp.*, 2010, pp. 547–550.

[39] S. E. Alper and T. Akin, “A Single-Crystal Silicon Symmetrical and Decoupled MEMS Gyroscope on an Insulating Substrate,” *J. Microelectromech. Syst.*, vol. 14, no. 4, pp. 707–717, 2005.

[40] S. E. Alper and T. Akin, “Symmetrical and decoupled nickel microgyroscope on insulating substrate,” in *The 17th European Conference on Solid-State Transducers,* 2004, vol. 115, pp. 336–350.

[41] R. C. Meier, J. Höfflin, and V. Badilita, “Resonance A fully MEMS-compatible process for 3D high aspect ratio micro coils obtained with an automatic wire bonder,” *J. Micromech. Microeng.,* vol. 20, no. 15021, 2009.

[42] F. Y. Yamaner, S. Olçum, H. K. Oğuz, A. Bozkurt, H. Köymen, and A. Atalar, “High-power CMUTs: design and experimental verification.,” *IEEE Trans. Ultrason. Ferroelectr. Freq. Control,* vol. 59, no. 6, pp. 1276–84, Jun. 2012.

[43] P. C. Eccardt, K. Niederer, and B. Fischer, “Micromachined transducers for ultrasound applications,” *1997 IEEE Ultrason. Symp. Proceedings. An Int. Symp. (Cat. No.97CH36118),* vol. 2, pp. 1609–1618, 1997.

[44] D. F. Lemmerhirt *et al.*, “A 32 x 32 capacitive micromachined ultrasonic transducer array manufactured in standard CMOS.,” *IEEE Trans. Ultrason. Ferroelect., Freq. Contr.,* vol. 59, no. 7, pp. 1521–36, 2012.

[45] Y. Tsuji, M. Kupnik, B. T. Khuri-yakub, and A. P. Flow, “Low Temperature Process for CMUT Fabrication with Wafer Bonding Technique,” in *Proc. IEEE Ultrason. Symp.,* 2011, pp. 1–4.

[46] I. O. Wygant *et al.*, “Integration of 2D CMUT arrays with front-end electronics for

volumetric ultrasound imaging.," *IEEE Trans. Ultrason. Ferroelect., Freq. Contr.*, vol. 55, no. 2, pp. 327–42, Feb. 2008.

[47] R. Wodnicki *et al.*, "Multi-row linear cMUT array using cMUTs and multiplexing electronics," in *Proc. IEEE Ultrason. Symp.*, 2009, pp. 2696–2699.

[48] A. Nikoozadeh *et al.*, "Forward-Looking Intracardiac Imaging Catheters Using Fully Integrated CMUT Arrays," in *Proc. IEEE Ultrason. Symp.*, 2010, vol. 2, pp. 770–773.

[49] S. Weichel *et al.*, "Low-temperature anodic bonding to silicon nitride," *Sensors Actuators, A Phys.*, vol. 82, no. 1, pp. 249–253, 2000.

[50] G. W. Hsieh, C. H. Tsai, and W. C. Lin, "Anodic bonding of glass and silicon wafers with an intermediate silicon nitride film and its application to batch fabrication of SPM tip arrays," *Microelectronics J.*, vol. 36, no. 7, pp. 678–682, 2005.

[51] T. M. H. Lee, D. H. Y. Lee, C. Y. N. Liaw, A. I. K. Lao, and I. M. Hsing, "Detailed characterization of anodic bonding process between glass and thin-film coated silicon substrates," *Sensors Actuators, A Phys.*, vol. 86, no. 1–2, pp. 103–107, 2000.

[52] S. Mack, H. Baumann, U. Gösele, H. Werner, and R. Schlögl, "Analysis of Bonding-Related Gas Enclosure in Micromachined Cavities Sealed by Silicon Wafer Bonding," *J. Electrochem. Soc.*, vol. 144, no. 3, pp. 1106–1111, 1997.

[53] I. Ladabaum *et al.*, "Silicon substrate ringing in microfabricated ultrasonic transducers,"

in *Proc. IEEE Ultrason. Symp.*, 2000, no. C.

[54] F. Y. Yamaner, X. Zhang, and O. Oralkan, "Fabrication of anodically bonded capacitive micromachined ultrasonic transducers with vacuum-sealed cavities," in *IEEE International Ultrasonics Symposium*, 2014, pp. 604–607.

[55] X. Zhang, F. Y. Yamaner, O. Adelegan, and Ö. Oralkan, "Design of high-frequency broadband CMUT arrays," in *Proc. IEEE Ultrason. Symp.*, 2015, pp. 1–4.

[56] P. C. Eccardt, K. Niederer, and B. Fischer, "Micromachined transducers for ultrasound applications," in *Proc. IEEE Ultrason. Symp.*, 1997, vol. 2, pp. 1609–1618.

[57] C. Daft, S. Calmes, D. da Graca, K. Patel, P. Wagner, and I. Ladabaum, "Microfabricated ultrasonic transducers monolithically integrated with high voltage electronics," in *Proc. IEEE Ultrason. Symp.*, 2004, vol. 1, pp. 493–496.

[58] R. A. Noble *et al.*, "Low-temperature micromachined cMUTs with fully-integrated analogue front-end electronics," in *Proc. IEEE Ultrason. Symp.*, 2002, vol. 0, no. c, pp. 1045–1050.

[59] A. Nikoozadeh *et al.*, "Forward-looking volumetric intracardiac imaging using a fully integrated CMUT ring array," in *Proc. IEEE Ultrason. Symp.*, 2009, pp. 511–514.

[60] C. H. Cheng, A. S. Ergun, and B. T. Khuri-Yakub, "Electrical through-wafer interconnects with sub-picofarad parasitic capacitance," in *Proc. IEEE Int. Conf. MEMS*,

2001, pp. 18–21.

[61] M. Topper *et al.*, “3-D Thin film interposer based on TGV (Through Glass Vias): An alternative to Si-interposer,” in *Proc. IEEE Electr. Comp. Tech. Conf.*, 2010, pp. 66–73.

[62] M. Esashi, “Wafer level packaging of MEMS,” *J. Micromech. Microeng.*, vol. 18, no. 7, p. 73001, 2008.

[63] N. Toan, S. Hahng, Y. Song, and T. Ono, “Fabrication of vacuum-sealed capacitive micromachined ultrasonic transducer arrays using glass reflow process,” *Micromachines*, vol. 7, no. 5, p. 76, 2016.

[64] R. Mukhiya *et al.*, “Fabrication of capacitive micromachined ultrasonic transducer arrays with isolation-trenches using anodic wafer bonding,” *IEEE Sensors J.*, vol. 15, no. 9, pp. 5177–5184, 2015.

[65] S. Takahashi, K. Horiuchi, K. Tatsukoshi, M. Ono, N. Imajo, and T. Mobley, “Development of through glass via (TGV) formation technology using electrical discharging for 2.5/3D integrated packaging,” in *Proc. IEEE Electr. Comp. Tech. Conf.*, 2013, pp. 348–352.

[66] Q. Yan, H. Lei, Y. Chen, Y. Zhu, and Research, “Preparation and polishing performances of α-Al2O3 /Fe2O3 composite particles,” in *Proc. CIST2008 & ITS-IFToMM2008*, 2007, vol. 14, no. 5, pp. 945–950.

[67] R. Keusseyan and T. Mobley, “Hermetically sealed glass packages,” in *Proc. IMAPS Int. Symp. on Microelectronics*, 2015, pp. 375–378.

[68] X. Zhang, F. Y. Yamaner, and Ö. Oralkan, “Fabrication of capacitive micromachined ultrasonic transducers with through-glass-via interconnects,” in *Proc. IEEE Ultrason. Symp.*, 2015, pp. 1–4.

[69] S. Enderling *et al.*, “Sheet resistance measurement of non-standard cleanroom materials using suspended Greek cross test structures,” *IEEE Trans. Semicond. Manuf.*, vol. 19, no. 1, pp. 2–9, 2006.

[70] G. G. Yaralioglu, S. A. Ergun, and B. T. Khuri-Yakub, “Finite-element analysis of capacitive micromachined ultrasonic transducers,” *IEEE Trans. Ultrason. Ferroelect., Freq. Contr.*, vol. 52, no. 12, pp. 2185–2198, 2005.

[71] X. Zhang, F. Y. Yamaner, and Ö. Oralkan, “Fabrication of Vacuum-Sealed Capacitive Micromachined Ultrasonic Transducers With Through-Glass-Via Interconnects Using Anodic Bonding,” *J. Microelectromechanical Syst.*, vol. 26, no. 1, pp. 226–234, Feb. 2017.

[72] J. Camacho and C. Fritsch, “Protection circuits for ultrasound applications,” *IEEE Trans. Ultrason. Ferroelect., Freq. Contr.*, vol. 55, no. 5, pp. 1160–1164, 2008.

[73] N. C. Chaggares, R. K. Tang, A. N. Sinclair, F. S. Foster, K. Haraieciwz, and B. Starkoski, “Protection circuitry and time resolution in high frequency ultrasonic NDE,”

*Proc. IEEE Ultrason. Symp.*, vol. 1, pp. 819–822, 1999.

[74] R. E. Mihailovich *et al.*, “MEM relay for reconfigurable RF circuits,” *IEEE Microw. Wirel. Components Lett.*, vol. 11, no. 2, pp. 53–55, 2001.

[75] J. J. Yao and M. F. Chang, “A Surface Micromachined Miniature Switch For Telecommunications Applications With Signal Frequencies From DC Up To 4 Ghz,” *Proc. Int. Solid-State Sensors Actuators Conf. - TRANSDUCERS ’95*, vol. 2, pp. 384–387, 1995.

[76] G. M. Rebeiz and J. B. Muldavin, “RF MEMS switches and switch circuits,” *IEEE Microw. Mag.*, vol. 2, no. 4, pp. 59–71, 2001.

[77] F. Y. Yamaner, “Finite element and equivalent circuit modeling of capacitive micromachined ultrasonic transducer (cMUT),” *Master Thesis*, 2006.

[78] W. D. Greason, “Analysis of the effect of ESD on the operation of MEMS,” *IEEE Trans. Ind. Appl.*, vol. 45, no. 6, pp. 2185–2191, 2009.

[79] X. Zhang, X. Wu, O. J. Adelegan, F. Y. Yamaner, and Ö. Oralkan, “A MEMS T/R switch embedded in CMUT structure for ultrasound imaging frontends,” *IEEE Trans. Ultrason. Ferroelect., Freq. Contr.*, pp. 1–4, 2017.

[80] X. Zhang, O. J. Adelegan, F. Y. Yamaner, and Ö. Oralkan, “A fast-switching (1.35-µs), low-control-voltage (2.5-V) MEMS T/R switch monolithically integrated with a

capacitive micromachined ultrasonic transducer (CMUT)," *J. Microelectromechanical Syst.*, accepted, 2017.

[81] A. Fortini, M. I. Mendelev, S. Buldyrev, and D. Srolovitz, "Asperity contacts at the nanoscale: Comparison of Ru and Au," *J. Appl. Phys.*, vol. 104, no. 7, 2008.

[82] H. Kwon *et al.*, "Investigation of the electrical contact behaviors in Au-to-Au thin-film contacts for RF MEMS switches," *J. Micromech. Microeng.*, vol. 18, p. 105010, 2008.

[83] Z. Yang, D. J. Lichtenwalner, A. S. Morris, J. Krim, and A. I. Kingon, "Comparison of Au and Au-Ni alloys as contact materials for MEMS switches," *J. Microelectromech. Syst.*, vol. 18, no. 2, pp. 287–295, 2009.

[84] J. Dhennin, F. Courtade, C. Dieppedal, P. Pons, X. Lafontan, and R. Plana, "Characterization of gold/gold, gold/ruthenium, and ruthenium/ruthenium ohmic contacts in MEMS switches improved by a novel methodology," *J. Micro/Nanolithography, MEMS, MOEMS*, vol. 9, no. 4, p. 41102, 2010.

[85] R. Fisher *et al.*, "Reconfigurable arrays for portable ultrasound," in *Proc. IEEE Ultrason. Symp.*, 2005, vol. 1, no. c, pp. 495–499.

[86] J. Chen, M. Wang, J. C. Cheng, Y. H. Wang, P. C. Li, and X. Cheng, "A photoacoustic imager with light illumination through an infrared-transparent silicon CMUT array," *IEEE Trans. Ultrason. Ferroelect., Freq. Contr.*, vol. 59, no. 4, pp. 766–775, 2012.

[87] G. Brodie, Y. Qiu, S. Cochran, G. Spalding, and M. MacDonald, “Optically transparent piezoelectric transducer for ultrasonic particle manipulation,” *IEEE Trans. Ultrason. Ferroelectr. Freq. Control*, vol. 61, no. 3, pp. 389–391, 2014.

[88] H. Kim *et al.*, “Electrical, optical, and structural properties of indium–tin–oxide thin films for organic light-emitting devices,” *J. Appl. Phys.*, vol. 86, no. 11, p. 6451, 1999.

[89] M. Xu and L. V Wang, “Photoacoustic imaging in biomedicine,” *Rev. Sci. Instrum.*, vol. 77, no. 41101, 2006.

[90] L. V Wang and J. Yao, “A practical guide to photoacoustic tomography in the life sciences,” *Nat. Methods*, vol. 13, no. 8, pp. 627–638, 2016.

[91] S. Vaithilingam *et al.*, “Three-dimensional photoacoustic imaging using a two-dimensional CMUT array,” *IEEE Trans. Ultrason. Ferroelect., Freq. Contr.*, vol. 56, no. 11, pp. 2411–2419, 2009.

[92] X. Wu, J. Lunsford Sanders, Ö. Oralkan, and D. N. Stephens, “Photoacoustic-Imaging-Based Temperature Monitoring for High-Intensity Focused Ultrasound Therapy,” in *Proc IEEE Eng. Med. Biol. Soc*, 2016, pp. 3235–3238.

[93] S. A. Ermilov *et al.*, “Development of laser optoacoustic and ultrasonic imaging system for breast cancer utilizing handheld array probes,” in *Proc. SPIE*, 2009, vol. 7177, p. 717703.

[94] H. F. Zhang, K. Maslov, G. Stoica, and L. V. Wang, “Functional photoacoustic microscopy for high-resolution and noninvasive in vivo imaging,” *Nat. Biotechnol.*, vol. 24, no. 7, pp. 848–851, 2006.

[95] Y. Asao *et al.*, “Photoacoustic mammography capable of simultaneously acquiring photoacoustic and ultrasound images,” *J. Biomed. Opt.*, vol. 21, no. 11, p. 116009, 2016.

[96] S. Manohar, A. Kharine, J. C. G. van Hespen, W. Steenbergen, and T. G. van Leeuwen, “The Twente Photoacoustic Mammoscope: system overview and performance.,” *Phys. Med. Biol.*, vol. 50, no. 11, pp. 2543–57, 2005.

[97] B. Chen, F. Chu, X. Liu, Y. Li, J. Rong, and H. Jiang, “AlN-based piezoelectric micromachined ultrasonic transducer for photoacoustic imaging,” *Appl. Phys. Lett.*, vol. 103, no. 3, pp. 1–4, 2013.

[98] A. Nikoozadeh *et al.*, “Photoacoustic imaging using a 9F microLinear CMUT ICE catheter,” in *Proc. IEEE Ultrason. Symp.*, 2012, pp. 24–27.

[99] A. Nikoozadeh *et al.*, “An integrated Ring CMUT array for endoscopic ultrasound and photoacoustic imaging,” in *Proc. IEEE Ultrason. Symp.*, 2013, pp. 1178–1181.

[100] J. J. Niederhauser, M. Jaeger, M. Hejazi, H. Keppner, and M. Frenz, “Transparent ITO coated PVDF transducer for optoacoustic depth profiling,” *Opt. Commun.*, vol. 253, no. 4–6, pp. 401–406, 2005.

[101] X. Chen, R. Chen, Z. Chen, J. Chen, K. K. Shung, and Q. Zhou, "Transparent lead lanthanum zirconate titanate (PLZT) ceramic fibers for high-frequency ultrasonic transducer applications," *Ceram. Int.*, vol. 42, no. 16, pp. 18554–18559, 2016.

[102] P. V. Chitnis, O. Aristizábal, E. Filoux, A. Sampathkumar, J. Mamou, and J. Ketterling, "Coherence-Weighted Synthetic Focusing Applied to Photoacoustic Imaging Using a High-Frequency Annular-Array Transducer.," *Ultras. Imaging*, vol. 38, no. 1, pp. 32–43, 2016.

[103] E. Zhang, J. Laufer, and P. Beard, "Backward-mode multiwavelength photoacoustic scanner using a planar Fabry Perot polymer film ultrasound sensor for high resolution three-dimensional imaging of biological tissues," *Appl. Opt.*, vol. 47, no. 4, pp. 561–577, 2008.

[104] E. Zhang and J. Laufer, "In vivo high-resolution 3D photoacoustic imaging of superficial vascular anatomy," *Phys. Med. Biol.*, vol. 54, 2009.

[105] M. Karaman, L. Pai-Chi, and M. O'Donnell, "Synthetic aperture imaging for small scale systems," *IEEE Trans. Ultrason. Ferroelect., Freq. Contr.*, vol. 42, no. 3, pp. 429–442, May 1995.

[106] P. C. Li and M. L. Li, "Adaptive imaging using the generalized coherence factor," *IEEE Trans. Ultrason. Ferroelect., Freq. Contr.*, vol. 50, no. 2, pp. 128–141, 2003.

[107] M. Xu and L. V. Wang, "Universal back-projection algorithm for photoacoustic computed

tomography," *Phys. Rev. E*, vol. 71, no. 1, pp. 1–7, 2005.

[108] O. Ratib, "OSIRIX: An Open Source Platform for Advanced Multimodality Medical Imaging," in *4th Int. Conf. Information and Communications Technology*, 2006, pp. 1–2.

[109] D. Sette *et al.*, "Transparent piezoelectric transducers for large area ultrasonic actuators," in *Proc. IEEE Int. Conf. MEMS,* 2017, pp. 793–796.

[110] G. Luo *et al.*, "High fill factor piezoelectric micromachined ultrasonic transducers on transparent substrates," *TRANSDUCERS - Int. Solid-State Sensors, Actuators Microsystems Conf.*, pp. 1053–1056, 2017.

[111] S. Chen, "Optical Microring Resonators for Photoacoustic Imaging and Detection," *PhD Thesis, Univ. Michigan,* 2012.

[112] E. Z. Zhang, "2D backward-mode photoacoustic imaging system for NIR (650-1200nm) spectroscopic biomedical applications," in *Proc. SPIE*, 2006, vol. 6086, p. 60860H–60860H–8.

[113] D.-C. Pang and Y.-H. Chiang, "A transparent capacitive micromachined ultrasonic transducer (CMUT) array for finger hover-sensing dial pads," *TRANSDUCERS - Int. Solid-State Sensors, Actuators Microsystems Conf.*, pp. 2171–2174, 2017.

[114] A. I. Chen, L. L. P. Wong, S. Na, Z. Li, M. Macecek, and J. T. W. Yeow, "Fabrication of a Curved Row–Column Addressed Capacitive Micromachined Ultrasonic Transducer

Array," *J. Microelectromechanical Syst.*, vol. 25, no. 4, pp. 675–682, 2016.

[115] X. Zhang, X. Wu, O. J. Adelegan, F. Y. Yamaner, and Ö. Oralkan, "Photoacoustic Imaging With Backside Illumination Through a CMUT With Improved Transparency," *IEEE Trans. Ultrason. Ferroelectr. Freq. Control*, DOI 10.1109/TUFFC.2017.2774283, 2017. (IEEE early access article available)

[116] X. Zhang, O. J. Adelegan, F. Y. Yamaner, and Ö. Oralkan, "An optically transparent capacitive micromachined ultrasonic transducer (CMUT) fabricated using SU-8 or BCB adhesive wafer bonding," in *Proc. IEEE Ultrason. Symp.*, 2017, pp. 1–4.

## APPENDICES

# Appendix A: Finite element modeling of the MEMS switch (ANSYS 17.2)

## Static Analysis

```
finish
/clear,nostart

/prep7

sthick=2.15  !thickness of the plate
snthick=0.20  !nitride thickness
smthick=0.185   !top metal thickness
slength=100   !side of the square
plength=80    !plate side length
blength=13   !bump length
bwidth=6     !bump width

eps=5.7     !insulation layer permittivity
fg=snthick/eps  !insulation layer effective gap for the nitride
gap=0.220
tb=0.070          !bump thickness
gapt=gap+fg   !total effective gap


gaprf=gap   !rf line gap
gapclos=tb

et,1,solid45


mp,dens,1,2328e-18  !silicon's density
mp,ex,1,148e3       !silicon's young's modulus
mp,nuxy,1,0.177     !silicon's possion's ratio

mp,dens,2,3187e-18  !si3n4's density
mp,ex,2,296e3       !si3n4's young's modulus
mp,nuxy,2,0.27      !si3n4's possion's ratio

mp,dens,3,19300e-18 !au's density
mp,ex,3,79e3        !au's young's modulus
mp,nuxy,3,0.44      !au's possion's ratio

!!!!!!!!geometry!!!!!!!!!!
block,-slength/2,slength/2,-slength/2,slength/2,0,sthick
block,-slength/2,slength/2,-slength/2,slength/2,sthick,smthick+sthick
block,-slength/2,slength/2,-slength/2,slength/2,0,-snthick
block,-blength/2,blength/2,-slength/2,slength/2,-snthick,smthick+sthick
block,-slength/2,slength/2,-bwidth/2,bwidth/2,-snthick,smthick+sthick
block,-blength/2,blength/2,-bwidth/2,bwidth/2,-snthick-tb,smthick+sthick
vovlap,all
vglue,all

vsel,s,loc,z,0,sthick   !material type 1, element type 1
vatt,1,,1,0
```

```
vsel,s,loc,z,0,-snthick
vatt,2,,1,0

vsel,s,loc,z,sthick,sthick+smthick
vatt,3,,1,0

vsel,s,loc,z,-snthick,-tb-snthick
vatt,3,,1,0

allsel
!!!!!!!!!mesh!!!!!!!!!

esize,1
vmesh,all

allsel

!!!!!!!!!boundary conditions!!!!!!!!!

nsel,s,loc,z,-snthick
nsel,r,loc,x,-plength/2,+plength/2
nsel,r,loc,y,-plength/2,+plength/2
cm,underpnodes,node

allsel
nsel,s,loc,z,-snthick
cmsel,u,underpnodes,node
d,all,uy,0
d,all,ux,0
d,all,uz,0

allsel

nsel,s,loc,z,-snthick-tb
cm,rf,node

allsel

nsel,s,loc,z,-snthick
nsel,r,loc,x,-plength/2+3,plength/2-3
nsel,r,loc,y,-plength/2+3,-bwidth/2-5
cm,side1,node
allsel

nsel,s,loc,z,-snthick
nsel,r,loc,x,-plength/2+3,plength/2-3
nsel,r,loc,y,bwidth/2+5,plength/2-3
cm,side2,node
allsel

emtgen,'side1','trans1','gnd1','uz',-gapt,fg,1      !generate trans126 elements: emtgen,
emtgen,'rf','trans3','gnd3','uz',-gaprf,gapclos,1
emtgen,'side2','trans2','gnd2','uz',-gapt,fg,1
```

```
d,gnd1,ux,0          !define dof boundary conditions
d,gnd1,uy,0
d,gnd1,uz,0
d,gnd1,volt,0

d,gnd2,ux,0          !define dof boundary conditions
d,gnd2,uy,0
d,gnd2,uz,0
d,gnd2,volt,0

d,gnd3,ux,0          !define dof boundary conditions
d,gnd3,uy,0
d,gnd3,uz,0
d,gnd3,volt,0
allsel
! !!!!!!!!!!slove!!!!!!!!!!!!!!!!
*dim,mydata,array,241,2,0.5 !define array: 49 rows, 10 columns, 1 plane
*do,i,0,9,1
vdc=i*10
/solu
antyp,0     ! static analysis type

allsel
esel,s,type,,1      !select type 1 elements
nsel,r,ext         !reselect its external node
nsel,r,loc,z,sthick+smthick     !reselect its nodes on the surface
sf,all,pres,0.1
allsel
neqit,50
d,side1,volt,vdc
d,rf,volt,0
d,side2,volt,vdc
allsel
solve
finish
/post1      !/post1 General Postproc; /post26 TimeHist Postproc
path,cross,2,30,240
ppath,1,, -slength/2, 0,0
ppath,2,, slength/2, 0,0
pdef,deflect,u,sum,avg
plpath,deflect
allsel

*get, test, path, 0, nval
*do,k,1,test,1
*get, mydata(k,1,1), path, 0, item,s,pathpt,k
*get, mydata(k,2,1), path, 0, item,deflect,pathpt,k
*enddo

*cfopen,deflect_path_%i%,txt
*vwrite,mydata(1,1,1),mydata(1,2,1)
(e24.5,'     ',e24.5)
*cfclose
finish
*enddo
/eof
```

**Transient Analysis (with a control signal rising edge of 300 ns and fall edge of 300 ns)**

```
finish
/clear,nostart

/prep7

sthick=2.15  !thickness of the plate
snthick=0.20  !nitride thickness
smthick=0.185   !top metal thickness
slength=100   !side of the square
plength=80    !plate side length
blength=13   !bump length
bwidth=6     !bump width

rfluid=300      !may need 1000=1.5*1500/2

eps=5.7     !insulation layer permittivity
fg=snthick/eps  !insulation layer effective gap for the nitride
gap=0.220
tb=0.07          !bump thickness
gapt=gap+fg   !total effective gap

gaprf=gap   !rf line gap
gapclos=tb

et,1,solid45
et,2,fluid30,,1  !fluid-structure-interface (fsi) off
et,3,fluid30,,0  !fsi on
et,4,fluid130    !absorping boundary
r,4,rfluid,0,0,smthick+sthick    !radius of absorping boundary

mp,dens,1,2328e-18  !silicon's density
mp,ex,1,148e3       !silicon's young's modulus
mp,nuxy,1,0.177     !silicon's possion's ratio

mp,dens,2,3187e-18  !si3n4's density
mp,ex,2,296e3       !si3n4's young's modulus
mp,nuxy,2,0.27      !si3n4's possion's ratio

mp,dens,3,19300e-18 !au's density
mp,ex,3,79e3        !au's young's modulus
mp,nuxy,3,0.44      !au's possion's ratio

mp,dens,4,1000e-18  !fluid's properties
mp,sonc,4,1500e6

!!!!!!!geometry!!!!!!!!!
block,-slength/2,slength/2,-slength/2,slength/2,0,sthick
block,-slength/2,slength/2,-slength/2,slength/2,sthick,smthick+sthick
block,-slength/2,slength/2,-slength/2,slength/2,0,-snthick
block,-blength/2,blength/2,-slength/2,slength/2,-snthick,smthick+sthick
block,-slength/2,slength/2,-bwidth/2,bwidth/2,-snthick,smthick+sthick
block,0,slength/2,-slength/2,slength/2,-snthick,smthick+sthick
block,-blength/2,blength/2,-bwidth/2,bwidth/2,-snthick-tb,smthick+sthick
block,0,-blength/2,-bwidth/2,bwidth/2,-snthick-tb,smthick+sthick
wpof,,,smthick+sthick
wprota,0,90,0
```

```
wprota,0,90,0
sphere,0,rfluid,0,180
vovlap,all
vglue,all
wprota,0,-90,0
wpof,,,-(smthick+sthick)

vsel,s,loc,z,0,sthick   !material type 1, element type 1
vatt,1,,1,0

vsel,s,loc,z,0,-snthick
vatt,2,,1,0

vsel,s,loc,z,sthick,sthick+smthick
vatt,3,,1,0

vsel,s,loc,z,-snthick,-tb-snthick
vatt,3,,1,0

vsel,s,loc,z,smthick+sthick,10000
vatt,4,,2

allsel

!!!!!!!!!mesh!!!!!!!!!

vsel,s,loc,z,-(snthick+tb),sthick+smthick
esize,2
vmesh,all

vsel,s,loc,z,sthick+smthick,10000
esize,30        !wavelength/20
mshape,1,3d
mshkey,0
vmesh,all

allsel

!!!!!!!fluid-structure interface!!!!!!!!!!!!
esel,s,type,,1
nsel,s,ext
nsel,r,loc,z,sthick+smthick
esel,s,type,,2
esln,r
emodif,all,type,3
sf,all,fsi,1
allsel

!!!!!!!absorbing boundary!!!!!!!!!!!!
wpof,,,smthick+sthick
csys,wp
WPSTYL, , , , , , 2
nsel,s,loc,x,rfluid,rfluid+1
csys,0
type,4
real,4
```

```
mat,4
esurf           !absorbing boundary fluid
WPSTYL, , , , , ,0
wpof,,,-smthick-sthick
csys,0

!!!!!!!!!boundary conditions!!!!!!!!!
nsel,s,loc,z,-snthick
nsel,r,loc,x,-plength/2,+plength/2
nsel,r,loc,y,-plength/2,+plength/2
cm,underpnodes,node

allsel
nsel,s,loc,z,-snthick
cmsel,u,underpnodes,node
d,all,uy,0
d,all,ux,0
d,all,uz,0
allsel

nsel,s,loc,z,-snthick-tb
cm,rf,node

!cmsel,s,rf,node   !selecting nodes named rf
allsel

nsel,s,loc,z,-snthick
nsel,r,loc,x,-plength/2+3,plength/2-3
nsel,r,loc,y,-plength/2+3,-bwidth/2-5
cm,side1,node
allsel

nsel,s,loc,z,-snthick
nsel,r,loc,x,-plength/2+3,plength/2-3
nsel,r,loc,y,bwidth/2+5,plength/2-3
cm,side2,node
allsel

emtgen,'side1','trans1','gnd1','uz',-gapt,fg,1     !generate trans126 elements: emtgen,
emtgen,'rf','trans3','gnd3','uz',-gaprf,gapclos,1
emtgen,'side2','trans2','gnd2','uz',-gapt,fg,1

d,gnd1,ux,0          !define dof boundary conditions
d,gnd1,uy,0
d,gnd1,uz,0
d,gnd1,volt,0

d,gnd2,ux,0          !define dof boundary conditions
d,gnd2,uy,0
d,gnd2,uz,0
d,gnd2,volt,0

d,gnd3,ux,0          !define dof boundary conditions
d,gnd3,uy,0
d,gnd3,uz,0
d,gnd3,volt,0
```

```
allsel
vdc=69.5
vp=2.5
t_start=1000e-9
t_step=20e-9

/solu
antype,trans
trnopt,full
autots,on

timint,off

d,rf,volt,0
d,side1,volt,vdc
d,side2,volt,vdc
allsel
esel,s,type,,1
nsel,r,ext
nsel,r,loc,z,sthick+smthick
sf,all,pres,0.1
allsel

time,t_start
nsubst,2    !Nonzero initial displacement and zero initial velocity
kbc,1       !
outres,all,all
allsel
solve

!!!!!!!!!!!!!!!!!!!!!!!!!!!!!!!!!!!!!!!!!
*do,aa,1,100,1
timint,on
t1= t_start+t_step*aa
time,t1

allsel
esel,s,type,,1
nsel,s,loc,z,0
nsel,r,loc,x,0,slength
nsel,r,loc,y,0,slength
sf,all,pres,0.1
allsel
autots,on
kbc,0
neqit,50

d,side1,volt,vdc
d,side2,volt,vdc
d,rf,volt,0

deltim,2.5e-9,1e-9 , ,on
outres,all,all
alls
solve
*enddo
```

```
*do,aa,1,15,1

t2=t1+t_step*aa

timint,on
time,t2
d,side1,volt,vdc+(aa*vp/15)
d,side2,volt,vdc+(aa*vp/15)
d,rf,volt,0

allsel
esel,s,type,,1
nsel,r,ext
nsel,r,loc,z,sthick+smthick
sf,all,pres,0.1
allsel
autots,on
kbc,0
neqit,50 !iterations of 50


deltim,2.5e-9,1e-9 , ,on  !specifies the time step sizes to be used for this load step.
outres,all,all  !controls the solution data written to the database.
allsel
solve
*enddo
!!!!!!!!!!!!!!!!!!!!!!!!!!!!!!!!!!!!!!!!!
*do,aa,1,100,1

timint,on
t3= t2+t_step*aa
time,t3

allsel
esel,s,type,,1
nsel,s,loc,z,0
nsel,r,loc,x,0,slength
nsel,r,loc,y,0,slength
sf,all,pres,0.1
allsel
autots,on
kbc,0
neqit,50

d,side1,volt,vdc+vp
d,side2,volt,vdc+vp
d,rf,volt,0


deltim,2.5e-9,1e-9 , ,on
outres,all,all
alls
solve
*enddo
```

```
*do,aa,1,15,1
timint,on
t4= t3+t_step*aa
time,t4
allsel
esel,s,type,,1
nsel,s,loc,z,0
nsel,r,loc,x,0,slength
nsel,r,loc,y,0,slength
sf,all,pres,0.1
allsel
autots,on
kbc,0
neqit,50
d,side1,volt,vdc+vp-(aa*vp/15)
d,side2,volt,vdc+vp-(aa*vp/15)
d,rf,volt,0
deltim,2.5e-9,1e-9 , ,on
outres,all,all
alls
solve
*enddo

*do,aa,1,200,1
timint,on
t5= t4+t_step*aa
time,t5
allsel
esel,s,type,,1
nsel,s,loc,z,0
nsel,r,loc,x,0,slength
nsel,r,loc,y,0,slength
sf,all,pres,0.1
allsel
autots,on
kbc,0
neqit,50
d,side1,volt,vdc
d,side2,volt,vdc
d,rf,volt,0
deltim,2.5e-9,1e-9 , ,on
outres,all,all
alls
solve
*enddo
finish

/post26
nsol,2,8345,volt,, volt_5,
nsol,3,node(0,0,0),u,z

*dim,results,table,1002,3
vget,results(1,0),1
vget,results(1,1),2
vget,results(1,2),3
finish
/post1

*cfopen,up300fall300_jarvan,csv
*vwrite,results(1,0),results(1,1),results(1,2)
(e24.5,',',e24.5,',',e24.5)
*cfclose
finish
/eof
```

# Appendix B: Image reconstruction code for backward-PAI with universal back-projection and coherence factor weighting for graphite target (MATLAB 2017a)

```
clear;
close all;

fileName = 'C: \M2.csv';
ascan = csvread(fileName);

fc = 1.2709e6;
flow = 0.15e6;
fcutoff = 4.5e6;
fs = 50e6;
c = 1490;
lamda = c/fc;

stepSize = 0.3e-3;
scanDistance = 12.9e-3;

numberOfElements = round(scanDistance/stepSize+1);
elementPosition = -0.5*scanDistance+(0:(numberOfElements-1))*stepSize;

imagingStartDepth = 0e-3;
imagingEndDepth = 32e-3;
imagingWidth = scanDistance;

delta = lamda/16;

numberOfPixelsX = ceil(imagingWidth/delta);
numberOfPixelsZ = ceil((imagingEndDepth-imagingStartDepth)/delta);

pixelPositionsX = -0.5*imagingWidth + (0:(numberOfPixelsX-1))*delta;
pixelPositionsZ = imagingStartDepth + (0:(numberOfPixelsZ-1))*delta;

numberOfSamples = size(ascan,2);
ascanBackProj = ascan;
for m = 1:numberOfElements;
    for n = 2:numberOfSamples
        ascanBackProj(m,n) = ascan(m,n) - (n-1)*(ascan(m,n)-ascan(m,n-1));
    end
end

numberOfZeros = 50;
lpf = fir1(2*numberOfZeros,[flow/(fs/2),fcutoff/(fs/2)]);

filteredAscan = zeros(size(ascan,1),size(ascan,2)+numberOfZeros*2);
```

```
filteredAscanBackProj = zeros(size(ascan,1),size(ascan,2)+numberOfZeros*2);
for m = 1:numberOfElements;
    temp = conv(lpf,ascan(m,:));
    filteredAscan(m,:) = hilbert(temp);

    temp = conv(lpf,ascanBackProj(m,:));
    filteredAscanBackProj(m,:) = hilbert(temp);
end

hahahaha = 15*pi/2/99;
threshold = cos(hahahaha);

pic = zeros(numberOfPixelsZ,numberOfPixelsX);
coherence = zeros(numberOfPixelsZ,numberOfPixelsX);

picBackProj = zeros(numberOfPixelsZ,numberOfPixelsX);
coherenceBackProj = zeros(numberOfPixelsZ,numberOfPixelsX);

for m = 1:numberOfPixelsX
    for n = 1:numberOfPixelsZ
        x = pixelPositionsX(m);
        z = pixelPositionsZ(n);

        pixelValue = 0;
        b = 0;
        ha = 0;

        pixelValueBackProj = 0;
        bBackProj = 0;

        for l = 1:numberOfElements
            elementPos = elementPosition(l);
            distance = norm([x,z]-[elementPos,0]);
            cosValue = z/distance;
            if(cosValue<threshold)continue;end
            index = round(distance/c*fs)+numberOfZeros;
            if(index<=size(filteredAscan,2))
                pixelValue = pixelValue + filteredAscan(l,index);
                b = b + abs(filteredAscan(l,index))^2;

                pixelValueBackProj = pixelValueBackProj + filteredAscanBackProj(l,index);
                bBackProj = bBackProj + abs(filteredAscanBackProj(l,index))^2;

                ha = ha+1;
            end
        end
        pic(n,m) = pixelValue;
        coherence(n,m) = abs(pixelValue)^2/b/ha;
```

```
        picBackProj(n,m) = pixelValueBackProj;
        coherenceBackProj(n,m) = abs(pixelValueBackProj)^2/bBackProj/ha;
    end
end

dyn = 40;


envRealData = abs(pic);
envRealDataCF = envRealData.*coherence;

envRealDatadB = 20*log10(envRealData/max(max(envRealData)));
envRealDatadB(envRealDatadB<-dyn) = -dyn;

envRealDataCFdB = 20*log10(envRealDataCF/max(max(envRealDataCF)));
envRealDataCFdB(envRealDataCFdB<-dyn) = -dyn;

envRealDataBackProj = abs(picBackProj);
envRealDataBackProjCF = envRealDataBackProj.*coherenceBackProj;

envRealDataBackProjdB = 20*log10(envRealDataBackProj/max(max(envRealDataBackProj)));
envRealDataBackProjdB(envRealDataBackProjdB<-dyn) = -dyn;

envRealDataBackProjCFdB = 20*log10(envRealDataBackProjCF/max(max(envRealDataBackProjCF)));
envRealDataBackProjCFdB(envRealDataBackProjCFdB<-dyn) = -dyn;

[m1,n1] = find(envRealDatadB==0);
disp(['In real data, the 0dB pixel is at X:',num2str(pixelPositionsX(n1)*1000),'mm,
Z:',num2str(pixelPositionsZ(m1)*1000),'mm']);

[m2,n2] = find(envRealDataCFdB==0);
disp(['In real data with coherence factor, the 0dB pixel is at
X:',num2str(pixelPositionsX(n2)*1000),'mm, Z:',num2str(pixelPositionsZ(m2)*1000),'mm']);

[m3,n3] = find(envRealDataBackProjdB==0);
disp(['In real data with back projection, the 0dB pixel is at
X:',num2str(pixelPositionsX(n3)*1000),'mm, Z:',num2str(pixelPositionsZ(m3)*1000),'mm']);

[m4,n4] = find(envRealDataBackProjCFdB==0);
disp(['In real data with back projection with coherence factor, the 0dB pixel is at
X:',num2str(pixelPositionsX(n4)*1000),'mm, Z:',num2str(pixelPositionsZ(m4)*1000),'mm']);

figure(1)
imagesc(pixelPositionsX*10^3,pixelPositionsZ*10^3,envRealDatadB,[-dyn,0]);
xlabel('mm','fontsize',12,'fontweight','bold');
ylabel('mm','fontsize',12,'fontweight','bold');
title(['Real data: angle:
',num2str(hahahaha/pi*180),'^{\circ}'],'fontsize',12,'fontweight','bold');
set(gca,'linewidth',1.5,'fontsize',12,'fontweight','bold');
```

```
colormap('gray');
h = colorbar;
set(h,'fontsize',12,'fontweight','bold');
truesize;

figure(2)
imagesc(pixelPositionsX*10^3,pixelPositionsZ*10^3,envRealDataCFdB,[-dyn,0]);
xlabel('mm','fontsize',12,'fontweight','bold');
ylabel('mm','fontsize',12,'fontweight','bold');
title(['Real data with coherence factor: angle:
',num2str(hahahaha/pi*180),'^{\circ}'],'fontsize',12,'fontweight','bold');
set(gca,'linewidth',1.5,'fontsize',12,'fontweight','bold');
colormap('gray');
h = colorbar;
set(h,'fontsize',12,'fontweight','bold');
truesize;

figure(3)
imagesc(pixelPositionsX*10^3,pixelPositionsZ*10^3,envRealDataBackProjdB,[-dyn,0]);
xlabel('mm','fontsize',12,'fontweight','bold');
ylabel('mm','fontsize',12,'fontweight','bold');
title(['Real data with back projection: angle:
',num2str(hahahaha/pi*180),'^{\circ}'],'fontsize',12,'fontweight','bold');
set(gca,'linewidth',1.5,'fontsize',12,'fontweight','bold');
colormap('gray');
h = colorbar;
set(h,'fontsize',12,'fontweight','bold');
truesize;

figure(4)
imagesc(pixelPositionsX*10^3,pixelPositionsZ*10^3,envRealDataBackProjCFdB,[-dyn,0]);
xlabel('mm','fontsize',12,'fontweight','bold');
ylabel('mm','fontsize',12,'fontweight','bold');
title(['Real data with back projection and coherence factor: angle:
',num2str(hahahaha/pi*180),'^{\circ}'],'fontsize',12,'fontweight','bold');
set(gca,'linewidth',1.5,'fontsize',12,'fontweight','bold');
colormap('gray');
h = colorbar;
set(h,'fontsize',12,'fontweight','bold');
truesize;
```